\documentclass[12pt]{article}

\usepackage{amssymb,amsmath}
\usepackage{graphicx}

\usepackage{epsfig}

\usepackage{cite}

\numberwithin{equation}{section}

\newcommand{\be}{\begin{equation}}
\newcommand{\ee}{\end{equation}}
\newcommand{\bea}{\begin{eqnarray}}
\newcommand{\eea}{\end{eqnarray}}
\newcommand{\Dlt}{\Delta}
\newcommand{\dlt}{\delta}
\newcommand{\prt}{\partial}
\newcommand{\br}{{\bf r}}
\newcommand{\ba}{{\bf a}}
\newcommand{\bu}{{\bf u}}
\newcommand{\bk}{{\bf k}}
\newcommand{\bfe}{{\bf e}}

\newcommand{\bp}{{\bf p}}
\newcommand{\bP}{{\bf P}}
\newcommand{\bD}{{\bf D}}
\newcommand{\bv}{{\bf v}}
\newcommand{\bS}{{\bf S}}

\newcommand{\bt}{\beta}
\newcommand{\vp}{\varphi}
\newcommand{\ep}{\varepsilon}
\newcommand{\al}{\alpha}
\newcommand{\ra}{\rightarrow}
\newcommand{\sgm}{\sigma}

\newcommand{\gm}{\gamma}
\newcommand{\om}{\omega}
\newcommand{\Om}{\Omega}
\newcommand{\Gm}{\Gamma}
\newcommand{\dgr}{\dagger}
\newcommand{\lbd}{\lambda}
\newcommand{\Lbd}{\Lambda}

\newcommand{\cH}{{\cal H}}

\newcommand{\rgl}{\rangle}
\newcommand{\lgl}{\langle}

\begin{document}

\begin{center}

{\Large {\bf Effects of symmetry breaking in finite quantum systems} \\ [5mm]

J.L. Birman$^1$, R.G. Nazmitdinov$^{2,3}$, and V.I. Yukalov$^3$ } \\ [3mm]

{\it $^1$Department of Physics, City College, \\
City University of New York, New York, NY 10031, USA \\ [2mm]

$^2$Departament de Fisica, \\
Universitat de les Illes Balears,Palma de Mallorca 07122, Spain \\ [2mm]

$^3$Bogolubov Laboratory of Theoretical Physics, \\
Joint Institute for Nuclear Research, Dubna 141980, Russia}

\end{center}

\vskip 3cm

\begin{abstract}

The review considers the peculiarities of symmetry breaking and symmetry
transformations and the related physical effects in finite quantum systems.
Some types of symmetry in finite systems can be broken only asymptotically.
However, with a sufficiently large number of particles, crossover transitions
become sharp, so that symmetry breaking happens similarly to that in
macroscopic systems. This concerns, in particular, global gauge symmetry
breaking, related to Bose-Einstein condensation and superconductivity, or
isotropy breaking, related to the generation of quantum vortices, and the
stratification in multicomponent mixtures. A special type of symmetry
transformation, characteristic only for finite systems, is the change of
shape symmetry. These phenomena are illustrated by the examples of several
typical mesoscopic systems, such as trapped atoms, quantum dots, atomic
nuclei, and metallic grains. The specific features of the review are:
(i) the emphasis on the peculiarities of the symmetry breaking in finite
mesoscopic systems; (ii) the analysis of common properties of physically
different finite quantum systems; (iii) the manifestations of symmetry
breaking in the spectra of collective excitations in finite quantum systems.
The analysis of these features allows for the better understanding of the
intimate relation between the type of symmetry and other physical properties
of quantum systems. This also makes it possible to predict new effects by
employing the analogies between finite quantum systems of different physical
nature.
\end{abstract}

\vskip 1cm
{\bf keywords}:
Symmetry breaking,  Finite quantum systems, Trapped atoms,
Quantum dots, Atomic nuclei, Metallic grains

\newpage

{\bf Contents}

\vskip 5mm

1. {\bf Preamble}

\vskip 3mm
2. {\bf Symmetry breaking}

\vskip 2mm
2.1. {\it Symmetry in macroscopic systems}

2.1.1. Symmetry of thermodynamic phases

2.1.2. Landau phase-transition theory

2.1.3. Symmetry and conservation laws

2.1.4. Thermodynamic limit and symmetry

2.1.5. Methods of symmetry breaking

\vskip 2mm
2.2. {\it Symmetry in finite systems}

2.2.1. Asymptotic symmetry breaking

2.2.2. Effective thermodynamic limit

2.2.3. Fluctuations and thermodynamic stability

2.2.4. Equilibration in quantum systems

2.2.5. Convenience of symmetry breaking

\vskip 2mm
2.3. {\it Geometric symmetry transformations}

2.3.1. Role of shape symmetry

2.3.2. Geometric shape transitions

2.3.3. Geometric orientation transitions

\vskip 3mm
3. {\bf Trapped atoms}

\vskip 2mm
3.1. {\it Bose–Einstein condensation}

3.1.1. Gauge symmetry breaking

3.1.2. Representative statistical ensemble

3.1.3. Equation for condensate function

3.1.4. Size and shape instability

3.1.5. Elementary collective excitations

\vskip 2mm
3.2. {\it Atomic fluctuations and stability}

3.2.1. Uniform ideal gas

3.2.2. Trapped ideal gas

3.2.3. Harmonically trapped gas

3.2.4. Interacting nonuniform systems

3.2.5. Importance of symmetry breaking

\vskip 2mm
3.3. {\it Nonequilibrium symmetry breaking}

3.3.1. Dynamic instability of motion

3.3.2. Stratification in multicomponent mixtures

3.3.3. Generation of quantum vortices

3.3.4. Topological coherent modes

3.3.5. Nonequilibrium crossover transitions

\vskip 2mm
3.4. {\it Pairing of fermionic atoms}

3.4.1. Harmonically trapped fermions

3.4.2. Pair formation and superfluidity

3.4.3. Pairing and symmetry breaking

\vskip 2mm
3.5. {\it On condensation of unconserved quasiparticles}

3.5.1. Conserved versus unconserved particles

3.5.2. No condensation of self-consistent phonons

3.5.3. No condensation of generic bogolons

3.5.4. No condensation of equilibrium magnons

3.5.5. Condensation of auxiliary quasiparticles

\vskip 3mm
4. {\bf Quantum dots}

\vskip 2mm
4.1. {\it Basic features}

\vskip 2mm
4.2. {\it Shell effects in a simple model}

4.2.1. Magnetic field and shapes

4.2.2. Magnetic properties

\vskip 2mm
4.3. {\it Two-electron quantum dot: a new paradigm in mesoscopic physics}

4.3.1. Hidden symmetries in a two-electron quantum dot

4.3.2. Center-of-mass and relative-motion Hamiltonians

4.3.3. Classical dynamics and quantum spectra

\vskip 2mm
4.4. {\it Dimensionality effects in ground-state transitions of two-electron quantum dots }

4.4.1. First singlet-triplet transition in a two-electron quantum dot

4.4.2. Topological transitions in a two-electron quantum dot

4.4.3. Effective charge

\vskip 2mm
4.5. {\it Symmetry breaking: mean field and beyond}

4.5.1. Theoretical approaches

4.5.2. Hartree-Fock approximation

4.5.3. Roto-vibrational model

4.5.4. Geometric transformation in a two-electron quantum dot

4.5.5. Shell structure and classical limit

4.5.6. Ground state in random phase approximation

\vskip 3mm
5. {\bf Atomic nuclei }

\vskip 3mm
5.1. {\it Signatures of symmetry breaking in nuclear structure}

\vskip 2mm
5.2. {\it Nuclear structure models and symmetry breaking phenomena}

5.2.1. Geometric collective model

5.2.2. Algebraic approach: Interaction boson model

5.2.3. Microscopic models with effective interactions

5.2.4. Cranking approach

\vskip 2mm
5.3. {\it Symmetry breaking in rotating nuclei }

5.3.1. Symmetries

5.3.2. Shape transitions in rotating nuclei

5.3.3. Collective excitations as indicator of symmetry breaking

\vskip 3mm
6. {\bf Metallic grains }

\vskip 2mm
6.1. {\it General properties }

\vskip 2mm
6.2. {\it Pairing effects and shell structure}

6.2.1. Theoretical approaches

6.2.2. Effects of magnetic field

6.2.3. Shell effects

\vskip 3mm
7. {\bf Summary}

\vskip 2mm
{\it Acknowledgments}

\vskip 3mm
{\bf Appendix}. Two-dimensional harmonic oscillator in a perpendicular magnetic field

\vskip 3mm
{\bf References}

\newpage

\section{Preamble}

\label{sec:I}

Finite quantum systems are now intensively studied, since many of their
specific properties promise a variety of applications in science and
technology. Two basic features of these systems are encoded in the names
{\it quantum} and {\it finite}. The first adjective implies that the
correct description of such systems requires the use of quantum theory.
And the second adjective distinguishes these objects from macroscopic
systems. Generally, the finite systems, to be considered in the review,
can be called {\it mesoscopic}. This means that their typical sizes are
intermediate between microscopic sizes, related to separate quantum
particles, and macroscopic sizes characterizing large statistical systems,
such as condensed matter.

Being intermediate between microscopic and macroscopic sizes, mesoscopic
systems exhibit the combination of properties of both their limiting cases.
On the one hand, they inherit many physical properties of large statistical
systems, while, from another side, they possess a number of properties,
caused by their finiteness, which are absent in large systems. Examples
of finite quantum systems are trapped atoms, quantum dots, atomic nuclei,
and metallic grains. Though the physical nature of these systems is very
different, they possess a variety of common properties caused by the fact
that all of them are quantum and finite.

One should not confuse spontaneous symmetry breaking occurring under phase
transitions in bulk matter and the more general notion of symmetry changes
that can happen in finite systems. In the latter, it is not the finite-size
corrections to extensive quantities, which are usually small, that are of
interest, but the symmetry changes that result not in corrections, but
in the effects principally distinguishing finite systems from bulk matter. For
instance, though traps can house quite a number of atoms, but the spectrum of
elementary excitations in traps is principally different from that in infinite
condensed matter. The long-wave spectrum of trapped atoms is discrete and
strongly dependent on the trap shape and its symmetry. These effects are
described in Sec. 3.1. Moreover, despite that a trap can contain a number of
atoms, the trap symmetry is crucially important for defining the stability of
the system as a whole in the case of dipolar or attractive interactions, as
is explained in Sec. 3.2. So, it would be wrong to identify trapped atoms
with an infinite system in thermodynamic limit.

Throughout the review, we stress the similarity of physical effects in
different finite systems. Thus, the long-wave spectrum of collective
excitations for trapped atoms is practically the same as that for atomic
nuclei and similar to the spectrum for quantum dots. The description of
rotating Bose-Einstein condensates is analogous to that of rotating nuclei
or quantum dots in magnetic fields. The pairing effects for trapped fermionic
atoms are almost the same as these effects in atomic nuclei or metallic grains.
Geometric shape transitions happen for atomic nuclei as well as for trapped
atoms. We repeatedly emphasize the common physical features and analogous
mathematical treatment of different finite systems.

The general feature of the finite systems, considered in the review, is
their mesoscopic size, being intermediate between microscopic and macroscopic
scales. Such systems possess the combination of the properties of both their
limiting cases. From one side, they can inherit symmetry properties, related
to large systems. But from another side, they can exhibit symmetry changes
typical of only finite systems, e.g., shape and orientation transitions
accompanied by symmetry variations.

Thus, trapped atoms may carry features typical of many-particle systems and
demonstrate Bose-Einstein condensation or superconducting fermion paring. The
typical size of a trapped atomic cloud can contain from just a few up to
millions of atoms. The size of a quantum dot can be comparable to that of
a trapped atomic cloud, though the number of particles, as is considered in
the review, is varied between two to the order of ten. The recent progress
in semiconductor technology has made it feasible to fabricate and probe such
confined systems at different values of magnetic field and at low temperatures,
close to those typical of trapped atoms. Quantum dots in an external magnetic
field, atomic nuclei, and rotating trapped atoms share a lot of common physics
as well as mathematics, despite rather different numbers of constituent
particles.

It is remarkable that different finite systems, housing quite different
numbers of particles, nevertheless, can exhibit similar physical features.
For examples, shell effects exist in quantum dots with a small number of
electrons, as well as in atomic nuclei with many nucleons, and even in
metallic grains with a rather large number of particles. Shell effects are
caused by the degeneracy of quantum spectra, which, for quantum dots,
can be controlled by external magnetic fields. Contrary to this, the
similar phenomenon can only be observed in rotating nuclei but not
controlled. Therefore, the comparison of analogous effects in different
systems may help to shed light on the physics of atomic nuclei, quantum
dots, and metallic grains in external magnetic fields.

Magic numbers, associated with the degeneracy of quantum spectrum, define
the shape symmetry and shape changes from spherical to deformed shapes,
as is discussed for quantum dots in Sec. 4.2.1 and 4.5.5, and for rotating
nuclei in Sec. 5.1. Such a symmetry breaking phenomenon, occurring in
very different systems, is due to the finite number of the constituent
particles.

The main aim of this review is to consider the specific symmetry properties
of mesoscopic quantum systems, emphasizing their similarities. In
Sec. \ref{sec:II}, we briefly recall the symmetry properties of macroscopic
systems and discuss their difference from those of finite systems.
The characteristic features, due to the system finiteness, are described on
the general level, which stresses their common origin. In the following
sections, these typical properties are illustrated for several particular
finite systems: for trapped atoms in Sec. \ref{sec:III}, for quantum dots
in Sec. \ref{sec:IV}, for atomic nuclei in Sec. \ref{sec:V}, and for metallic
grains in Sec. \ref{sec:VI}. Section \ref{sec:VII} concludes, summarizing
the main points.

In many places, we mention exact mathematical results and theorems. In
doing this, we do not overload the review by the corresponding proofs
and we do not enumerate all conditions required for the validity of the
theorems. We remember the saying of Arnold \cite{Arnold_1999} comparing
mathematics and physics: "mathematics have always played absolutely
subordinate role, like that of orthography or even calligraphy in poetry".
At the same time, we know that understanding mathematical facts in a number
of situations is crucial. Therefore, mentioning mathematical results, we
always provide the references, where all details can be found.

Throughout the paper, when this does not yield confusion, the system of
units is used where the Planck constant $\hbar =1$ and the Boltzmann
constant $k_B = 1$.

\section{Symmetry Breaking}
\label{sec:II}

The mathematically rigorous meaning of symmetry is usually associated
with infinite systems. Therefore, we start this section with the basic
notions of symmetry as applied to such macroscopic systems and then pass
to finite systems, explaining what they inherit from the former and what
appears to be principally different.

\subsection{Symmetry in macroscopic systems}
\label{subsec:II.A}

\subsubsection{Symmetry of thermodynamic phases}
\label{subsubsec:II.A.1}

A physical system, composed of the same elements, say, atoms or
molecules, can exhibit, under special conditions, several thermodynamic
phases differing by their properties, among which one of the most
important is the system symmetry. Thus, water can be gas, liquid, or
solid in different regions of the pressure-temperature plane. Likewise,
a solid can show phases differing in their crystalline structure. Magnetic
materials can be in paramagnetic or magnetized states. Ferroelectrics can
be in ferroelectric phase, having dipole moment, or in paraelectric phase,
without such a dipole moment. Liquid helium-4 can be superfluid or normal.
Each thermodynamic phase, as a rule, can be characterized by its symmetry.
The symmetry property implies invariance under some transformations. The
group of the related transformations forms a symmetry group. There are
many good books on group theory, such as the classical monographs
\cite{Wigner_1959,Hamermesh_1964}. Various applications of group theory
to physical systems can be found in several books, e.g.,
\cite{Wigner_1970,Birman_1974,Louck_2008}.

One distinguishes the symmetry of the system Hamiltonian, or Lagrangian,
from the symmetry of the order parameter. The system can be described by
the same Hamiltonian, or the same Lagrangian, invariant under a general
group of transformations, while its thermodynamic phases are characterized
by the order parameters associated with some subgroups of the general group.

The order parameters are the quantities specifying thermodynamic phases
and qualitatively distinguishing between the latter. For instance,
magnetization $\bf M$ is the order parameter distinguishing ferromagnet,
with nonzero $\bf M$, from paramagnet, with ${\bf M} \equiv 0$. Similarly,
dipole moment $\bf P$ distinguishes ferroelectric, with nonzero $\bf P$,
from paraelectric, with ${\bf P} \equiv 0$. A particle density
$\rho (\bf r)$ distinguishes between a crystal, with the density
$\rho({\bf r}+{\bf a})=\rho (\bf r)$, which is periodic with respect to
lattice vectors $\bf a$, and a liquid, with the uniform density
$\rho (\br) = \rho = const$. Different crystalline structures are
distinguished by the periodicity of the particle density over different
lattice vectors. Generally, an order parameter can be a scalar, vector,
or a tensor. Order parameters for particular substances have been
appearing in different mean-field theories, such as the Weiss \cite{Weiss_1907}
mean-field theory of ferromagnetism. The order parameter, as a general
concept, was defined by Landau in a series of papers starting with
\cite{Landau_1937}. All these papers are reprinted in \cite{Landau_1967}.

The qualitative change of the order parameter, in particular, of its
symmetry, is associated with phase transitions. The common classification
of phase transitions, accepted nowadays, distinguishes two main types of
them. {\it First-order} phase transitions are characterized by a discontinuous
jump of the order parameter at the transition point, for instance, as
density in the crystal-liquid transition. {\it Second-order} phase
transitions, also called critical phenomena, correspond to a continuous
change of the order parameter at the critical point between an ordered
phase, where it is non-zero, and a disordered phase, where it is zero.
The typical case is the ferromagnet-paramagnet phase transition, with
a continuous change of magnetization. There is also the third kind of
transitions, called {\it crossover}, when the order parameter varies
continuously, but does not become exactly zero in the region corresponding
to an almost disordered phase. This can be illustrated by the behavior of
magnetization of a ferromagnet in the presence of an external magnetic
field. Then the magnetization in the paramagnetic region is nonzero.
Phase transitions in macroscopic systems are described in textbooks on
statistical mechanics, e.g., \cite{Huang_1963,Isihara_1971,LandauLif_1980}.
There are several books entirely devoted to phase transitions, e.g.,
\cite{Fisher_1965,Stanley_1971,Ma_1976,YukalovShu_1990}.

In finite systems, the majority of transitions are crossovers. However,
when the number of particles in the finite system is sufficiently large,
the crossover is so sharp that it becomes practically non-distinguishable
from either first or second order phase transition.

When the system Hamiltonian enjoys a symmetry characterized by some
general symmetry group, but the system is stable for a phase with the
order parameter whose symmetry corresponds to a subgroup of the general
group, this is termed {\it spontaneous symmetry breaking}. The majority
of phase transitions are accompanied by such spontaneous breaking of
symmetry. Also, the symmetry breaking at phase transitions is related
to the change of state entanglement
\cite{Keyl_2002,Vedral_2002,GalindoMar_2002} and of entanglement
production \cite{Yukalov_2003a,Yukalov_2003b}.

\subsubsection{Landau phase-transition theory}
\label{subsubsec:II.A.2}

A very simple and general way of describing spontaneous symmetry
breaking is provided by the Landau theory of phase transitions
\cite{Landau_1937,Landau_1967}. Landau theory is based on the assumption
that the system free energy $F$ is an analytic function of the order
parameter, and hence, can be expanded in powers of the latter. This,
taking for illustration a scalar order parameter $\eta$, yields
\be
\label{2.1}
F(\eta) = F(0) + a_2 \eta^2 + a_3 \eta^3 + a_4 \eta^4 \;  ,
\ee
where $a_2$ and $a_3$ can be of different sign, while $a_4 > 0$. The
structure of the free energy is prescribed by the symmetry properties
of the system. Minimizing the free energy with respect to the order
parameter defines the stable physical state.

When there is inversion symmetry, so that $F(-\eta) = F(\eta)$, then
$a_3 = 0$. In this case, for $a_2 > 0$, a disordered thermodynamic phase
is stable, corresponding to zero order parameter $\eta = 0$. But for
$a_2 < 0$, an ordered phase is stable with $\eta \neq 0$. The
order-disorder phase transition is continuous, that is, of second
order. When there is no inversion symmetry, and $a_3 \neq 0$, then
the order-disorder phase transition occurs discontinuously, being a
first-order phase transition.

The classification of phase transitions, based on the behaviour of
the order parameter, onto first-order discontinuous and second-order
continuous transitions, following from the Landau scheme, is now widely
accepted, replacing the cumbersome Ehrenfest \cite{Ehrenfest_1959}
classification, based on the behaviour of derivatives of thermodynamic
potentials with respect to thermodynamic variables.

The Landau approach can be used for finite as well as for zero
temperatures. The order parameter can also be a vector or a tensor.
The theory can be applied for single-component and for multicomponent
systems. Among the merits of Landau theory is its ability to predict
symmetry changes occurring at phase transitions. An efficient symmetry
group analysis has been developed for this purpose
\cite{JaricBir_1977,Michel_1980,Birman_1982,BelitzKirkVoj_2005}.

The form of the free energy (\ref{2.1}) assumes that a uniform
equilibrium system is treated, since the order parameter does not
depend on spatial variables and time. The generalization to the
nonuniform and, generally, nonequilibrium, case is done by the
Ginzburg-Landau \cite{Landau_1967,GinzburgLan_1950,Ginzburg_2004}
approach, where the order parameter $\eta = \eta({\bf r}, t)$ is
considered as a function varying in space and time. The Ginzburg-Landau
functional
\be
\label{2.2}
G[\eta] = \int \left [ F(\eta) + D (\nabla\eta)^2
\right ] \; d\br
\ee
generalizes the Landau free energy (\ref{2.1}). The order parameter
function $\eta({\bf r}, t)$ is defined through the variational
derivative of $G[\eta]$ over this function.

The Landau approach, as any mean-field approximation, becomes
ineffective in the vicinity of a critical point. Thus, the theory
predicts universal critical indices of the mean-field type,
independently of the system details and dimensionality. The reason
for the failure of Landau theory to correctly describe the
peculiarities of the critical behaviour is caused by its inadequate
treatment of fluctuations of the order parameter around its mean value.
In Landau theory, these fluctuations are assumed to be Gaussian in all
cases. The deviations of these fluctuations from the Gaussian character
are usually stronger for lower dimensionalities and for order parameters
with fewer components \cite{BelitzKirkVoj_2005}.

Even though Landau theory is designed to be applied to continuous phase
transitions, ironically, it cannot even provide a sufficient condition
whether the transition is continuous. It has been observed in some
cases that Landau theory predicts a continuous phase transition, while
in reality the transition is discontinuous due to fluctuations.

Fluctuations need to be taken into account beyond the Gaussian
approximation in order to obtain the correct critical behaviour.
This is done by applying renormalization group approach to
Ginzburg-Landau-Wilson functional
\cite{Ma_1976,WilsonKog_1974,Fisher_1974,Wilson_1975,Barber_1977,Hu_1982}.
This theory makes it possible to calculate accurate values of critical
indices. In the frame of this approach, in order for a phase transition
to be continuous, there should exist a stable renormalization-group
fixed point, and the transition region should be within the attraction
domain of that fixed point.

\subsubsection{Symmetry and conservation laws}
\label{subsubsec:II.A.3}

The existence of symmetries in a system is always related to some
conservation laws. For example, let us consider the conservation
of momentum. For quantum systems, this implies that the operator of
momentum $\hat{\bf P}$ commutes with the system Hamiltonian:
\be
\label{2.3}
[\hat\bP , H ] = 0 \; , \qquad
\hat\bP \equiv \int \psi^\dgr(\br) (-i\nabla)\psi(\br) \; d\br \; .
\ee
Here and in what follows, the field operator $\psi({\bf r}, t)$
is assumed to be in the Heisenberg representation, but the explicit
dependence on time, for brevity, is not shown, when this does not lead
to confusion. Under condition (\ref{2.3}), the Hamiltonian is invariant
with respect to the translation transformations characterized by the
operators
\be
\label{2.4}
\hat T(\br) = \exp ( i \hat\bP \cdot \br ) \; ,
\qquad \hat T^+(\br) = \exp ( -i \hat\bP \cdot \br ) \; ,
\ee
forming a unitary group, so that
\be
\label{2.5}
\hat T^+(\br) H \hat T(\br) = H \;  .
\ee
This is called {\it translation symmetry} or {\it translation
invariance}, which is just another side of the momentum conservation
(\ref{2.3}).

For an equilibrium system, under translation symmetry, one has
$$
 \left\lgl \prod_{i=1}^m \psi^\dgr(\br_i+\br) \;
\prod_{j=1}^n \psi(\br_j+\br) \right \rgl =
$$
\be
\label{2.6}
= \left \lgl \prod_{i=1}^m \psi^\dgr(\br_i)
\prod_{j=1}^n \psi(\br_j) \right \rgl \; ,
\ee
with ${\bf r}$ being an arbitrary spatial vector. In particular, the
density does not depend on the spatial variable:
\be
\label{2.7}
 \rho(\br) \equiv \lgl \psi^\dgr(\br) \psi(\br) \rgl =
\lgl \psi^\dgr(0) \psi(0) \rgl \;  ,
\ee
which means that the system is always uniform. That is, the order
parameter is trivial, corresponding to a disordered phase. Under
these conditions, there can be no periodic crystalline matter.

Another example is the spin conservation, when the spin operator
$\hat{\bf S}$ commutes with the system Hamiltonian,
\be
\label{2.8}
[ \hat\bS, H ] = 0 \;  ,
\ee
as it happens for the Heisenberg model. Under condition (\ref{2.8}),
there is {\it spin-rotation symmetry}. Then, for any equilibrium
system, the average spin is exactly zero,
\be
\label{2.9}
 \lgl \hat\bS \rgl = 0 \; ,
\ee
hence, magnetic moment is always zero. This means that magnetism is
absent.

One more example is the conservation of the number of particles, when
the number operator $\hat{N}$ commutes with the system Hamiltonian:
\be
\label{2.10}
 [\hat N, H ] = 0 \; , \qquad
\hat N \equiv \int \psi^\dgr(\br) \psi(\br) \; d\br \;  .
\ee
Then the Hamiltonian is invariant under the global gauge transformations
\be
\label{2.11}
 \hat U_\vp = e^{i\vp\hat N} \; , \qquad
\hat U_\vp^+ = e^{-i\vp\hat N} \; ,
\ee
where $\varphi$ is a real number, so that
\be
\label{2.12}
 \hat U_\vp^+ H \hat U_\vp = H \; .
\ee
Transformations (\ref{2.11}) form the gauge group that is the unitary
group $U(1)$. Under this {\it global gauge symmetry}, one gets
\be
\label{2.13}
\left \lgl \prod_{i=1}^m \psi^\dgr(\br_i) \;
\prod_{j=1}^n \psi(\br_j) \right \rgl = 0 \qquad (m \neq n) \;  .
\ee
Consequently, under this condition, there can be no anomalous averages,
hence, no Bose-Einstein condensation and no superconductivity.

The above relations between conservation laws, symmetry properties, and
order parameters are exact; their derivation can be found in literature
\cite{YukalovShu_1990,Bogolubov_1970}.

\subsubsection{Thermodynamic limit and symmetry}
\label{subsubsec:II.A.4}

As we know, phase transitions are usually accompanied by spontaneous
symmetry breaking. However, the relations of the previous subsection
are valid for any finite system, with the number of particles $N$ in
a given volume $V$. Strictly speaking, no symmetry breaking of the
type considered above can occur in a finite system. This can happen
only in thermodynamic limit that corresponds to the limiting
procedure
\be
\label{2.14}
 N \ra \infty \; , \qquad V \ra \infty \; , \qquad
\frac{N}{V} \ra const \; .
\ee
To give a mathematically accurate definition of spontaneous symmetry
breaking, one has to consider thermodynamic limit (\ref{2.14}). This
is a principal point to be kept in mind.

The structure of the space of microstates and of statistical states can
be principally different for finite and infinite systems. Suppose the
space of microstates for a finite system is a Hilbert space ${\cal H}$
that is invariant with respect to a symmetry group $G$. And let the
observable quantities are given by the statistical averages
$\lgl\hat{A}\rgl$ of self-adjoint operators from the algebra of local
observables defined on $\mathcal{H}$. In thermodynamic limit, the unique
space $\mathcal{H}$ can split into a direct sum of spaces,
\be
\label{2.15}
 \cH \ra \bigoplus_\nu \cH_\nu \; ,
\ee
where each space $\mathcal{H}_\nu$ is invariant only with respect to a
subgroup $G_\nu$ of the total group $G$. Respectively, the observables
become the convex linear combinations
\be
\label{2.16}
\lgl \hat A \rgl \ra \sum_\nu \lbd_\nu \lgl \hat A \rgl_\nu
\ee
over ergodic states \cite{Emch_1972,BratteliRob_1979}.

The underlying idea of symmetry breaking is to incorporate in the
process of taking thermodynamic limit such constraints that would lead
not to the state combinations (\ref{2.16}), but to a desired particular
pure state $\lgl\hat{A}\rgl_\nu$.

\subsubsection{Methods of symmetry breaking}
\label{subsubsec:II.A.5}

When the system Hamiltonian is invariant under transformations related
to some symmetry group, the correct description of an ordered phase
requires that the system symmetry be somehow broken. There exist several
methods realizing the procedure of spontaneous symmetry breaking.

The oldest such a method is the use of the mean-field approximation
when the order parameter, with the necessary symmetry properties, is
explicitly introduced into the system Hamiltonian. The Weiss
mean-field theory of ferromagnets \cite{Weiss_1907} and Landau
mean-field theory of phase transitions \cite{Landau_1937,Landau_1967}
are the well known examples of this kind.

Another way of symmetry breaking is by imposing external potentials with
the symmetry lower than that of the Hamiltonian. Kirkwood \cite{Kirkwood_1965}
suggested this method for lifting the degeneracy connected with the
momentum conservation that results in the constant density (\ref{2.7}).
According to Kirkwood, in order to describe crystals, one has to impose
external forces fixing in space the crystal position and orientation.
Then the ambiguity in the definition of density is removed and the latter
acquires the periodicity of a crystalline phase \cite{Kirkwood_1965}.

Bogolubov \cite{Bogolubov_1970,Bogolubov_1967} developed a general
mathematical methodology of symmetry breaking by means of infinitesimal
external sources. In this method, one adds to the system Hamiltonian an
additional term lowering the symmetry. Thus, instead of the initial
Hamiltonian $H$, one forms \cite{Bogolubov_1970,Bogolubov_1967}
\be
\label{2.17}
 H_\ep \equiv H + \ep \hat \Gm \; ,
\ee
where the operator source $\hat{\Gamma}$ has a lower symmetry than
$H$ and $\lgl\hat\Gm\rgl\propto N$. Observable quantities, corresponding
to operators $\hat{A}$ from the algebra of local observables are defined
through the double limit
\be
\label{2.18}
 \lgl \hat A \rgl \equiv \lim_{\ep\ra 0} \;
\lim_{N\ra\infty} \lgl \hat A\rgl_\ep \;  ,
\ee
where the statistical average $\langle \hat{A} \rangle_\epsilon$ is
calculated with Hamiltonian (\ref{2.17}). The average in the left-hand
side of Eq. (\ref{2.18}) is called {\it quasiaverage}.

By this construction, it is clear that the role of the infinitesimal
source is just to break the symmetry, being removed afterwards. It is
crucially important that the limits here do not commute. The
thermodynamic limit has to be necessarily accomplished before the
limit $\ep\ra 0$. This is connected with the fact that spontaneous
symmetry breaking can occur only in thermodynamic limit, but cannot
happen in a finite system. The described way of symmetry breaking is
termed the {\it method of Bogolubov quasiaverages}.

The double limit (\ref{2.18}) can be replaced by a single limit in the
{\it method of thermodynamic sources}, or {\it method of thermodynamic
quasiaverages}, \cite{Yukalov_1981,Yukalov_1991a}. Then, instead of
Hamiltonian (\ref{2.17}), one introduces the Hamiltonian
\be
\label{2.19}
H_N \equiv H + \frac{1}{N^\al} \; \hat\Gm \qquad
( 0 < \al < 1) \;  ,
\ee
with the same source term $\hat{\Gamma}$, but with $1/N^\al$ instead
of $\ep$. The observables are defined through the single thermodynamic
limit
\be
\label{2.20}
\lgl \hat A \rgl \equiv \lim_{N\ra\infty} \lgl \hat A\rgl_N \; ,
\ee
where the average $\lgl\hat{A}\rgl_N$ is calculated with Hamiltonian
(\ref{2.19}). The power $\alpha \in (0,1)$ in Eq. (\ref{2.19}) is such
that the value $1/N^\alpha$ would tend to zero slightly slower than
$1/N$. This guarantees that the Hamiltonian symmetry be broken. If the
symmetry breaking source would go to zero faster, it would not be able
to break the symmetry.

In the process of symmetry breaking, one passes from the initial state
of microstates $\cH$ to a state $\cH_\nu$ corresponding to a phase with
broken symmetry. Therefore, the straightforward way of symmetry breaking
would be by calculating the averages not over the total space $\cH$, but
directly over the restricted space $\cH_\nu$. So that the statistical
states, characterized by a statistical operator $\hat{\rho}$, would be
given by the averages
\be
\label{2.21}
\lgl \hat A \rgl_\nu = {\rm Tr}_{\cH_\nu} \hat\rho \hat A \; ,
\ee
involving the trace over $\mathcal{H}_\nu$. This way is called the
{\it method of restricted trace} \cite{Brout_1965}.

As is clear, any method of symmetry breaking requires to impose some
additional constraints. The ways of imposing the latter can be different.
Except the methods described above, it is possible to invoke other ways,
such as asymptotically weakly breaking commutation relations for operators
or complimenting the equations of motion for correlation functions and
Green functions by additional symmetry conditions. Different methods of
symmetry breaking are discussed in review articles
\cite{BogolubovShuYuk_1984,Yukalov_1991b}.

\subsection{Symmetry in finite systems}
\label{subsec:II.B}

\subsubsection{Asymptotic symmetry breaking}
\label{subsubsec:II.B.1}

As is explained above, the concept of symmetry breaking can be rigorously
defined only for macroscopic systems in thermodynamic limit. Then in what
sense could one mean symmetry breaking in finite systems? It is evident
that for small systems consisting of just a few particles, it is impossible
to correctly define the notion of symmetry related to macroscopic systems.
However, our aim here is to study {\it mesoscopic} systems, whose number
of particles $N$ is large, though finite. For $N \gg 1$, it is possible
to define {\it asymptotic symmetry}, in the sense that it is approximate,
but becomes exact in the limit $N \ra \infty$.

In this asymptotic sense, it is admissible to consider phase transitions
in finite systems. In these systems, as we know, there are no such strictly
defined phase transitions as melting, magnetization, Bose-Einstein
condensation, or superconductivity. But there are the related crossover
transitions, which become more and more sharp as the number of particles
increases. The thermodynamic characteristics, such as susceptibilities, in
the case of a crossover, do not diverge, as at the critical point of a
second-order phase transition. But the peaks of these characteristics, at
the critical point, grow larger with increasing $N$ and can become divergent
in the limit of $N \ra \infty$. In that sense, one can define asymptotic
phase transitions and the related asymptotic symmetry changes.

How large should be the number of particles $N$ for imitating well
thermodynamic phases and phase transitions? As many computer simulations
show, the systems of about 100 (or even 10) particles already form a kind
of thermodynamic phases with the related approximate symmetries and also
demonstrate the symmetry changes associated with phase transitions
\cite{Strandburg_1988}.

\subsubsection{Effective thermodynamic limit}
\label{subsubsec:II.B.2}

To check what type of asymptotic symmetry is realized in the system, one
needs to resort to thermodynamic limit. The definition of thermodynamic
limit (\ref{2.14}) requires that the number of particles $N$ be given and
the system volume $V$ be fixed. The number of particles is understood as
the average number $N = \langle \hat{N} \rangle$. Therefore limit
(\ref{2.14}) is well defined even if the number of particles in
microscopic reactions is not conserved, provided that the average number
$N$ is given.

A complication in understanding thermodynamic limit can arise when the
system volume is not fixed. This happens when the system of finite number
of particles is confined in a potential that extends to infinity. For
instance, $N$ atoms, ions, or other particles are trapped in a confining
potential, e.g., a harmonic potential. How then could one understand
thermodynamic limit?

It is possible to give a general definition of thermodynamic limit
as follows \cite{Yukalov_2005a,Yukalov_2009}. Let $A_N$ be an
{\it extensive} observable quantity for a system with $N$ particles.
Then an effective thermodynamic limit is defined as
\be
\label{2.22}
 N \ra \infty \; , \qquad A_N \ra \infty \; , \qquad
\frac{A_N}{N} \ra const \; .
\ee
Recall that a quantity is extensive if it is proportional to $N$ for
large $N \gg 1$. Limit (\ref{2.22}) does not depend on what observable
is taken, provided it is an extensive observable. When the system
volume $V$ is prescribed, then an extensive observable quantity is also
proportional to $V$. In the latter case, the definition of thermodynamic
limit (\ref{2.22}) reduces to the standard form (\ref{2.14}). The effective
thermodynamic limit (\ref{2.22}) shows how one should vary the parameters
of a trapping potential under the condition $N \ra \infty$.

\subsubsection{Fluctuations and thermodynamic stability}
\label{subsubsec:II.B.3}

For macroscopic systems, there are the well known conditions that are
required for thermodynamic phases to be stable, see, e.g.,
\cite{LandauLif_1980,Kubo_1968}. Generally, such conditions are
expressed as inequalities for susceptibilities that are defined to be
positive and finite.

It is a common understanding that phase transitions occur when one phase
becomes unstable, because of which the system changes to another phase.
Respectively, phase transition lines are called stability boundaries
\cite{LandauLif_1980,Kubo_1968,Blinc_1974,White_1979}.

In order that an observable quantity would be measurable, it is necessary
that its fluctuations would be smaller than the observable itself, which
is just the requirement for the validity of the law of large numbers
\cite{LandauLif_1980}. A susceptibility characterizes the fluctuations of an
observable quantity and is represented through the variance of the related
operator of the observable. Therefore, for an extensive observable, the ratio
of the related variance, hence, susceptibility, over the number of particles
has to be finite, provided the system is in equilibrium.

This is in agreement with the general definition of stable states as such
states at which a system, being perturbed, returns to its initial state
\cite{Scott_2005}. However, at a critical point, the relaxation time becomes
infinite, which means that the system can never relax to the unperturbed state
\cite{Fisher_1965}.

Also, one should distinguish real physical systems from cartoon models. This,
for instance, concerns the case of the XY model in two dimensions, whose
magnetic susceptibility diverges below the Kosterlitz-Thouless phase transition
\cite{Thouless_1973,Kosterlitz_1974}.

In real life, purely two-dimensional systems do not exist. Any, even the
thinnest membrane, always possesses a finite, though maybe small, thickness.
And the three-dimensional XY model enjoys a finite magnetic susceptibility
everywhere except the critical point. The crossover from a three-dimensional
XY model to a two-dimensional one can be done by reducing one of the
spatial dimensions to zero. This crossover has been studied in several papers.
A detailed analysis of such a limit can be found in \cite{Janke_1990}. It is
shown that, when reducing the role of one of the spatial dimensionalities, some
of the observable quantities of a three-dimensional XY model, for instance,
energy and specific heat, do tend to the expressions that would correspond to
the two-dimensional XY model. But the susceptibility is always finite below the
critical point, and becomes infinite only in the unphysical limit of purely
two-dimensional case.

So, the susceptibilities of real equilibrium systems, outside of the critical
points, are always finite, which is confirmed by all existing experiments.

Susceptibilities are proportional to the variances of some operators of
observables. It is therefore possible to represent the general form of
stability conditions as follows. Let $\hat{A}$ be an operator from the algebra
of local observables. The operator variance
\be
\label{2.23}
{\rm var}(\hat A) \equiv \lgl \hat A^2 \rgl -
\lgl \hat A \rgl^2 \;
\ee
characterizes the fluctuations of the related observable. By definition,
the variance is non-negative, that is, either positive or zero. For
macroscopic systems, extensive observables can be measured when their
fluctuations do not diverge, in thermodynamic limit, faster than $N$.
Thus, a statistical system is {\it thermodynamically stable} when the
fluctuations of its observables are {\it thermodynamically normal},
such that
\be
\label{2.24}
0 \leq \frac{{\rm var}(\hat A)}{N} < \infty
\ee
for any $N$, including thermodynamic limit.

As examples of susceptibilities that are expressed through operator
variances, we can mention specific heat, expressed through the variance
of internal energy, and isothermal compressibility, expressed through
the variance of the number operator. Susceptibilities can diverge only
at the points of phase transitions, where the system becomes unstable,
which results in the change of thermodynamic phases. Condition
(\ref{2.24}) is to be valid everywhere outside of the phase transition
points.

The same condition (\ref{2.24}) should hold for finite systems. The
sole difference is that the latter may require the use of thermodynamic
limit in the general form (\ref{2.22}). More detailed discussion of the
relation between the fluctuations of observables and stability conditions
can be found in classical books \cite{LandauLif_1980,Kubo_1968} and in
recent literature
\cite{Yukalov_2004,Yukalov_2005b,Yukalov_2005c,Yukalov_2005d}.

\subsubsection{Equilibration in quantum systems}
\label{subsubsec:II.B.4}

When ascribing a type of symmetry to a thermodynamic phase, one tacitly
assumes an {\it equilibrium} thermodynamic phase. However, the notion of
equilibrium for finite systems is essentially more complicated than that
for macroscopic systems. The pivotal problem is whether a finite quantum
system could equilibrate at all.

Let us consider an observable quantity, corresponding to the average
\be
\label{2.25}
\lgl \hat A(t) \rgl \equiv {\rm Tr} \hat\rho(t) \hat A
\ee
of an operator $\hat{A}$ from the algebra of local observables. For an
isolated system, the evolution of the statistical operator is given by
the equation
\be
\label{2.26}
 \hat\rho(t) = \hat U(t) \hat\rho(0) \hat U^+(t) \; ,
\ee
with the evolution operator $\hat{U}(t) = exp( - iHt)$. Taking in the
observable quantity (\ref{2.25}) the trace over the eigenfunctions of
the system Hamiltonian yields
\be
\label{2.27}
 \lgl \hat A(t) \rgl = \sum_{mn} \rho_{mn}(0) A_{nm}(t) \; ,
\ee
where
\be
\label{2.28}
 A_{nm}(t) \equiv A_{nm} e^{i(E_n-E_m)t} \;  ,
\ee
with $E_n$ being the corresponding eigenvalues of $H$. The existence
of an equilibrium time-independent state presupposes the existence of
the time average
\be
\label{2.29}
\overline A \equiv \lim_{\tau\ra\infty} \; \frac{1}{\tau}
\int_0^\tau \lgl \hat A(t) \rgl dt \;  .
\ee
This, assuming that the Hamiltonian spectrum is nondegenerate, can be
written as
\be
\label{2.30}
 \overline A = \sum_n \rho_{nn}(0) A_{nn}  .
\ee

However, it is evident that Eq. (\ref{2.27}) is a quasiperiodic function
that cannot tend to a time-independent stationary state, such as Eq.
(\ref{2.30}). Any given value of observable (\ref{2.27}) will be
reproduced after a {\it recurrence time} \cite{Kac_1957,Penrose_1979}.
Then the pivotal question arises whether it is admissible in principle
to ascribe to finite quantum systems some equilibrium thermodynamic
states and related symmetries. Or, in other words, how could a finite
quantum system equilibrate?

The simplest way of understanding how this could happen is by assuming
that the treated system is connected with a thermostat
\cite{LandauLif_1980}. Even if there is no well defined thermostat, we
have to remember that there are no absolutely isolated systems, but
each system is immersed into its surrounding, being always a subsystem
of a larger system. That is, each given system is always influenced by
its surrounding \cite{Penrose_1979,Lebowitz_1993}. This could be the
influence of uncontrollable random perturbations during the preparation
period or during the system lifetime. Moreover, the notion of an absolute
isolation as such is self-contradictory, since in order to check that the
system is really isolated during a period of time, it is necessary to
realize a series of measurements proving this, hence, influencing the
system by measurement procedures \cite{Yukalov_1970,Yukalov_1971}. During
sufficiently long time, even very weak random perturbations can lead
to drastic changes in the system properties
\cite{Yukalov_1997a,Yukalov_2002,Yukalov_2003c}. Incorporating into
the system appropriate randomness, with an additional supposition that
the variances of operators, representing observable quantities, are
sufficiently small, it is possible to achieve equilibration in quantum
systems \cite{Deutsch_1991,Srednicki_1999,Reimann_2008,RigolDunOlsh_2008}.
Bogolubov \cite{Bogolubov_1962} also mentioned that for rigorously proving
the existence of equilibration in quantum systems it may be necessary to
invoke thermodynamic limit, similarly to the use of this limit for
rigorously defining the system symmetry and symmetry breaking
\cite{Bogolubov_1962}. For confined systems, thermodynamic limit should
be understood in the general sense (\ref{2.22}).

Even if one assumes an idealized situation of an isolated system, the
latter spends on average most of the time in a quasiequilibrium state
that can be described by a Gibbs ensemble called the representative
Gibbs ensemble
\cite{Yukalov_1991b,Gibbs_1931,Tolman_1940,terHaar_1955,Jaynes_1957}.
Generally, finite quantum systems can equilibrate to stationary states
that live sufficiently long for accomplishing with them the desired
measurements. More references on the equilibration of finite quantum
systems can be found in the recent review articles
\cite{Reimann_2010,Dziarmaga_2010,Polkovnikov_2011,Yukalov_2011a}.

\subsubsection{Convenience of symmetry breaking}
\label{subsubsec:II.B.5}

Sometimes one says that, since finite systems do not experience exact
symmetry breaking, there is no necessity of such a breaking at all,
but it is possible to deal with the system of the same symmetry for
any thermodynamic parameters, treating a phase transition just as a
sharp crossover. As a justification of this statement, one says that
in computer simulations, such as Monte Carlo, one does not break any
symmetry but can observe a kind of phase transitions.

This is certainly correct for computer simulations that always deal
with a finite number of particles. Then, instead of first or second
order phase transitions, one always observes crossovers. But, as has
been stressed above, for a large number of particles $N \gg 1$ such
crossovers become extremely sharp, closely imitating phase transitions.

One more complication in computer simulations is the necessity of
imposing additional constraints in order to get nontrivial order
parameters, as far as for finite systems these parameters, as is
explained above, are always the same due to conservation laws.

Moreover, computer simulations, accomplished for a finite system, are
always complimented by finite-size scaling allowing for an extension
of the finite-volume results to effectively infinite system. Only
with such a scaling, accounting for finite-size corrections, one is
able to extract correct system characteristics, especially in the
vicinity of phase transitions.

When one is interested in developing an analytical theory for finite
systems exhibiting phase transitions, then, trying to describe the
latter without symmetry breaking, results in extremely complicated
calculations that are often even not accomplishable. While, taking
into account symmetry breaking greatly simplifies calculations.
Owing to the fact that, for a large number of particles, crossovers
become so sharp that are practically non-distinguishable from phase
transitions, it is reasonable to invoke symmetry breaking for
describing ordered phases.

Thus, though for finite systems one could employ computer simulations
without symmetry breaking, but this is extremely unpractical for
analytical description of systems with phase transitions. Dealing with
mesoscopic systems, for which $N \gg 1$, it is a wise idea to
explicitly employ symmetry breaking, understanding that, strictly
speaking, it is asymptotic symmetry breaking, as is defined above.

\subsection{Geometric symmetry transformations}
\label{subsec:II.C}

\subsubsection{Role of shape symmetry}
\label{subsubsec:II.C.1}

There exists a type of symmetry transformation that is specific for finite
systems. This is shape symmetry breaking, when a finite system, under
varying external or internal parameters, exhibits the change of its shape.
This change can be spontaneous, in the sense that the shape form is not
imposed from outside, but the system acquires the chosen form because it
is energetically profitable.

In those cases when the finiteness of the system becomes important, the
system shape may also play an important role. This role becomes crucial
in several situations.

First of all, a system, composed of the same kind of particles, can be
either stable or unstable depending on its shape and size. For instance,
this concerns trapped atoms with attractive interactions and atoms with
dipolar forces, which will be considered in the following section. Atomic
nuclei is another example, where the shape is defined by the internal
properties.

When the system shape is not prescribed by external potentials, but can
be self-organized, then the effect of spontaneous shape symmetry
breaking can happen, as for rotating atomic nuclei.

Another characteristic of a finite system, where its shape starts playing
role, is the spectrum of collective excitations. Elementary collective
excitations in a finite system display two principally different types of
behavior. For short wavelengths of such excitations, such that the
wavelength $\lambda \ll L$ is much shorter than the characteristic
system size $L$, respectively, when the excitation wave vector $k$ is
such that $kL \gg 1$, then these excitations are of the same kind as
they would be in an infinite system. The spectrum of these short-wavelength
excitations is usually continuous with respect to the wave vector. The form
of the spectrum changes under phase transitions accompanied by symmetry
breaking \cite{BelitzKirkVoj_2005}.

But, as soon as the wavelength is comparable or larger than the system
size, hence $\lambda > L$, then the excitation spectrum becomes discrete
and strongly dependent on the system shape. Then the shape symmetry is
as important as the symmetry of the thermodynamic phase.

Examples, illustrating the influence of the system shape on the properties
of collective excitations, will be given below for several finite quantum
systems, such as trapped atoms, quantum dots, and atomic nuclei. Another
example is provided by the behavior of spin waves in finite magnetic
samples, which has been studied by Mills et al. \cite{Arias_2005,Mills_2005,
Mills_2006,Costa_2006,Arias_2007}.

Finally, in finite quantum systems there appear quantized collective
excitations of the type that does not exist in macroscopic systems. This
is because particles are bounded inside a finite system. The energy spectrum
of bounded particles is discrete. The eigenfunctions, corresponding to these
discrete energy states, form collective topological modes. The latter are
termed topological, since the related wave functions and, respectively,
particle densities display different spatial shapes, with different numbers
of zero. The topological modes are principally different from elementary
collective excitations. The latter correspond to small oscillations around
a given topological mode and are described by linearized equations. While
the topological modes are described by nonlinear equations. Explicit
illustrations of these points will be given in the following sections.

\subsubsection{Geometric shape transitions}
\label{subsubsec:II.C.2}

Probably, the first model of shape transitions was suggested by Jahn and Teller
for molecules. The Jahn-Teller effect, sometimes also known as Jahn-Teller
distortion, describes the geometrical distortion of non-linear molecules under
certain situations. This electronic effect was described by Jahn and Teller
using group theory. They showed that orbital non-linear spatially degenerate
molecules cannot be stable \cite{Jahn_1937}. This essentially states that any
non-linear molecule with a spatially degenerate electronic ground state will
undergo a geometrical distortion that removes that degeneracy, because the
distortion lowers the overall energy of the complex. This effect has been found
to occur for a number of substances in chemistry and solid-state physics
\cite{Bersuker_2006}. Similar shape transitions happen for some nuclei
\cite{pcej}, for which different phenomenological approaches and models
have been suggested. Because of the occurrence of such geometric distortions
in a wide variety of substances and because of the special role of shape
transitions for finite quantum systems, we delineate below the main ideas
of the effect.

It is important to stress that in their original paper \cite{Jahn_1937} Jahn
and Teller studied a molecule, that is, exactly a finite system, but not an
infinite crystal. The Jahn-Teller distortion of a finite system necessarily
results in its shape change, usually accompanied by a change of shape symmetry.
The application of the Jahn-Teller effect to describing the shape and symmetry
changes in molecules has been expounded in many textbooks, for instance, in
\cite{LandauLifshitz_2003, Petrashen_2009}.

And only later the Jahn-Teller effect started being considered for infinite
crystals. In both these cases, the physics of the effect is the same, being
related to the Jahn-Teller distortion, which, for bulk crystals, leads to the
symmetry variation of the crystalline lattice, while for finite systems, to
their shape change. The Jahn-Teller effect for finite systems, such as
molecules and clusters, has been intensively studied and is described in
voluminous literature.

There exists the well known term {\it finite Jahn-Teller systems}. Numerous
articles are devoted to symmetry changes in such finite Jahn-Teller systems,
as molecules and clusters, where the molecule or cluster distortion changes
their shape symmetry, which is called {\it geometric transformation}
\cite{O'Brien_1993,Qiu_1997,Alberto_2004,Ziegler_2005}.

The shape variations of molecules, caused by the distortion of their
structure, are described in a number of textbooks \cite{Wells_1984,
Cotton_1999,Hollman_2001,Clayden_2012}.

The common physical origin of the Jahn-Teller effect in finite and infinite
systems is expounded in detail in numerous literature, for instance, in the
books \cite{Englman_1972,Kaplan_2001}.

As can be inferred from the cited literature, two principal mechanisms
govern the physics of finite Jahn-Teller systems: {\it tunneling of
particles between different orbitals} and the {\it coupling of these particles
with phonons} \cite{Bersuker_2006,pcej,O'Brien_1993,Qiu_1997,Alberto_2004,
Ziegler_2005}. It is exactly these mechanisms that lead to the system distortion.
The fact that such a distortion results in the changes in the symmetry shapes
of molecules is rather evident and numerous examples can be found in the cited
literature. As the simplest illustration, imagine that a finite system enjoys
cubic structure. The distortion of the atomic locations in the cube can easily
result in the structures possessing either orthorhombic, or tetragonal, or
rhombohedrical, or triclinic symmetry
\cite{Wells_1984,Cotton_1999,Hollman_2001,Clayden_2012}.

In order to show how the distortion arises, due to the basic mechanisms of
{\it orbital tunneling} and {\it atom-phonon coupling}, let us consider a
system of $N$ particles, say atoms, that interact with each other through an
interaction potential $\Phi(\bf r)$ and are confined in a finite volume by means
of a confining potential $U(\bf r)$. The related microscopic Hamiltonian is
$$
\hat H = \int \psi^\dgr(\br) \left [ -\;
\frac{\nabla^2}{2m} + U(\br) \right ] \psi(\br) \; d\br \; +
$$
\be
\label{S.1}
 + \;
\frac{1}{2} \int \psi^\dgr(\br) \psi^\dgr(\br') \Phi(\br-\br')
\psi(\br') \psi(\br) \; d\br d\br' \; ,
\ee
where $\psi(\br)$ are the field operators satisfying either Bose or Fermi
commutation relations. The particle statistics is not important for what
follows. Assume that, as it is common for chemical molecules or solid
clusters, there are in the system preferable atomic locations denoted by
a set $\{\br_j\}$, with $j = 1,2,\ldots,N$. Due to the confining nature
of the potential $U(\bf r)$, the eigenproblem
\be
\label{S.2}
 \left [ -\; \frac{\nabla^2}{2m} + U(\br) \right ] \psi_n(\br-\br_j) =
E_{nj} \psi_n(\br -\br_j)
\ee
defines the localized atomic orbitals $\psi_n({\bf r} - {\bf r}_j)$ with
a discrete spectrum labeled by a multi-index $n$ pertaining to a discrete
manifold. The orbitals with different indices $n$ possess different spatial
symmetries. The field operators can be expanded over the localized orbitals
as
\be
\label{S.3}
 \psi(\br) = \sum_{nj} c_{nj} \psi_n(\br-\br_j) \;  ,
\ee
with the operators $c_{nj}$ satisfying the no-double-occupancy constraint
\be
\label{S.4}
 \sum_n c_{nj}^\dgr c_{nj} = 1 \; , \qquad
c_{mj} c_{nj} = 0 \;  .
\ee

As is stressed above, one of the main features of finite Jahn-Teller
systems is the existence of orbital tunneling. Assuming that temperature
is sufficiently low allows one to take into account only a few orbitals,
with the lowest energy levels of the discrete spectrum $\{E_{nj}\}$. For
instance, we can take two lowest levels labeled by $n = 1,2$. This makes
it straightforward to invoke the operator transformations
$$
c_{1j}^\dgr c_{1j} = \frac{1}{2} + S_j^x \; , \qquad
c_{2j}^\dgr c_{2j} = \frac{1}{2} - S_j^x \; ,
$$
\be
\label{S.5}
c_{1j}^\dgr c_{2j} = S_j^z - i S_j^y \; , \qquad
c_{2j}^\dgr c_{1j} = S_j^z + i S_j^y \; ,
\ee
in which the operators $S_j^\alpha$ satisfy the spin commutation relations,
independently from the statistics of the atomic operators $c_{nj}$. Since
the operators $S_j^\alpha$ do not need to describe real spins, but obey the
spin algebra, they are termed {\it pseudospin} operators. The inverse
transformations read as
$$
 S_j^x = \frac{1}{2} \left ( c_{1j}^\dgr c_{1j} -
c_{2j}^\dgr c_{2j} \right ) \; , \qquad
S_j^y = \frac{i}{2} \left ( c_{1j}^\dgr c_{2j} -
c_{2j}^\dgr c_{1j} \right ) \; ,
$$
\be
\label{S.6}
S_j^z = \frac{1}{2} \left ( c_{1j}^\dgr c_{2j} +
c_{2j}^\dgr c_{1j} \right ) \; .
\ee
Such a pseudospin representation is often invoked for finite-level systems
of different physical nature, e.g., for double-well optical lattices
\cite{YukalovYukalova_2008,YukalovYukalova_2009}.

Then expansion (\ref{S.3}) is substituted into Hamiltonian (\ref{S.1}).
The arising matrix elements of the interaction potential are denoted by
$A({\bf r}_{ij}), B({\bf r}_{ij})$, and $C({\bf r}_{ij})$, where
${\bf r}_{ij} \equiv {\bf r}_i - {\bf r}_j$. One introduces the
notation for the average potential energy per atom
\be
\label{S.7}
  E_0 \equiv \frac{1}{2N}
\sum_{nj} \lgl nj \; | \; U \; | \; nj \rgl \; ,
\ee
where $|nj\rgl$ is the corresponding orbital $\psi_{nj}$, for average
kinetic energy per particle
\be
\label{S.8}
 \frac{p_j^2}{2m} \equiv \frac{1}{2} \sum_n
\left \lgl nj \; \left | \left ( -\; \frac{\nabla^2}{2m} \right )
\right | \; nj \right \rgl \; ,
\ee
and for the orbital tunneling frequency
\be
\label{S.9}
 \Om_j \equiv E_{2j} - E_{1j} + \sum_i C(\br_{ij}) \;  .
\ee
As a result, Hamiltonian (\ref{S.1}) reduces to
$$
\hat H = E_0 N + \sum_j \left ( \frac{p_j^2}{2m} \; - \;
\Om_j S_j^x \right ) \; +
$$
\be
\label{S.10}
 + \; \sum_{i\neq j} \left [ \frac{1}{2} \; A(\br_{ij}) +
B(\br_{ij}) S_i^x S_j^x - I(\br_{ij}) S_i^z S_j^z \right ] \;  .
\ee

The physical meaning of the pseudospin operators is as follows. The
operator $S_j^x$ describes tunneling between the atomic orbitals,
$S_j^y$ corresponds to the internal Josephson current between the orbitals,
and $S_j^z$ characterizes interorbital coupling.

As is stressed above, the second basic mechanism leading to the atomic
distortion and, hence, to shape changes of finite systems, is the atom-phonon
coupling \cite{Bersuker_2006,pcej,O'Brien_1993,Qiu_1997,Alberto_2004,
Ziegler_2005}. Taking account of the phonon degrees of freedom is a known
procedure, because of which we delineate below only its general scheme.

To introduce vibrational degrees of freedom, one defines atomic
deviations ${\bf u}_j$ from a fixed location ${\bf a}_j$ by the relation
$$
 \br_{j} = \ba_j + \bu_j \;  .
$$
Assuming that the deviations are small, one expands the terms of
Hamiltonian (\ref{S.10}) in powers of the deviations up to the second
order. Then one represents the operators ${\bf p}_j$ and ${\bf u}_j$ by
means of a canonical transformation introducing the phonon operators $b_{ks}$,
labelled by a quasimomentum $k$ and a polarization index $s$. The phonon
and pseudospin degrees of freedom are decoupled by means of a self-consistent
approximation based on the Bogolubov variational principle, after which it
becomes possible to diagonalize the phonon part of the Hamiltonian and to
calculate the average distortion $\lgl {\bf u}_j \rgl$ of a $j$-th atom.
The explicit elaboration of all details for this procedure can be found in
the books \cite{Blinc_1974,Bruce_1981,Tyablikov_1995}. As a result of this
procedure, one gets the average distortion vector $\lgl {\bf u}_j \rgl$
with the components
$$
 \lgl u_i^\al \rgl = \sum_j \gm_{ij}^\al C_j +
\sum_{jf} \dlt_{ijf}^\al C_j C_f \;  ,
$$
which connects the distortion with the average
$$
 C_j \equiv \lgl S_j^z \rgl \;  ,
$$
playing the role of an order parameter. The coefficients $\gamma_{ij}^\alpha$
and $\delta_{ijf}^\alpha$ are expressed through the derivatives over
$r_i^\alpha$ of the interaction $I({\bf r}_{ij})$ from Hamiltonian (\ref{S.10})
(see details in \cite{Blinc_1974,Bruce_1981,Tyablikov_1995}).

The system Hamiltonian, being invariant under the inversion $S_j^z \ra - S_j^z$,
allows for the existence of a trivial solution with the zero order parameter
$C_j \equiv 0$, which corresponds to a symmetric state having no distortion.
However, the symmetric state, under certain conditions, is unstable against
the appearance of a nonzero order parameter $C_j \neq 0$. In that case, there
happens geometric distortion of the sample, characterized by the distortion
vector with the observable components given by the average of
$\lgl {\bf u}_j^\alpha \rgl$. The arising nontrivial order parameter can lead
to the geometric distortion and the change of system shape. This change occurs spontaneously owing to the fact that the distorted shape is more energetically
favorable than the symmetric shape. This situation, as is explained above, is
typical for the Jahn-Teller effect in finite systems.

As an illustration, we can again imagine a finite system enjoying cubic shape
structure. Then a distortion of atomic locations can easily lead to the
structures possessing either orthorhombic, or tetragonal, or rhombohedrical,
or triclinic symmetry \cite{Wells_1984,Cotton_1999,Hollman_2001, Clayden_2012}.

\subsubsection{Geometric orientation transitions}
\label{subsubsec:II.C.5}

The fact that finite systems demonstrate geometric transformations, even when
the system is as small as being composed of two particles, can be illustrated
by the following example of the orientational transformation in a dimer formed
by two monomers represented by their dipole vectors ${\bf D}_1$ and ${\bf D}_2$.
Dimer structures are very common in biophysics and biochemistry
\cite{Mathews_1990}. The Hamiltonian of the dimer can be written in the form
\be
\label{G.1}
H = J \bD_1 \cdot \bD_2 - U \left ( D_1^z + D_2^z \right ) \;  ,
\ee
where $J$ characterizes the interaction strength and $U$ describes an external
field acting on the dimer. Each vector ${\bf D}_i$ is described by the set
\be
\label{G.2}
 \bD_i = \left \{ D\gm_i^x , \; D\gm_i^y , \; D\gm_i^z \right \} \; ,
\ee
in which the orientation factors
\be
\label{G.3}
\gm_i^x = \sin\vartheta_i \cos\vp_i \; , \qquad
\gm_i^y = \sin\vartheta_i \sin\vp_i \; , \qquad
\gm_i^z = \cos\vartheta_i  \qquad ( i = 1,2 )
\ee
are expressed through the spherical angles $\vartheta_i$ and $\varphi_i$.

The spatial orientation of the vectors ${\bf D}_1$ and ${\bf D}_2$ are
defined by minimizing the dimer energy over the spatial angles under the
normalization constraint
\be
\label{G.4}
 \left ( \gm_i^x \right )^2 + \left ( \gm_i^y \right )^2 +
\left ( \gm_i^z \right )^2 = 1\; .
\ee
Accomplishing the minimization, we find that there exists a critical value
of the external field
\be
\label{G.5}
 U_c \equiv 2J D \;  ,
\ee
separating two different types of orientation.

At the values of the field below $U_c$, we have
\be
\label{G.6}
\gm_1^x + \gm_2^x = \gm_1^y + \gm_2^y =  0 \; , \qquad
\gm_1^z  = \gm_2^z  = \frac{U}{U_c} \qquad ( U < U_c ) \; ,
\ee
which gives the angles
\be
\label{G.7}
 \vartheta_1 = \vartheta_2 = \arccos \;  \frac{U}{U_c} \; , \qquad
\vp_1 - \vp_2 = \pi \; .
\ee
The related energy is
\be
\label{G.8}
 E = - J D^2 \left [ 1 + 2 \left ( \frac{U}{U_c} \right )^2 \right ]
\qquad ( U < U_c ) \;  .
\ee

This shows that the vectors ${\bf D}_i$ deviate from the axis $z$ by the same angle,
$\vartheta_i$, while their components on the $x-y$ plane are counter-aligned.
The direction of the transverse components is not defined, though this
degeneracy is lifted by an external infinitesimal field. Under the fixed
vectors ${\bf D}_i$, the dimer possesses the symmetry of rotation around the
$z$-axis with respect to the angles that are integers of $\pi$.

When the external field values are larger than $U_c$, the energy minimization
yields
\be
\label{G.9}
\gm_1^x = \gm_2^x = \gm_1^y = \gm_2^y = 0 \; , \qquad
\gm_1^z = \gm_2^z = 1 \; , \qquad  \vartheta_1 = \vartheta_2 = 0
\qquad ( U > U_c ) \; ,
\ee
with the energy
\be
\label{G.10}
 E =  J D^2 \left ( 1  -  4 \; \frac{U}{U_c} \right ) \; .
\ee

Then both vectors ${\bf D}_i$ are aligned along the axis $z$. Hence the dimer
enjoys the rotation symmetry with respect to an arbitrary angle $\varphi$.

In that way, at the critical field $U_c$, there happens the orientation
transition. When lowering the field $U$, the system transforms from the
state of the high rotational symmetry around the axis $z$ to the broken
symmetry state, having the lower $\pi$-rotational symmetry.

This example demonstrates that finite systems exhibit the type of
transformations that are absent for infinite systems. These are shape and
orientation transitions accompanied by symmetry changes. A similar
orientation transition, as will be shown below, occurs in a two-electron
quantum dot.

\section{Trapped Atoms}
\label{sec:III}

Trapped particles can form finite quantum systems of different sizes,
from just a few particles to millions of them. The possibility of varying
the number of particles as well as other properties of trapped systems
makes the latter the objects of high importance for both theoretical
studies as well as for a variety of applications. Particles can be charged
or neutral. Respectively, there are two rather distinct systems: trapped
ions and trapped atoms. Here, we concentrate on the physics of neutral
trapped atoms. Physics of trapped ions is a different topic requiring a
separate consideration \cite{Blaum_2006}. We analyze theoretical aspects
related to symmetry breaking in the systems of trapped atoms. But we do
not touch experimental methods of trapping atoms, whose description can
be found in literature \cite{Balykin_2000,Letokhov_2007}.

\subsection{Bose-Einstein condensation}
\label{subsec:III.A}

\subsubsection{Gauge symmetry breaking}
\label{subsubsec:III.A.1}

If trapped atoms are bosons, it is possible to cool them down reaching
Bose-Einstein condensation predicted by Bose \cite{Bose_1924} and
Einstein \cite{Einstein_1924}. Though superfluidity in liquid $^4$He,
since London, has been assumed to be accompanied by Bose-Einstein
condensation \cite{London_1954}, the measurement of the condensate
fraction has been a complicated experiment task \cite{WirthHallock_1987}.
This is connected with strong interactions between helium atoms and the
resulting strong condensate depletion. In superfluid helium, the condensate
fraction at zero temperature is only about $10\%$. This is why the direct
observation of Bose-Einstein condensation of trapped atoms has become such
an important experimental achievement. This phase transition was
demonstrated in 1995 almost simultaneously in three groups, condensing
the atoms of $^{87}$Rb \cite{Anderson_1995}, $^{23}$Na \cite{Davis_1995},
and $^7$Li \cite{Bradley_1995}. By the present time, Bose-Einstein
condensation has been achieved for the following trapped atoms: $^1$H,
$^4$He, $^7$Li, $^{23}$Na, $^{39}$K, $^{40}$Ca, $^{41}$K, $^{52}$Cr,
$^{84}$Sr, $^{85}$Rb, $^{86}$Sr, $^{87}$Rb, $^{88}$Sr, $^{133}$Cs,
$^{170}$Yb, and $^{174}$Yb. Bosonic atomic molecules can also be condensed,
such as boson-formed molecules: $^{23}$Na$_2$, $^{85}$Rb$_2$, $^{87}$Rb$_2$,
and $^{133}$Cs$_2$, fermion-formed molecules: $^6$Li$_2$ and
$^{40}$K$_2$, and heteronuclear boson-boson molecules: $^{85}$Rb$^{87}$Rb
and $^{41}$K$^{87}$Rb.

This phenomenon has been intensively studied both experimentally and
theoretically. Numerous references can be found in the books
\cite{Pitaevskii_2003,Pethick_2008} and review articles
\cite{Yukalov_2009,Yukalov_2004,Dalfovo_1999,Courteille_2001,Andersen_2004,
Bongs_2004,YukalovGirardeau_2005,Posazhennikova_2006,Morsch_2006,Moseley_2008,
Proukakis_2008,Yukalov_2011b}.

With regard to symmetry breaking under Bose-Einstein condensation, one can
meet in literature contradictory statements. This phase transition is
associated with the global gauge symmetry breaking. The global gauge symmetry
corresponds to the group $U(1)$. The order parameter, characterizing this
phase transition, is the average of the field operator $\langle \psi\rangle$.
In the normal phase, the system Hamiltonian is invariant under the gauge
transformations (\ref{2.11}), which results in Eq. (\ref{2.13}). Hence,
the order parameter is zero, which is a particular case of Eq. (\ref{2.13}).
The order parameter becomes nonzero in the Bose-condensed phase, which
requires gauge symmetry breaking. However, in literature, it is possible to
meet statements that Bose-Einstein condensation does not necessarily require
the gauge symmetry breaking. In order to be precise, we shall recall below
rigorous mathematical theorems proving that {\it global gauge symmetry breaking
is the necessary and sufficient condition for Bose-Einstein condensation}.
As has been explained above, such rigorous mathematical facts correspond to
thermodynamic limit. For finite systems, one has to keep in mind asymptotic
symmetry breaking, as is formulated in Sec. \ref{sec:II}.

In a {\it finite} system, the field operator of spinless bosons can be expanded
over a basis $\{\varphi_k(\bf r)\}$, where $k$ is a quantum multi-index.
Then the field operator can be decomposed into two terms
\be
\label{3.1}
\psi(\br) = \sum_k a_k \vp_k(\br) =
\psi_0(\br) + \psi_1(\br)  \;  ,
\ee
the first term, representing a quasicondensate, characterized by the
index $k_0$, and the second term, corresponding to uncondensed
particles,
\be
\label{3.2}
 \psi_0(\br) \equiv a_0 \vp_0(\br) \; , \qquad
\psi_1(\br) \equiv \sum_{k\neq k_0} a_k \vp_k(\br) \; .
\ee
Here $a_0 \equiv a_{k_0}$ and $\vp_0 \equiv \vp_{k_0}$. By definition,
{\it quasicondensate} is what would be condensate in thermodynamic limit.
The index $k_0$ here is not specified, keeping in mind that condensation
can happen in either uniform or nonuniform systems. For uniform systems,
$k$ becomes the wave vector and $k_0 = 0$.

It is of principal importance to stress that the expansion basis
in Eq. (\ref{3.1}) is in no way arbitrary, but it has to be the
{\it natural basis} that is composed of {\it natural orbitals},
which are the eigenfunctions of the first-order density matrix
\cite{PenroseOnsager_1956,ColemanYukalov_2000}. Only in this case,
the quasicondensate will become real condensate in thermodynamic limit.

Also, it is necessary to emphasize that, for a {\it finite} system, the
parts $\psi_0$ and $\psi_1$ are not separate operators but just two parts
of one field operator $\psi$, their commutation relations being
$$
[\psi_0(\br),\psi_0^\dgr(\br') ] =
\vp_0(\br) \vp_0^*(\br') \; ,
$$
\be
\label{3.3}
[\psi_1(\br),\psi_1^\dgr(\br') ] =
\sum_{k\neq k_0} \vp_k(\br) \vp_k^*(\br') \;.
\ee
All operators of observables are defined on the Fock space
$\mathcal{F}(\psi)$ generated by the field operator $\psi$
\cite{Berezin_1966}.

The fact that spontaneous gauge symmetry breaking yields Bose-Einstein
condensation is rather straightforward. Let us define the quasiaverages
as in Eq. (\ref{2.18}), breaking the symmetry of the system Hamiltonian
by an infinitesimal source, as in Eq. (\ref{2.17}). Spontaneous gauge
symmetry breaking implies that
\be
\label{3.4}
 \lim_{\ep\ra 0} \; \lim_{N\ra\infty} \;
\frac{| \lgl a_0\rgl_\ep|^2}{N} \; > \; 0 \; .
\ee
When the latter holds true, then, from the Cauchy-Schwarz inequality
$|\lgl a_0\rgl_\ep|^2\leq \lgl a_0^\dgr a_0\rgl_\ep$, it follows that
\be
\label{3.5}
 \lim_{\ep\ra 0} \; \lim_{N\ra\infty} \;
\frac{\lgl a_0^\dgr a_0 \rgl_\ep}{N} \; > \; 0 \;  ,
\ee
which means Bose-Einstein condensation.

Ginibre \cite{Ginibre_1968} proved the theorem showing that the
thermodynamic potential of a system with gauge symmetry breaking acquires,
in thermodynamic limit, the form, where the quasicondensate field operator
$\psi_0$ is replaced by the condensate order parameter $\eta$ defined as
the minimizer of the thermodynamic potential \cite{Ginibre_1968}. Bogolubov
showed that the same replacement, in thermodynamic limit, of $\psi_0$ by
$\eta$ is valid for all correlation functions given by the averages with
broken gauge symmetry \cite{Bogolubov_1970}. Therefore, spontaneous gauge
symmetry breaking is a {\it sufficient condition} for condensation.

The necessary condition was, first, proved by Roepstorff \cite{Roepstorff_1978}.
Recently, the proof was polished by several authors \cite{Suto_2005,Lieb_2005}.
According to the Roepstorff theorem, the inequality
\be
\label{3.6}
\lim_{N\ra\infty} \;
\frac{\lgl a_0^\dgr a_0 \rgl}{N} \leq
\lim_{\ep\ra 0} \; \lim_{N\ra\infty} \;
\frac{|\lgl a_0\rgl_\ep|^2}{N}
\ee
is valid, where the left-hand side is defined without gauge symmetry
breaking. This means that Bose-Einstein condensation necessarily
results in gauge symmetry breaking \cite{Roepstorff_1978}.

Concluding, it is a rigorous mathematical fact that: {\it Spontaneous
breaking of global gauge symmetry is the necessary and sufficient
condition for Bose-Einstein condensation}. For finite systems, it is
to be understood in the sense of asymptotic symmetry breaking. A more
detailed discussion can be found in the review article \cite{Yukalov_2007}.

\subsubsection{Representative statistical ensemble}
\label{subsubsec:III.A.2}

A system with broken symmetry requires to be correctly described, for
which purpose, one has to employ representative statistical ensembles.
A statistical ensemble, by definition, is a pair
$\{ {\mathcal F}, \hat{\rho}\}$ of the space of microstates ${\mathcal F}$
and a statistical operator $\hat{\rho}$ describing the system. The
temporal evolution of the statistical operator is governed by the evolution
operator, with the system Hamiltonian $H$ being the evolution generator.
This can be denoted as the dependence $\hat{\rho}(H,t)$. For
equilibrium and quasi-equilibrium systems, the form of the statistical
operator is prescribed by the principal of minimal information, that is,
by minimizing the information entropy under given additional constraints
\cite{Jaynes_1957}.

It is important that the statistical ensemble should uniquely define
the considered statistical system, for which it is necessary to include
in the definition all conditions and constraints characterizing the system
\cite{Gibbs_1931,Gibbs_1928}. Such an ensemble, correctly defining the
system, is called {\it representative} \cite{Tolman_1940,terHaar_1955}.
In the case of a system with condensate, one has to accurately take into
account the global gauge symmetry breaking and the method by which this is
realized.

Treating a system with Bose-Einstein condensate, one usually accepts
one of the following two ways:

(i) One possibility is to work with a {\it finite system}, having one field
operator $\psi$ generating the Fock space ${\mathcal F}(\psi)$. The system
evolution is described by a grand Hamiltonian that is the evolution generator.
In defining the grand Hamiltonian, one takes into account physical constraints
associated with the system. One such a constraint is the definition of
internal energy as the average of the energy Hamiltonian $\hat{H}$. Another
constraint is the definition of the number of particles as the average
of the number operator $\hat{N}$. And one more constraint is the
necessity of breaking the symmetry of the Hamiltonian by an infinitesimal
source, as in Eq. (\ref{2.17}). Then the grand Hamiltonian reads as
\be
\label{3.7}
H_\ep = \hat H - \mu \hat N + \ep \hat\Gm \;   ,
\ee
which is defined on the Fock space ${\mathcal F}(\psi)$. Using this
approach, based on the statistical ensemble
$\{{\mathcal F}(\psi), \hat{\rho}(H_\varepsilon,t)\}$, it is necessary
to make calculations for a finite system, employing the commutation
relations (\ref{3.3}). At the end, one passes to thermodynamic limit
in order to break gauge symmetry, as in Eq. (\ref{2.18}). This way,
though being in principle admissible \cite{Leggett_2006}, is extremely
cumbersome and can be followed only by invoking heavy numerical calculations,
for example, by using Monte Carlo numerical simulations.

(ii) Another approach, developed by Bogolubov \cite{Bogolubov_1947} suggests
to break the global gauge symmetry explicitly by introducing the so-called
{\it Bogolubov shift} of the field operator, replacing the operator
$\psi({\bf r})$ by the field operator
\be
\label{3.8}
 \hat\psi(\br) = \eta(\br) + \psi_1(\br) \; ,
\ee
in which $\eta({\bf r})$ is the {condensate wave function} and
$\psi_1({\bf r})$ is the field operator of uncondensed particles. For
the latter, the standard Bose commutation relations are valid,
\be
\label{3.9}
\left [ \psi_1(\br) , \psi_1^\dgr(\br') \right ] =
\dlt(\br-\br') \; ,
\ee
which makes their use very convenient
\cite{Bogolubov_1970,Bogolubov_1967,Bogolubov_1947}.

The Bogolubov shift (\ref{3.8}) is a canonical transformation that is
correctly defined for any Bose-condensed system. This transformation
does not require that $\psi_1$ be in any sense small. It is important
to emphasize that, after this canonical transformation, the operators
of observables are defined on the Fock space ${\mathcal F}(\psi_1)$
generated by the field operator $\psi_1$. The spaces ${\mathcal F}(\psi)$
and ${\mathcal F}(\psi_1)$ are asymptotically (as $N$ tends to infinity)
orthogonal, thus, realizing unitary nonequivalent operator
representations with the unitary nonequivalent representations for
commutation relations \cite{Umezawa_1982,Yukalov_2006a}.

Realizing the Bogolubov shift, one gets two independent field variables,
$\eta$ and $\psi_1$ that are orthogonal to each other,
\be
\label{3.10}
 \int \eta^*(\br) \psi_1(\br) \; d\br = 0 \;  .
\ee
Respectively, in addition to the definition of the internal energy, there
are two normalization conditions, for the number of condensed particles,
\be
\label{3.11}
 N_0 = \int | \eta(\br)|^2 \; d\br \; ,
\ee
and for the number of uncondensed particles,
\be
\label{3.12}
N_1 = \lgl \hat N_1 \rgl \; , \qquad
\hat N_1 \equiv
\int \psi_1^\dgr(\br) \psi_1(\br) \; d\br \;  .
\ee
There is also the restriction classifying the uncondensed particles
as normal particles satisfying the condition
\be
\label{3.13}
 \lgl \psi_1(\br) \rgl = 0 \; .
\ee
The latter condition can be rewritten in the standard form of a
statistical average
\be
\label{3.14}
\lgl \hat\Lbd \rgl = 0 \; , \qquad
\hat\Lbd \equiv \int [\lbd(\br) \psi_1^\dgr(\br) +
\lbd^*(\br) \psi_1(\br) ] \; d\br \;  ,
\ee
where $\lambda({\bf r})$ is a complex function playing the role of
the Lagrange multiplier guaranteeing the validity of Eq. (\ref{3.14}).
In that way, the grand Hamiltonian acquires the form
\be
\label{3.15}
 H = \hat H - \mu_0 N_0 - \mu_1 N_1 -\hat\Lbd \; ,
\ee
with the Lagrange multipliers $\mu_0$ and $\mu_1$ conserving the
normalization conditions (\ref{3.11}) and (\ref{3.12}).

This form of the grand Hamiltonian makes it possible to define a
representative ensemble providing a completely self-consistent
description of any Bose-condensed system
\cite{Yukalov_2005d,Yukalov_2006b,Yukalov_2006c,Yukalov_2008}.
In order to stress that the field variables $\eta$ and $\psi_1$ are
independent, it is possible to define the grand Hamiltonian on the
composite Fock space ${\mathcal F}_0(\eta) \bigotimes {\mathcal F}(\psi_1)$
\cite{Yukalov_2006a}. Generally, the Lagrange multipliers $\mu_0$ and
$\mu_1$ are different, though they can coincide in some particular
cases, as in the Bogolubov approximation \cite{Bogolubov_1947},
corresponding to low temperatures and asymptotically weak interactions.
The existence of two Lagrange multipliers resolves the Hohenberg-Martin
dilemma \cite{HohenbergMartin_1965}, making the theory conserving and
the spectrum gapless, as is prescribed by the Hugenholtz-Pines relation
\cite{HugenholtzPines_1959}.

Both these ways, described above, are equivalent as far as the values of
observable quantities do not depend on the used operator representation
\cite{Bogolubov_1970,Emch_1972,BratteliRob_1979,Umezawa_1982}.

The method of breaking the global gauge symmetry by means of the
Bogolubov shift (\ref{3.8}) seems to be more convenient than the
method of infinitesimal sources (\ref{3.7}) because the commutation
relations (\ref{3.9}) are much simpler than those in Eq. (\ref{3.3}).
But, breaking the gauge symmetry with the Bogolubov shift requires to
be accurate, taking into account those constraints that uniquely define
this method. If these constraints are not taken into account, the
description will become not self-consistent and plagued with internal
defects, such as the appearance of a gap in the spectrum, breaking of
general thermodynamic relations, and the loss of the system stability.

The equations of motion are obtained from Hamiltonian (\ref{3.15}) in
the standard way, by means of variational derivatives. The equation
for the {\it condensate function} reads as
\be
\label{3.16}
i\; \frac{\prt}{\prt t} \; \eta(\br,t) = \left \lgl
\frac{\dlt H}{\dlt\eta^*(\br,t)} \right \rgl \;  .
\ee
And the equation for the operators of uncondensed particles is given
by the variational derivative
\be
\label{3.17}
i\; \frac{\prt}{\prt t} \; \psi_1(\br,t) =
\frac{\dlt H}{\dlt\psi_1^\dgr(\br,t)} \; .
\ee

Note that the variational equations are equivalent to the Heisenberg
equations of motion \cite{Yukalov_2011b,Yukalov_2011c}. For an
equilibrium system, the condensate-function equation (\ref{3.16})
reduces to the minimization of the Hamiltonian with respect to this
function, which constitutes a necessary stability condition.

\subsubsection{Equation for condensate function}
\label{subsubsec:III.A.3}

The global gauge symmetry is broken in the system if and only if there
exists a nonzero solution of the condensate-function equation (\ref{3.16}).
This equation, therefore, is of principal importance.

To explicitly illustrate the form of the equation for the condensate
function, let us accept for the energy Hamiltonian the standard expression
$$
\hat H = \int \hat\psi^\dgr(\br) \left ( -\; \frac{\nabla^2}{2m} +
U \right ) \hat\psi(\br) \; d\br \; +
$$
\be
\label{3.18}
 +  \; \frac{1}{2} \int \hat\psi^\dgr(\br) \hat\psi^\dgr(\br')
\Phi(\br-\br')\hat\psi(\br')\hat\psi(\br) \; d\br d\br' \; ,
\ee
in which $\Phi({\bf r})$ is a pair interaction potential and
$U = U({\bf r}, t)$ is an external potential. Then the condensate-function
equation (\ref{3.16}) yields
$$
i\; \frac{\prt}{\prt t} \; \eta(\br,t) = \left ( -\;
\frac{\nabla^2}{2m} + U - \mu_0 \right ) \eta(\br,t) \; +
$$
\be
\label{3.19}
+\; \int \Phi(\br-\br')
\lgl \hat\psi^\dgr(\br') \hat\psi(\br') \hat\psi(\br) \rgl d\br' \; .
\ee
This is the exact equation for the condensate function. Substituting
here the Bogolubov shift (\ref{3.8}), we meet the following densities:
the condensate density
\be
\label{3.20}
\rho_0(\br) = | \eta(\br)|^2 \; ,
\ee
the density of uncondensed atoms
\be
\label{3.21}
 \rho_1(\br) = \lgl \psi_1^\dgr(\br) \psi_1(\br) \rgl \; ,
\ee
and the total density
\be
\label{3.22}
\rho(\br) = \rho_0(\br) + \rho_1(\br) \;  .
\ee
One also needs the notation for the density matrix
\be
\label{3.23}
 \rho_1(\br,\br') = \lgl \psi_1^\dgr(\br') \psi_1(\br) \rgl \; ,
\ee
for the pair anomalous average
\be
\label{3.24}
 \sgm_1(\br,\br') = \lgl \psi_1(\br') \psi_1(\br) \rgl \; ,
\ee
and for the triple anomalous average
\be
\label{3.25}
 \al(\br,\br') = \lgl \psi_1^\dgr(\br') \psi_1(\br')
\psi_1(\br) \rgl \; .
\ee
The averages (\ref{3.24}) and (\ref{3.25}) are called anomalous, since
they appear only when gauge symmetry is broken, while under preserved
gauge symmetry, they are exactly zero, in view of Eq. (\ref{2.13}).

Introducing a self-energy operator
\be
\label{3.26}
  \hat\Sigma(\br) = \hat\Sigma_N(\br) +
\hat\Sigma_A(\br) \; ,
\ee
consisting of the sum of the normal operator, defined by the equation
\be
\label{3.27}
 \hat\Sigma_N(\br) \eta(\br) \equiv \int \Phi(\br-\br')
[ \rho(\br') \eta(\br) + \rho_1(\br,\br')\eta(\br')
] \; d\br' \;  ,
\ee
and of the anomalous operator, defined by the relation
\be
\label{3.28}
 \hat\Sigma_A(\br)\eta(\br)  \equiv \int \Phi(\br-\br')
[ \sgm_1(\br,\br') \eta^*(\br') + \al(\br,\br')] \; d\br' \;  ,
\ee
we can rewrite Eq. (\ref{3.19}) as
\be
\label{3.29}
 i\; \frac{\prt}{\prt t} \; \eta(\br,t) = \left ( -\;
\frac{\nabla^2}{2m} + U -\mu_0 \right ) \eta(\br,t) +
\hat\Sigma(\br) \eta(\br,t) \;  .
\ee
The statistical ensemble $\{{\mathcal F}(\psi_1), \hat{\rho}(H,t)\}$
provides a correct description of thermodynamics for Bose-condensed
systems at finite temperatures and arbitrary interactions
\cite{YukalovKleinert_2006,YukalovYukalova_2006,YukalovYukalova_2007}.

In the particular case of dilute gas, for which the interaction
radius is much shorter than the mean inter-atomic distance, the
interactions are modelled by the local potential
\be
\label{3.30}
 \Phi(\br) = \Phi_0\dlt(\br) \; , \qquad
\Phi_0 \equiv 4\pi \; \frac{a_s}{m} \;  ,
\ee
where $a_s$ is s-wave scattering length. At zero temperature and
very weak interactions, when practically all atoms are condensed,
so that $\hat\Sigma_N(\br)=\Phi_0 \rho_0(\br)$ and $\hat\Sigma_A(\br)=0$,
equation (\ref{3.29}) reduces to the Gross-Pitaevskii equation
\cite{Gross_1957,Gross_1958,Gross_1961,Pitaevskii_1961}
\be
\label{3.31}
 i \; \frac{\prt}{\prt t} \; \eta(\br,t) = \left [ -\;
\frac{\nabla^2}{2m} + U - \mu_0 + \Phi_0 |\eta(\br,t)|^2
\right ] \eta(\br,t) \; .
\ee
Gross interpreted this equation as the equation for a system of atoms
all of which are in the same state \cite{Gross_1957,Gross_1958,Gross_1961},
that is, in a coherent state \cite{Klauder_1985}.

\subsubsection{Size and shape instability}
\label{subsubsec:III.A.4}

Bose-Einstein condensate and, respectively, the occurrence of broken
gauge symmetry, can happen only in stable systems. The stability of
finite systems essentially depends on their sizes and shapes. Usually,
the {\it size-shape instability} appears when particle interactions
are either attractive or anisotropic, containing an attractive part.

In the case of isotropic interactions, the stability of Bose-condensed
systems is connected with the sign of the integral pair interaction
$$
  \Phi_0 = \int \Phi(\br) \; d\br = 4\pi \; \frac{a_s}{m} \; ,
$$
where $\Phi(\br)$ is a pair interaction potential. A uniform
system of atoms with effective attractive interactions, for which
$\Phi_0$ is negative, is known to be unstable \cite{terHaar_1977}.
However, a finite system of trapped atoms can be stable, when
their positive kinetic energy compensates the negative potential
energy. This imposes the limitation on the maximal number of atoms
$N_{max}$ that could form a stable finite system. This number has
been calculated numerically for harmonic traps, employing the
Gross-Pitaevskii equation (\ref{3.31}) by several authors, as
reviewed in \cite{Dalfovo_1999}. An analytical formula has also
been derived \cite{YukalovYukalova_2005} in the form
\be
\label{3.32}
 N_{max} = \sqrt{\frac{\pi}{2}} \;
\frac{l_x l_y l_z}{|a_s|(l_x^2+l_y^2+l_z^2)} \;  ,
\ee
in which $a_s < 0$ is attractive scattering length,
$l_\alpha \equiv 1/\sqrt{m \omega_\alpha}$ is a characteristic trap
length, $\alpha = x,y,z$, and $\omega_\alpha$ is a trap frequency.
The maximal number of atoms essentially depends on the trap size and
its anisotropy. The trap symmetry plays an important role. For the
purpose of housing more atoms, with the given scattering length, the
spherical shape seems to be more favorable, as compared to anisotropic
traps, elongated in one or two directions.

Typical anisotropic interactions are the dipole interactions, such as
exist for $^{52}$Cr atoms \cite{Griesmaier_2005} which possess
comparatively large magnetic dipole moments $\mu_0 = 6\mu_B$. The
dipolar interactions, for dipoles polarized along the $z$-axis, are
\be
\label{3.33}
 D(\br) = \frac{\mu_0^2}{r^3} \; ( 1 - 3\cos^2\vartheta) \; ,
\ee
where $r \equiv |{\bf r}|$, with ${\bf r}$ being a vector between the
interacting atoms, and $\vartheta$ is the angle between ${\bf r}$ and
the dipole orientation $z$. The same form of the dipole interactions
exists for atoms with electric dipoles, for which the magnetic moment
$\mu_0$ should be replaced by the electric dipole moment $d_0$
\cite{Baranov_2008}.

Strictly speaking, atomic interactions are composed of the sum of
the contact interaction (\ref{3.30}) and the dipolar interaction
(\ref{3.33}). The magnitudes of these interactions can be varied in
experiment. Dipolar interactions can be varied by applying magnetic
or electric fields, while contact interactions can be varied by means
of the Feshbach resonance techniques
\cite{Timmermans_1999,DuineStoof_2004,Kohler_2006,Yurovsky_2008}.

For the case of pure dipolar interactions (\ref{3.33}), a spatially
homogeneous Bose-condensed gas is unstable, similarly to the
homogeneous Bose gas with attractive isotropic interactions. For a
cigar shaped trap, with the aspect ratio $\omega_z/ \omega_\perp \ll 1$,
the effective dipole interactions are attractive and the gas can be
stable only for the number of atoms less than the maximal number
$N_{max}$. For a pancake shaped trap, with the aspect ratio
$\omega_z/ \omega_\perp \gg 1$, there also exists the maximal number
of atoms $N_{max}$, when the system becomes unstable, but this maximal
number is much larger than that for the cigar shaped traps. The value
of the maximal number of atoms depends on the trap shape and on the
average interaction
$$
D_0 = \frac{1}{N} \int D(\br-\br')
\rho_0(\br) \rho_0(\br') \; d\br d\br' \; .
$$
The pancake-shaped systems with dipolar interactions are always more
stable than the cigar-shaped ones.

\subsubsection{Elementary collective excitations}
\label{subsubsec:III.A.5}

Finiteness of quantum systems influences their spectra of collective
excitations. At zero temperature and very weak interactions, collective
excitations are usually described by considering small deviations from
the condensate function and linearizing the Gross-Pitaevskii equation
(\ref{3.31}). At finite temperatures and interactions, the correct
description of collective excitations requires to consider the whole
system Hamiltonian, whose diagonal part defines the spectrum of
excitations \cite{Yukalov_2008}. Employing the Hartree-Fock-Bogolubov
approximation, taking, for simplicity, the local interaction potential
(\ref{3.30}), and using the notation
$$
\hat\om(\br) \equiv -\; \frac{\nabla^2}{2m} + U(\br) - \mu_1
+ 2 \Phi_0 \rho(\br) \; ,
$$
\be
\label{3.34}
\Dlt(\br) \equiv \Phi_0 [ \rho_0(\br) + \sgm_1(\br) ] \; ,
\ee
where $\sigma_1({\bf r})$ is the diagonal anomalous average,
Eq. (\ref{3.24}), results in the Bogolubov equations for collective
excitations
$$
\hat\om(\br) u_k(\br) + \Dlt(\br) v_k(\br) = \ep_k u_k(\br) \; ,
$$
\be
\label{3.35}
 \hat\om(\br) v_k(\br) + \Dlt^*(\br) u_k(\br) =
-\ep_k v_k(\br) \;  .
\ee
Generally, these equations for trapped atoms can be solved only
numerically. Analytical solutions are available only in some
particular cases.

Collective excitations, whose wavelength is much shorter than the
system size and, respectively, the excitation frequency $\omega$
is much larger than the characteristic trapping frequency $\omega_0$,
behave as bulk excitations that are continuous and can be described
in the local density approximation
\cite{Pitaevskii_2003,Pethick_2008,Courteille_2001,Giorgini_1997}.
In this approximation, Eqs. (\ref{3.35}) yield the local Bogolubov
spectrum
\be
\label{3.36}
\ep(\bk,\br) = \sqrt{ c^2(\br) k^2 +
\left ( \frac{k^2}{2m} \right )^2 } \;   ,
\ee
where the local sound velocity is defined by the equation
\be
\label{3.37}
mc^2(\br) = [\; \rho_0(\br) + \sgm_1(\br) \;] \;\Phi_0 \; .
\ee
This form of the Bogolubov spectrum is in good agreement with
experiment for short-wave excitations \cite{Ozeri_2005}.

Collective excitations can also be defined as the poles of Green
functions. For systems with broken gauge symmetry, the poles of
single-particle and two-particle Green functions coincide
\cite{Gavoret_1964}. In uniform systems, the spectrum of collective
excitations is gapless, which is required by the Hugenholtz-Pines
relation \cite{HugenholtzPines_1959} and the condition of condensate
existence \cite{Yukalov_2009}. Such gapless excitations correspond
to massless Goldstone modes \cite{Goldstone_1961} arising under
spontaneous breaking of a global continuous symmetry
\cite{BelitzKirkVoj_2005}. Collective excitations with the Bogolubov
spectrum are called {\it bogolons}.

In finite quantum systems, those excitations, whose wavelength is
of the order or larger than the system size and, equivalently, for
which $\omega \lesssim \omega_0$, become discrete and essentially
dependent on the system shape symmetry. At zero temperature and
asymptotically weak interactions, such low-frequency excitations can
be found from hydrodynamic equations derived from the Gross-Pitaevskii
equation, with invoking the Thomas-Fermi approximation. Thus, for a
harmonic spherical trap the discretized spectrum is \cite{Stringari_1996}
\be
\label{3.38}
\om_{nl} = \om_0 \sqrt{2n^2 + 2nl + 3n + l} \; ,
\ee
where $n$ and $l$ are radial and orbital quantum numbers, respectively.
The excitations with $n=0$ are called {\it surface} excitations. A
particular case is the {\it dipolar mode} $(n=0, l=1)$. The excitations
with $n \neq 0$ are termed {\it compressional}. The case with $n=1, l=0$
corresponds to {\it breathing mode}.

For cigar-shape harmonic traps $(\omega_z \ll \omega_\perp)$, the
low-frequency spectrum is \cite{Fliesser_1997,Stringari_1998}
\be
\label{3.39}
 \om_n = \frac{\om_z}{2} \; \sqrt{n(n+1)} \; .
\ee
Particular cases are the {\it dipole mode} $(n=1)$ and the
{\it quadrupole mode} $(n=2)$.

For disk-shape harmonic traps $(\omega_z \gg \omega_\perp)$, the spectrum
of the lowest modes becomes \cite{Stringari_1998}
\be
\label{3.40}
 \om_{nm} = \om_\perp \;
\sqrt{\frac{4}{3}\; n^2 + \frac{4}{3}\; nm + 2n + m} \;  ,
\ee
where $m$ is azimuthal quantum number.

In a cylindrically symmetric trap, there is an excitation for which
$m = \pm l$ and the spectrum is
\be
\label{3.41}
 \om_l = \om_\perp\; \sqrt{l} \; .
\ee
In the case of $l=2$, these excitations are similar to quadrupole modes
$K=1^+$ well known for atomic nuclei \cite{BM}. Analogous excitations also
exist in anisotropic traps, where, for $l = 2$, they have the frequencies
$\sqrt{\omega_x^2 + \omega_y^2}$, $\sqrt{\omega_y^2 + \omega_z^2}$, and
$\sqrt{\omega_z^2 + \omega_x^2}$ \cite{susr}. In literature on trapped atoms,
they are called scissor modes, keeping in mind the modes in nuclei that
describe out-of-phase oscillations of neutron and proton densities, which
resembles the opening and closing of a pair of scissors \cite{Rich}.

The occurrence of collective excitations with discretized spectrum is a
specific feature of finite quantum systems. In such systems, short-wave
excitations demonstrate the properties of bulk excitations, with continuous
spectra. While the long-wave excitations display discrete spectra, whose
form essentially depends on the system shape.

\subsection{Atomic fluctuations and stability}
\label{subsec:III.B}

\subsubsection{Uniform ideal gas }
\label{subsubsec:III.B.1}

Equilibrium Bose-condensed systems can actually exist only when they
are thermodynamically stable. This implies the validity of the stability
conditions discussed in Sec. \ref{subsec:II.B}. The meaning of these
conditions is straightforward, telling us that the fluctuations of
observable quantities must be thermodynamically normal. In the other
case, anomalous thermodynamic fluctuations destroy the equilibrium
state of the system.

As a simple example, it is possible to mention the case of the ideal
Bose gas in a $d$-dimensional box of volume $V$. The condensation
temperature for this gas is
\be
\label{3.42}
 T_c = \frac{2\pi}{m} \; \left [
\; \frac{\rho}{\zeta(d/2)} \; \right ]^{2/d} \;  ,
\ee
where $\rho$ is average particle density. Due to the properties of
the Riemann zeta function $\zeta(\cdot)$, a finite transition
temperature exists only for $d > 2$. This is the manifestation of the
general feature of the absence of phase transitions in low-dimensional
systems with continuous symmetry \cite{MerminWagner_1966}.

However, the finiteness of the transition temperature does not yet
mean that below it the system can really exist, being stable. It is
necessary to check the stability conditions for large numbers of
particles \cite{Yukalov_2009,Yukalov_2005d}. The fluctuations of energy
satisfy the stability condition (\ref{2.24}). But we need also to study
particle fluctuations. The operator for the total number of particles
in a Bose-condensed system,
\be
\label{3.43}
\hat N = \hat N_0 + \hat N_1
\ee
is the sum of the terms corresponding to condensed, $\hat{N}_0$, and
uncondensed, $\hat{N}_1$, particles.

Here it is necessary to remember that, as is explained above, there are
two ways of dealing with Bose-condensed systems.

If one separates the field operator into the sum $\psi = \psi_0 + \psi_1$
of two operators defined in the Fock space ${\cal F}(\psi)$, generated by the
field operator $\psi$, then one has to calculate the variance of the operator
$\hat{N}_0$ representing the number of condensed particles as well as the
variance of the number operator of uncondensed particles \cite{Lifshitz_1980}.

However, when one employs the Bogolubov shift of the field operator by making
the canonical transformation $\psi = \eta + \psi_1$, where $\eta$ is the
condensate wave function, one has to work in the Fock space ${\cal F}(\psi_1)$
generated by $\psi_1$. In this representation, the number of condensed atoms
is a nonoperator quantity $N_0 = \int |\eta|^2 d \bf r$ that defines the
normalization condition for the condensate wave function. The variance of a
nonoperator quantity, as is evident, is identically zero.

Of course, both these ways give the same value for the compressibility that
is the observable quantity quantifying particle fluctuations. The observable
quantities do not depend on the used operator representation
\cite{Emch_1972,BratteliRob_1979}.

If the system gauge symmetry is broken by means of the Bogolubov
shift (\ref{3.8}), then ${\rm var}(N_0)$ is identically zero. In
that case,
\be
\label{3.44}
 {\rm var}(\hat N) = {\rm var}(\hat N_1 ) \; .
\ee

In three-dimensional space, one has
\be
\label{3.45}
 {\rm var}(\hat N_1 ) = \left ( \frac{mT}{\pi}
\right )^2 V^{4/3} \;  .
\ee
The reduced variance ${\rm var}(\hat N_1 )/N$ is proportional to the
isothermal compressibility, that has to be finite in thermodynamic limit.
But from the above equation it follows that the compressibility diverges
as $N^{1/3}$, for large $N$. This divergence contradicts the stability
condition (\ref{2.24}) and means that the ideal uniform Bose-condensed
gas is unstable.

The fact that divergent compressibility of the ideal Bose gas implies its
instability is easy to understand. If the compressibility of a system is
infinite, then the sound velocity is zero and the system structure factor
is infinite. Certainly, no one physical system displays such abnormal
behavior. As is well known, the general stability condition of any
equilibrium system requires that $\partial P/ \partial V < 0$. This
necessary condition is broken for the ideal Bose gas, for which this
derivative becomes zero, as a result of which the compressibility diverges,
implying instability \cite{Lifshitz_1980}. The instability of the degenerate
Bose gas is widely accepted and is explained in modern textbooks
\cite{Wasserman_2011}.

Moreover, to understand that a system with infinite compressibility is
unstable, there is even no need to address textbooks, but it is sufficient
to have just common sense. Really, infinite compressibility, being
proportional to $\partial V/ \partial P$, means that any arbitrarily weak
noise in pressure would immediately result in the system explosion.

\subsubsection{Trapped ideal gas}
\label{subsubsec:III.B.2}

In order to check whether the ideal Bose gas could be stabilized being
confined inside a trap, let us consider the most often used trap shape
described by the power-law potential
\be
\label{3.46}
 U(\br) = \sum_{\al=1}^d \; \frac{\om_\al}{2} \; \left |\;
\frac{r_\al}{l_\al} \; \right |^{n_\al} \; ,
\ee
where $\om_\al\equiv 1/m l_\al^2$ are the characteristic
trap frequencies.

An important quantity, defining the confining power of the trapping
potential is the {\it confining dimension} \cite{Yukalov_2005a}
\be
\label{3.47}
 s \equiv \frac{d}{2} + \sum_{\al=1}^d \; \frac{1}{n_\al} \; .
\ee

For ideal gases confined in the power-law trapping potentials, it
is convenient to use the generalized quasi-classical approximation
\cite{Yukalov_2005a} yielding the results that are in agreement with
the quantum consideration \cite{KetterleDruten_1996,Klunder_2009}.
In this approximation, the condensation temperature is
\be
\label{3.48}
T_c = \left [\; \frac{N}{Bg_s(1)} \; \right ]^{1/s} \; ,
\ee
where the generalized Bose function
\be
\label{3.49}
g_s(z) \equiv \frac{1}{\Gm(s)} \; \int_{u_0}^\infty \;
\frac{zu^{s-1}}{e^u-z} \; du
\ee
is introduced and the notations
$$
B \equiv \frac{2^s}{\pi^{d/2}} \; \prod_{\al=1}^d
\frac{\Gm(1+1/n_\al)}{\om_\al^{1/2+1/n_\al}} \; ,
$$
\be
\label{3.50}
 u_0 \equiv \frac{\om_0}{2T} \; , \qquad \om_0 \equiv
\left ( \prod_{\al=1}^d \om_\al \right )^{1/d}
\ee
are used. In the limit of $u_0 \ra 0$, one returns to the standard
semiclassical approximation \cite{Courteille_2001,BagnatoKleppner_1991}.

Considering the effective thermodynamic limit (\ref{2.22}), we can take
the internal energy $E_N$, which, for the Bose-condensed system, is
\be
\label{3.51}
 E_N = B s g_{1+s}(1) T^{1+s} \; .
\ee
Then the thermodynamic limit (\ref{2.22}) becomes
\be
\label{3.52}
 N \ra \infty \; , \qquad B \ra \infty \; , \qquad
\frac{B}{N} \ra const \; .
\ee
For the usual case of unipower trapping potentials, with $n_\al=n$,
limit (\ref{3.52}) reduces to
\be
\label{3.53}
  N \ra \infty \; , \qquad \om_0 \ra 0 \; , \qquad
N\om_0^s \ra const \;  .
\ee

The variance of the particle number, below $T_c$, reads as
\be
\label{3.54}
 {\rm var}(\hat N) = \frac{g_{s-1}(1)}{g_s(1)} \;
\left ( \frac{T}{T_c} \right )^s N \; .
\ee
Taking into account the effective limit (\ref{3.52}), we find, at
large $N$, the following behavior
$$
\frac{ {\rm var}(\hat N)}{N} \propto N \qquad
( s = 1 ) \; ,
$$
$$
\frac{{\rm var}(\hat N)}{N} \propto N^{(2/s)-1} \qquad
( 1 < s <2 ) \; ,
$$
$$
\frac{{\rm var}(\hat N)}{N} \propto  \ln N \qquad
( s = 2 ) \; ,
$$
\be
\label{3.55}
 \frac{{\rm var}(\hat N)}{N} \propto const \qquad
( s > 2 ) \;  .
\ee
This tells us that the ideal Bose-condensed gas can be stabilized only
in the traps with the confining dimension $s > 2$.

\subsubsection{Harmonically trapped gas}
\label{subsubsec:III.B.3}

The most common shape of traps is harmonic, when $n_\alpha = 2$.
The condensation temperatures formally exist for all dimensions,
being
$$
T_c = \frac{N\om_0}{\ln(2N)} \qquad (d=1) \; ,
$$
\be
\label{3.56}
 T_c = \om_0 \left [ \frac{N}{\zeta(d)} \right ]^{1/d}
\qquad ( d \geq 2) \; .
\ee
This is contrary to the uniform case, where a finite $T_c$ exists
only for $d = 3$.

However, again, we have to remember that the formal existence of a
transition temperature does not yet guarantee that the system could
be stable \cite{Yukalov_2005a}. It is necessary to check the stability
conditions

Particle fluctuations are described by the variances
$$
{\rm var}(\hat N) = 2 \left ( \frac{T}{\om_0} \right )^2 \qquad
(d = 1) \; ,
$$
$$
{\rm var}(\hat N) =  \left ( \frac{T}{\om_0} \right )^2
\ln \left ( \frac{2T}{\om_0} \right ) \qquad
(d = 2) \; ,
$$
\be
\label{3.57}
{\rm var}(\hat N) = \frac{\pi^2 N}{6\zeta(3)}
\left ( \frac{T}{T_c} \right )^3  \qquad (d = 3) \;   ,
\ee
which, for large $N$, yield
$$
\frac{{\rm var}(\hat N)}{N} \propto N \qquad
( d = 1 ) \; ,
$$
$$
\frac{{\rm var}(\hat N)}{N} \propto \ln N \qquad
( d = 2 ) \; ,
$$
\be
\label{3.58}
\frac{{\rm var}(\hat N)}{N} \propto const \qquad
( d = 3 ) \;   .
\ee
Therefore, only the three-dimensional harmonic trap stabilizes the
ideal Bose-condensed gas.

Of course, it could be possible to say that the stability condition
(\ref{2.24}) is formally satisfied for a finite number of particles $N$.
Thence, it would be admissible to assume that the ideal gas can form
a kind of, maybe metastable, quasicondensate in one- and two-dimensional
harmonic traps. However, to be classified as quasicondensate, it should
allow for ascribing the notion of asymptotic symmetry breaking, which
requires the validity of the stability condition (\ref{2.24}) at
asymptotically large $N$.

Taking into account interactions does not stabilize harmonically trapped
Bose-condensed gas in low dimensions. According to the theorems, proved
in Refs. \cite{Mullin_1997,Carrasquilla_2012}, there can be no Bose-Einstein
condensate in harmonically trapped interacting Bose gas for
dimensionalities $d=1$ and $d=2$.

\subsubsection{Interacting nonuniform systems}
\label{subsubsec:III.B.4}

To consider particle fluctuations for interacting systems, it is
convenient to use the following expression for the variance:
\be
\label{3.59}
 {\rm var}(\hat N) = N + \int \rho(\br) \rho(\br')
[ g(\br,\br') - 1 ] \; d\br d \br' \; ,
\ee
in which $g(\br,\br')$ is the pair correlation function.
This expression (\ref{3.59}) is valid for any system of $N$ particles.
Using the Hartree-Fock-Bogolubov and local-density approximations
gives \cite{Yukalov_2010}
\be
\label{3.60}
 {\rm var}(\hat N) = \frac{T}{m} \int
\frac{\rho(\br)}{c^2(\br)} \; d\br \;  .
\ee

Equation (\ref{3.60}) holds true for three-dimensional nonuniform
Bose-condensed systems in the presence of arbitrary external potentials.
The integral here is proportional to $N$, provided that the sound
velocity is nonzero. Therefore, the stability condition (\ref{2.24}) is
satisfied.

\subsubsection{Importance of symmetry breaking}
\label{subsubsec:III.B.5}

In this subsection, we emphasize that forgetting about the occurrence
of gauge symmetry breaking in Bose-condensed systems is not as innocent,
as one could think.

First of all, the appearance of Bose-Einstein condensate as such
necessarily requires the global gauge symmetry breaking, as is explained
above. Strictly speaking, if there is no gauge symmetry breaking, there
can be no condensate.

If, forgetting the necessity of the symmetry breaking, one invokes a
conserving-symmetry approach, then there arise unphysical divergences
making calculations meaningless. For example, in the Hartree-Fock
approximation, that preserves the gauge symmetry, the variance for the
number of particles becomes thermodynamically anomalous,
${\rm var}{\hat{N}} \propto N^2$. Then the stability condition (\ref{2.24})
is not valid, which implies that the system is unstable. However, the
appearance of this instability has no physical meaning, being caused
just by the error of forgetting to break the gauge symmetry. This
fictitious instability is immediately removed as soon as the gauge
symmetry for the condensed phase is broken \cite{terHaar_1977}. Similar
unphysical divergences in the Hartree-Fock approximation occur for some
other observable quantities, but become removed under gauge symmetry
breaking \cite{HohenbergMartin_1965}. This means that the
approximations, preserving gauge symmetry, are principally inappropriate
for describing systems with Bose-Einstein condensate. In particular, the
standard Hartree-Fock approximation, preserving gauge symmetry, is
inapplicable for such systems.

It is worth recalling that thermodynamically anomalous particle
fluctuations imply the divergence of isothermic compressibility that
is proportional to ${\rm var}(\hat{N})/N$ \cite{Yukalov_2005a,Yukalov_2005b,
Yukalov_2005c,Yukalov_2005d}. This quantity ${\rm var}(\hat{N})/N$ is also
proportional to the structure factor that is known to be finite for
real systems with Bose condensate \cite{Ozeri_2005}.

The occurrence of the global gauge symmetry breaking is necessary and
sufficient not merely for the condensate existence, but also for the
appearance of the anomalous averages, as in (\ref{3.24}). Both, the
condensate itself and the anomalous averages, either exist together or
are together zero, in line with Eq. (\ref{2.13}). Direct calculations
\cite{YukalovKleinert_2006,YukalovYukalova_2006,YukalovYukalova_2007}
show that the anomalous average is always either larger than the density
of uncondensed atoms or larger than the density of condensed atoms, so
practically can never be neglected, as has been suggested in the Shohno
model \cite{Shohno_1964}. Moreover, omitting the anomalous averages
makes the system unstable and results in the appearance of unphysical
divergences, as it happens for the discussed above situation when gauge
symmetry breaking is forgotten. The correct description of Bose-condensed
systems necessarily requires to accurately take into account the
anomalous averages and the related anomalous Green functions
\cite{Beliaev_1958,Popov_1983,Popov_1987}.

\subsection{Nonequilibrium symmetry breaking}
\label{subsec:III.C}

\subsubsection{Dynamic instability of motion}
\label{subsubsec:III.C.1}

The macroscopic occupation of a single quantum state by Bose-condensed
atoms has clear similarities with conventional laser systems. Therefore
it has been natural to try to realize directed beams of Bose-condensed
atoms, which would imitate optical laser beams. Such devices that could
be used for forming coherent atomic beams are called {\it atom lasers}.
Two conditions are the most important for the latter, the atomic beam
coherence  \cite{Wiseman_1997,Kneer_1998} and well defined directionality
\cite{Yukalov_1997b}. The coherent atomic beam, propagating with a given
velocity, should be similar to a coherent photon ray with a given
momentum \cite{RehlerEberly_1971}. The standard way of creating atomic
beams is to form, first, a trapped ensemble of Bose-condensed atoms and
then to direct these atoms out of the trap by an output coupler. There
are many types of the output couplers and extensive literature on atom
lasers, which can be found in the review articles
\cite{Courteille_2001,ParkinsWalls_1998}.

It is not our aim here to discuss the details of how a moving beam of
coherent atoms could be created. But our goal is to recall that the beam
coherence and, hence, the properties of the Bose-Einstein condensate in
the beam, including its superfluidity, depend on the velocity of atomic
motion. Increasing the velocity of moving atoms above a critical velocity,
one can destroy the superfluid condensate, as is well known from the Landau
criterion \cite{Landau_1967}. When the Bose condensate is destroyed, the
system passes to the normal state. The destruction of the moving Bose
condensate is, actually, a kind of a quantum phase transition. For a finite
system, the Landau criterion can be generalized by taking account of the
system spatial nonuniformity.

Suppose that the flow of Bose-condensed atoms moves with a constant
velocity ${\bf v}$, being described by the Gross-Pitaevskii equation.
The latter is invariant with respect to the Galilean transformation
\be
\label{3.61}
\eta_v(\br,t) = \eta(\br-\bv t,t) \exp \left ( i m \bv \cdot
\br - i \; \frac{mv^2}{2} \; t \right ) \;  ,
\ee
where $v \equiv |{\bf v}|$. Considering small deviations from a
stationary solution  $\eta({\bf r} - {\bf v}t)$ in the form
\be
\label{3.62}
 \eta(\br-\bv t,t) = \eta(\br-\bv t) +
u_v(\br) e^{-i\om t} + v_v^*(\br) e^{i\om t} \; ,
\ee
we substitute this form into the Gross-Pitaevskii equation. Then, by the
standard procedure of the linearization of the Gross-Pitaevskii equation
with respect to the small deviations, we come to a variant of the Bogolubov
equations for the moving gas. In the local-density approximation, this
gives the spectrum of collective excitations for the moving system
\be
\label{3.63}
\ep_v(\bk,\br) = \bv \cdot \bk +
\sqrt{ c^2(\br) k^2 + \left ( \frac{k^2}{2m} \right )^2 } \; .
\ee
When all flow is Bose-condensed, the local sound velocity is given
by the relation
\be
\label{3.64}
 mc^2(\br) = \rho_0(\br) \Phi_0 \; ,
\ee
which is a particular form of Eq. (\ref{3.37}).

The flow is everywhere stable if the excitation spectrum is everywhere
positive. This requires the validity of the stability condition
\be
\label{3.65}
v < v_c \equiv \min_{\bk,\br} \; \frac{\ep(\bk,\br)}{k} =
\min_\br c(\br) \;  .
\ee
This condition is a slight generalization of the Landau criterion
of superfluidity \cite{Landau_1967} to a nonuniform system treated in
the local-density approximation. When atoms move with a low velocity,
smaller than $v_c$, the system is in Bose-condensed state. However, as
soon as the velocity exceeds the critical value $v_c$, there arise
exponentially diverging solutions leading to the system instability.
As a result, the unstable system has to pass to the normal state,
without condensate and with no superfluidity.

\subsubsection{Stratification in multicomponent mixtures}
\label{subsubsec:III.C.2}

In a system, consisting of several components of atomic condensates,
there can occur a phase transition accompanied by spontaneous breaking
of translational symmetry. Then a uniform mixture of atomic species
separates into different spatial parts, with each part filled by only
one kind of atoms. This phase transition is called {\it stratification}
\cite{Nepomnyashchy_1974,Yukalov_1980}. Conditions, when such a
stratification arises, depend on atomic parameters and their
interactions.

Let us have a mixture of several bosonic species, enumerated by an
index $i = 1,2,\ldots$. Respectively, atomic masses of the species are
denoted as $m_i$ and the interaction potentials between the $i$-th and
$j$-th components are $\Phi_{ij}({\bf r})$. Keeping in mind the
local-density approximation, we need the average interaction parameters
\be
\label{3.66}
\Phi_{ij} \equiv \int \Phi_{ij}(\br) \; d\br \; .
\ee
The diagonal terms $\Phi_{ii}$ are assumed to be positive in order that
each of the components of Bose atoms would be stable \cite{terHaar_1977}.
Two components can be uniformly mixed at zero temperature, provided that
the stability condition
\be
\label{3.67}
\Phi_{ij} < \sqrt{\Phi_{ii} \Phi_{jj}}
\ee
holds. The condition can be obtained from the minimization of the system
energy \cite{Pethick_2008}. If this condition does not hold, the system
stratifies into separated spatial regions of single-component subsystems,
so that the translational symmetry becomes broken.

The same condition (\ref{3.67}) can be derived by analyzing the stability
of the spectrum of collective excitations
\cite{Nepomnyashchy_1974,Yukalov_1980}. For the mixture of two
Bose-condensed components, the spectrum splits into two branches, given
by the equation
$$
 \ep_\pm^2(\br,\bk) = \frac{1}{2} \left \{ \ep_1^2(\bk,\br) +
\ep_2^2(\bk,\br) \pm \right.
$$
\be
\label{3.68}
\pm \left.
\sqrt{ [\ep_1^2(\bk,\br)-\ep_2^2(\bk,\br)]^2+
4\ep_{12}^4(\bk,\br) } \right \} \; ,
\ee
where the notations
$$
\ep_{ij}^2(\bk,\br) = c_{ij}^2(\br) k^2 +
\dlt_{ij}\left ( \frac{k^2}{2m_i} \right )^2 \; ,
$$
\be
\label{3.69}
c_{ij}^2(\br) = \sqrt{ \frac{\rho_i(\br)\rho_j(\br)}{m_i m_j} } \;
\Phi_{ij} \; ,
\ee
and $\varepsilon_i \equiv \varepsilon_{ii}$ are used. The upper part
of the spectrum, $\ep_{+}({\bf k},{\bf r})$, describes the density wave
propagating through the mixture, while the lower part, $\ep_{-}(\bk,\br)$,
corresponds to the oscillations of the components with respect to each
other. The mixture stratifies when the lower spectrum goes to zero.

When there are two components moving with the velocities ${\bf v}_i$,
then the spectrum of collective excitations is defined by the equation
\be
\label{3.70}
\prod_{i=1}^2 \left [ ( \om - \bv_i \cdot \bk )^2 -
\ep_i^2(\bk,\br) \right ] = \ep_{12}^4(\bk,\br) \; .
\ee
Then, even if the mixture has been stable in the immovable state,
it becomes stratified as soon as the relative velocity of the second
component through the first one reaches the critical value
\cite{YukalovYukalova_2004}
$$
v_c = \min_\br \;
\frac{\sqrt{c_1^2(\br)c_2^2(\br)-c_{12}^4(\br)}}{c_1(\br)} =
$$
\be
\label{3.71}
= \min_\br \; \sqrt{ \frac{\rho_2(\br)}{m_2\Phi_{11}}
\left ( \Phi_{11} \Phi_{22} - \Phi_{12}^2 \right ) } \; ,
\ee
where $c_i \equiv c_{ii}$.

In uniform macroscopic systems, the stratification occurs simultaneously
in the whole volume. But in finite systems, that are nonuniform, the
stratification may start in a part of the system and then proliferate to
other parts by a kind of shock waves producing turbulence \cite{Takeuchi_2010}.
This means that in finite systems, the stratification is easier achieved
than in large uniform systems.

\subsubsection{Generation of quantum vortices}
\label{subsubsec:III.C.3}

Breaking of translational symmetry also happens when there appear
quantum vortices. The straightforward way of creating vortices is by
means of rotation. Suppose that the system is rotated with a constant
angular frequency $\vec{\om}$. The single-particle Hamiltonian, in the
Schr\"{o}dinger representation, in the frame rotating with the same
angular frequency, has the form \cite{LandauLifshitz_1960}
\be
\label{3.72}
\hat H_{rot}(\br) = \frac{\hat\bp^2}{2m} \; - \;
\vec{\om} \cdot \hat{\bf L} + U(\br) \; ,
\ee
with the momentum operator $\hat{\bf p} \equiv - i \nabla$
and angular momentum  $\hat{\bf L} \equiv {\bf r} \times \hat{\bf p}$.
Here $U({\bf r})$ is a trapping potential. This Hamiltonian can
also be represented as
\be
\label{3.73}
 \hat H_{rot}(\br) = \frac{(\hat\bp-m\bv)^2}{2m} \; - \;
\frac{mv^2}{2} + U(\br) \; ,
\ee
with the linear velocity $\bv \equiv\vec{\om} \times \br$.
This expression shows that the motion of a neutral particle in a rotating
system is analogous to the motion of a charged particle in an external
magnetic field, with a vector potential ${\bf A}$ and magnetic field
${\bf B}$ given by the relations
$$
\frac{e}{c} \; {\bf A} \equiv m\bv = m (\vec{\om}\times\br ) \; ,
\qquad
{\bf B} \equiv \vec{\nabla} \times {\bf A} =
\frac{2m}{e}\; \vec{\om} \; .
$$

It is common to choose the direction of the angular frequency along the
$z$- axis, so that $\vec{\omega} = \omega {\bf e}_z$. If the trapping
potential is harmonic,
\be
\label{3.74}
U(\br) = \frac{m}{2} \; \om_\perp^2 \left ( x^2 + y^2 \right ) +
\frac{m}{2} \; \om_z^2 z^2 \; ,
\ee
then Hamiltonian Eq. (\ref{3.73}) reads as
$$
\hat H_{rot}(\br) = \frac{(\hat\bp - m\bv)^2}{2m} \; +
$$
\be
\label{3.75}
+ \; \frac{m}{2} \left ( \om_\perp^2 - \om^2 \right ) \left ( x^2 +
y^2 \right ) + \frac{m}{2} \; \om_z^2 z^2 \; .
\ee
The problem is equivalent to that describing the motion of a
charged particle in a constant magnetic field in the presence of a
harmonic potential \cite{Fock,LandauLifshitz_2003}. The eigenvalues of
Hamiltonian (\ref{3.75}) are
\be
\label{3.76}
E_{nlk} = ( 2n + |l| + 1) \om_\perp + \left ( k +
\frac{1}{2} \right ) \om_z - l\om \;  ,
\ee
where $n = 0,1,2,\ldots$ is the radial quantum number,
$l=0,\pm 1,\pm 2,\ldots$ is the azimuthal quantum number, sometimes
also called winding number or circulation quantum, and $k =0,1,2,\ldots$
is the axial quantum number. The spectrum reduces to the set of Landau
levels if either $\om=\om_\perp$, $l\geq 0$ or $\om=-\om_\perp$, $l\leq 0$.

The many-particle system in the rotating frame is characterized by the
energy Hamiltonian in the Heisenberg representation
$$
\hat H = \int \hat\psi^\dgr(\br) \hat H_{rot}(\br)
\hat\psi(\br) \; d\br  \; +
$$
\be
\label{3.77}
 + \; \frac{1}{2} \int \hat\psi^\dgr(\br) \hat\psi(\br')
\Phi(\br-\br') \hat\psi(\br') \hat\psi(\br) \; d\br d\br' \; .
\ee
Substituting here the Bogolubov shift of the field operator, from the
equation of motion, one obtains the equation for the condensate wave
function, as is explained in Sec. 3. This equation, for the case
of weak interactions and temperatures, reduces to the Gross-Pitaevskii
equation.

The first vortex, with the winding number $l = 1$, appears when
the angular frequency reaches a critical value $\omega_c$. An explicit
expression of this value depends on the physical situation and the
approximation involved. When the Bose system is trapped in a cylindrically
symmetric harmonic potential, fully condensed, being described by the
Gross-Pitaevskii equation, and allows for the Thomas-Fermi approximation,
then the critical rotation frequency is
\be
\label{3.78}
 \om_c = \frac{5}{2mR^2} \;
\ln \left ( 0.7 \; \frac{R}{\xi} \right ) \;  ,
\ee
where $\xi = 1/\sqrt{2 m \rho_0 \Phi_0}$ is the healing length, $R$,
Thomas-Fermi radius, and $\rho_0 = \rho(0)$ is the condensate density
at the trap center in the absence of rotation \cite{Pethick_2008}.

The vortices with the higher winding numbers $l > 1$ require larger
critical frequencies. But such vortices have larger energies and are
unstable with respect to the decay into the vortices with lower
winding numbers \cite{Courteille_2001,Pethick_2008}. Increasing the
rotation frequency generates, first, a small number of vortices and
subsequently an array of many vortices forming a triangular lattice
\cite{ButtsRokhsar_1999}, with all the vortices having a single quantum
of circulation $l = 1$. This vortex lattice is analogous to the
triangular Abrikosov lattice of magnetic vortices in superconductors
\cite{Abrikosov_1957}.

In experiment, vortices can be created in several ways. The
straightforward method is by stirring the cloud of trapped atoms with
a laser spoon. It is also possible to apply radio-frequency field for
transferring atoms between hyperfine states, from a non-rotating state
to a rotating state. There is a method of creating vortices that does
not rely on mechanical stirring, but employs adiabatic transfer
between hyperfine states, with the formation of spatially dependent
geometric Berry phase \cite{Berry_1984}.

More details on the physics of vortices in Bose-condensed systems can
be found in review articles \cite{Bloch_2008,Cooper_2008,Fetter_2009,Saar}.
The mathematics of describing vortices in trapped atoms are similar to
that of characterizing quantum dots in magnetic fields and rotating nuclei.

\subsubsection{Topological coherent modes}
\label{subsubsec:III.C.4}

Quantum vortices can be generated in macroscopic systems, such as
$^4$He, as well as in trapped gases. But in finite quantum systems
there can be generated a larger variety of nonlinear excitations
breaking translational and rotational symmetry, of which the
vortices are just a particular type. These are the
{\it topological coherent modes} \cite{YYB_1997} that are the
stationary solutions of the condensate-function equation (\ref{3.29}).
For example, in the case of the fully Bose-condensed gas, when the
Gross-Pitaevskii equation (\ref{3.31}) is applicable, the stationary
solutions are given by the eigenproblem
\be
\label{3.79}
\left [ -\; \frac{\nabla^2}{2m} + U(\br) \right ] \eta_n(\br)
+ \Phi_0 | \eta_n(\br)|^2 \eta_n(\br) = E_n \eta_n(\br) \;  ,
\ee
where $n$ is a quantum multi-index enumerating the modes and
$\mu_0 = {\rm min}_n E_n$.

The stationary nonlinear Schr\"{o}dinger equation is the notion that
is widely accepted. It enjoys the form $H_{NLS} \eta_n = E_n \eta_n$,
with the $H_{NLS}$ being the nonlinear Schr\"{o}dinger Hamiltonian. This is
the standard form of an eigenproblem equation. By the common definition,
the eigenvalue of a Hamiltonian is termed energy.

Note the principal difference between these coherent modes, defined
by the nonlinear Schr\"{o}dinger equation (\ref{3.79}), and the
elementary collective excitations, described by the linear Bogolubov
equations (\ref{3.35}). The modes are called coherent, since they
correspond to the condensate function that characterizes the
coherent part of the Bose system. And they are termed topological,
since different solutions of Eq. (\ref{3.79}) have different spatial
topology, with different numbers of zeros. Each coherent mode
corresponds to a stationary nonground-state condensate.

In a finite system, where atoms are confined by a trapping potential,
the spectrum of the eigenvalues of Eq. (\ref{3.79}) is discrete. This
makes the principal difference of the situation in finite systems
from that of uniform macroscopic systems, where the spectrum is
continuous. When the spectrum is discrete, it is possible to transfer
atoms between different energy levels, thus, transferring atoms between
different topological modes. Such a transfer can be effectively
realized by means of the corresponding resonance fields
\cite{YYB_1997,YYB_2002,YMY_2004}. The situation is similar to the
case of resonance atoms, whose electrons can be transferred between
their energy levels by resonance electromagnetic fields. The latter
resonance transitions are well studied in the optical and infrared
regions \cite{MandelWolf_1995} and even are considered for the gamma
region \cite{Baldwin_1981}. Another similarity is with spin systems
that allow for realizing transitions between different discrete Zeeman
levels \cite{Slichter_1980}. The main feature of the considered case
of Bose-condensed atoms is that the energy levels $E_n$, given by
Eq. (\ref{3.79}), correspond to many-particle states, but not to
single-electron or single-spin states. Consequently, the related
eigenproblem is nonlinear, including atomic interactions. The role of
a resonant object is played by the whole trapped Bose system, but not
by a single particle.

Trapped atoms can be transferred between their energy levels $E_n$,
say between the levels $E_1$ and $E_2$, by two main techniques. One
method \cite{YYB_1997} is by modulating the trapping potential with a
temporal variation
\be
\label{3.80}
V(\br,t) = V_1(\br) \cos(\om t) + V_2(\br) \sin(\om t) \;  ,
\ee
tuning the modulation frequency $\omega$ close to the resonance
with the desired transition frequency $\omega_{12} \equiv E_2 - E_1$,
so that the resonance condition
\be
\label{3.81}
 \om = \om_{12}
\ee
be approximately satisfied. The modulating field is added to the
trapping potential $U({\bf r})$.

The other method \cite{Ramos_2008,YukalovBagnato_2009} is by modulating
atomic interactions, which can be realized by Feshbach resonance techniques,
so that the atomic interaction becomes alternating,
\be
\label{3.82}
\Phi(t) = \Phi_0 + \Phi_1 \cos(\om t) + \Phi_2\sin(\om t) \; ,
\ee
with the frequency $\omega$ tuned to the resonance condition (\ref{3.81}).

The solution to the time-dependent equation (\ref{3.31}) can be presented
as the expansion over the coherent modes,
\be
\label{3.83}
 \eta(\br,t) = \sum_n c_n(t) \eta_n(\br) e^{-i\om_n t} \; ,
\ee
with $\omega_n = E_n - E_0$. Fractional mode populations are given by the
expressions
\be
\label{3.84}
p_n(t) \equiv | c_n(t)|^2 \;  .
\ee
The solution of the corresponding time-dependent nonlinear Schr\"odinger
equation for the condensate function can be done by invoking averaging
methods \cite{YYB_1997,YYB_2002} or numerically
\cite{YMY_2004,Adhik_2003,Adhik_2004}.

If the driving frequency of the alternating pumping field is tuned close
to the resonance condition (\ref{3.81}), then, for generating the coherent
modes, it is sufficient to have a rather weak amplitude of the pumping
field \cite{YYB_1997,YYB_2002}. Increasing the amplitude of this field
makes it possible to generate coherent modes under the wider class of
resonance conditions \cite{YMY_2004}. Thus, the modes are generated under
the conditions of {\it harmonic generation}
\be
\label{3.85}
n\om = \om_{12} \qquad (n = 1,2,\ldots) \;  .
\ee
If there are two modulating fields, with the frequencies $\omega_1$ and
$\omega_2$, coherent modes can be generated under the condition of
{\it parametric conversion}
\be
\label{3.86}
 \om_1 \pm \om_2 = \om_{12}
\ee
or the condition of {\it combined resonance}
\be
\label{3.87}
 n_1 \om_1 + n_2 \om_2 = \om_{12} \qquad
( n_i = \pm 1, \pm 2,\ldots ) \; .
\ee
When there are several modulating fields, with the frequencies $\om_i$,
coherent modes can be generated under the condition of
{\it generalized resonance}
\be
\label{3.88}
 \sum_i n_i \om_i = \om_{12} \qquad
( n_i = \pm 1, \pm 2, \ldots) \; .
\ee

The topological coherent modes, representing nonground-state condensates,
possess a variety of interesting properties, such as interference patterns
and interference current, mode locking, dynamical phase transitions and
critical phenomena, chaotic motion, atomic squeezing, entanglement
production, and Ramsey fringes \cite{YYB_1997,YYB_2002,YMY_2004,Ramos_2007}.
Examples of the coherent modes are the well known quantum vortices and the
dipole mode that was generated in experiment \cite{Williams_2000}.

\subsubsection{Nonequilibrium crossover transitions}
\label{subsubsec:III.C.5}

Among all topological coherent modes, the basic vortex with the unit
circulation number $l=1$ plays a special role. This is caused by the
fact that the transition frequencies $\omega_n \equiv E_n - E_0$,
characterizing the transitions from the ground state to an excited
state with the energy $E_n$, depend on the parameter $\alpha g$, where
$\alpha \equiv \omega_z / \omega_\perp$ is the trap aspect ratio and
$g \equiv 4\pi N a_s/l_\perp$ is the dimensionless interaction parameter.
The transition frequencies for all modes, except the basic vortex mode,
grow with this parameter as $(\alpha g)^{2/5}$. But the transition
frequency of the vortex, vice versa, diminishes with this parameter
\cite{Courteille_2001,YYB_2002}. For instance, the critical frequency
(\ref{3.78}) depends on this parameter as
\be
\label{3.89}
 \om_c = \frac{0.9}{(\al g)^{2/5}} \;
\ln ( 0.8 \al g) \; .
\ee
Therefore, at large parameter $\alpha g$, the energy of the basic vortex
mode is the lowest among all coherent modes. Therefore, all other modes,
except the basic vortex mode, are energetically unstable and, being
created, decay into the basic vortices with unit circulation.

This unique feature of the basic vortex suggests that, by modulating
either the trapping potential with the additional alternating field
(\ref{3.80}) or the atomic interaction, as in Eq. (\ref{3.82}), it is
admissible to generate multiple vortices and anti-vortices. For
sufficiently large modulation amplitude and/or sufficiently long
modulation time, a great number of vortices can be created. Contrary
to the case of rotation, where only vortices are generated, in the
considered case of modulation, the vortices are created in pairs with
anti-vortices. And, instead of forming a triangular vortex lattice, as
in the case of rotation, in the process of the alternating modulation,
vortices and anti-vortices form a random vortex tangle.

Such a random vortex tangle is a typical feature of {\it quantum turbulence}.
The latter has been widely studied for the case of turbulent superfluids
$^3$He and $^4$He, and quantum gases, as is reviewed in literature
\cite{Sonin_1987,Svistunov_1995,NemirovskiiFiszdon_1995,VinenNiemela_2002,
Vinen_2006,Tsubota_2008,Tsubota_2010,Tsubota_2012}. Quantum turbulence of
trapped atoms has been observed in experiments
\cite{Henn_2009,Shiozaki_2011,Seman_2011}.

The state of the system, produced by means of the suggested modulation,
depends on both, the modulation amplitude and modulation time. This is
because both these characteristics are responsible for the amount of
energy injected into the system. Let the total Hamiltonian consist of
two terms, $H = H_0 + \hat{V}(t)$, where the first term does not depend
on time, while the second term, $\hat{V}(t)$ accomplishes the modulation
during the time $t_{mod}$. Then the amount of energy per atom, injected
into the system by the modulating term, can be defined as
\be
\label{3.90}
 E_{inj} \equiv \frac{1}{N} \int_0^{t_{mod}} \left | \left \lgl
\frac{\prt\hat V(t)}{\prt t} \right \rgl \right | \; dt \;  .
\ee
The injected energy plays the role of an effective temperature. If the
effective amplitude of the modulating field is $V_{mod}$ and the
modulation frequency is $\omega$, then the injected energy (\ref{3.90})
can be estimated as $E_{inj} = \om t_{mod} V_{mod}$.

The overall picture, under increasing the injected energy (\ref{3.90}),
is as follows. Let, the system be prepared in the Bose-condensed superfluid
state without vortices. This state can be called {\it regular superfluid}.
Switching on pumping, first, produces a small number of vortices, which
forms what can be called the state of {\it vortex superfluid}. Increasing
the injected energy (\ref{3.90}) generates a random tangle of multiple
vortices forming the state of {\it turbulent superfluid}.

Since the influence of the injected energy is analogous to that of
temperature, increasing $E_{inj}$ depletes the amount of condensed atoms,
transferring them to the normal, uncondensed, fraction. When the normal
fraction of atoms becomes substantial, the system stratifies into the
pieces of Bose-condensed droplets separated by the regions of normal
fluid. Inside each of the droplets, there exists coherence, but
different droplets are not coherent with each other. The locations of
the coherent droplets are random. Their shapes also are random, with
the surfaces that can be fractal \cite{Mandelbrot_1983}. Such a state
is equivalent to a heterophase system consisting of a random mixture
of different phases \cite{Yukalov_1991b,Yukalov_het}. Hence, it can be
called {\it heterophase fluid}. This heterophase state, consisting of
coherent, Bose-condensed, and incoherent, normal, spatially separated
regions, could also be called nonequilibrium Bose glass or granular
condensate. The formation of this heterophase mixture could be called
granulation.

Increasing further the injected energy depletes the condensate more and
more, finally destroying all coherence in the system, making the whole
fluid normal. But, since there is a strong external pumping field, the
fluid is strongly nonequilibrium, demonstrating chaotic oscillations,
because of which the state can be named {\it chaotic fluid}. This state
reminds classical turbulence \cite{LandauLifshitz_1987}, such as occurs in
plasma \cite{terHaar_1981,Zakharov_1985}.

The general sequence of the nonequilibrium states, arising under the
pumping of the system of trapped Bose atoms, is shown in Fig. 1.
Transitions from one state to another are not genuine phase
transitions, but rather are crossovers. The sequence of the described
states, under the rising amount of the injected energy, was observed in
experiment \cite{Shiozaki_2011,Seman_2011}. All the states, starting from
the regular superfluid to the granulated heterophase fluid were demonstrated
and studied. Only the last stage of chaotic fluid was not achieved.

\begin{figure}[t]
\centerline{
\includegraphics[width=6cm]{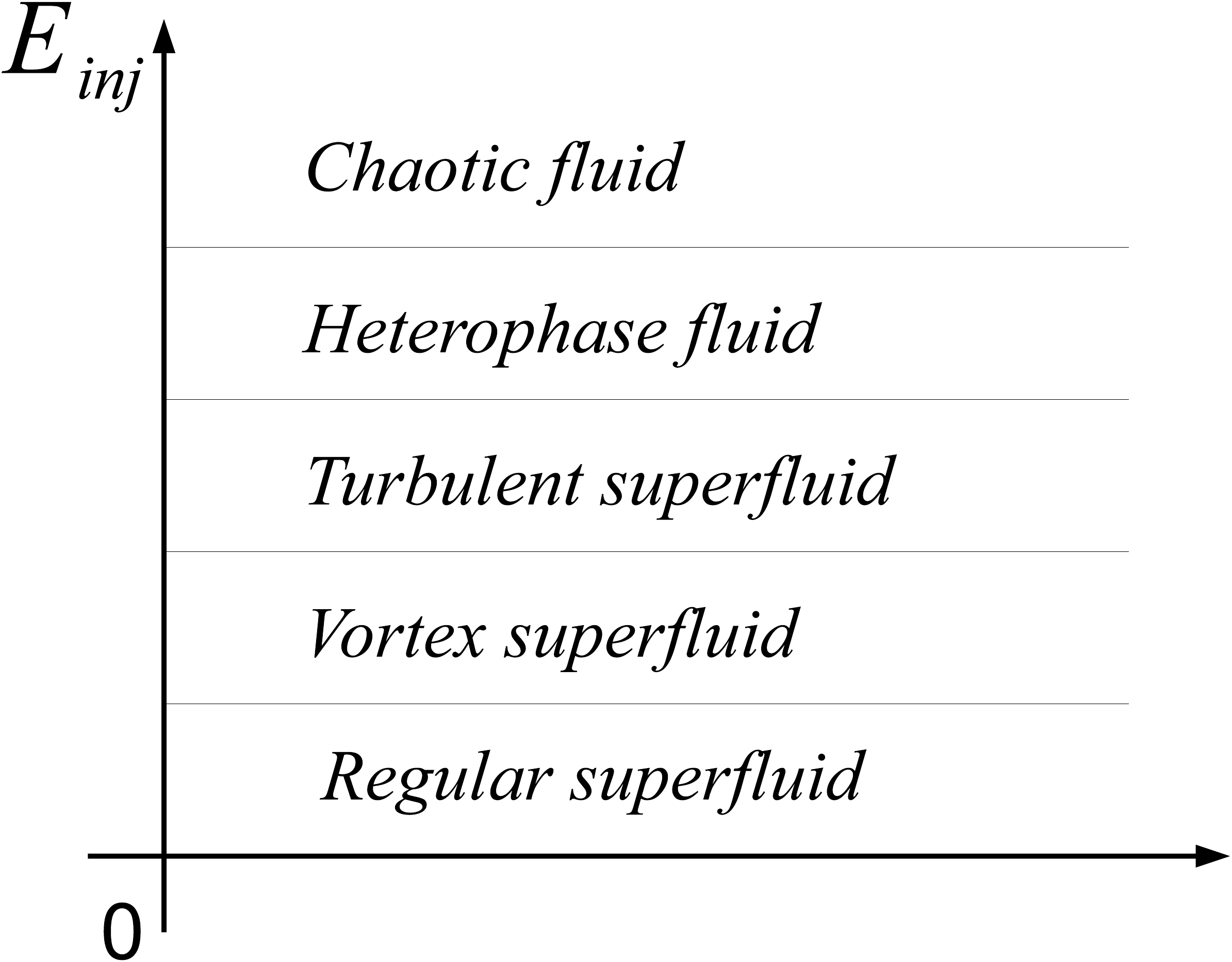}}
\caption{\label{fig:1} Qualitative scheme of the sequence of states for a
trapped Bose system subject to the pumping of an alternating modulating
field, with the increasing injected energy $E_{inj}$.}

\end{figure}

\subsection{Pairing of fermionic atoms}
\label{subsec:III.D}

\subsubsection{Harmonically trapped fermions}
\label{subsubsec:III.D.1}

Several species of neutral Fermi atoms have been trapped and cooled
down to low temperatures corresponding to quantum degenerate regime.
These are, the alkali atoms potassium ($^{40}$K) and lithium ($^6$Li)
and also $^3$He and $^{173}$Yb. The physics of ultracold degenerate
trapped fermions has been discussed in the reviews
\cite{Giorgini_2008,Ketterle_2008,Chevy_2010}. Here, we emphasize only
some basic facts concerning these gases, concentrating on their symmetry
properties.

Trapped fermions provide a unique opportunity of studying different
states of the same atomic system by varying atomic interactions, which
can be done by means of Feshbach resonance techniques \cite{Chin_2010}.
By these techniques, the scattering length can be varied by an external
magnetic field in a very wide range between negative and positive values.
The scattering length, as a function of the magnetic field $B$, is given
by the expression
\be
\label{3.91}
a_s(B) = a_s \left ( 1 + \frac{\Dlt_B}{B-B_0} \right ) \; ,
\ee
in which $B_0$ is the resonance magnetic field, $\Delta_B$ is the
resonance width, and $a_s$ is the off-resonant scattering length.

The principal difference between bosons and fermions is that Bose
gases with large scattering lengths experience strongly enhanced
inelastic collisions, which prevents the experiments to reach the
strongly interacting regime \cite{Stenger_1999}. While Fermi gases,
due to Pauli suppression effect, can be remarkably stable against
inelastic decay, allowing for the formation of stable molecular
quantum gases \cite{Ketterle_2008}.

One usually considers harmonically trapped Fermi gases with spin $1/2$.
Their Fermi energy is
\be
\label{3.92}
 E_F = \om_0 (3N)^{1/3} \; ,
\ee
where $\omega_0 \equiv (\omega_x \omega_y \omega_z)^{1/3}$ and $N$ is
the total number of fermions in the trap. The Fermi wave number $k_F$
is defined through the relation
\be
\label{3.93}
 \frac{k_F^2}{2m} = E_F \; ,
\ee
which gives
\be
\label{3.94}
k_F = \frac{\sqrt{2}}{l_0} \; (3N)^{1/6} \; .
\ee
The atomic density at the center of the trap is
\be
\label{3.95}
 \rho(0) = \frac{k_F^3}{3\pi^2} \; .
\ee
Similar relations are valid for fermions in atomic nuclei, that is,
neutrons and protons.

\subsubsection{Pair formation and superfluidity}
\label{subsubsec:III.D.2}

Dilute gases, whose interaction radius is much shorter than the
mean interatomic distance, are described by the contact interaction
potential (\ref{3.30}). The interaction regimes are characterized by
the dimensionless parameter $1/(k_F a_s)$. In the limit
$1/(k_F a_s)\ll -1$, fermions form Cooper pairs with the standard
Bardeen-Cooper-Schrieffer (BCS) ground state \cite{Bardeen_1957}. The
Cooper pairs are formed by two atoms, with opposite momenta and spins,
on the surface of the Fermi sphere. The pairing energy, i.e. the gap,
is small compared with the Fermi energy, and the Cooper pair size
greatly exceeds the typical inter-atomic spacing. The ground state
is equivalent to the BCS superconducting state, but, since the atoms
are neutral, the state is termed superfluid.

In the limit $1/(k_F a_s) \gg 1$, the fermions form bosonic molecules,
with the ground state being Bose-Einstein condensate (BEC). The molecular
binding energy is large compared with all other energies, and the molecular
size is small compared with typical interatomic spacing. The physics of
these small molecules is understood as that of composite bosonic atoms
forming superfluid state.

The crossover region $ -1 < 1/(k_F a_s) < 1$ corresponds to the strongly
interacting regime, where the pairs are no longer pure Cooper pairs or
pure bosonic molecules. Their binding energy is comparable to the Fermi
energy, and their size is about the inter-atomic spacing. The ground state
is also superfluid, but such that both bosonic and fermionic degrees of
freedom are important, which is called {\it resonance superfluidity}
\cite{Timmermans_2001,Holland_2001}.

Weakly bound molecules, made of fermionic atoms near a Feshbach
resonance, are called {\it dimers}. The elastic atom-molecule collisions
\cite{Skorniakov_1957,Petrov_2003} are described by the effective
scattering length $a_{am}$, and molecule-molecule collisions
\cite{Petrov_2004}, by the scattering length $a_{mm}$, which are
expressed through the atom-atom scattering length $a_s$ by the relations
\cite{Petrov_2003,Petrov_2004}
\be
\label{3.96}
a_{am} = 1.18 a_s \; , \qquad a_{mm} = 0.60 a_s \;  .
\ee

The ground state at $T=0$, throughout the whole range of interactions,
is formed by pairs, either Cooper pairs or bosonic molecules, and it is
superfluid for all $k_F a_s$. But the critical temperature of superfluid
transition $T_c$, generally, does not coincide with the temperature of
molecule formation $T^*$. In the framework of BCS theory, there is no
difference between $T_c$ and $T^*$, which means that the formation of
Cooper pairs and superfluid transition occur simultaneously. But on the
BEC side, molecules are formed at much higher temperatures than
Bose-Einstein condensation occurs. Thus, $T_c \leq T^*$.

In the BCS limit, the characteristic temperatures are given
\cite{Ketterle_2008} by the known expressions
$$
T_c = 0.277 E_F \exp \left ( - \; \frac{\pi}{2k_F|a_s|}
\right ) \; ,
$$
\be
\label{3.97}
 T^* = T_c \qquad \left ( \frac{1}{k_F a_s} \ll - 1
\right ) \; .
\ee
And in the BEC limit, the temperatures are defined \cite{Perali_2004}
by the equations
$$
12 \left ( \frac{T^*}{E_F} \right )^3 = \exp \left \{
\frac{2E_F}{T^*(k_F a_s)^2} \right \} \; ,
$$
\be
\label{3.98}
 T_c = 0.518 E_F \qquad \left ( \frac{1}{k_F a_s} \gg 1
\right ) \; .
\ee
The transition between the BCS and BEC limits is a gradual crossover.

\begin{figure}[t]
\centerline{
\includegraphics[width=7.0cm]{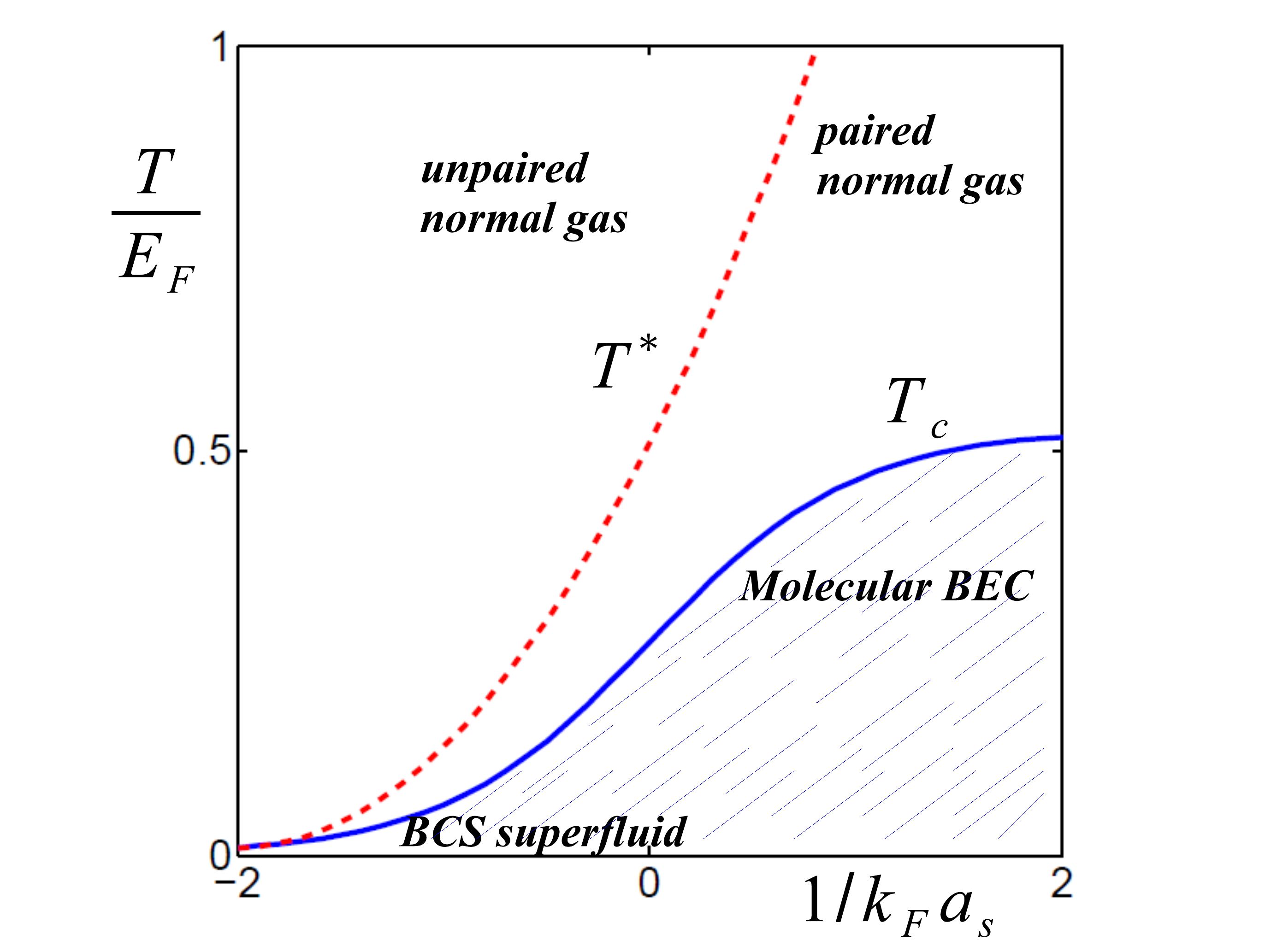} }
\caption{\label{fig:2}(Color online) Schematic phase diagram of the BCS-BEC crossover.
The critical temperature $T_c$ (solid line) shows the phase transition
between the normal phase, with unbroken gauge symmetry, and superfluid
phase, with broken gauge symmetry. The pairing temperature $T^*$ (dashed
line) marks the onset of pairing inside the normal phase. The superfluid
region with broken gauge symmetry is shadowed.}
\end{figure}

\subsubsection{Pairing and symmetry breaking}
\label{subsubsec:III.D.3}

In the region of temperatures $T_c < T < T^*$, pairs are formed, however,
gauge symmetry is not yet broken. The gauge symmetry becomes broken only
at temperatures below $T_c$. The difference from the case of Bose-Einstein
condensation of bosonic atoms, is that the breaking of gauge symmetry for
fermions is accompanied by the appearance of only {\it even} anomalous
averages, such as
\be
\label{3.99}
\lgl \psi_\sgm(\br) \psi_{-\sgm}(\br') \rgl \neq 0 \;  ,
\ee
where $\psi_\sigma({\bf r})$ is a Fermi field operator for a spin component
$\sigma$ \cite{Belitz_1994}. But odd anomalous averages are zero, e.g.,
\be
\label{3.100}
\lgl \psi_\sgm(\br) \rgl = 0 \; ,
\ee
which is related to the conservation of spin.

The schematic phase diagram illustrating the BCS-BEC crossover and the regions
of broken and unbroken gauge symmetry is presented in Fig. 2. This diagram
concerns atoms with the standard Hamiltonian including only contact interactions,
characterized by $s$-wave scattering length. No dipole or spinor interactions
are included. Recently, there have also appeared publications studying effective
spin-orbit interactions of trapped atoms \cite{Stanescu_2007,Stanescu_2008}.
Such additional interactions, of course, can change the whole picture.

\subsection{On condensation of unconserved quasiparticles}
\label{subsec:III.E}

\subsubsection{Conserved versus unconserved particles}
\label{subsubsec:III.E.1}

In all sections above, atomic systems are considered, where the total
average number of atoms $N$ has been fixed. In that sense the atoms, as such,
are stable and conserved. There are also particles or quasiparticles whose
total number is not fixed but can be varied by creating quasiparticles with
applying external fields. While without external supporting fields, in an
equilibrium system, the quasiparticles decay, so that their total number is
not conserved. Examples of such quasiparticle excitations are excitons and
polaritons.

Unconserved quasiparticles, whose number is not fixed, cannot form
Bose-Einstein condensates in equilibrium systems. The chemical potential
of unconserved quasiparticles is zero. When diminishing temperature, their
density decreases, so that Bose condensation does not occur.

However, in nonequilibrium systems, subject to the action of external
time-dependent fields, the density of unconserved quasiparticles can be
supported at sufficiently high level, and condensation can be feasible. For
example, there happens nonequilibrium condensation of excitons
\cite{Keldysh_1968,Kohn_1970,Butov_2003,Eisenstein_2004},
polaritons \cite{Deng_2002,Keeling_2007,Szymanska_2007}, and photons
\cite{Klaers_2010}.

Another type of unconserved quasiparticles is represented by elementary
collective excitations, such as phonons (density fluctuations), bogolons
(density fluctuations in a Bose-condensed system), and magnons
(spin fluctuations). An important question is whether such collective
excitations could form Bose-Einstein condensates by themselves, and,
if so, under what conditions? As is explained above, Bose condensation
is necessarily accompanied by the global gauge symmetry breaking. So,
the question is whether this breaking can occur in a subsystem of
collective excitations?

\subsubsection{No condensation of self-consistent phonons}
\label{subsubsec:III.E.2}

Phonons are introduced by defining, first, deviations ${\bf u}_j$ from a
given spatial point $\ba_j$, and then representing these deviations
through phonon operators $b_{ks}$. If the location $\ba_j$ is fixed, then
formally phonons can start condensing at the point of a phase transition.
But, if the vector $\ba_j$ is defined as the average location, then there
is no phonon condensation, but there is the variation of this location $\ba_j$.
The latter self-consistent procedure can be generalized to nonequilibrium
situations, when atomic positions depend on time, so that one could write
\be
\label{C.1}
\br_j(t) = \ba_j(t) + \bu_j(t) \;   ,
\ee
with the self-consistency condition
\be
\label{C.2}
\ba_j(t) = \lgl \br_j(t) \rgl
\ee
definition the vector $\ba_j$ as the average location. Owing to this
self-consistent way of introducing deviations, one has
\be
\label{C.3}
 \lgl \bu_j(t) \rgl = 0 \;  .
\ee
Consequently, the corresponding phonon operators are such that
\be
\label{C.4}
 \lgl b_{ks}(t) \rgl = 0 \;  ,
\ee
which implies the absence of Bose condensation of phonons.

In this way, if one introduces the deviation ${\bf u}_j$, counted from
a given fixed location $\ba_j$, then at the phase transition point there
can arise phonon Bose condensation implying the necessity of changing
the definition of the location vectors $\ba_j$, which is equivalent to
the change of the spatial system structure. However, if one imposes the
self-consistency condition (\ref{C.2}), then the phonons, defined in
such as self-consistent way, cannot condense, whether in equilibrium or
in nonequilibrium systems. This also concerns phase transition points,
where there occurs a sharp variation of the mean locations $\ba_j$.

\subsubsection{No condensation of generic bogolons}
\label{subsubsec:III.E.3}

Bogolons are quasiparticles characterizing elementary collective excitations
in a Bose-condensed system. They are introduced by means of the Bogolubov
shift that, for an arbitrary nonequilibrium system, reads as
\be
\label{C.5}
 \hat\psi(\br,t) = \eta(\br,t) + \psi_1(\br,t) \;  ,
\ee
where the condensate wave function is the system order parameter
\be
\label{C.6}
 \eta(\br,t) = \lgl \hat\psi(\br,t) \rgl \;  .
\ee
Recall that the Bogolubov shift (\ref{C.5}) is not an approximation, but an
exact canonical transformation that does not require that $\psi_1$ be in any
sense small.

Because of this definition, one has the exact equality
\be
\label{C.7}
 \lgl \psi_1(\br,t) \rgl = 0 \;  .
\ee
The bogolon operators $b_k(t)$ are introduced through the Bogolubov
canonical transformation
\be
\label{C.8}
\psi_1(\br,t) =
\sum_k \left [ u_k(\br) b_k(t) + v_k^*(\br) b_k^\dgr(t) \right ]\; ,
\ee
in which $k$ is a multi-index. In view of Eqs. (\ref{C.7}) and (\ref{C.8}),
one has
\be
\label{C.9}
\lgl b_k(t) \rgl = 0 \;  ,
\ee
which tells us that the gauge symmetry for bogolons cannot be broken.

The self-consistent definition of bogolons shows that they cannot form
Bose condensate in any system, whether equilibrium or not, similar to
the absence of condensation of self-consistently defined phonons. This
should be of no surprise, since bogolons, similarly to phonons,
characterize density fluctuations, albeit in a Bose-condensed system.
In that system, conserved atoms condense, yielding the order parameter
(\ref{C.6}), while unconserved bogolons cannot condense, possessing
only the trivial parameter (\ref{C.9}).

\subsubsection{No condensation of equilibrium magnons}
\label{subsubsec:III.E.4}

Between Bose systems and spin systems, there exists a direct analogy
\cite{Matsubara_1956,Batyev_1984}. Therefore, on should expect that
the elementary collective excitations of spin systems, that is, magnons,
should not be able to condense, at least in equilibrium systems, as has
been stressed by Mills \cite{Mills_2007}. Despite this, many authors
state that there can occur condensation of magnons in equilibrium. Below
we show that this is impossible, provided magnons are correctly defined.

Let us consider a system of spins ${\bf S}_j$, enumerated by the index
$j=1,2,\ldots,N$. And let the system of coordinates be chosen so that the
mean spin be directed along the $z$-axis. Magnon operators $b_j$ are
introduced by means of the Holstein-Primakoff \cite{Holstein_1940}
representation
\be
\label{C.10}
S_j^+ = \sqrt{2S - b_j^\dgr b_j} \; b_j \; , \qquad
S_j^- = b_j^\dgr \;\sqrt{2S - b_j^\dgr b_j} \; , \qquad
S_j^z = S - b_j^\dgr b_j \; ,
\ee
where $S$ is the spin value. As is emphasized by Holstein and Primakoff,
this representation is valid under the condition that the magnetization
is directed along the $z$-axis,
\be
\label{C.11}
\lgl \bS_j \rgl = \lgl S_j^z \rgl \bfe_z \;   ,
\ee
and that it is close to the saturation value, so that
\be
\label{C.12}
 \frac{|\; S - \lgl S_j^z \rgl \; |}{2S} \; \ll \; 1 \;  .
\ee
Condition (\ref{C.11}) tells us that there is no transverse magnetization,
\be
\label{C.13}
 \lgl S_j^\pm \rgl = 0 \;  ,
\ee
while inequality (\ref{C.12}) shows that the magnon density must be small,
\be
\label{C.14}
 \frac{\lgl b_j^\dgr b_j\rgl }{2S} \; \ll \; 1 \;  .
\ee
This is a necessary condition for interpreting the roots of operators in Eq.
(\ref{C.10}) as expansions in Taylor series in powers of $b_j^\dgr b_j$.
Note that the functions of operators are standardly defined as expansions
over these operators. In the present case, the expansion over $b_j^\dgr b_j$,
implies the smallness of the eigenvalues of $b_j^\dgr b_j/2S$, which is
a sufficient condition for the validity of condition (\ref{C.14}).

In the lowest order, one has
$$
 S_j^+ \cong \sqrt{2S} \; b_j \; , \qquad
S_j^- \cong \sqrt{2S} \; b_j^\dgr \;  .
$$
In that way, by definition (\ref{C.13}), we have
\be
\label{C.15}
 \lgl b_j \rgl = 0 \;  ,
\ee
implying that magnons cannot condense in an equilibrium system.

Conversely, from the absence of magnon Bose condensation, that is from the
equation $\lgl b_j \rgl = 0$, it immediately follows that
$\lgl S_j^+ \rgl = 0$, hence $\lgl {\bf S}_j \rgl = \lgl S_j^z \rgl {\bf e}_z$,
corresponding to the mean spin direction along the $z$-axis.

It is, certainly, not necessary to choose the mean spin being directed
along the $z$-axis, but it is just convenient. It is, of course, possible
to consider the case of the mean spin directed arbitrarily and to write down
the Holstein-Primakoff transformation for this case, as has been done by
R\"{u}ckriegel et al. \cite{Ruckriegel_2012}. In the general case of a mean
spin $\lgl {\bf S}_j \rgl$ directed arbitrarily, the spin components
$\{S_j^1$, $S_j^2$, $S_j^3\}$ are the projections of the spin operator
${\bf S}_j$ on the axes defined by the unit mutually orthogonal vectors
${\bf e}^1$, ${\bf e}^2$, and ${\bf e}^3$, with the latter vector being
$$
 \bfe_j^3 = \frac{\lgl \bS_j \rgl }{|\lgl \bS_j \rgl | } \;  .
$$
The spin operator is represented as the sum
$$
 \bS_j = \frac{1}{2} \left ( S_j^+ \bfe_j^- + S_j^- \bfe_j^+ \right )
+ S_j^3 \bfe_j^3 \;  ,
$$
in which
$$
 S_j^\pm \equiv S_j^1 \pm i S_j^2 \; , \qquad \bfe_j^\pm \equiv
\bfe_j^1 \pm  i\bfe_j^2  .
$$
Then the components $S_j^{\pm}$ are given by the same transformations as
above and $S_j^3$, as the component $S_j^z$.

According to this construction, one has
$$
 \lgl \bS_j \rgl = \lgl S_j^3 \rgl \bfe_j^3 \;  ,
$$
from where, as before, it follows that
$$
 \lgl \bS_j^\pm \rgl  = 0 \; , \qquad \lgl b_j \rgl = 0 \; ,
$$
implying the absence of magnon condensation. This conclusion is clear, since
physics does not depend on the choice of coordinates.

When one meets a formal magnon condensation in equilibrium, this just
means that there appears a transverse spin component and there occurs
magnetization rotation. Then, in order that the Holstein-Primakoff
representation would be valid, one needs to redirect the $z$-axis
along the new magnetization direction. The situation is completely
analogous to the case of formal phonon condensation that simply
signifies the occurrence of a phase transition and the necessity to
redefine the system ground state.

Some authors assume that magnon condensation could be possible in
nonequilibrium systems, where condensation could be achieved by
parametric pumping of a magnetic material, such as yttrium-iron-garnet,
by an external pumping field, whose frequency would play the role
of an effective nonzero chemical potential determining the magnon
density \cite{Demokritov_2006,Demidov_2008,Volovik_2008,Bunkov_2010}.
In the frame of this assumption, spin systems, prepared in a strongly
nonequilibrium state, being coupled to a resonator, forming feedback field,
as a result of which exhibiting fast coherent relaxation accompanied by
spin superradiance \cite{YukalovPRL_1995,YukalovPRB_1996,YukalovPPN_2004,
YukalovPRB_2005,YukalovLPL_2011}, could also be related to nonequilibrium
magnon condensation, since their relaxation is connected with the
formation of large transverse coherently rotating magnetization. However,
for strongly nonequilibrium systems, the introduction of magnons does not
look to be clear.

\subsubsection{Condensation of auxiliary quasiparticles}
\label{subsubsec:III.E.5}

Accomplishing operator transformations, one may introduce auxiliary
quasiparticles that do not necessarily correspond to physical objects,
but rather serve as convenient mathematical tools. Nevertheless, one
may consider the formal occurrence of Bose-Einstein condensation of such
auxiliary quasiparticles.

A very often employed transformation, convenient in describing dimerized
magnets is the bond-operator representation \cite{Sachdev_1990}. One
considers a system of spin dimers, consisting of two spins one-half in
spatial locations ${\bf r}_j$, with $j = 1,2,\ldots,N$. A spin dimer is
represented through auxiliary quasiparticle operators $s_j$ and
$t_{j \alpha}$, where $\alpha = x,y,z$. The former operator characterizes
singlet states, so it can be assumed to be related to quasiparticles
{\it singletons}. And the operators $t_{j \alpha}$ correspond to triplet
states, hence the related quasiparticles can be termed {\it triplons}.
Each dimer consists of two spins connected by bonds, one of the spins
being represented as
\be
\label{C.16}
S_{1j}^\al = \frac{1}{2} \left ( s_j^\dgr t_{j\al} +
t_{j\al}^\dgr s_j -
i \sum_{\bt \gm} \ep_{\al\bt\gm} t_{j\bt}^\dgr t_{j\gm} \right )
\ee
and the second spin operator, as
\be
\label{C.17}
S_{2j}^\al = -\; \frac{1}{2} \left ( s_j^\dgr t_{j\al} +
t_{j\al}^\dgr s_j +
i \sum_{\bt \gm} \ep_{\al\bt\gm} t_{j\bt}^\dgr t_{j\gm} \right )   .
\ee
Here $\varepsilon_{\alpha \beta \gamma}$ is the completely antisymmetric
unit tensor.

The introduced quasiparticles have two very important features. First,
their statistics are not prescribed, so that they can be treated either
as bosons or as fermions. Treating them as bosons is an arbitrary
assumption. Second, by their definition, they must satisfy the
{\it no-double-occupancy constraint}
\be
\label{C.18}
 s_j^\dgr s_j + \sum_\al t_{j\al}^\dgr t_{j\al} = N \;  .
\ee
This means that these quasiparticles are conserved:
$$
\sum_j \lgl s_j^\dgr s_j \rgl \; + \;
\sum_\al \lgl t_{j\al}^\dgr t_{j\al} \rgl = N \;   .
$$
Since they are conserved, and assuming that they are bosons, they
are formally allowed to exhibit Bose-Einstein condensation.

One should not confuse triplons with magnons. The former are auxiliary
quasiparticles that can arbitrarily be defined as bosons; they are
conserved, hence, allowing for equilibrium condensation. On the contrary,
magnons are well defined bosons that are not conserved, thus, not
allowing for equilibrium condensation \cite{Mills_2007}.

Another type of widely used auxiliary quasiparticles are introduced
through the transformation
\be
\label{C.19}
S_j^\al = \frac{1}{2}
\sum_{\bt\gm} a_{j\bt}^\dgr \sgm_{\bt\gm}^\al a_{j\gm} \;   ,
\ee
where $\sigma^\alpha_{\beta \gamma}$ is the element of the Pauli matrix.
The unipolarity condition is imposed:
\be
\label{C.20}
 \sum_\al a_{j\al}^\dgr a_{j\al} = 2S \; .
\ee

Again, the quasiparticles, represented by the operators $a_{j\gm}$,
can be of any statistics, being either bosons or fermions. Bogolubov
\cite{Bogolubov_1949} considered the case of fermions, while Schwinger
\cite{Schwinger_1952} treated these quasiparticles as bosons, because
of which they acquired the name of Schwinger bosons.

In view of condition (\ref{C.20}), the quasiparticles are conserved:
$$
\sum_{j\al} \lgl  a_{j\al}^\dgr a_{j\al} \rgl = 2 S N \;  .
$$
Being treated as bosons and being conserved, they can condense in
equilibrium. But again, the Schwinger bosons have nothing to do with
magnons.

The general conclusion of this section, is that conserved particles
can form equilibrium as well as nonequilibrium Bose-Einstein
condensates. But unconserved  quasiparticles cannot condense in
equilibrium, though, in some cases, they can condense in nonequilibrium
states. The latter, for instance, concerns Bose-Einstein condensation of
excitons and polaritons.

\section{Quantum Dots}
\label{sec:IV}

The development of semiconductor technology has made it possible the
confinement of a finite number of electrons in a localized three-dimensional
(3D) space of a few hundreds Angstroms \cite{Kas,Ash96}.
This mesoscopic system, which is made up of artificially trapped electrons
between a few layers of various semiconductors and called quantum dot (QD),
opens new avenues in the study of the interplay between quantum and
classical behaviour at low--dimensional scale.
The quantum dot is formed by removing the electrons
outside the dot region with external gates (lateral dot),
or by etching out the material outside the dot region (vertical dot)
\cite{Jac,kou,tihn}. The dot is connected to its
environment by electrostatic barriers, the so called source and drain
contacts, and  gates to which one can apply a voltage $V_g$.
In order to observe quantum effects, QDs are cooled down to well
below 1 K.

In vertical QDs there is a strong screening of Coulomb
interactions in contrast to lateral ones \cite{kou}. This results
in strong quantum effects of the confinement potential on the dynamics of
confined electrons. The main effects discussed in the present review are
directly relevant to vertical dots.

The smaller the quantum dot, the larger the prevalence of quantum effects
upon the static and dynamic properties of the system.
Almost all parameters of QDs,-- size, strength of a
confining potential, number of electrons, coupling between dots,
dielectric environment, the shape of tunnelling barriers, as well
as external parameters, such as temperature and magnetic,
electrical and/or electro-magnetic fields, -- can be varied in a
controlled way. It is precisely to stress this controllability that the names
{\em artificial atoms} and {\em quantum dots} have been coined.
Therefore, QDs can be considered as a tiny laboratory
allowing direct investigation of fundamental
properties of charge and spin correlations at the atomic scale
\cite{Jac,kou,Chak,mak,RM,yan,Saar}. Another strong motivation
for studying the properties of QDs is due to a rapid development of
the field of quantum computing, since the entangled states of electrons
confined in a quantum dot may give a natural realization
of a quantum bit or "qubit" \cite{kou2,bit,qcom2}. It is expected that QDs could
lead to novel device applications in fields such as quantum cryptography,
quantum computing, optics and optoelectronics, information
storage, biology (fluorescent labelling of cellular targets).

For small QDs, where the
number of electrons is well defined $(N\leq 30)$, the mean free path of the
electrons at Fermi energy ($\lambda_F\sim 100$ nm) appears to be larger or
comparable with the diameter of the dot ($d\sim 10-100$ nm) \cite{kou3}.
It seems therefore natural to assume that the properties of
the electron states in QDs close to the Fermi level
should be determined by the effective mean-field potential of
the "artificial atom", produced by non-trivial interplay of the external confinement
governed by gate voltage and electron-electron interaction.
However, the atom--quantum dot analogy should not
be carried too far: unlike electrons in an isolated atom, carriers
in semiconductor QDs interact strongly with lattice
vibrations and could be strongly influenced by defect, surface, or
interface states. In contrast to real atoms, for which the
confining Coulomb potential is well known, the forces that keep
the carriers in self-organized traps are difficult to
estimate from first principles. The exact shape and composition of
the traps often are not well known and depend on the growth procedure;
in addition, complications are introduced by the complex band
structure of the strained material and, in some cases, by the effect of
piezoelectric forces.
A good assessment of the effective confining
potential inside the dot can be obtained from a combined study of
the ground-state and excitation energies. Ground-state energies
are investigated by capacitance spectroscopy or by single electron tunnelling
spectroscopy \cite{kou,tihn}. Far-infrared spectroscopy is used to study the
excitations of $N$-electron states in the dots (see below).  A rich information
about the intrinsic structure of QDs and correlations effects is expected
under the influence of the applied magnetic field.

The electron states of few-electron quantum dots subjected
to a strong magnetic field have been studied extensively
in various experiments. The electrodynamic response
(far-infrared spectroscopy) of QDs
is expected to be dominated by the many-body effects
produced by confined and interacting electrons.
Sikorski and Merkt \cite{1} found experimentally, however,
the surprising result that the resonance frequencies in the magneto-optical
spectrum are independent of the number of electrons in the QD.
In these systems, which have been experimentally realized,
the extension in the $x-y$ - plane is much larger than in
the $z$-direction. Based on the assumption that the extension in the
$z$-direction can be effectively considered zero, a good description of the
far-infrared resonance frequencies has been obtained \cite{1} within
a two-dimensional ($2D$) harmonic oscillator model
in the presence of a magnetic field \cite{Fock,Darwin}.
This result was interpreted as a consequence of
Kohn's theorem \cite{Kohn} which is applied for a parabolic
potential \cite{BJH,Li}. The proof is based on the identities
\bea
&&{\sum}_{j=1}^N {\bf x}_j^2=
N{\bf X}^2+{\sum}_{i<j}({\bf x}_i-{\bf x}_j)^2/N\; ,\nonumber\\
&&{\sum}_{j=1}^N {\bf p}_j^2=
{\bf P}^2/N+{\sum}_{i<j}({\bf p}_i-{\bf p}_j)^2/N\nonumber  \; ,
\eea
where the center-of-mass coordinate and momentum are ${\bf X}={\sum}_j{\bf x}_j/N$
and ${\bf P}={\sum}_j{\bf p}_j$, respectively.
This implies that, for the parabolic confinement, the total Hamiltonian $H_N$ can be
separated into the center-of-mass motion term, the spin term, and the relative motion
term that contains the electron-electron interactions.
The wave length of the external laser field far exceeds the average dot diameter
and, therefore, can be well approximated  by a dipole electric term only.
Since the radiation of an external
electric dipole field couples only to the center-of-mass motion and does not affect
the relative motion, the dipole resonance frequencies
should be exactly the same as those of the non-interacting system with the
parabolic confinement and, therefore, be independent from the
electron-electron interaction. The more complicated resonance structure,
observed by \cite{h1,h2}, raised,
however, the question on the validity of Kohn's theorem for QDs. In order to
describe the experimental data, it has been assumed that there is a deviation of
the confining potential from the parabolic form, and different phenomenological
corrections have been introduced \cite{Gud,Pfan}.
Considering  external gates and surrounding of a two-electron QD as
the image charge, it was shown \cite{Zhaug} that the effective potential
has, indeed, anharmonic corrections to the parabolic potential. However, their
contribution becomes less important with the increase of the
magnetic field strength.

Recent single-electron capacitance spectroscopy experiments
in vertical QDs \cite{e1,ni2,ni} provide
another strong evidence in favour of the parabolic potential as an effective
confinement potential in small QDs. In these experiments,
shell structure phenomena have been clearly observed.
In particular, the energy needed to place the extra electron (addition energy)
into a vertical QD at zero magnetic field
has characteristic maxima which correspond to
the sequence of magic numbers (due to a complete filling of shells)
of a $2D$ harmonic oscillator. The energy gap between
filled shells is approximately $\hbar \omega_0$, where $\hbar \omega_0$ is
the lateral confinement energy of the $2D$ harmonic oscillator.
In fact, these atomic-like features, when the confining energy is
comparable to or larger than the interaction energy, have been
predicted independently in several publications \cite{Mac1,Mac2,PL,W}.
Indeed, for a small dot size and a small number of electrons,
the confinement energy becomes prevalent over the Coulomb energy.
This has been confirmed for the $3D$
parabolic potential for two interacting electrons \cite{Din} and
in the Hartree-Fock approach for $N\leq 12$ electrons \cite{fuj}.
In small QDs with the incomplete shell, produced
by the parabolic potential, the Hund's rule prior the first experiment in
1996 \cite{e1} was predicted \cite{ser,fuj,wojs}.

These experimental and theoretical studies lead to the conclusion that,
indeed, for a few-electron small QDs the parabolic
potential is a good approximation for the effective confinement.
Note that for a typical voltage $\sim 1$V applied to
the gates, the confining potential in small QDs is of the order of
1 eV deep, which is large compared to a few meV of the confining frequency.
Hence, the electron wave function is localized close to the minimum
of the well which can always be approximated by a parabolic potential.
This approximation becomes especially attractive for comparing symmetry
breaking effects in QDs in a perpendicular magnetic field with rotating BEC
and rotating nuclei. In this review we will pay a special attention to the
manifestations of symmetry breaking due to the {\it three-dimensional} nature
of small QDs, in contrast to reviews \cite{RM, Saar, yan} which are
devoted to $2D$ approaches to QDs.

The simplest description of finite quantum systems of interacting
fermions is based on the idea that their interactions create
an effective potential in which particles are assumed to move independently.
For finite Fermi systems, like nuclei and metallic clusters, the bunching of
single-particle levels known as shells \cite{BM,Heer,Brack,mclus} is one consequence
of this description, since the mean free path is comparable with the size of
the system. A remarkable stability is found in nuclei and metallic clusters
at magic numbers that correspond to closed shells in an effective potential.
According to accepted wisdom, strong shell effects is the manifestations of
a high degree of symmetry of the effective potential of a quantum
many-body system \cite{BM}.

The rearrangement of the intrinsic structure of small QDs under the
perpendicular magnetic field can be traced within a
simple shell model \cite{PL}. For instance, the model describes the effect of
symmetry breaking of the mean field due to the magnetic field and the number
of electrons. For pedagogical purposes, in order to illustrate some basic
features of structural properties of small QDs, we discuss this model
in Sec. 4.2. In fact, this discussion will help us to understand some common
properties of rotating nuclei and rotating trapped atoms.
In Sec. 4.3, we trace the
dynamical effects of the confinement strength, the magnetic strength, the
Coulomb repulsion, and their mutual interplay in the model for a
two-electron QD. We show that at a particular strength of
the magnetic field the nonlinear dynamics of two-electron QD becomes
separable. Sec. 4.4 is devoted to the comparison of theoretical and
experimental results for the ground state energies of two-electron quantum
dots in a perpendicular magnetic field.
In Sec. 4.5, we discuss the symmetry breaking phenomena in N-electron
quantum dots in a mean field and a random phase approximations.

\subsection{Basic features}
\label{subsec:IV.A}

For the analysis of experimental data, several approximations are commonly used.
The underlying lattice of the semiconductor material
is taken into account by using the effective mass for the conduction electrons,
and a static dielectric constant reducing the Coulomb repulsion.
As it was mentioned above, an effective trapping potential in
small QDs with a few electrons is quite well approximated by a parabolic confinement.
The ground state energy of the dot is calculated assuming that the dot is isolated.
This approximation is well justified, when the tunnelling
between the QD and an external source and drain is relatively weak. Using these
approximations, one can study the influence of magnetic field on the electron
spectrum of the dot. Hereafter, the magnetic field is assumed to be perpendicular
to the plane $x-y$ of the electron motion.

Thus, the Hamiltonian of an isolated quantum dot with $N$ electrons
in a perpendicular magnetic field reads as
\begin{eqnarray}
\label{hamr}
&&H = H_0+H_{\it int}=\sum_j^N h_j+
\sum_{i>j=1}^N\frac{k}{\vert{\mathbf r}_i \!- {\mathbf r}_j\vert} \; , \\
&&h_j=\bigg[ \frac{1}{2m^*\!}\,
\Big({\bf p}_j - \frac{e}{c} {\mathbf A}_j \Big)^{\! 2}
+ U({\mathbf r}_j) + \mu^*{\sigma}_z( j)B \bigg],\nonumber
\end{eqnarray}
where $k = e^2/4\pi\varepsilon_0\varepsilon_r$ .
Here, $e$, $m^*$, $\varepsilon_0$ and
$\varepsilon_r$ are the unit charge, effective electron mass, vacuum
and relative dielectric constants of a semiconductor,
respectively. The confining potential is
approximated by a three-dimensional harmonic oscillator potential (HO)
\be
U({\mathbf r}) = m^* [\omega_x^2x^2 \!+ \omega_y^2y^2 \! + \omega_z^2z^2]/2,
\ee
where $\hbar\omega_i \,( i= x,y,z)$ are the energy scales of confinement in
the $x,y,z$-directions, respectively, and  $\sigma_z$ is a Pauli matrix.

In a quantum dot, the natural unit of length is
$a^*=\hbar^2/[m^*(e^2/\varepsilon)]$, with $(4\pi\varepsilon_0=1)$.
In what follows (if not mentioned otherwise), we use in numerical examples
the effective mass $m^*=0.067 m_e$ and the
dielectric constant $\varepsilon_r \approx 12$,  which are typical for GaAs.
In this case, the length unit $a^*\approx 180 a_0$, where the Bohr radius
$a_0=\hbar^2/(m_ee^2)=5.29\times 10^{-2}$ nm. The energy unit in this case is
$E^*=e^2/(\varepsilon_r a^*)=e^2/a_0\times (a_0/\varepsilon_r a^*)\approx 12$ meV,
where the atomic unit of energy $E_0=e^2/a_0=27.2$ eV.

The effective spin magnetic moment
is $\mu ^*=g_L\mu _B$ with
$\mu_B=|e|\hbar/2m_e c \approx 5.79\times 10^{-2}$meV/T (T is Tesla).
The effective mass determines the orbital magnetic moment $\mu _B^{{\rm eff}}$
for electrons through the expression of the Larmor frequency
$\hbar\omega_L=\mu_{B}Bm_e/m^{*}=\mu _B^{{\rm eff}}B$ and
leads to $\mu _B^{{\rm eff}}\approx 15\mu _B$. Evidently, the orbital magnetic
moment is much stronger than the effective spin magnetic moment
(with the effective Lande factor $|g_L|=0.44$).
If the magnetic length  $\ell_B=\sqrt{\hbar c/(eB)}$ equals the unit length
$a^*$, one extracts the magnetic unit strength $B^*=(a_0/a^*)^2B_0$, where
$B_0=e^3m_e^2c/\hbar^3=E_0/(2\mu_B)\approx 2.35\times 10^5$T corresponds to
free electrons.
For GaAs one obtains $B^*\approx 7.2$ T. In other words, for magnetic fields that
are available in the laboratory, one can study various phenomena related to orbital
magnetism that can occur only in neutron stars.

\subsection{Shell effects in a simple model}

\subsubsection{Magnetic field and shapes}
\label{subsubsec:IV.A.1}

The effect of an external homogeneous magnetic field on a three-dimensional ($3D$)
harmonic oscillator potential can be taken into account exactly, irrespectively to
the direction of the field \cite{PL,HeNa97}.
For a perpendicular magnetic field, we choose
the vector potential with
a gauge ${\mathbf A} = \frac{1}{2} {\mathbf B}
\times {\mathbf r} = \frac{1}{2}B(-y, x,0)$.
In this case, the electronic spectrum, generated by the Hamiltonian
(\ref{hamr}) without interaction,
is determined by the sum $H_0=\sum_i^N h_i$ of the single-particle harmonic
oscillator Hamiltonians $h = h_0+h_z$ where
\be
\label{rh}
h_0=\frac{{p_x}^2+{p_y}^2}{2m^*}+
\frac{m^{*}}{2}\sum_{i=1}^2\omega_i^2x_i^2
-\omega_L l_z \; .
\ee
Here, for a perpendicular magnetic field, we have
\be
\omega_1^2=\omega_x^2+\omega_L^2 \;,\omega_2^2=\omega_y^2+\omega_L^2 \;,
\ee
and $\omega_L= |e|B/(2m^{*}c)$. Since the orbital momentum $l_z$
\be
l_z = xp_y-yp_x
\ee
couples lateral variables, the dynamics in $z$ direction is determined by
one-dimensional harmonic oscillator
$h_z={p}_z^2/2m^*+m^{*}\omega_z^2z^2/2$.

Before proceeding further, a few remarks are in order. First, the
external field is the dominant part of the mean field, hence the effective confining
potential should reflect the main features of this field. Yet it
must also contain the effect of the interplay
between the external fields which are governed by
the charges in the adjacent layers and gates and the magnetic
field. Due to these considerations, we assume that the confining
potential should also take into account the changes
that affect the properties of the single-electron states owing
to a variation of the homogeneous magnetic field, as well as
the slab thickness. We suppose that the system adjusts itself to the influence
of the magnetic field, under the given particle number.
Minimizing $E_{{\rm tot}}=\langle H_0\rangle$ connects the magnetic field strength
with the related shape of the confining effective potential, which is defined by
the oscillator frequencies.
In other words, for a given magnetic field, we need to find the minimum
of $E_{{\rm tot}}$ with respect to the variation of the oscillator frequencies.
In this way, we accommodate the effect of the interplay between
external fields, the external confinement and
the magnetic field.
Here, the Pauli principle is essential as it limits
the accessible quantum configuration space for electrons.
The variation cannot be unrestricted as the
confining potential contains a fixed number of electrons.
Assuming that the electron density area does not change, we introduce
a fixed volume constraint which translates into the subsidiary condition
$\omega _x\omega _y\omega _z=\omega _0^3$ with $\omega _0$ fixed. Denoting
the Lagrange multiplier by $\lambda$, we solve the variational problem
\be
\delta (\langle g|H_0|g\rangle -\lambda \omega _x\omega _y\omega _z)=0 \;,
\label{var}
\ee
where $|g\rangle $ denotes the ground state. From Eq. (\ref{var}), after the
differentiation with respect to the
frequencies and using the Feynman's theorem \cite{Fey}
\be
{d\over d\omega _k}\langle g|H_0|g\rangle =
\langle g|{dH_0\over d\omega _k}|g\rangle  \; ,
\label{feyn}
\ee
we get the useful condition
\be
\omega _x^2\langle g|x^2|g\rangle=\omega _y^2\langle g|y^2|g\rangle
=\omega _z^2\langle g|z^2|g\rangle \; ,
\label{cond}
\ee
which must be obeyed at the minimum of $E_{{\rm tot}}$.
We restrict ourselves to the consideration of
a thin slab with large lengths in two dimensions.
This is achieved by varying only $\omega _x$ and $\omega _y$ in the
minimization procedure while keeping $\omega _z$ fixed at a value, which
is several time (e.g., five times) larger than the other two frequencies.
In this case, only the condition
$\omega _x^2\langle g| x^2|g\rangle=\omega _y^2\langle g|y^2|g\rangle $
is to be fulfilled. Choosing different
(fixed) values of $\omega _z$ allows one to study the dependence of the
results on the slab thickness.

Since the electron interaction is crucial only for partially
filled electronic shells \cite{kou,RM},
we deal in this section mainly with closed shells.
This case corresponds to the quantum limit $\hbar\omega_0>k/l_0$,
where $k/l_0$ is the typical Coulomb energy and
$l_0=(\hbar/m^*\omega_0)^{1/2}$ is the effective oscillator length.
In fact, for small dots, where large gaps between closed shells occur
\cite{Mac1,Mac2,W,Zozu},
the electron interaction plays the role
of a weak perturbation, which, in the first approximation, can be neglected.
But even in the regime
$\hbar \omega_0< k/l_0$, an essentially larger additional energy is needed
for the addition of an electron to a closed shell \cite{HiWi}.
We do not take into account the effect of finite temperature; this is
appropriate for experiments which are performed at temperatures
$k_BT\ll \hbar\omega_0$, with $\hbar\omega_0 \sim 2-5$ meV being the mean
level spacing. Indeed, for experiments \cite{e1,ni2} a typical temperature
is estimated to be below $\sim 100$ mK ($\sim 0.008$ meV). In the following,
we use meV for the energy units and T for magnetic field strength.

\begin{figure}[th]
\centerline{
\includegraphics[width=0.45\textwidth,clip=]{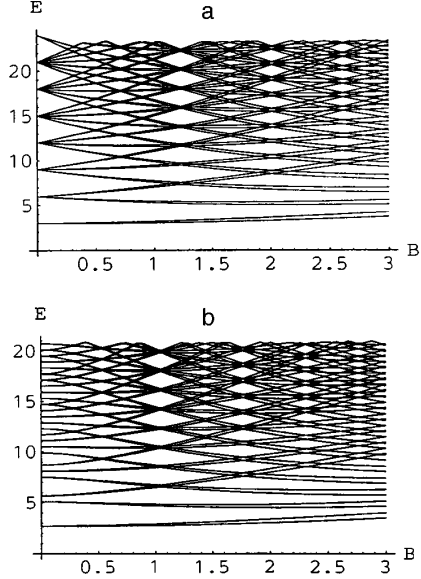} }
\caption{Single-particle spectra as a function of the magnetic field strength.
Spectra are displayed for: (a) a plain isotropic ($\omega _x=\omega _y$)
two-dimensional oscillator;
(b) a deformed two-dimensional oscillator.
From \cite{om}.}
\label{sheff}
\end{figure}

Evidently, for the dot with a fixed
$\omega_z\gg \omega_{x,y}$, with the number of electrons $N\sim 20$,
electrons occupy the states with $n_z=0$.
As a result, shell effects are determined
by the ratio of the eigenmodes $\Omega _{\pm }$ in the lateral plane
(see Appendix \ref{appa}).
The shell structure occurs whenever the ratio of the two eigenmodes
$\Omega _{\pm }$ of the Hamiltonian $H_0$
is a rational number with a small numerator and denominator.
Closed shells are particularly pronounced if
the ratio is equal to one (for $B=0$) or two (for
$B\approx 1.23$ ), or three (for $B\approx 2.01$ ), and are lesser
pronounced if the ratio is 3/2 (for $B=0.72$) or 5/2 (for $B=1.65$)
for a circular case $\omega_x=\omega_y$ (see Fig.\ref{sheff}a).
For illustration, we use for the spin splitting
the value $2\mu_B$ instead of the correct $\mu^*$ in the Figures;
the discussions and conclusions are based on the correct value.
The values given here for $B$ depend on $m^*$ and $\omega _{x,y}$.
As a consequence, a material with an even smaller effective mass $m^*$ would
show these effects for a correspondingly smaller magnetic field.
For $B=0$ the magic numbers (including spin) turn out to be the
usual sequence of the $2D$ isotropic oscillator numbers, that is,
$2,6,12,20,\ldots $. For $B\approx 1.23$,
a new shell structure arises {\em as if}  the confining potential would be a
deformed harmonic oscillator without magnetic field.
The magic numbers are
$2,4,8,12,18,24,\ldots $, which are just the numbers obtained for the
$2D$ oscillator, where $\omega _> = 2\omega _<$. Here $\omega _>$ and
$\omega _<$ denote the larger and smaller values of the frequencies.
Similarly, for $B\approx 2.01$ the magic numbers
$2,4,6,10,14,18,24,\ldots $ appear, which correspond to $\omega _>=3\omega _<$.
If one starts with a deformed mean field
$\omega _x=(1-\beta )\omega _y$ with $\beta >0$,
the degeneracies (shell structure), lifted at $B=0$,
re-occur at higher values for $B$.
In Fig. \ref{sheff}b, we display such an example
referring to $\beta =0.2$. Shell structures are restored by a magnetic field
in an isolated QD that
does not give rise to magic numbers at zero field strength due to deformation.
Thus, the choice of $\beta =0.5$ would shift the pattern, found for
$B\approx 1.23$ in Fig. \ref{sheff}a, to the value $B=0$.
Closed shells are obtained for the values of $B$ and $\beta $, which yield
$\Omega _+/\Omega _-=l=1,2,3,\ldots $.

The ground state energy of a QD, as a function of magnetic
field, can be probed very elegantly by single-electron capacitance
spectroscopy \cite{as} or by single-electron tunnelling
spectroscopy \cite{su,Sch-al95}.
At low temperature $\sim 100$ mK, a large electrostatic charging energy
prevents the flow of current and, therefore, the dot has a fixed number of electrons.
Applying a gate voltage to the contacts
brings the electro-chemical potential of the contacts in resonance
with the energy $\mu(N)$ that is necessary for adding the $N$-th electron,
tunnelling through the barrier, into the dot with $N-1$ electrons.
Indeed, it is the shell structure caused by the effective mean field,
which produces the maxima that are observed experimentally in the addition
energy
\be
\Delta\mu=\mu (N)- \mu (N-1)
\label{aden}
\ee
for $N=2,6,12$ electrons \cite{e1},
where $\mu(N) = E(N)-E(N-1)$ is an electrochemical potential and
$E(N)$ is the total ground state energy of an $N$-electron dot.

In order to shed light on this phenomenon let us calculate $\Delta\mu$
in a constant-interaction (CI) model that provides an approximate description of the
electronic states of QDs \cite{kou}.
In the CI model the total ground state energy of an $N$-electron dot is
\be
E(N)=[e(N-N_0)-C_gV_g]^2/2C+\sum_i^N\varepsilon_i \; ,
\label{ci}
\ee
where $N=N_0$ for the gate voltage $V_g=0$. The term $C_gV_g$ represents the
charge (a continuous variable) induced on the dot by a gate voltage $V_g$, through
the gate capacitance $C_g$. It is assumed that the Coulomb interactions of
an electron on the dot with all other electrons, in and outside the dot, are
parametrized by a constant total capacitance $C$.
The total capacitance between the dot and the source,
drain, and gate is $C=C_s+C_d+C_g$. The quantum contribution is determined by the sum
over all occupied single-particle energies $\varepsilon_i$, which depend on
the magnetic field. In the CI model it is assumed that the single-particle spectrum
is calculated for non-interacting electrons.
Electrons can flow from the source
(left) to the drain (right) through a transport (bias) window
$eV_{\rm sd}$, when $\mu_{\rm left}>\mu_{\rm dot}(N)>\mu_{\rm right}$
(with $-|e|V_{\rm sd}=\mu_{\rm left}-\mu_{\rm right}$). With the aid of
Eq.(\ref{ci}) one obtains the addition energy
\be
\Delta\mu=\varepsilon_{N}-\varepsilon_{N-1}+ e^2/C \; ,
\ee
where $\varepsilon_N$ is the highest filled single-particle state for an
$N$-electron dot and
$e^2/C$  is the classical electrostatic energy.
In the CI model the addition energies of single electrons are periodic in $e^2/C$,
since the difference $\varepsilon_{N}-\varepsilon_{N-1}$ is usually neglected.
In reality, however, it is the fluctuations (shell effects) of the difference that
matter, at least for small QDs.

A similar effect is known in nuclear physics and for metallic clusters.
For specific numbers of fermions, these systems are particularly stable.
For example, the synthesis of superheavy nuclei is guided by a predicted shell
structure for these systems (see, e.g., \cite{oga,prl_stav}). From the theoretical
point of view, the shell effects, due to the single-particle motion, create
fluctuations in the total potential energy that is dominated by the bulk
energy, which is the classical liquid drop energy \cite{NR,Heer,Brack}.
The analogy goes even further: in an isolated small
QD, the external magnetic
field acts like the rotation of a nucleus, thus creating a new shell structure;
in this way superdeformation (axis ratio 2:1) has been established
for rotating nuclei owing to the shell gaps in the single-particle
spectrum \cite{NR}.

Various shapes of the QD can be obtained by the energy minimization
\cite{HeNa97}. In this context, it is worth noting that, under particular
values of the magnetic field, where a pronounced shell structure occurs,
the energy minimum would be obtained for axially-symmetric (or circular) dots,
if the particle number were chosen to be equal to one of the magic numbers.
Deviations from these magic numbers usually give
rise to deformed shapes at the energy minimum. To what extent these 'spontaneous'
deformations actually occur
is the subject of more detailed experimental investigations.
The far-infrared spectroscopy in a small isolated QD
could be a useful tool to provide pertinent data.

The question arises as to what extent the discussed findings depend on the
particular choice of the mean field.  The Coulomb interaction lowers the
electron levels for the increasing magnetic quantum number $|m|$ in the
parabolic potential \cite{Din,taver}
and in a hard-wall potential of finite height \cite{PS96}.
One may add to the Hamiltonian $H_0$ (\ref{hamr}) the term
$-\gamma \hbar \omega_L L^2$, where $L$ is the dimensionless $z$-component of
the orbital momentum operator, to produce a similar result \cite{HeNa98}.
Indeed, for $\gamma >0$, the additional term
mimics the Coulomb interaction effect in the Coulomb blockade regime of
deformed QDs \cite{HHW}, as well as the surface effect.
In particular, the bunching of single-particle levels
with high orbital momenta comes
from the presence of the surface in the Woods-Saxon potential which could
be suitable for modelling the surface effects in QDs.
As a consequence, this term has the effect of interpolating between the
oscillator and the square-well single-particle spectra.
This behaviour is well known in nuclear systems \cite{NR} and
recently has been used for explaining the dominance of prolate deformed shapes
in contrast to oblate shapes for small finite quantum systems such as
nuclei \cite{ham09}.

For $\omega_x=\omega_y\ne \omega _z$ and $\gamma \ne 0$,
the combined Hamiltonian $H'=H_0 - \gamma \hbar \omega_L L^2$
is nonintegrable \cite{HeNa94,HeNa98}, and the
level crossings encountered in Figs.\ref{sheff} are replaced by the avoided level crossings.
The shell structure, which prevails for
$\gamma =0$ throughout
the spectrum at $B\approx 1.23$ or $B\approx 2.01$, is therefore disturbed to an
increasing extent with increasing shell number.
But even for $\gamma \leq 0.1$,
for sufficiently low electron numbers, virtually any
binding potential will produce the patterns found for the harmonic
oscillator.

\subsubsection{Magnetic properties}
\label{subsubsec:IV.A.2}

Orbital magnetism of an ensemble of QDs  was discussed
for noninteracting electrons \cite{AGI,alt,Opp,RUJ,terra}, but little attention
was paid to the shell structure of an individual dot.
When the magnetic field is changed continuously for a QD with a fixed
electron number, the ground state will undergo a rearrangement at the
values of $B$, where level crossings occur (see discussion above).
In fact, it leads to strong variation in the magnetization \cite{yos,Chak} and
should be observable also in the magnetic susceptibility
$\chi=-\partial E_{{\rm tot}}^2/\partial B^2$ \cite{ser}, since
it is proportional to the second derivative
of the total energy with respect to the field strength.

Note that if one replaces $B\Rightarrow \omega_L$, the susceptibility
will be rescaled by a constant factor. However this replacement establishes the
obvious link between the susceptibility and the nuclear dynamical moment
of inertia ${\cal J}^{(2)}=-d^2E/d\Omega^2$,
where $\Omega$ is a rotational frequency of rotating nuclei. We will return
to this point in Sec. 5.

We now focus on the special cases which give rise to the pronounced shell
structure, that is, when the ratio
$\Omega_+/\Omega_-=l=1,2,3,\ldots $.
To avoid cumbersome expressions, we analyze in detail the
circular shape ($\omega_x=\omega_y=\omega_0$, $\langle z^2 \rangle=0$)
for which the eigenmodes, Eq.(\ref{mod}), become
$\Omega_{\pm} = (\Omega \pm \omega_L )$, with
$\Omega = \sqrt{\omega_0^2+\omega_L^2}$.
In this case, the total energy for the closed QD (see Appendix \ref{appa}) is
\be
E_{\rm tot}=\Omega_{+}\Sigma_{+}+\Omega_{-}\Sigma_{-}-
\mu^*B<S_z>+E_z \; ,
\label{hoen}
\ee
with $\sum_{\pm } = \sum _j^N(n_{\pm }+1/2)_j$ and the shell number
$N_{\rm sh}=n_{+}+n_{-}$ $(n_z=0)$.
For the magnetization $M=-dE_{{\rm tot}}/dB$,
taking into account that, after the differentiation of the total
energy (\ref{hoen}), $\Omega_{+}=l\Omega_{-}$ $(\hbar=1)$, we have
\be
\label{mag1}
M= \mu_B^{{\rm eff}}(1-{\omega_L \over \Omega})
({\sum}_{-}-l {\sum}_{+})- \mu^*<S_z>
\ee

Let us consider a QD with $g_L=0$, which results in
zero spin contribution, since $ \mu^*=0$. From the orbital motion we obtain
for the susceptibility
\be
\label{suc}
\chi= - d^2E_{\rm tot}/dB^2=
- \frac{{\mu_B^{{\rm eff}}}^2}{\Omega} (\frac{\omega_0}{\Omega})^2
({\sum}_{+}+{\sum}_{-})
\ee
It follows from Eq. (\ref{suc}) that, for a completely filled shell,
the magnetization due to the orbital motion leads to
diamagnetic behavior. For zero magnetic field ($l=1$) the
system is paramagnetic and the magnetization vanishes
($\sum_- = \sum_+$). The value $l=2$ is reached for $B\approx 1.23$.
When calculating $\sum_-$ and $\sum_+$, we have to
distinguish between the cases, where the shell number $N_{\it sh}$ of the
last filled shell is even or odd. With all shells filled
from the bottom we find:

(i) For the last filled shell even number:
\bea
{\sum}_+ &= &(N_{\it sh}+2)[(N_{\it sh}+2)^2+2]/12\;,\\
{\sum}_- &=&(N_{\it sh}+1)(N_{\it sh}+2)(N_{\it sh}+3)/6\;,
\eea
which implies
\be
M=-\mu_B^{{\rm eff}}(1-{\omega_L / \Omega})(N_{\it sh}+2)/2\;;
\ee

(ii) For the last filled shell odd number:
\be
{\sum}_+ = {\sum}_-/2 =
(N_{\it sh}+1)(N_{\it sh}+2)(N_{\it sh}+3)/12\;,
\ee
which, in turn, implies $M=0.$

Therefore, if $\Omega_+/\Omega_-=2$, the orbital magnetization vanishes
for the magic numbers
$4,12,24,\ldots $, while it leads to diamagnetism for the magic numbers
$2,8,18,\ldots $. A similar picture is obtained for ${\Omega_+}/\Omega{-}=3$,
which happens at $B\approx 2.01$: for each third filled shell number
(magic numbers $6,18,\ldots $) the magnetization is zero.
Since the presented results are due to shell effects, they do not depend
on the assumption $\omega _x/\omega _y=1$, which was made to facilitate the
discussion. The crucial point is the relation
$\Omega_+/\Omega_-=l=1,2,3,\ldots $, which can be obtained for a variety
of combinations of the magnetic field strength and the ratio
$\omega _x/\omega _y$.
Whenever the appropriate combination of the field strength and deformation is
chosen to yield, say, $l=2$, the above discussion is valid.

\subsection{Two-electron quantum dot: a new paradigm in mesoscopic physics}
\label{subsec:IV.C}

Two-electron QDs
have drawn a great deal of experimental and theoretical attention
in recent years. Progress in semiconductor technology has made it possible
to fabricate and probe such confined system at different values of magnetic
field \cite{kou,ni,tihn}.
In particular, one observes transitions between the states that
can be characterized by different quantum numbers $m$ and a total spin $S$
of the Fock-Darwin states \cite{as}. These transitions have been earlier predicted
in the $2D$ approximation: in
numerical calculations \cite{gs1} and in a perturbative approach
in the limit of high magnetic fields \cite{Wag}.
The experiments  stimulated numerous theoretical
studies of two-electron QDs \cite{mak,yan}. Indeed, a competition between
a confining potential, approximated quite well by the HO, and the repulsive
electron-electron interaction produces a rich variety of phenomena.
Being a simplest non-trivial system, a two-electron QD  poses, however,
a significant challenge to theorists.

It is well known that there is a restricted class of exactly solvable
problems in quantum mechanics. Such examples serve as paradigms for
illustrating fundamental principles or/and new methods in the respective fields.
This is especially important for finite systems.
In fact, a two-electron QD becomes a testing--ground for different
quantum-mechanical approaches and experimental techniques
that could provide highly accurate data for this system \cite{Ash96,kou,tihn}.
Therefore, two-electron systems play an
important role in understanding electron correlation
effects, since their eigenstates can be obtained very accurately, or in some
cases, exactly.

For example, Pfannkuche {\it et al.} \cite{Pfan1}  compared the results of
the Hartree, Hartree-Fock, and  of the exact treatment of a $2D$ two-electron
QD and found that the exchange effects are very important.
The most popular model to study the electronic exchange-correlation energy
in the density functional theory is the Hookean two-electron atom. The basic
Hookean atom is formed by two electrons interacting by the Coulomb potential
but bound to a nucleus by a harmonic potential that mimics a nuclear-electron
attraction. For certain values of the confinement strength
$(\omega_x=\omega_y)$, there exist
exact solutions for the $2D$ Hookean atom ground state \cite{kai,taut}.
Turbiner \cite{tur} was able to show that
some analytical solutions for this problem occur due to a hidden
$sl_2$ algebraic structure. This model can be equally
viewed as a model of a $2D$ quantum dot.
At zero magnetic field, the $2D$ problem is integrable in two cases:
$\omega_x:\omega_y=1,2$.
For the circular symmetry, the problem of relative motion becomes separable
in polar coordinates. Although a closed-form solution for the eigenfunctions
was obtained with the aid of power series methods \cite{china}, the exact
energies were calculated numerically.
The separability in the case of $\omega_x:\omega_y=2$ \cite{drou}
provided a basis for algebraic solutions for certain values of $\omega_x$ \cite{pol}.
For the Hookean atom placed in a perpendicular magnetic field,
Taut found analytical solutions for a $2D$ case at particular values
of the magnetic field for the circular case \cite{taut2}.
The circular dot at arbitrary values of the magnetic field was
studied in various
approaches in order to find a closed-form solution \cite{DiNa99,loz,kand1,grokr}.
Analytical solutions, developed for a $3D$ model of a QD \cite{Din,zhtr},
provide the explicit completion
of five integrable cases \cite{hs}.
Discovery of the closed--form
solutions stimulated additional efforts for understanding the validity of
the density functional theory at large magnetic fields. In the case of $3D$
Hookean atom in an external magnetic field,
the comparison with the exact solutions demonstrated clearly the major
qualitative failures of several widely used approximate density functional
theory exchange-correlation energy functionals \cite{edf2}.

The major aspect of Sec. \ref{subsec:IV.C} is to demonstrate that the hidden
symmetries could be observed in a two-electron QD with a $3D$ effective parabolic
confinement under a tunable perpendicular magnetic field. Note that these
symmetries have been overlooked in a plain quantum-mechanical models. It is,
therefore, necessary to focus our analysis on the nonlinear classical dynamics
of such finite systems.
At certain conditions, the motion becomes integrable, which indicates the
existence of symmetries in the quantum spectrum.

\subsubsection{Hidden symmetries in a two-electron quantum dot}
\label{subsubsec:IV.C.1}

A three-dimensional harmonic oscillator, with the frequencies in rational
ratios (RHO), and a Coulomb system are the benchmarks for the hidden symmetries,
which account for the accidental degeneracies of their quantum spectra,
e.g., \cite{LandauLifshitz_2003}.
For the isotropic case, the hidden symmetries
define the $so(n)$ symmetry algebra for the oscillator, and the $so(n+1)$ algebra
for Coulomb systems,
where $n$ is a dimension. In both, the classical and quantum cases,
the transformation $r=R^2$, where $r$ and $R$ denote the radial coordinates
of Coulomb and oscillator systems, respectively, converts the $(n+1)$-dimensional
radial Coulomb problem to a $2n$-dimensional radial oscillator. In three cases,
$n=1,2,4$, one can establish a complete correspondence between the Coulomb and
oscillator systems with the
aid of Bohlin (or Levi-Cevita) \cite{boh,lc}, Kustaanheimo-Stiefel \cite{ks},
and Hurwitz \cite{hur} transformations, respectively. For the $n$-dimensional
case, it was found \cite{zeng} that there exists a simple relationship between
the energy and eigenstates
of the hydrogen atom and those of the HO by means of $su(1,1)$ algebra.

The degeneracies of the HO
model, which occur when the frequency ratio is a rational number, have been the
subject of several investigations in the $2D$ case \cite{hill,Dem} and in
the $3D$ case \cite{DuZa,bragre,vend,mavi,cimc,nadob} in various fields.
For example, these degeneracies result in the appearance of spherical and
superdeformed magic gaps and magic numbers in nuclear systems \cite{BM,NR}.
As is well known, the invariance group of the HO is SU(3) \cite{El}, while
that of the hydrogen atom is $O(4)$ \cite{bi}.

If the HO and the Coulomb potential are combined, most of the symmetries are
expected to be broken. Nevertheless, in particular cases, the Coulomb (Kepler)
system and the RHO may have common  symmetries, as it was already noticed a
long time ago \cite{hill}. Jauch and Hill \cite{hill} could not find, however,
a physical application for this phenomenon. These symmetries were rediscovered
in the analysis of laser--cooled ions in a Paul trap \cite{blum} and
of the hydrogen atom in the generalized van der Waals potential \cite{alh}.

\subsubsection{Center-of-mass and relative-motion Hamiltonians}
\label{subsec:IV.C.2}

Let us consider the Hamiltonian (\ref{hamr}) of a two-electron QD in a magnetic
field with the confining frequencies $\omega_x=\omega_y=\omega_0\neq\omega_z$.
In the present analysis, we neglect the spin interaction (the Zeeman term), since
the corresponding energy is small compared to the confinement
and the Coulomb energies and is not important for our discussion.

By introducing the relative and center-of-mass (CM) coordinates
$$
\mathbf{r} = \mathbf{r}_1-\mathbf{r}_2 \;,  \qquad \mathbf{R} =
\frac{1}{2}(\mathbf{r}_1+\mathbf{r}_2) \; ,
$$
the Hamiltonian (\ref{hamr}) can be separated into the CM and relative motion
terms due to the Kohn theorem \cite{Kohn}: $H = H_{\rm cm} + H_{\rm rel}$,
where
\bea
H_{\rm cm}&=&\frac{{\mathbf P}^2}{2M^*\!}
+\frac{M^*\!}{2}\Big[\,\omega_\rho^2\,(X^2\!\! +\!Y^2) +
\omega_z^2\,Z^2\Big]-\omega_L L_z \; \\
H_{\rm rel}&=&\frac{{\mathbf p}^2}{2\mu}+
\frac{\mu}{2}\Big[\,\omega_\rho^2\,(x^2\!\!
+\!y^2)+
\omega_z^2\,z^2\Big]- \omega_L l_z + \frac{k}{r}  \; .
\label{relham0}
\eea
Here $M^* = 2m^*$ and $\mu = m^*/2$ are the total and reduced masses,
$\omega_L$ is the Larmor frequency and $L_z$ and $l_z$ are the
$z$-projections of the angular
momenta for the CM and relative motions, respectively.
The effective confinement frequency for the $\rho$-coordinate,
$\omega_{\rho}=(\omega_{L}^{2}+\omega_{0}^{2})^{1/2}$,
depends, through $\omega_{L}$, on the magnetic field. In this way, the
magnetic field can be used to control the effective lateral
confinement frequency of the QD for a fixed value of the
vertical confinement,
i.e., for a fixed ratio $\omega_z/\omega_\rho$.

The total two-electron wave function
$\Psi(\mathbf{r}_1,\mathbf{r}_2) = \psi(\mathbf{r}_1,\mathbf{r}_2)
\chi(\sigma_1,\sigma_2)$ is a product of the orbital
$\psi(\mathbf{r}_1,\mathbf{r}_2)$ and spin
$\chi(\sigma_1,\sigma_2)$ wave functions.
Due to the Kohn theorem, the orbital wave function is factorized
as a product of the CM and the relative motion wave functions
$\psi(\mathbf{r}_1,\mathbf{r}_2) =
\psi_\mathrm{CM}(\mathbf{R})\,\psi_\mathrm{rel}(\mathbf{r})$.
The parity of $\psi_\mathrm{rel} (\mathbf{r})$ is a good
quantum number, as well as the magnetic quantum number $m$, since
$\ell_z$ is the integral of motion.
The CM eigenfunction is a product of the
Fock-Darwin state (\ref{fd}) (the eigenstate of a single electron in an isotropic
2D harmonic oscillator potential in a perpendicular
magnetic field) \cite{Fock} in the $(X,Y)$-plane
and the oscillator function in the $Z$-direction
(both sets for a particle of mass $M^*$).

The CM eigenenergies are the sum of Fock-Darwin levels
and oscillator levels in the $z$-direction,
\begin{equation}
E_{\rm cm} = \hbar\omega_\rho\,(2N+\vert M\vert+1) + \hbar\omega_z\,(N_z+1/2)
-\omega_L M  \; .
\label{cm_energy}
\end{equation}
Here $N = 0,1,...$ is the radial quantum number, $M = 0,\pm2,\pm3,...$ is
the azimuthal number, and $N_z = 0,1,2,...$ is the quantum number for the
CM excitations in the $z$-direction.

In the following, we concentrate on the dynamics associated with $H_{\rm rel}$.
For our analysis, it is convenient to use the
cylindrical {\it scaled} coordinates,
$$
\tilde\rho = \rho/l_0, {\tilde p}_{\rho} = p_{\rho}l_0/\hbar \; ,
$$
$$
\tilde z = z/l_0, {\tilde p}_{z} = p_z l_0/\hbar \; ,
$$
where $l_{0} = (\hbar/\mu\omega_0)^{1/2}$ is the characteristic
length of the confinement potential with the reduced mass $\mu$. The
strength parameter $k$ of the Coulomb repulsion goes over to
$\lambda = k/(\hbar \omega_0 l_0)$.
Although our consideration is general, for a numerical demonstration,
we choose the values: $\lambda = 1.5$ (GaAs);
$\hbar \omega_0 \approx 2.8$ meV and $\omega_z/\omega_0 = 2.5$.
Hereafter, for the sake of simplicity,
we drop the tilde, i.e. for the scaled variables we use the same symbols as
before scaling.

In these variables, the
Hamiltonian for the relative motion takes the form
(in units of $\hbar \omega_0$)
\be
h \equiv \frac{H_{\rm rel}}{\hbar\omega_0} = \frac{1}{2}\bigg(
p_\rho^2 + \frac{m^2}{\rho^2} + p_z^2 +
{\tilde\omega_{\!\rho}}^2 \rho^2 +
{\tilde\omega_z}^2 z^2\bigg) +
\frac{\lambda}{r} - {\tilde\omega_L} m \; ,
\label{relham}
\ee
where $r = (\rho^2+z^2)^{1/2}$,
$\tilde\omega \equiv \omega/\omega_0$, $m = l_z/\hbar$.

Since the Coulomb interaction
($k \Rightarrow \lambda\neq 0$) couples the motions in $\rho$ and
$z$-directions,
the eigenfunctions of the Hamiltonian
for relative motion (\ref{relham}) are expanded in the basis of
the Fock-Darwin states $\Phi_{n,m}(\rho, \varphi)$ and
oscillator functions in the $z$-direction,
$\phi_{n_z}(z)$ (for a particle of mass $\mu$), i.e.
\begin{equation}
\psi_\mathrm{rel}(\mathbf{r}) = \sum_{n,n_z} c_{n,n_z}^{(m)}
\Phi_{n,m}(\rho,\varphi)\,\phi_{n_z}(z).
\label{expansion_3d}
\end{equation}
For non-interacting electrons the ground state is described by the
wave function $\psi_\mathrm{rel} = \Phi_{0,0}\,\phi_0$. For
interacting electrons, however, the ground state (in the form
(\ref{expansion_3d})) evolves from $m = 0$ to higher values of $m$
as the magnetic field strength increases. Since the quantum number
$m$ and the total spin are related by the expression
\begin{equation}
\label{spM}
S =\frac{1}{2}[1 - (-1)^m]\,,
\end{equation}
this evolution leads to the
singlet-triplet (S-T) transitions (see below Sec.\ref{subsec:IV.D.1}).
Note that if the Zeeman term is included, the
splitting (with $g^* < 0$) lowers the energy of the $M_S =1$
component of the triplet states, while leaving the singlet states
unchanged. As a consequence, the ground state is characterized by
$M_S = S$.

\subsubsection{Classical dynamics and quantum spectra}
\label{subsec:IV.C.3}

Due to the axial symmetry of the system the $\varphi$-motion is
separated from the motion in the $\rho, z$-plane.
Considering classical dynamics, we associate the Hamiltonian with the energy of
our system. Evidently, the energy is the integral of motion, since we analyze
the autonomous system. Besides the
energy $(\epsilon\equiv h)$, the $z$-component of the angular momentum
$l_z$ is an integral of motion.
Therefore, the magnetic quantum number $m$ is always a
good quantum number.
Since the Hamiltonian (\ref{relham})
is invariant under the reflection with respect to the origin, the parity $\pi$
is also a good quantum number.

The classical trajectories can be obtained by solving (numerically)
the Hamilton equations for a fixed energy \cite{hs}.
Although the motion in $\varphi$ is separated
from the motion in the $\rho, z$-plane,
the problem is in general non-integrable,
since the Coulomb term couples the $\rho$ and $z$-coordinates.

\begin{figure}[t]
\centerline{
\includegraphics[width=0.45\textwidth,clip=]{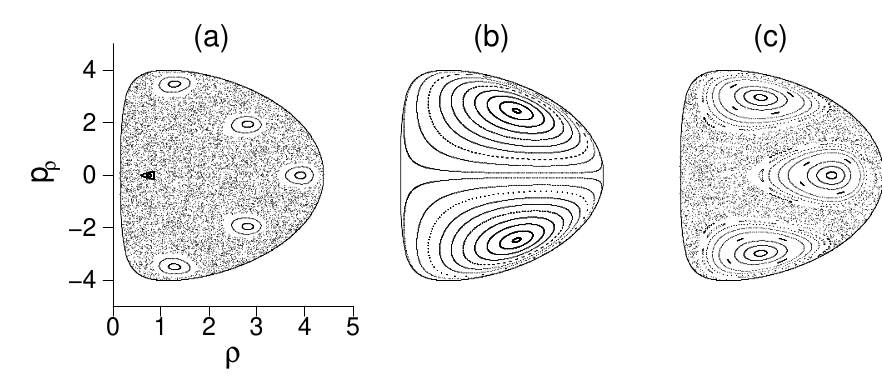} }
\caption{Poincar\'e sections $z = 0$, $p_z > 0$
for the relative motion ($\lambda=1.5$, $\epsilon = 10$, $m = 0$), with:
(a) $\omega_z/\omega_\rho = 5/2$,
(b) $\omega_z/\omega_\rho = 2$ and
(c) $\omega_z/\omega_\rho = 3/2$.
The section (b) indicates that for the ratio
$\omega_z/\omega_\rho = 2$ the system is integrable.
From \cite{hs}.}
\label{fig4}
\end{figure}

The examination of the Poincar\'e sections by varying
the parameter $\omega_z/\omega_\rho$ (see, e.g., Fig.~\ref{fig4})
in the interval $(1/10,10)$, with a small step, indicates
that there are five integrable cases.
The trivial cases
are $\omega_z/\omega_\rho \to 0$ and $\omega_z/\omega_\rho \to \infty$, which
correspond to $1D$ vertical and $2D$ circular QDs, respectively.
The non-trivial cases are
$\omega_z/\omega_\rho = 1/2,1,2$.
These results hold for any strength of the Coulomb
interaction and agree with
the results for the Paul trap \cite{blum}.
Below, we discuss the non-trivial cases only.
The typical trajectories in
cylindrical coordinates are shown in Figs.~\ref{clor}a,c.

The results, obtained with the aid of the
Poincar\'e sections, are invariant under the
coordinate transformation.
On the other hand, the integrability is a necessary condition for the
existence of a coordinate system, in which the motion can be separated.
In turn, the analogous quantum mechanical system would be characterized
by a complete set of quantum numbers.

\begin{itemize}
\item{The case $\omega_\rho=\omega_z$}

At the value $\omega_L^\prime = (\omega_z^2 - \omega_0^2)^{1/2}$,
the magnetic field gives rise to the spherical symmetry
$(\omega_z/\omega_\rho=1)$ in an axially symmetric QD
(with $\omega_z > \omega_0$) \cite{NSR}.
In this case, Hamiltonian (\ref{relham})
is separable in (scaled) spherical coordinates
\be
h = \frac{p_r^2}{2} + \frac{({\bf l}/\hbar)^2}{2r^2} +
\frac{{\tilde\omega_z}^2 r^2}{2\,} + \frac{\lambda}{r} -
{\tilde\omega_L^\prime}m
\label{relsph}
\ee
and the dynamics is integrable. The additional integral of motion is
the square of the total angular momentum ${\bf l}^2$.

Due to the separability of  Hamiltonian (\ref{relsph}) in spherical
coordinates, the corresponding eigenfunctions can be written in the form
\be
\psi({\mathbf r}) = \frac{\phi_{lm}(r)}{r}\, Y_{lm}(\vartheta, \varphi) \; .
\label{psisph}
\ee
The functions $\phi_{lm}(r)$ are the solutions of the radial equation
\be
\left[-\frac{d^2}{d r^2} + \frac{l(l\!+\!1)}{r^2} + \tilde\omega_z^2 r^2 +
\frac{2\lambda}{r} - 2({\tilde\omega_L}^\prime m \!+\! \epsilon) \right]\!
\phi_{lm}(r) = 0 \; ,
\label{reqsph}
\ee
where $l$ and $m$ are the orbital and magnetic quantum numbers, respectively.
Eq.~(\ref{reqsph}) can be solved numerically.
Hence, good quantum numbers for this case are $(n_r,l,m)$, where
the radial quantum number $n_r = 0,1,2,...$ enumerates the radial functions
$\phi_{n_r,l,m}(r)$ within each $(l,m)$-manifold. For the spherical case,
$\pi=(-1)^l$. Thus, the magnetic field, reducing the
$O(4)$ symmetry, creates the dynamical symmetry $O(4)\supset O(3)$
(see also Alhassid {\it et al.} \cite{alh}).

In this case, it is straightforward to use
a semiclassical quantization of Hamiltonian (\ref{relsph}) to calculate
the spectrum. The procedure reduces to the WKB quantization of $r$-motion,
due to the separability of the problem in spherical coordinates.
The momentum $p_r$, determined from Eq.~(\ref{relsph}), enters the action
integral
\be
I_r = \frac{\hbar}{2\pi}\oint p_r\,dr =
\frac{\hbar}{\pi}\int_{r_\mathrm{min}}^{r_\mathrm{max}} |p_r|\,dr \; ,
\label{actsph}
\ee
with the turning points $r_\mathrm{min}$, $r_\mathrm{max}$ as
the positive roots of equation $p_r(r) = 0$.
The WKB quantization conditions
\begin{eqnarray}
&&I_r(\epsilon) = \hbar\,(n_r + \hbox{$\frac{1}{2}$}),
\quad  n_r = 0, 1, ...,\nonumber \\
&&\vert\,{\bf l}\,\vert = \hbar\,(l + \hbox{$\frac{1}{2}$}),
\quad l = 0, 1, ..., \\
&&m = 0, \pm 1, ..., \pm l\nonumber
\end{eqnarray}
determine the energy levels. For non-interacting electrons ($\lambda = 0$),
the analytical calculation of the action integral leads to the
(quantum mechanically exact) eigen-energies (\ref{fce}).
For $\lambda \neq 0$, one can calculate the action integral (\ref{actsph})
numerically with a few iterations to determine the eigenvalues.
The results for the spherically symmetric case, obtained by the WKB
approach and for the cases discussed below, can be found in \cite{NSR}.

The restoration of the rotational symmetry of the electronic states by the
magnetic field for noninteracting electrons was discussed in
Sec. \ref{subsubsec:IV.A.1}. This phenomenon
was also recognized in the results for {\it interacting} electrons in
self-assembled QDs \cite{W}. It was interpreted as an approximate symmetry that
had survived from the noninteracting case due to the dominance of the
confinement energy over a relatively small Coulomb interaction energy. However,
as it is clear from the form of Eq. (\ref{relsph}), the symmetry is not
approximate but {\it exact} even for
strongly interacting electrons, because the radial electron-electron repulsion
does not break the rotational symmetry.

\item{The case $\omega_z=2\omega_\rho$ and $\omega_z=\omega_\rho/2$.}

The spherical coordinates
are a particular limit of the spheroidal (elliptic)
coordinates well suitable for the analysis
of the Coulomb systems, e.g., \cite{Kom}.
Therefore, to search for separability in other integrable cases,
it is convenient to use the spheroidal coordinates
$(\xi,\eta,\varphi)$, where $\xi = (r_1+r_2)/d$ and $\eta = (r_1-r_2)/d$.
In the {\it prolate} spheroidal coordinates
$r_1 = [\rho^2+(z+d)^2]^{1/2}$, $r_2 = r$.
The parameter $d\in(0,\infty)$ is the distance between two foci of the
coordinate system (with the origin at one of them).
In the limit $d\to 0$, the motion is separated when
$\omega_z/\omega_\rho=1$ (see Fig.\ref{clor}b). In this limit
$\xi\to\infty$, so that
$r=d\xi/2$ is finite, $\eta = \cos\vartheta$, and we obtain the spherical
coordinate system.

\begin{figure}[t]
\centerline{\psfig{figure=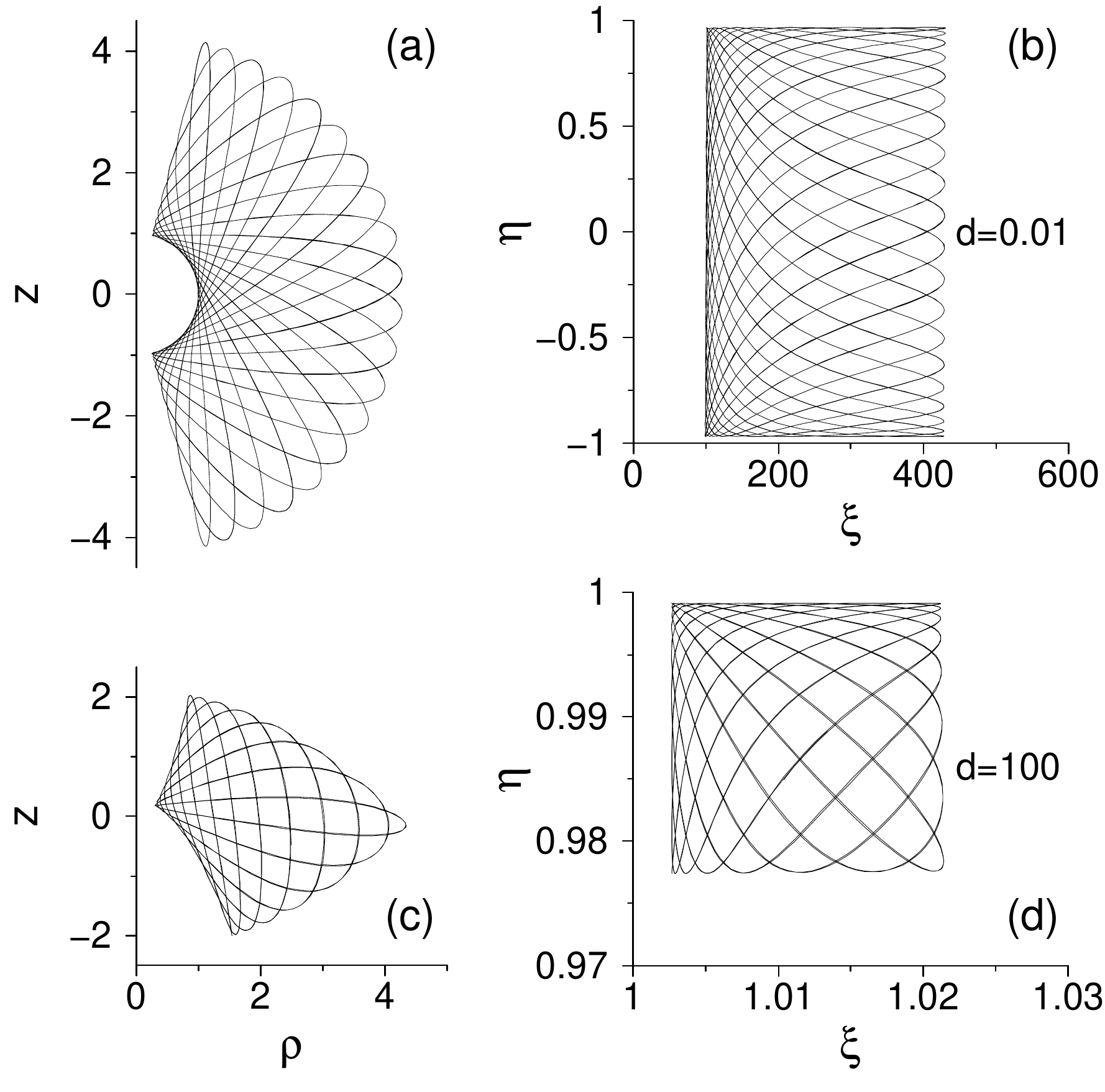,width=3.2in,clip=} }
\caption{
Typical trajectories ($\epsilon = 10$, $m = 1$) of the
relative motion at $\lambda=1.5$,
for $\omega_z/\omega_\rho = 1$ (a,b) and $\omega_z/\omega_\rho = 2$ (c,d),
are shown in cylindrical and prolate spheroidal coordinates, respectively.
From \cite{hs}.}
\label{clor}
\end{figure}

Let us turn to the case $\omega_z/\omega_\rho = 2$ which
occurs at the value of the magnetic field
$\omega_L^{\prime \prime} = (\omega_z^2/4 - \omega_0^2)^{1/2}$.
In the prolate  spheroidal coordinates the motion is
separated in the limit $d\to\infty$ (Fig.~\ref{clor}d).
In fact, at $d\to\infty$: $\xi\to 1$, $\eta\to 1$, so that
$\xi_1 = d(\xi-1)$, $\xi_2 = d(1-\eta)$ are finite, we obtain the
parabolic coordinate system $(\xi_1,\xi_2,\varphi)$, where
$\xi_{1,2} = r\pm z$. In these coordinates
Hamiltonian (\ref{relham}) has the form
\begin{eqnarray}
h \!&=&\! \frac{1}{\xi_1+\xi_2}\Bigg{[}\,2(\xi_1 p_{\xi_1}^2 +
\xi_2 p_{\xi_2}^2) +
\frac{m^2}{2}\left(\frac{1}{\xi_1} + \frac{1}{\xi_2} \right)\nonumber\\
&&+\frac{{\tilde\omega_z}^2}{8}\,(\xi_1^3 + \xi_2^3) + 2\lambda \,\Bigg{]}
-{\tilde\omega_L^{\prime \prime}}m
\label{relpar}
\end{eqnarray}
and the equation $(\xi_1+\xi_2)(h - \epsilon) = 0$ is separated into two
decoupled equations for $\xi_1$ and $\xi_2$ variables,
\be
2\xi_j p_{\xi_j}^2 + \frac{m^2}{2\xi_j} + \frac{{\tilde\omega_z}^2}{8}\xi_j^3 -
(\epsilon + {\tilde\omega_L^{\prime\prime}}m)\xi_j + \lambda = (-1)^j c,
\; j=1,2 \; .
\label{seppar}
\ee

Simple manipulations define the separation constant
\be
c = a_z - {\tilde\omega_{\!\rho}}^2\rho^2 z \; ,
\label{rln}
\ee
which is the desired third integral of motion.
Here $a_z$ is the $z$-component of the Runge-Lenz vector
\be
{\mathbf a} = {\mathbf p}\times{\mathbf l} +
\lambda\frac{\mathbf r}{r}\;,
\ee
which is a constant of motion for
the pure Coulomb system
(i.e., when $\omega_{\rho} = \omega_z = 0$) \cite{LandauLifshitz_2003}.
The quantum mechanical counterpart of the
integral of motion, Eq.~(\ref{rln}), does not commute
with the parity operator, and we should expect the degeneracy
of quantum levels.

Due to the separability of the motion in the
parabolic coordinate system,
the eigenfunctions of the corresponding Schr\"odinger
equation can be expressed in the form
$\psi({\mathbf r}) = f_1(\xi_1)\,f_2(\xi_2)\, e^{im\varphi}$,
where the functions $f_j$ are the solutions of the equations
\begin{eqnarray}
&&\frac{d}{d\xi_j}\bigg(\xi_j\frac{d f_j}{d\xi_j}\bigg)
\!\!-\!\! \frac{1}{4} \bigg[ \frac{m^2}{\xi_j}+
\frac{\tilde\omega_z^2}{4}\xi_j^3 -
2(\epsilon+\tilde\omega_L^{\prime\prime} m)\xi_j+\nonumber\\
&&+
2\lambda - (-1)^j 2c\, \bigg]\,f_j = 0,\quad j=1,2  \; .
\label{fgeqs}
\end{eqnarray}
Eqs.~(\ref{fgeqs}) can be solved numerically.
Let $n_1$ and $n_2$ be the nodal quantum numbers of the functions
$f_1$ and $f_2$, respectively.
Note that Eqs.~(\ref{fgeqs}) are coupled by the constants of motion and,
therefore, both functions depend on all three quantum numbers $(n_1,n_2,m)$.
The states $\vert n_1,n_2,m\rangle$ have, in the coordinate representation,
the explicit form
\be
\psi_{n_1,n_2,m}({\mathbf r}) =
f_{n_1}^{(n_2,m)}(\xi_1)\,f_{n_2}^{(n_1,m)}(\xi_2)\,
\frac{e^{{\mathrm i}m\varphi}}{\sqrt{2\pi}}  \; .
\label{psif1f2}
\ee
The simple product of these functions
has no definite parity. Since
${\mathbf r} \to -{\mathbf r} \Leftrightarrow \{\xi_1\to\xi_2, \xi_2\to\xi_1,
\varphi\to\varphi + \pi\}$, the even/odd eigenfunctions are
constructed as
\begin{eqnarray}
\psi^{(\pm)}_{N,k,m}({\mathbf r})=
\frac{e^{im\varphi}}{\sqrt{2}}\,
[f_{n_1}^{(n_2,m)}(\xi_1)\,f_{n_2}^{(n_1,m)}(\xi_2)
\!\pm\!
(-1)^m f_{n_2}^{(n_1,m)}(\xi_1)\,f_{n_1}^{(n_2,m)}(\xi_2)] \; ,
\label{psipar}
\end{eqnarray}
where $N = n_1+n_2$ and $k = \vert n_1-n_2\vert$.
These states
are the eigenfunctions of $h$, $l_z$, $c$, and the parity operator.
For $c > 0$, the eigenstates (\ref{psipar})
appear in doublets of different parity and, therefore, of a different
total spin.
For $c = 0$ in Eqs.~(\ref{fgeqs}), $f_1 = f_2$ and, obviously,
only the states with parity $\pi = (-1)^m$ exist. In this case, the
dynamical symmetry is $O(4)\supset O(2)\otimes O(2)$.

For the magnetic field
$\omega_L^{\prime\prime\prime}
\equiv (4\omega_z^2 - \omega_0^2)^{1/2}$, we obtain
the ratio $\omega_z/\omega_\rho = 1/2$.
Hamiltonian (\ref{relham}), expressed
in the {\it oblate} spheroidal coordinates
($r_1 = [z^2+(\rho+d)^2]^{1/2}$, $r_2 = r$),
is separable for $m = 0$ (at $d\to\infty$).
For $m \neq 0$, the term
$m^2/\rho^2$ and, consequently, Hamiltonian (\ref{relham}),
is not separable in these coordinates. For $m = 0$, the cases
$\omega_z/\omega_\rho = 1/2$ and $2$ are equivalent,
if we interchange the $\rho$ and $z$ coordinates and,
hence, the additional integral of motion is
$\vert a_\rho - {\tilde\omega_{\!z}}^2 z^2\rho \vert$.
For $m\neq 0$, the recipe \cite{blum} enables one to obtain
the following integral of motion
\be
C = [(a_\rho - {\tilde\omega_z}^2 z^2\rho)^2 + a_\varphi^2 +
4m^2\tilde\omega_z^2 r^2]^{1/2} \; ,
\label{secint}
\ee
where $a_\rho$ and $a_\varphi$ are the $\rho$ and $\varphi$
components of the Runge-Lenz vector, respectively. Due to the
existence of three independent integrals of motion, $h$, $m$ and $c$,
which are in involution, the dynamics for $m \neq 0$, although
non-separable, is integrable.
The further analysis for $m=0$ is similar to the
previous one and we omit it here. The corresponding dynamical symmetry
is $O(4)\supset O(3)\supset O(2)$.
\end{itemize}

\subsection{Dimensionality effects in ground-state transitions of
two-electron quantum dots}
\label{subsec:IV.D}

Experimental data, including transport measurements
and spin oscillations in the ground state in a perpendicular magnetic field
for two-electron QDs, can be explained by the interplay between
electron correlations, lateral confinement, and a
magnetic field.
Using a $2D$ quantum dot model, one is able to reproduce a general trend for
the first singlet-triplet transitions observed in two-electron QDs
in a perpendicular magnetic field.
However, the experimental positions of the singlet-triplet transition
points are systematically higher \cite{kou,ni}.
The ignorance of the third dimension is the most evident source
of the disagreement, especially, in vertical QDs \cite{Din,ron,bruce}.

\subsubsection{First singlet-triplet transition in a two-electron quantum dot}
\label{subsec:IV.D.1}

As is discussed above, the additional energy Eq. (\ref{aden}) is one of the
major quantities which should be calculated and compared with the available
experimental data.
Here we are concerned with
$\mu(1)$ and $\mu(2)$ only, which we calculate with the aid of
the model considered in Sec. \ref{subsec:IV.C}.
The first is simply the harmonic oscillator
energy for a single electron in the dot, $\mu(1) = E(1,B)$, where
$E(N,B)$ denotes the total energy of the QD with $N$
electrons under a magnetic field of strength $B$.
The latter can be split into contributions
from the relative and center-of-mass motion
$E_{CM}$, where $E_{CM} = E(1,B)$.
The addition energy (direct probe of electron correlation
in the dot) takes the form
$\Delta\mu \equiv \mu(2) -\mu(1) =  \hbar\omega_0\epsilon - E(1,B)\,$,
where $\epsilon$ is the relative energy determined by Hamiltonian (\ref{relham})
and $E(1,B)=\hbar\omega_{\rho}+\hbar\omega_z/2$.

\begin{figure}[hb]
\centerline{\psfig{figure=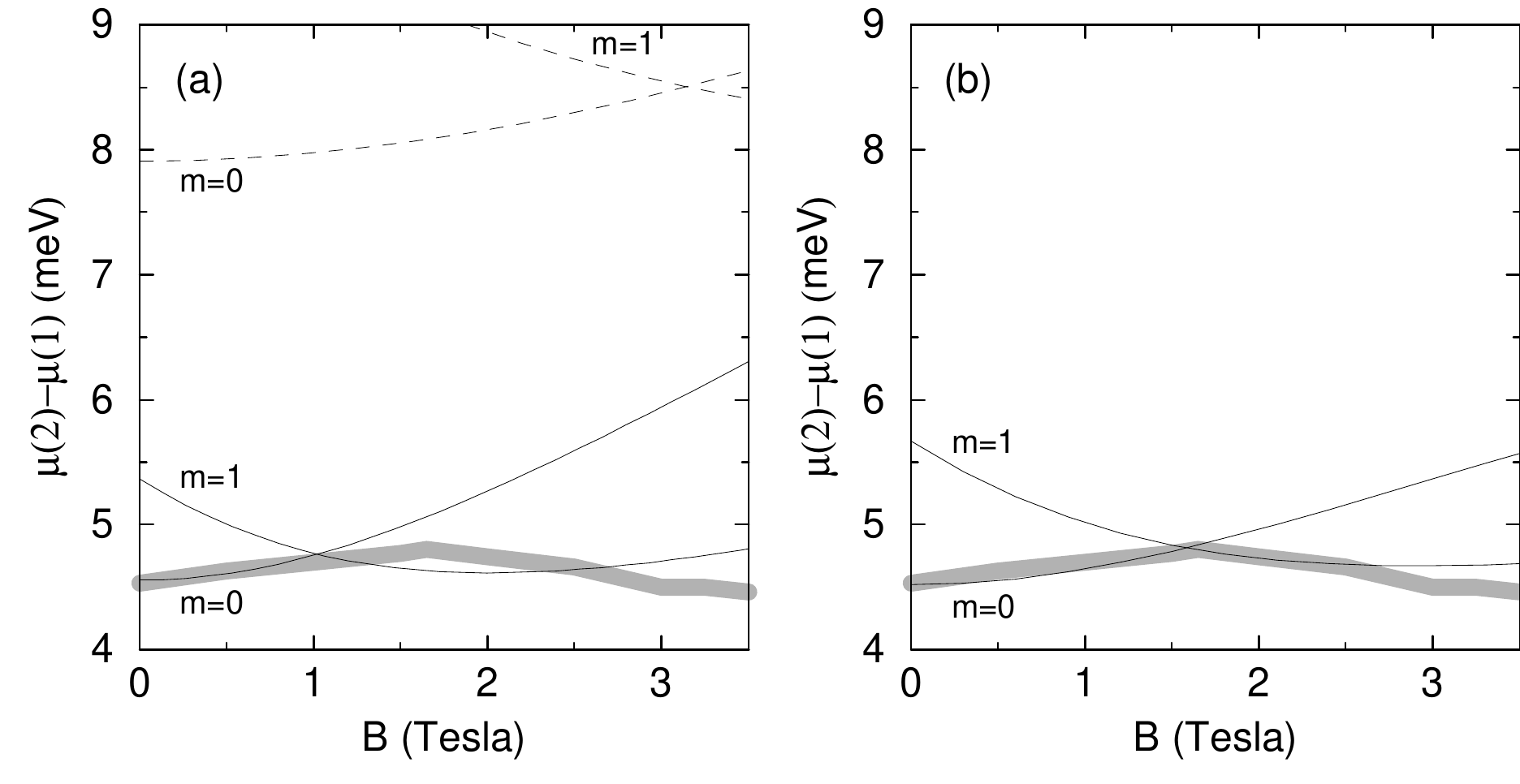,width=3.4in,clip=}}
\caption{The addition energy  $\Delta\mu$
(a shaded curve from experiment \cite{as}). Part (a) shows the theoretical
$\Delta\mu$
from a $2D$ quantum dot model, with $\hbar\omega_{0} = 5.4\,$meV
(dashed) and $\hbar\omega_{0} = 2.3\,$meV
(solid). Part (b) shows $\Delta\mu$ from a 3D model with
$\hbar\omega_{0} = 2.6\,$meV and $\omega_{z}/\omega_{0} = 2.4$ (solid).
From \cite{NSR}.
}
\label{fig44}
\end{figure}

In a number of papers (e.g., \cite{as,Sch-al95,Kou-al97,Pal-al94})
$\mu(1,B)$ has been used to estimate the confining frequency
$\hbar\omega_{0}$ in a $2D$ model of the QD. Indeed, with
$\hbar\omega_{0}=5.4\,$meV $(\hbar\omega_z=0)$,
one obtains a very satisfactory fit to $\mu(1)$ \cite{as}.
However, with this $\hbar\omega_{0}$, neither $\Delta\mu$
(which is by almost a factor 2 too large) nor the value for $B$, where the
first singlet-triplet transition occurs, are reproduced correctly, as is
obvious from Fig.~\ref{fig44}a.  It has been argued that for the increasing
magnetic field,
$\mu(N,B)$ might not follow the behaviour modelled by a simple QD with a
constant confining frequency, \cite{Sch-al95, Pal-al94}.
It was suggested \cite{NSR}
to extract $\hbar\omega_{0}$ from the difference of the chemical potentials
$\mu(2,0)-\mu(1,0)$ at zero magnetic field. Such a procedure, with
$\hbar\omega_{0}=2.3\,$meV $(\lambda=1.66)$, leads
to the first singlet-triplet transition $(m=0,S=0)\Rightarrow (m=1,S=1)$
at $B=1.02$ T (the Zeeman term is absent) (see Fig.\ref{fig44}a).
This value differs from
the experimental value of $B\approx 1.5$ T only by
about 30\%, contrary to
the difference of more than a factor of 2 with
$\hbar\omega_{0}=5.4\,$meV (dashed line).

The discrepancy of 30\% vanishes if one proceeds to a 3D description
of the QD. In this case, $\hbar\omega_{0} = 2.6\,$meV $(\lambda=1.56)$
is needed to
match $\mu(2,0)-\mu(1,0)$, only slightly different from the 2D case,
but the first singlet-triplet transition occurs now at $B = 1.59$ T
(see Fig.~\ref{fig44}b). If one includes the contribution from the
Zeeman energy
$E_Z=\frac{1}{2}\mu^*B[1-(-1)^m]$,
with $\mu^*=g_L\mu_B$ $(g_L=-0.44)$, then
this value reduces to $B = 1.52$ T in a good
agreement with the experiment. Of course, this agreement is achieved
by tuning a second parameter, available in the $3D$ case, namely
$\omega_{z}/\omega_{0}= 2.4$, i.e. the ratio of vertical to
lateral confinement. On the other hand, a rough estimate assuming
$\omega_z/\omega_0 \sim d_0/d_z$,
with the experimental value $d_z = 175\,$\AA, reveals a lateral size of
$d_0 \approx 420\,$\AA$\,$  which is of the correct order of
magnitude, although the exact lateral extension in
the experiment is not known \cite{as}. The analysis shows that, in contrast
to a $2D$ description, the $3D$
description provides a consistent way to describe the energy spectrum for small
$B$, the value of the magnetic field for the first singlet-triplet
transition, and the ratio of lateral to vertical extension of the dot.
The details of the recent study of excited states in two-electron
vertical QDs \cite{ni} confirm this point of view (see below).

It has been predicted that the ground state of an $N$-electron QD at
a high magnetic field, becomes the spin polarized maximum density droplet (MDD)
\cite{mac}. In the MDD case the single-particle orbitals in the lowest
Landau level become singly occupied. A detailed discussion of different
phases in $N$-electron 2D QDs in a magnetic field can be found in \cite{RM}.
The spin-polarized droplet of electrons in the lowest Landau level
has the lowest possible total angular momentum $L=N(N-1)/2$, compatible
with the Pauli principle. For a two-electron QD it is expected that the
MDD occurs after a first singlet-triplet transition.

\subsubsection{Topological transitions in a two-electron quantum dot}
\label{subsec:IV.D.2}

Theoretical calculations \cite{gs,gs1,haw,Wag,Din,DiNa99,PS96}
assert that after the first singlet-triplet transition the increase of
the magnetic field induces several ground state transitions to higher
orbital-angular and spin-angular momentum states.
This issue was addressed in the transport study of the correlated
two-electron states up to 8 T  and 10 T in a lateral \cite{ihn} and vertical
\cite{ni,ni2} QDs, respectively. It is quite difficult to detect the structure
of the ground states after the first singlet-triplet transition in a
lateral QD due to a strong suppression
of the tunnelling coupling between the QD and contacts. Altering the lateral
confinement strengths exhibits the transitions, beyond the first singlet-triplet
transition, reported in vertical QDs \cite{ni}. In fact, the variation of the
confining frequency with
{\it the same experimental setup} opens a remarkable opportunity
for a consistent study of effects of the magnetic field on electron
correlations.

Three  vertical QDs, with different lateral confinements, have been studied
in the experiment \cite{ni}.  In all samples, clear shell structure effects
for the electron numbers $N=2,6,...$ at $B=0$ T have been observed, implying
a high rotational symmetry.
Although there is a sufficiently small deviation from this symmetry in a sample,
classified by Nishi {\it et al.} \cite{ni} as C, a complete shell
filling for two and six electrons was observed.
Such a shell structure is generally associated with
a $2D$ harmonic oscillator $x-y$ confinement \cite{kou}.
However, a similar
shell structure is produced by a $3D$ axially symmetric
HO, if the confinement in the $z$-direction, with
$\omega_z=1.5\omega_0$, is only slightly larger than the lateral confinement
($\omega_x=\omega_y=\omega_0$). In this case, six electrons fill the lowest
two shells of the Fock-Darwin energy levels with $n_z=0$.
It was also found that the lateral confinement frequency, for the axially
symmetric QD, decreases with the increase of the electron number \cite{Mel},
since the screening in the lateral plane becomes stronger with large
electron numbers. In turn, this effectively increases the ratio
$\omega_z/\omega_0$, making the dot
effectively more "two-dimensional", since the
vertical confinement is fixed by the sample thickness. Indeed, the $N$-dependence
of the effective lateral frequency is observed in \cite{ni}.

Fitting the $B$-field dependence of the first and second Coulomb oscillation
peak positions to the lowest Fock-Darwin energy levels
of the $2D$ HO with the potential $m^*\omega_0^2r^2/2$,
Nishi {\it et al.} \cite{ni} estimated $\omega_0$
for all three samples A, B, and C. Although the general
trend in the experimental data is well reproduced by the $2D$ calculations,
the experimental positions of the singlet-triplet transition
points are systematically higher.
Different lateral confinements in the above experiment are achieved
by the variation of the electron density, without changing the sample thickness.

\begin{figure}[bt]
\centerline{
\includegraphics[height=0.22\textheight,clip]{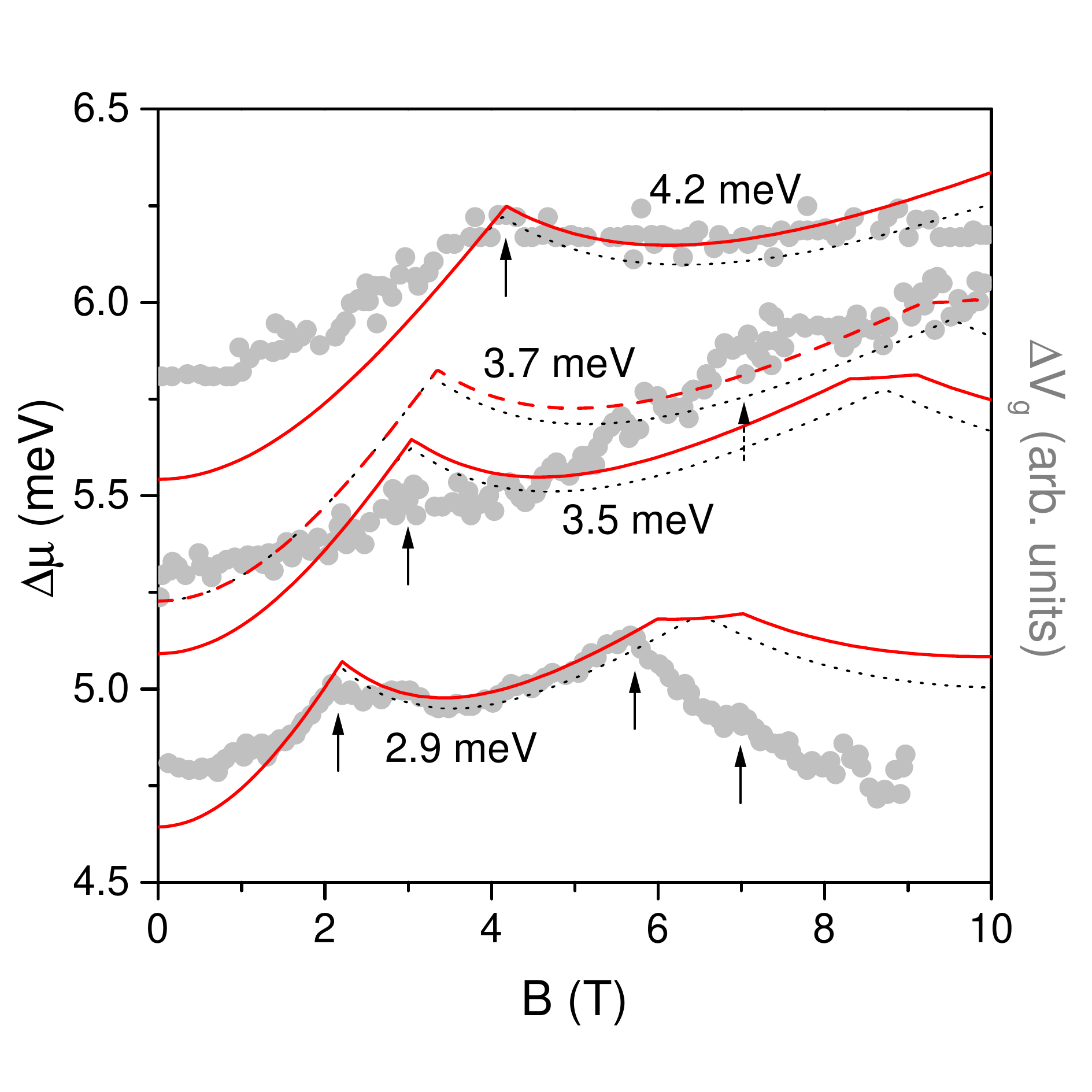} }
\caption{(Color online) The magnetic dependence of the addition energy
$\Delta \mu$ in two-electron QDs with lateral confinement
$\hbar\omega_0=4.2,3.7,3.5,2.9$ meV (the first, second, and fourth values
are experimental values for the samples A, B, and C, respectively \cite{ni}).
The confinement in the third ($z$) direction $\hbar\omega_z=8$ meV is fixed for all
samples. The results for $|g^*|=0.3(0.44)$ are connected by solid (dotted) line
for $\hbar\omega_0=4.2,3.5,2.9$ meV and by dashed (dotted) line for
$\hbar\omega_0=3.7$ meV.
The solid gray lines display the experimental
spacing $\Delta V_g$ as a function of $B$.
The arrows identify the position of experimental ground state
transitions \cite{ni}. From \cite{nen3}.}
\label{fig7}
\end{figure}

Using the "experimental" values for the lateral confinement and
the confinement frequency $\omega_z$ as a free parameter,
it was found, by means of the exact diagonalization of Hamiltonian (\ref{hamr})
for two interacting electrons,
that the value $\hbar\omega_z=8$ meV provides the best fit for
the positions of kinks in the addition energy (\ref{aden}) with the Zeeman
energy $E_Z$ \cite{nen3}. In Fig.\ref{fig7},
the magnetic dependence of the experimental spacing between
the first and the second Coulomb oscillation peaks
$\Delta V_g=V_g(2)-V_g(1)$ for the samples A--C is shown.
The spacing can be transformed to the addition energy $\Delta \mu$ \cite{ni,ni2}.
In the $\Delta V_g-B$ plot, the ground state transitions appear as upward
kinks and shoulders. It was found from the Zeeman splitting
at high magnetic fields that $|g^*|=0.3$ \cite{ni2}
and the addition energy is calculated with this and the bulk values.

The  experimental position of the first singlet-triplet transitions
at $B=4.2,3,2.3$ T in samples A, B, and C, respectively is reproduced quite
well in the calculations (see Fig. \ref{fig7}).
When the magnetic field is low, the difference between
the calculations with different $|g^*|$ factors is negligible.
Upon decreasing the lateral confinement $\hbar\omega_0$ from the
sample A to the sample C (the increase of the ratio $\omega_z/\omega_0$),
the Coulomb interaction becomes dominant in the
interplay between electron correlations and the confinement \cite{NSR}.
In turn, the smaller the lateral confinement at fixed thickness (the stronger
the electron correlations), the smaller the value of the magnetic field where
the singlet-triplet transitions or, in general, crossings between excited
states and the ground state, may occur.

There is no signature of the second crossing in the ground
state for the sample A at large $B$ (up to $10$ T). Here, the ratio
$\omega_z/\omega_0\approx 1.9$, and the effect of the
third dimension is the most visible: the confinement has a dominant role in the
electron dynamics, and a very high magnetic field is required to observe the next
transition in the ground state due to
electron correlations. Thus, the MDD phase survives until very high magnetic
fields ($B \sim 10$ T).

A second kink is observed at $B=7$ T in the sample B \cite{ni}.
The calculations with the "experimental" lateral confinement
$\hbar\omega_0=3.7$ meV produce the second kink at $B=9.5$ T, which is located
higher than the experimental value. The slight decrease of the lateral frequency
until $\hbar\omega_0=3.5$ meV shifts the second kink to $B=8.7$ T, improving
the agreement with
the experimental position of the first singlet-triplet transition as well.
In addition, the use of $|g^*|=0.3$ (instead of the bulk value) with
the latter frequency creates a plateau, which bears resemblance to the experimental
spacing $\Delta V_g$. However, there is no
full understanding of this kink. It seems,
there is an additional mechanism responsible for the second kink in the sample B.
\begin{figure}[b]
\centerline{\psfig{figure=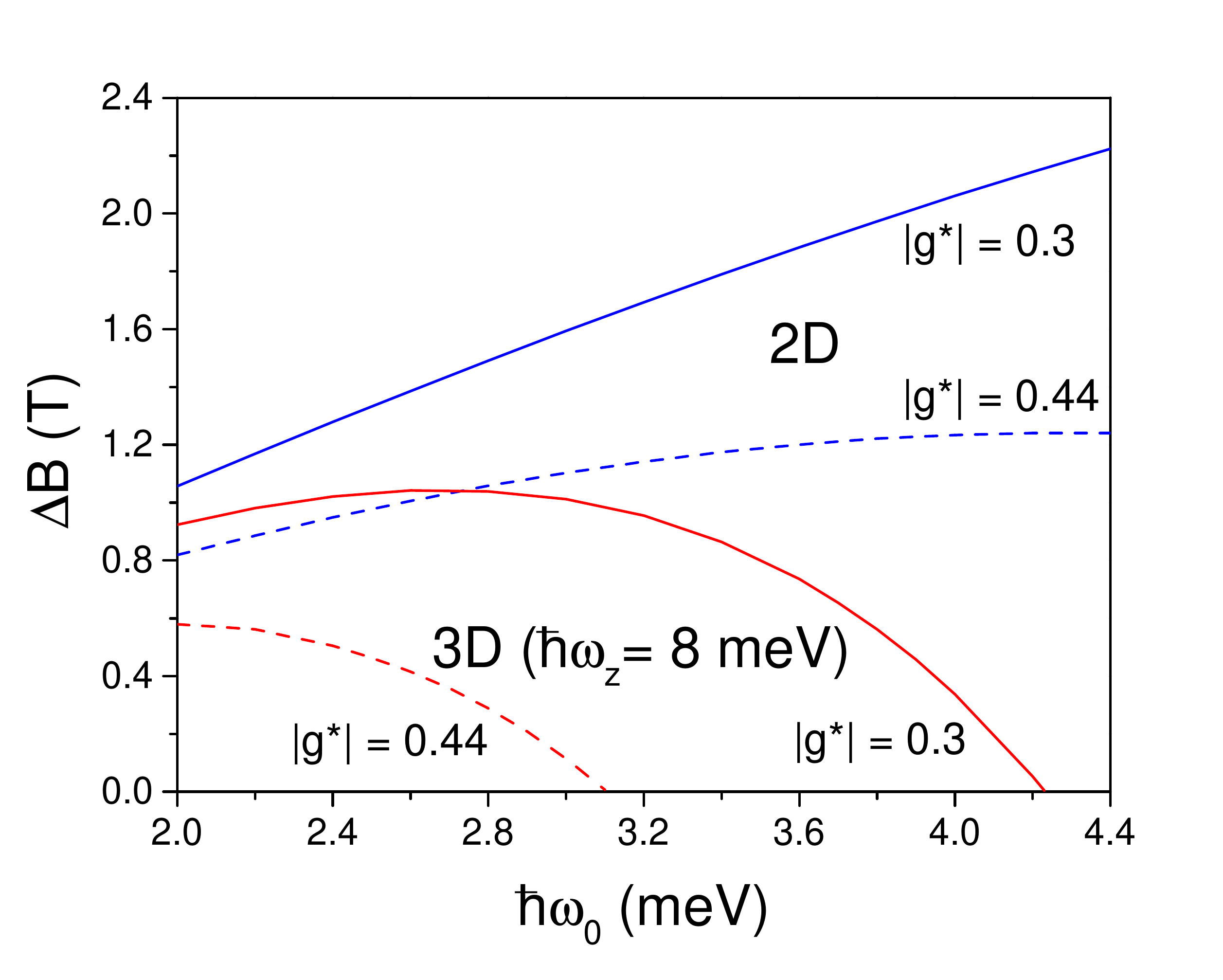,width=2.2in,clip=} }
\caption{(Color online) The interval $\Delta B$ where the singlet
state $(2,0)$ survives as a function of the lateral confinement for $2D$ and $3D$
calculations. The confinement in the third ($z$) direction
$\hbar\omega_z=8$ meV is fixed for the $3D$ calculations.
From \cite{nen3}.}
\label{fig9}
\end{figure}
In the sample C, the first experimental singlet-triplet
transition occurs at $B=2.3$ T, while the signatures of the second and the third
ones are observed at $B\approx 5.8, 7.1$ T, respectively.
The $2D$ calculations (with the "experimental" values
$\hbar\omega_0=2.9$ meV, $|g^*|=0.44$)
predict the first, second, and third singlet-triplet crossings at lower
magnetic fields: $B=1.7, 4.8, 5.8$ T, respectively.
The results can be improved to some degree with $|g^*|=0.3$.
To reproduce the data for $\Delta \mu$,
Nishi {\it et al.} \cite{ni} increased the lateral confinement
($\hbar\omega_0=3.5$ meV, $|g^*|=0.44$). As a result, the first, second, and
third singlet-triplet transitions occur at $B=2,6.3, 7.5$ T, respectively.
Evidently, $2D$ calculations overestimate
the importance of the Coulomb interaction. The increase of the lateral
confinement simply weakens the electron correlations in such calculations.
In contrast, the $3D$ calculations reproduce quite well the positions of all
crossings, with the "experimental" lateral confinement
$\hbar\omega_0=2.9$ meV at $B=2.3,5.8,7.1$ T (see discussion below).

One of the questions, addressed in the experiment \cite{ni}, is related to
a shoulder-like structure observed in a small range of values of the magnetic
field (see Fig. \ref{fig7}). This structure is identified as
the second singlet state $(2,0)$ that persists till the next crossing with the
triplet state $(3,1)$. According to Nishi {\it et al.} \cite{ni}, the ground
state transition from the triplet $(1,1)$ state to the singlet $(2,0)$ state
is associated with the collapse
of MDD state for $N=2$. Therefore, a question arises: at what conditions it
would be possible to avoid the collapse of the MDD phase (in general, to preserve
the spin-polarized state); i.e., at what conditions the singlet $(2,0)$ state
will never show up in the ground state.
In fact, the collapse of the MDD depends {\it crucially} on the value of the
lateral confinement. It was found that in the $2D$ consideration,
the $(2,0)$ state always exists for experimentally available lateral
confinement (see  Fig. \ref{fig9}).
Moreover, in this range of $\hbar\omega_0$, the $2D$ approach predicts the
monotonic increase for the interval of the values of the magnetic field
$\Delta B$, at which the second singlet
state survives with the increase of the lateral confinement.
In contrast, in the $3D$ calculations, the size of the interval
is a vanishing function of the lateral confinement for a fixed thickness
($\hbar\omega_z=8$ meV).
It is quite desirable, however, to measure this interval to draw
a definite conclusion.

As discussed above, the decrease of the confinement, at a fixed thickness,
increases the dominance of the electron correlations in the electron dynamics.
For a low enough electron density, Wigner \cite{W1} predicted that
electrons should localize, creating an ordered spatial structure
that breaks the complete translational symmetry of the homogeneous electron
gas. Therefore, the decrease of the confinement, related to the decrease of the
electron density \cite{Mel,ni}, creates the favourable conditions for the
onset of electron localization. According to a general wisdom, the Wigner
crystallization in QDs, whose localized states are referred to as
Wigner molecules \cite{Maks,mukn}, should occur at significantly larger
densities than in bulk. This is based on the argument that in QDs
potential-energy contributions can easily exceed the kinetic
terms and, therefore, electronic motion can be effectively quenched
by manipulating the external confinement and/or an applied magnetic field.

The dynamics of electrons in QDs is determined by the ratio
\begin{equation}
\label{RW}
 R_W=\bar{\ell_0}/a^*\equiv \sqrt{2}\lambda=k/\bar{\ell_0}\hbar \omega_0\,,
\end{equation}
where $\bar{\ell_0}=\sqrt{\hbar/m^*\omega_0}$ is the oscillator length
with the mass $m^*$ and $\hbar \omega_0$
can be determined from the volume conservation condition
(see Sec. \ref{subsubsec:IV.A.1}).
For a strong confinement, $R_W\rightarrow 0$, and the Coulomb energy can be
neglected. In this case, the interaction acts merely as a small perturbation.
In the opposite limit of a weak confining potential,
$\omega_0\rightarrow 0\Rightarrow R_W\rightarrow \infty$,
the interaction becomes arbitrarily strong and the system will undergo a phase
transition to the Wigner crystal.

For the QDs considered in the
experiments of Nishi {\it et al.} \cite{ni}, in the $2D$ approach for a
circular dot, $R_W\sim 3$. For a $2D$ two-electron QD,
it is predicted that the Wigner molecule can be formed for $R_W\sim 200$
at zero magnetic field \cite{YL}, or at very high magnetic field \cite{szaf}
for $\hbar\omega_0\sim 3$meV and small $R_W$,
such as in the discussed experiments.
In the $3D$ axially symmetric QDs the ratio between vertical and
lateral confinements (anisotropy) may, however, affect the formation of the
Wigner molecule.
\begin{figure}
\begin{center}
\includegraphics[height=0.52\textheight,clip]{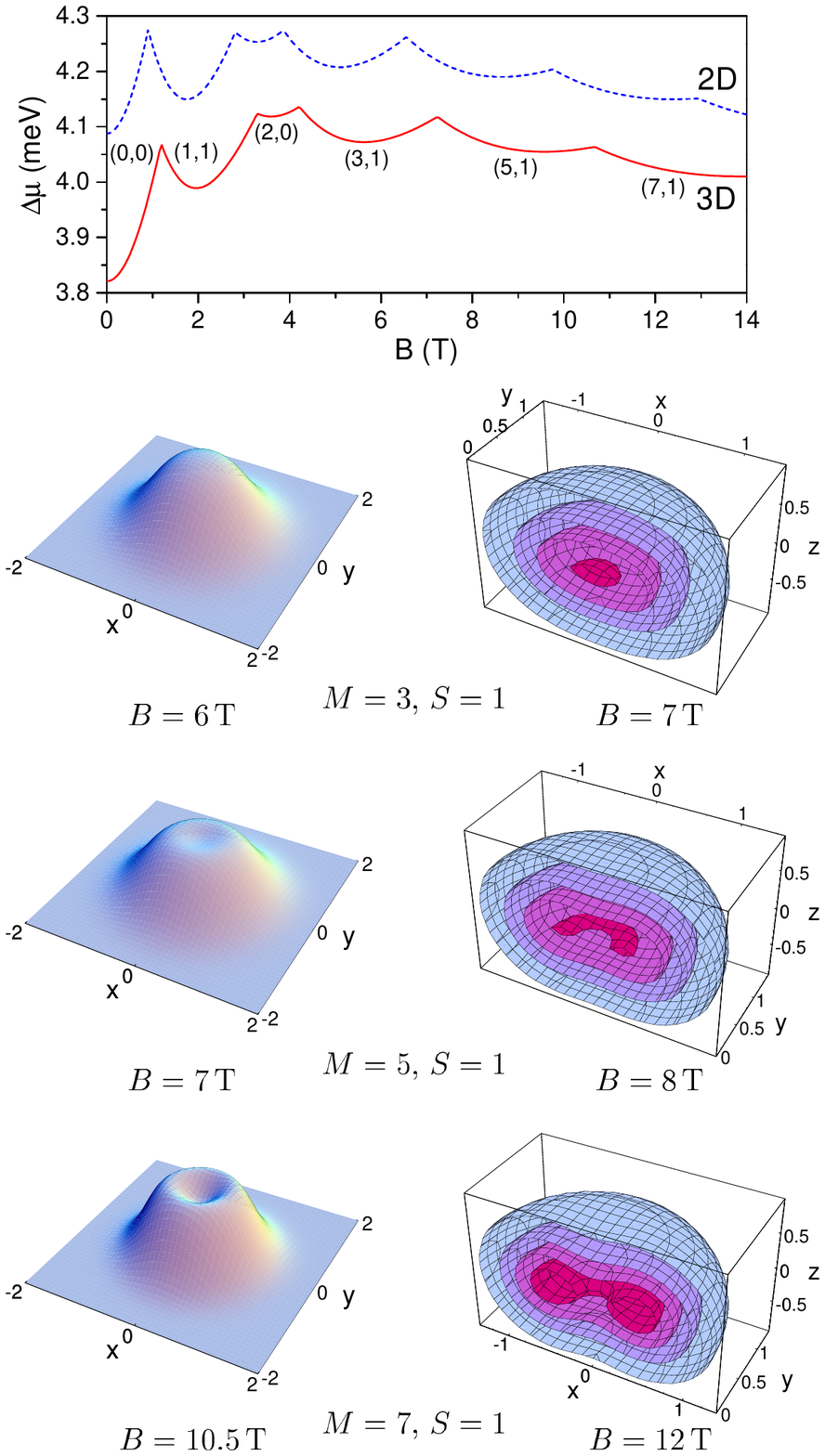}
\end{center}
\caption{(Color online)
Top: Magnetic dependence of the ground state
in $2D$ and $3D$ ($\hbar\omega_z=8$ meV) approaches for a lateral confinement
$\hbar\omega_0=2$ meV. The $2D$ (left) and $3D$ (right) electron densities are
displayed for different ground states (M,S) at corresponding magnetic fields.
The largest $3D$ density grows, with the increase of the magnetic field, from
the central small core over a ring to the torus. From \cite{nen3}.
}
\label{fig10}
\end{figure}
This problem can be studied by dint of the electron density
\be
n({\bf r}) = \int \left[\,|\Psi({\bf r},{\bf
r}^\prime)|^2+|\Psi({\bf r}^\prime,{\bf r})|^2 \right] {\rm d}{\bf
r}^\prime \; ,
\label{el-density}
\ee
when one electron is at a position $\bf r$, while another is located at
a position ${\bf r}^\prime$.
A criterion for the onset of the crystallization in two-electron QDs can be the
appearance of a local electron density minimum in the center of the dot
\cite{mukn,Cref}. For $2D$ QDs, this leads to the radial modulation in the
electron density, resulting in the formation of rings and roto-vibrational
spectra \cite{ton,yan04}.

The 3D analysis of the electron density
(see Fig. \ref{fig10}) indicates that, at $B > 7.25$ T, the
triplet state $(5,1)$ may be associated with the formation of the Wigner
molecule, in agreement with the above discussion. There is an evident
difference between the $2D$ and $3D$ approaches: the $2D$ calculations
predict the crystallization at the lower magnetic field ($\Delta B\sim 1$ T).
The further increase of the magnetic field leads
to the formation of a ring and a torus of the maximal density
in $2D$- and $3D$-densities, respectively. Notice that if the geometrical
differences are disregarded, $3D$ evolution of the ground state can be
approximately reproduced in $2D$ approach with the effective charge
concept \cite{NSR,sreen}.

\subsubsection{Effective charge}
\label{subsec:IV.D.3}

It seems evident that in QDs, the Coulomb interaction couples lateral and
vertical coordinates, and the problem, in general, is non-separable.
By means of the exact diagonalization
of 3D effective Hamiltonian, one can study
the effect of the vertical confinement on the energy spectrum \cite{Mel}.
This can be done, however, only for QDs
with a small number of electrons. Even in this case there are difficulties
related to the evaluation of $3D$ interaction matrix elements.

\begin{figure}
\centerline{
\includegraphics[width=0.55\textwidth]{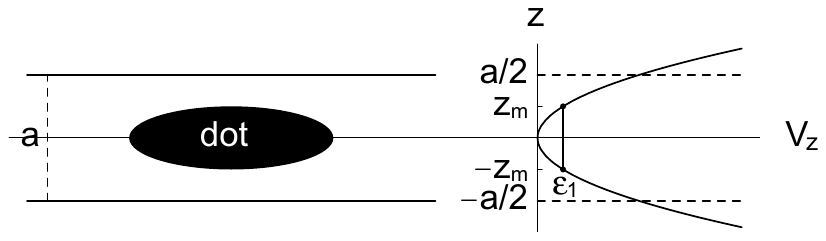} }
\caption{
Left: the localization of QD in the layer of the thickness $\mathrm{a}$.
Right: the schematic representation of the position of zero-point
motion in the parabolic confinement relative to the layer thickness.
From \cite{sreen}.}
\label{fig12}
\end{figure}

The thickness of QDs is much smaller in comparison with the
lateral extension. Therefore, the vertical confinement, with
$\hbar\omega_z$, is much stronger then the lateral confinement, with
$\hbar\omega_0$, and this fact is usually employed to justify
a $2D$ approach to QDs.
One can develop, however, the procedure that accounts for
the thickness of QDs \cite{sreen}.
Note that there is a nonzero contribution from the
vertical dynamics, since the energy level,
available for each of noninteracting electrons in the $z$-direction, is
$\varepsilon = \hbar\omega_z(n_z+1/2)$. For the lowest state
$n_z = 0 \Rightarrow \varepsilon_1 = \frac{1}{2}\,
\hbar\omega_z $. Because of the condition $V_z(\pm z_m)\equiv
m^*\omega_z^2 z_m^2/2 = \varepsilon_1 $, one defines the
turning points: $z_m = \sqrt{\hbar/(m^*\omega_z)}$.
One may assume that the distance between turning points should not exceed
the layer thickness, i.e., $2z_m \leq a$ (see Fig.\ref{fig12}).
Owing to this inequality, the lowest limit for the vertical confinement in
the layer of thickness $a$ is $\hbar \omega_z \geq 4\hbar^2/(m^*a^2)$.
For typical GaAs samples, with the
thickness $a$ between 10\,nm and 20\,nm, this estimate gives the
minimal value for $\hbar\omega_z$ between 45\,meV and 11\,meV,
respectively. These estimates provide a genuine cause for
the use of the adiabatic approach \cite{RRM} in the case of QDs, since
\be
T_z(=2\pi/\omega_z)\ll T_0(=2\pi/\omega_0)\;.
\ee
To lowest order, the adiabatic approach consists of averaging
the full $3D$ Hamiltonian (\ref{hamr}) over the angle-variables
$\theta_{z_i} = \omega_{z_i} t$ (fast variables) of the
unperturbed motion $(k = 0)$ of two electrons after rewriting the
$(z_i, p_{z_i})$ variables in terms of the action-angle variables
$(J_{z_i}, \theta_{z_i})$.
As a result, the dynamics effectively
decouples into an unperturbed motion in the vertical direction,
governed by the potential $\sum_i V(J_{z_i}, \theta_{z_i})$,
and the lateral motion governed by the effective potential
$V_\mathrm{eff}(\{x,y\};\{J_{z}\})$ that contains the memory on $z$ dynamics
through the integrals of motion $J_{z_i}$ \cite{sreen}.
The effective electron-electron interaction
affects, therefore, only the dynamics in the lateral plane,
where the confining potential is parabolic.
Hence, the effective Hamiltonian for a two-electron QD is
\be
H_\mathrm{eff} = H_0+ E_{z} + V_\mathrm{int}^\mathrm{eff}\;,
\ee
where $E_z=\sum_i\varepsilon_i$ and $\varepsilon_i$
is the electron energy of the unperturbed motion
in the vertical direction. The term $H_0=\sum_{i=1}^2h_0(i)$
consists of the contributions related only to the lateral
dynamics in the $x-y$-plane of noninteracting electrons (see Eq.(\ref{rh})).

The main idea of the procedure is based on the consideration
of the Coulomb term determined by the
effective value $k_\mathrm{eff}$ (effective charge) with the appropriate
consideration of the vertical ($z$) dynamics, i.e.,
\be
\label{efVC}
V_C = k/r_{12} \approx k_\mathrm{eff}/{\rho}\;,
\ee
where $\rho = [(x_1-x_2)^2 + (y_1-y_2)^2]$.
Thus, the procedure of evaluating the effective charge
consists of two steps: (i) the averaging of the Coulomb term $V_C(\rho,z)$
over the angle variables in the $z$-direction, which gives the effective
$2D$ potential
\be
V_\mathrm{int}^\mathrm{eff}(\rho) = k f(\rho)/\rho\;;
\ee
(ii) the calculation
of the mean value of the factor $f(\rho)$ upon the non-perturbed lateral
wave functions, i.e.,
\be
\label{ef2d}
k_\mathrm{eff} = k\langle f(\rho)\rangle \equiv \langle
\rho\,V_\mathrm{int} ^\mathrm{eff}(\rho)\rangle\;.
\ee
As a result, one has to solve only the Schr\"odinger
equation for the $2D$ effective Hamiltonian where the full charge $k$ is
replaced by $k_\mathrm{eff}$.
The effective charge can be calculated as a quantum-mechanical
mean value of the Coulomb term with the aid of the Fock-Darwin $|n_r,m\rangle$
and a one-dimensional harmonic oscillator $|n_z\rangle$ states:
\be
k_{\mathrm{eff}} =  \langle \rho\,V_C(\rho,z)\rangle  =
k\, \langle (1+z^2/\rho^2)^{-1/2}\rangle\;.
\ee
 Since the lateral extension exceeds the
thickness of the QDs by several times, one may suggest to consider
the ratio $(z/\rho)^2$ as a  small parameter of theory.
\begin{figure}[ht]
\centerline{
\includegraphics[height=0.28\textheight,clip]{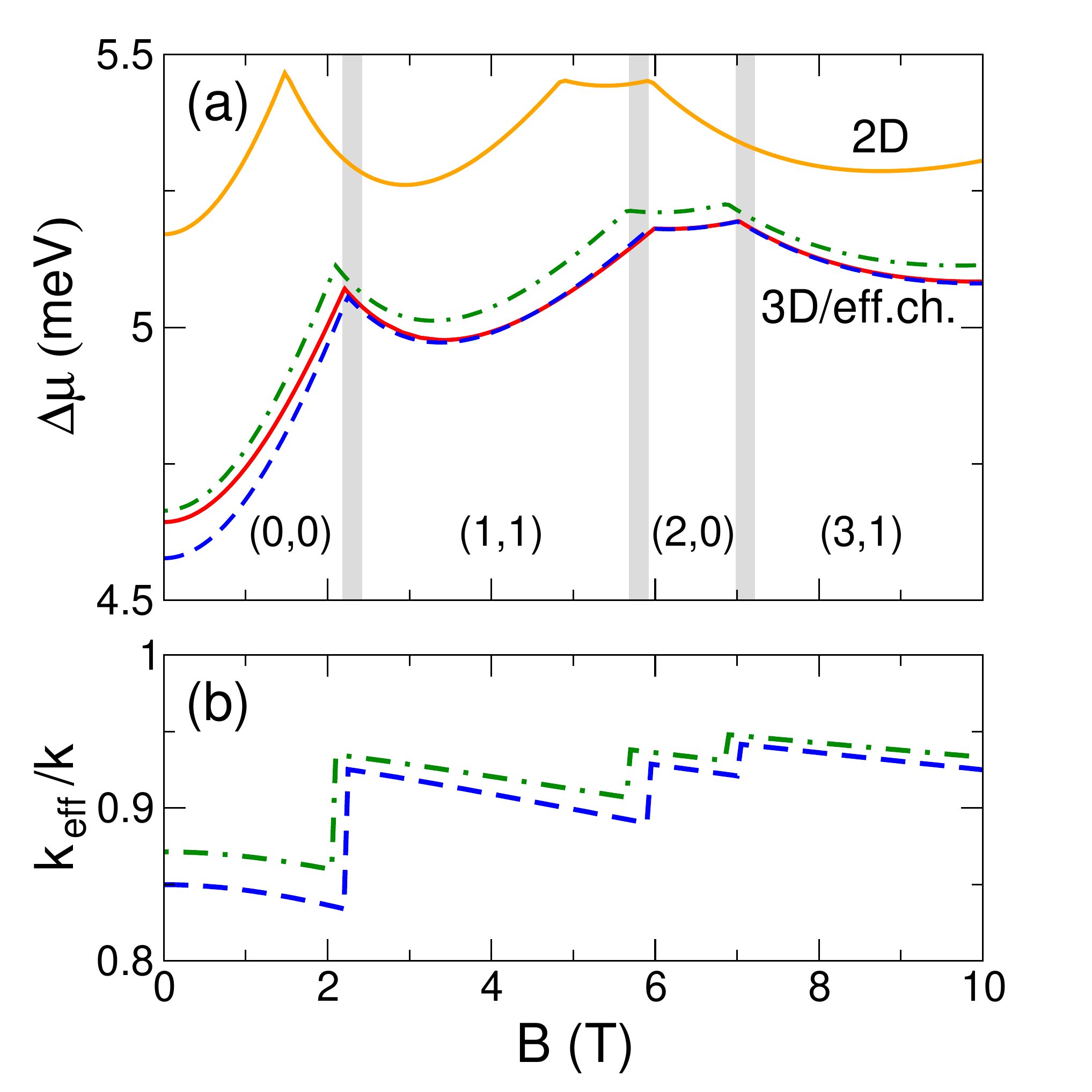} }
\caption{(Color online) (a)
The addition energy $\Delta\mu$ as a function
of the magnetic field in the parabolic model. The results of calculations
with a lateral confinement only
(the $2D$ approach, $\hbar\omega_0 = 2.9$\,meV, $|g^*| = 0.3$)
and full $3D$ approach \cite{nen3} ($\hbar\omega_z = 8$\,meV)
are connected by the thick (orange) and thin (red) lines, respectively.
The vertical gray lines indicate the position of the experimental
crossings between different ground states in a sample C \cite{ni}.
Ground states are labelled by
$(m,S)$, where $m$ and $S$ are
the quantum numbers of the operators $l_z$ and the total spin,
respectively.
The results based upon the adiabatic approximation, and
the plain quantum-mechanical averaging procedure, are connected
by dashed (blue) and dot-dashed (green) lines, respectively.
(b) The ratios $k_\mathrm{eff}/k$ as functions of the magnetic field,
based on the adiabatic approach and the plain quantum-mechanical averaging,
are connected by dashed (blue) and dot-dashed (green) lines, respectively.
From \cite{sreen}.}
\label{comef}
\end{figure}

In the considered cases of the parabolic, square, and triangular well
potentials \cite{sreen}, the value of the effective charge depends on
the (good) quantum number $m$ of the correlated
state. In particular, for the parabolic
confinement in the $z$-direction, the effective charge can be expressed in terms
of the Meijer $G$-function \cite{grad}
\begin{equation}
k_\mathrm{eff} = \frac{k}{\pi |m|!}\,
      G^{2,2}_{2,3}
  \biggl(
\frac{\omega_\rho}{\omega_z} \biggr|
 \begin{array}{c}
{1/2}\ \ {1/2}\\
0\,\,{m\!+\!1}\,\,0\\
\end{array} \biggr).
 \label{kaap}
\end{equation}
The screening due to the sample thickness is especially strong for the quantum states
with small values of the quantum number $m$. We recall that
these states determine the structure of ground state transitions
at small and intermediate values of the magnetic field.
Therefore, the screening provides a consistent way to
deal with the effect of thickness upon the position of the
singlet-triplet transitions \cite{ihn}. In particular, the screening should be
taken into account
for the analysis of the evolution of the energy difference between
singlet and triplet states in a magnetic field.
This energy is considered to be important
for the analysis of entanglement and concurrence in QDs  \cite{marcus}.

The comparison of the results with available experimental data
demonstrates a remarkable agreement and provides a
support to the validity of the approach (see Fig. \ref{comef}).
The results, based on the adiabatic approximation, are in a
better agreement with the full $3D$ calculations,
in contrast to those obtained with the aid of the plain quantum-mechanical
averaging procedure.
As discussed above, the adiabatic approach is based
on the effective separation of fast (vertical) and slow (lateral) dynamics,
with the subsequent averaging procedure. In contrast, the plain quantum-mechanical
averaging represents a type of perturbation theory based on the first order
contribution with respect to the ratio $z/\rho$ only. The higher-order terms may
improve the agreement at small magnetic field, since the vertical dynamics is
not negligible and affects the lateral dynamics.
The increase of the quantum number $m$, caused by the increase of the
magnetic field strength, reduces the orbital motion of electrons in the
vertical direction. The larger $m$, the stronger the centrifugal forces,
which induce the electron localization inside a plane, and,
therefore, the lesser importance of the vertical electron
dynamics. In the limit of strong magnetic field (large $m$) the dot becomes
rather a "two-dimensional" system. This explains the improvement of the accuracy
of the plain quantum-mechanical averaging procedures at
large $m$, i.e., for the ground states at high magnetic fields.
It follows that the $2D$
bare Coulomb potential becomes reliable in the $2D$ approaches,
for the analysis of the ground state evolution of QDs,
only at large magnetic fields.

\subsection{Symmetry breaking: mean field and beyond}
\label{subsec:IV.E.}
\subsubsection{Theoretical approaches}

The studies of different phases in QDs \cite{RM,yan} shed light on
a variety of aspects of strongly correlated {\it finite} systems, which
can be controlled experimentally.
This makes it possible to understand the common features of and the
principal differences between finite-size systems and conventional
condensed matter. One can  analyze the relation
between these systems, depending on the strength of electron-electron interactions
and of external fields. In fact, the
high controllability of QDs creates a remarkable opportunity to investigate
in detail novel strongly correlated phenomena in various experimental setups.

Several approaches, based on the parabolic
confining potential, including the exact diagonalization
method, a diffusion Monte Carlo (DMC) and a path integral Monte Carlo (PIMC),
density functional methods, have been applied with some degree of success
to analyze the ground state energies of the $N$-electron QD; see for the recent
reviews \cite{RM,yan,Saar}.
However, the exact diagonalization method is limited to relatively small
particle number, e.g., \cite{Pfan1,Pal-al94,eto,mik1,mik2,szaf04,ronci}.
The local-density approximation cannot properly describe the localized
states, due to the lack of exact cancellation of the direct and exchange Coulomb
interactions. The pros and cons of the density functional approach
related to the description of symmetry breaking phenomena are discussed in
\cite{Saar}.
PIMC  treats an interaction accurately \cite{eger,fil}.
However, it generates a thermal average of states with
different $L$ and $S$ quantum numbers, preserving only $S_z$ symmetry.
The energy for low-lying excitations becomes very small
in the low density limit (strong electron correlations), and, therefore,
the constraint on the temperature becomes extremely stringent.
DMC provides quite reliable results \cite{pedr,pedr1,ghos,umr}.
It contains some systematic "fixed-mode" error which are, however, smaller
than the systematic and statistical errors of PIMC.
A good introduction to the discussed methods, with applications to atomic
gases and quantum dots, can be found in \cite{elip}.
Among promising approaches, which could provide an
accurate treatment of correlation and spin effects for the ground and excited
states of QDs, we can mention the coupled cluster methods  \cite{17}.
The comparison of the coupled cluster method results with the available
analytical solutions \cite{Din,china,kai,taut} shows that this method takes
into account between 99\% and 91\% of the total energy from two to eight
electrons \cite{18}.

\subsubsection{Hartree-Fock approximation}

In many cases, a convenient starting point to treat finite systems is
a mean field description, like the Hartree-Fock (HF) approach
\cite{Rowe,Ring,BR86}.
In spite of its simplicity, the HF approach is extremely powerful.
The nonlinearity of the HF equations allows for spontaneous breaking of
fundamental symmetries of many-body Hamiltonians. In fact, this is a primary
mechanism that enables one to incorporate various correlations into a single
Slater determinant. Indeed, in a symmetry preserving
approach, the wave function of the system must have the same symmetry as the
many-body Hamiltonian. However, the most direct description of electron
localization can be obtained by using the unrestricted HF approximation with
broken rotational symmetry, when the wave function has a lower symmetry than
the Hamiltonian. The problem of the transition from a rotationally symmetric
solution to the so-called Wigner molecule \cite{Maks} and
crystal \cite{loz1,loz2,bol,bed}, where electrons occupy fixed spatial sites,
with the increase of the magnetic field, have been investigated
in the geometrically unrestricted Hartree-Fock approach
by M\"uller and Koonin \cite{mukn}. The mean field solutions with
broken rotational symmetry were interpreted as "intrinsic" states.
M\"uller and Koonin \cite{mukn} constructed the rotational states by
projecting the Hartree-Fock Slater
determinant onto eigenstates of a good angular momentum. These ideas,
borrowed from nuclear physics \cite{Ring}, resolved doubts about the validity
of the mean field approach. It is appropriate at this point to recall that
the assumption of deformed shape in the case of open-shell nuclei is essential
for the description of nuclear rotational spectra \cite{BM, sfrau,satw,sven}.
In fact, axially deformed shapes for QDs, with up to 40 electrons, have been
obtained  in the Hartree-Fock calculations with isotropic harmonic confinement
in all three spatial dimensions \cite{voblu}. The quadrupole moment
$q_{20}\sim \sum_{ij}\langle i|2z^2-x^2-y^2|j\rangle\rho_{ij}$, where
$\rho_{ij}$ is the Hartree-Fock density, vanishes for closed shells,
being nonzero for open-shell QDs.

In the simpler case of a two-electron $2D$ quantum dot in zero magnetic field,
Yannouleas and Landman \cite{YL} pointed out that
the excited-state energies of this system closely follow the
rotor sequence, when the repulsion-to-confinement ratio,
given by the Wigner parameter $R_W$, is large enough (${\sim}200$).
This was shown to be a proof of the crystallization of two electrons
in fixed positions in a rotating reference frame.
Quite remarkably, the hypothesized {\em rotating Wigner molecule}
fulfills at the same time the strict symmetry conditions of
quantum mechanics --circularity in this case-- and the obvious preference
for opposite positions, when repulsion is large enough.
This is the major difference from the above mentioned bulk case,
where the translation Hamiltonian symmetry is broken by the
crystallized state. For Wigner molecules, symmetries are preserved
in the laboratory frame and one must consider an intrinsic (rotating)
frame to notice the underlying deformation.
A similar situation is found
for particular states of two-electron atoms that have been much
investigated in physical chemistry (we address the reader to the
review paper \cite{Ber}).

Although the exact ground-state wave function of the two-electron artificial
atom can be obtained, at least numerically, it may seem
paradoxical that one needs also the excited states in order to ascertain
the existence of crystallization. In fact, this inability to
disentangle in a clear way the system intrinsic structure from its
full wave function can be taken as a weakness of the {\it ab initio},
symmetry preserving, approaches. In general, even in the cases where the exact
ground- and excited-state wave functions and energies are known, an intrinsic
deformation can only be inferred by comparing with the results for
simpler models, in which either symmetries are relaxed or the intrinsic
structure is imposed. A clear example of the former approach is given by the
unrestricted HF method for the ground state \cite{mukn,yan},
followed by the random-phase approximation (RPA) for excitations \cite{lor}.
On the contrary, the roto-vibrational model
for two electrons \cite{ton,yan04} could be included in the latter category.

One should be aware that when symmetries are relaxed, as in the Hartree-Fock
approach, artifacts or non-physical properties may appear \cite{RG}.
Therefore, a complete physical understanding requires
both exact results and model solutions. In this way, the system intrinsic
deformations are physically understood and, at the same time, artifacts can
be safely discarded. As an example, the exact solutions,
Hartree-Fock and RPA solutions were compared for two-electron
2D parabolic QD \cite{ton}.

\subsubsection{Roto-vibrational model}
\label{subsec:IV.E.3.}

In order to trace the evolution of the spectra from weak to strong
interaction and/or magnetic field, it is convenient
to introduce two dimensionless parameters:
\be
R_{\it mp} = \frac{k}{\ell_\Omega \hbar\Omega},\quad
W_{\it mp} = {\omega_c\over \Omega} \;.
\ee
Here, $\Omega=\sqrt{\omega_0^2+\omega_c^2/4}$, where
$\omega_c=eB/m^*c$ is the cyclotron frequency,
and the length $\ell_\Omega=\sqrt{\hbar/m^*\Omega}$ characterizes
the effective lateral confinement.
In the absence of a magnetic
field, $R_{\it mp}$ coincides with the Wigner parameter
$R_W$ (see Eq.(\ref{RW}), Sec. \ref{subsec:IV.D.2}).
Note also that $W_{\it mp}$ has a maximal value $W_{\it mp}=2$  that
corresponds to a zero confinement $\omega_0=0$.
With these dimensionless parameters, the Hamiltonian (\ref{hamr})
transforms into
\bea
\label{rv1}
{\cal H} &=&\frac{H}{\hbar\Omega}= \sum_{i=1}^N{\left[
-\frac{1}{2}\nabla^2 + \frac{1}{2} r^2+ \frac{W_{\it mp}}{2} \ell_z
\right]_i}+
R_{\it mp}\sum_{i>j=1}^{N}\frac{1}{r_{ij}}
+ \frac{g^* m^*}{2} W_{\it mp} S_z\; ,
\eea
where $r\rightarrow r/\ell_\Omega$. Note that we use the effective
electron-electron interaction, where the parameter $R_{\it mp}$ is associated
with the effective charge (a screening parameter), discussed in
Sec. \ref{subsec:IV.D.3}.

For the two-electron problem, introducing the standard
center of mass $(R,\Theta)$ and relative $(r,\theta)$ coordinates,
the Hamiltonian can be separated and, therefore, the
wave function factorizes. The center of mass problem is that of a
single particle in a harmonic potential and magnetic field, having an
analytic solution in terms of the Fock-Darwin orbitals and energies
(see Sec. \ref{subsec:IV.C.2}).
Focusing next on the relative problem, one introduces the wave function
$e^{im\theta} u_{nm}(r)/\sqrt{r}$ having good $\ell_z$ angular momentum ($m$)
and an additional quantum number $n$ whose
meaning will be clarified below.
The equation for the unknown $u_{nm}(r)$ reads
\begin{equation}
\label{eqrel}
u_{nm}''+\left[
\tilde\varepsilon^{\rm (rm)}_{nm}-\left(\frac{1}{4}r^2+\frac{R_{\it mp}}{r}
+ {m^2-1/4\over r^2}\right)
\right] u_{nm} = 0 \; ,
\end{equation}
where we have defined
$\tilde\varepsilon^{\rm (rm)}_{nm}=\varepsilon_{nm}^{\rm (rm)}-m W_{\it mp}/2$
in terms of the relative-motion energy $\varepsilon_{nm}^{\rm (rm)}$
and the parameter $W_{\it mp}$.

The above Eq. (\ref{eqrel}) resembles a Schr\"odinger one-dimensional equation
with an effective potential
\begin{equation}
\label{eq7}
V_{\it eff}(r)=\frac{1}{4}r^2+\frac{R_{\it mp}}{r}
+ {m^2-1/4\over r^2} \;
\end{equation}
that includes the rotational motion term $\sim m^2/r^2$ characterized
by the angular momentum quantum number $m$.
We can expect a rigid-rotor behaviour if $V_{\it eff}(r)$
has a deep minimum at a particular value $r=r_0$.
When this occurs, the situation resembles that of diatomic
molecules like H$_2$, where the potential well
for nuclear motion is described by the Morse potential \cite{BJ}.

The minimization of $V_{\it eff}(r)$ yields the rotor radius
from the equation
\begin{equation}
\label{eq8}
\frac{r_0}{2}-\frac{R_{\it mp}}{r_0^2}-{2(m^2-1/4)\over r_0^3} =0 \; .
\end{equation}
Neglecting the third contribution on the left-hand-side (the assumption
that is valid for large enough $r_0$),
one finds the asymptotic law $r_0\approx (2R_{\it mp})^{1/3}$.
Now, expanding up to
second order of deviations around $r_0$, we approximate
\begin{eqnarray}
V_{\it eff}(r) &\approx& V_{\it eff}(r_0) +
\frac{1}{2}\left(\frac{3}{2}+2{m^2-1/4\over r_0^4}\right)
(r-r_0)^2\nonumber\\
&=& {\it const.} + \frac{1}{2}k(r-r_0)^2 \; .
\end{eqnarray}
This result, after the substitution into Eq.\ (\ref{eqrel}),
leads to the analytical prediction
\begin{eqnarray}
\tilde\varepsilon^{\rm (rm)}_{nm} &=& \frac{1}{4}r_0^2+\frac{R_{\it mp}}{r_0}
+ {m^2-1/4\over r_0^2}
+\left(n+\frac{1}{2}\right)
\sqrt{3+4{m^2-1/4\over r_0^4}} \; .
\label{eq9}
\end{eqnarray}

Equation  (\ref{eq9}) has a clear physical interpretation.
It contains a rotor-like contribution, $\sim m^2/(2{\Im})$, with the moment
of inertia given by ${\Im}=r_0^2/2$, and a vibrational contribution,
characterized by the quantum number $n$.
The vibrational frequency
$\omega_{\it vib}=\sqrt{k/\mu}$ ($\mu=1/2$) is given
by the last square-root factor. Similarly to atomic molecules, there is a
roto-vibrational coupling, since the vibration frequency depends on $m$, and
also a centrifugal distortion, since $r_0$ also depends on $m$. For large
enough values of $R_{\it mp}$,
implying large $r_0$ and, therefore, small average densities,
the centrifugal distortion disappears and one has
$r_0\approx(2R_{\it mp})^{1/3}$ for all $m$'s.
In this limit, the rotational terms become negligible, as well as the
roto-vibrational ones. Thus, Eq.\ (\ref{eq9}) reduces to a simple
$m$-independent asymptotic expression
\begin{equation}
\label{asympt}
\tilde\varepsilon^{\rm (rm)}_{n} = \frac{3}{2^{4/3}} R_{\it mp}^{2/3}
+\sqrt{3}\left(n+\frac{1}{2}\right) \; .
\end{equation}

Adding magnetic field transforms the roto-vibrational energy to
\begin{eqnarray}
&&
\varepsilon^{\rm (rm)}_{nm}=\tilde\varepsilon^{\rm (rm)}_{nm}+
m W_{\it mp}/2\simeq {(m+W_{\it mp} r_0^2/4)^2\over r_0^2}+
\nonumber\\
&&
+\frac{R_{\it mp}}{r_0}+\left(n+\frac{1}{2}\right) \omega_{\it vib}-
\bigg[\frac{r_0 W_{\it mp}}{4}\bigg]^2\;,
\end{eqnarray}
in agreement with the expectations for two interacting
electrons in a strong magnetic field \cite{mape}.

The validity of the roto-vibrational model has been proved by the comparison
with the exact results \cite{ton}. As a matter of fact, for $R_{mp}>2$
the discrepancy for $\tilde\varepsilon_{00}$ is always below 2{\%},
even with the asymptotic expression Eq.\ (\ref{asympt}).
The roto-vibrational model allows one to determine
the crystallization onset from the criterion that the rotation and vibration
motions decouple when intrinsic-frame electron localization sets in.
On the contrary, when the coupling is strong,
the system could be represented by either {\it a vibrating rotor} or
{\it a rotating vibrator} and, therefore, the situation can not be clearly
resolved. It is also worth stressing that the roto-vibrational model describes
all possible excitations of the relative-motion problem. For this
particular system, this amounts to a description of all excitations,
since the center-of-mass and spin degrees of freedom can be
analytically integrated out. Indeed, with $B=0$ in the limit
$r_0\approx(2R_{\it mp})^{1/3}$, the model describes three basic excitations:
(i) the Kohn mode $\omega_0$ corresponding to the excitation of
the center-of-mass; (ii) the breathing mode with a frequency $\omega=\sqrt{3}$
corresponding to the vibration of the mean square radius; and
(iii) the frequency $\omega=0$ corresponding to the rotation of the circular
system as a whole. These modes are excitations of classical electrons in a
parabolic potential and are independent on $N$ \cite{vit}.

\begin{figure}[t]
\centerline{\psfig{figure=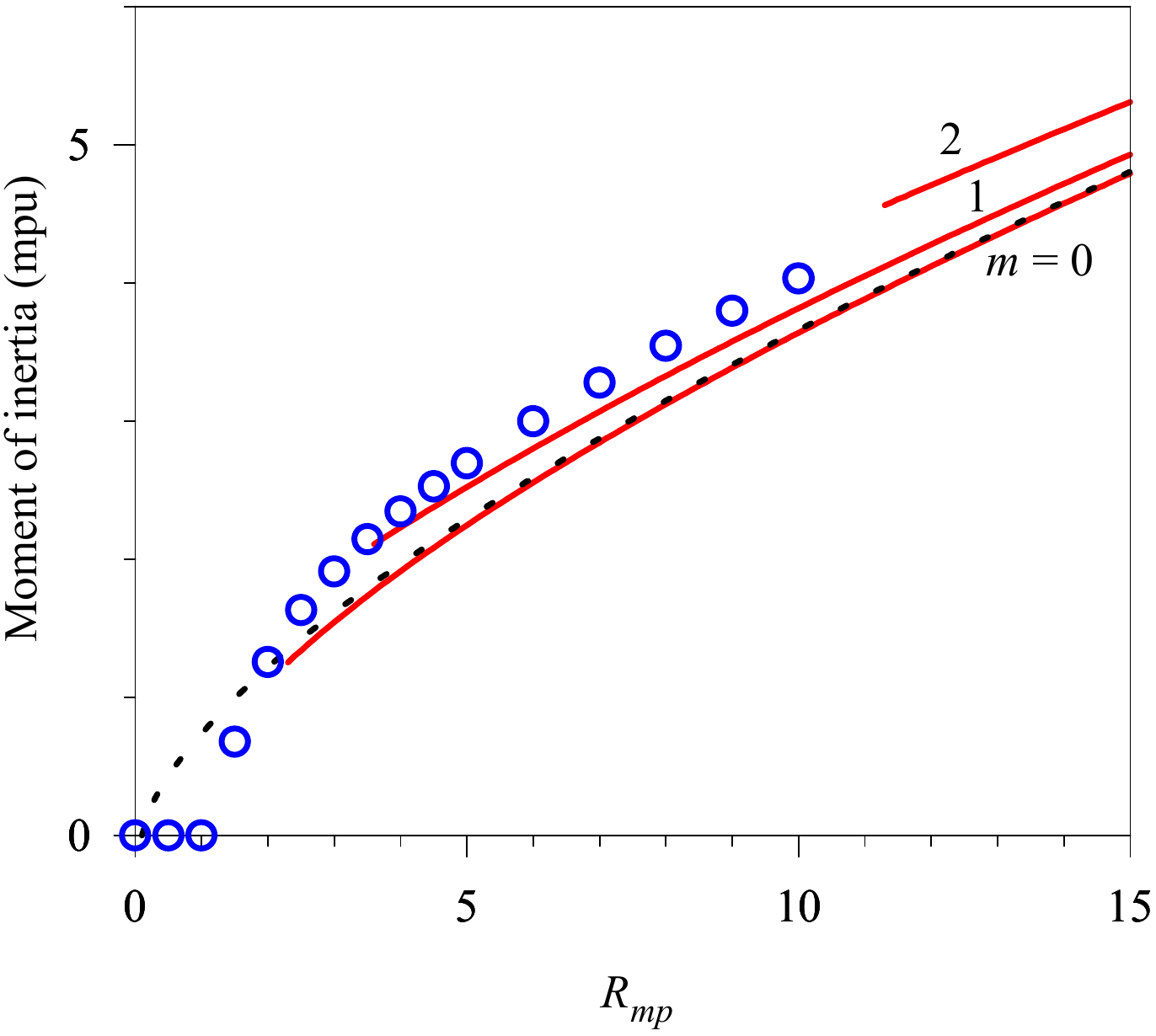,width=2.5in,clip=}}
\caption{(Color online) Moment of inertia, computed in the RPA
approximation (circles) at $W_{\it mp}=0$, as a function of the dimensionless
parameter $R_{\it mp}$. Solid lines show the evolution of the corresponding
values in the analytical model of Sec.\ref{subsec:IV.E.3.}
for different $m$-states. Each line starts at the
crystallization onset for the corresponding angular momentum, according to
the criterion $(\omega_{\it vib}-\sqrt{3})/\sqrt{3}> 0.01$.
The dashed line represents the asymptotic
value ${\Im}=r_0^2/2$ taking $r_0\approx(2R_{\it mp})^{1/3}$, to which all
solid lines converge at very high values of $R_{\it mp}$.
From \cite{lor}.}
\label{qdmoi}
\end{figure}

As will be discussed in Sec. \ref{subsec:IV.E.2}, the RPA determines the
moment of inertia associated with the collective rotation of a
deformed HF structure.
Fig. \ref{qdmoi} presents the evolution, with $R_{\it mp}$, at $W_{\it mp}=0$,
of the RPA moment of inertia ${\Im}_{RPA}$ (circles). For comparison,
the values computed through
the solution of Eq.\ (\ref{eq8}) of the roto-vibrational model,
${\Im}=r_0^2/2$, are also shown (solid lines).
Each ${\Im}$-line starts at the crystallization onset for the
corresponding angular momentum.
Note that ${\Im}_{RPA}$ remains zero until the
HF solution breaks the rotational symmetry at $R_{\it mp} \simeq 1$. After
this, it reasonably agrees with the exact values, somehow averaging
the exact results for different $m$.
The molecule stretching, yielding larger $r_0$ (${\Im}$),
as $m$ increases, is obviously outside RPA.
All ${\Im}$ values slowly converge to a
common result with increasing $R_{\it mp}$, i.e., to
an exact rigid-rotor behaviour.

\subsubsection{Geometric transformation in a two-electron quantum dot}
\label{subsec:IV.E.4.}
The roto-vibrational model demonstrates a transition from
the circular to the deformed shape due to the effect of strong interaction
and magnetic field in the 2D case. The question remains to answer about
the contribution of the third dimension. In particular,
how vibrational frequencies would evolve at
a shape transition in the 3D case?
These questions may shed light on the similarity and difference between
shape transitions and, in general,
quantum phase transitions (QPTs) in finite systems.

Quantum phase transitions in many-body systems, driven by
quantum fluctuations at zero temperature, attract a considerable
attention in recent years \cite{Sad}. They are recognized  as
abrupt changes of the ground state of a many-body system, with
varying a non-thermal control parameter (magnetic field, pressure
etc.) in the Hamiltonian of the system. Although the phases should
be characterized by different types of quantum correlations on
either side of a quantum critical point (actual transition point),
in some cases they cannot be distinguished by any local order
parameter. Particular examples are the integer and fractional
quantum Hall liquids \cite{Hal1,Hal2}, which cannot be understood in
terms of the traditional description of phases based on symmetry
breaking and order parameters. Nowadays, there is a growing
interest in using quantum entanglement measures for study such
transitions \cite{am} and, in general, quantum correlations in
many-body systems \cite{Yukalov_2003b,TMB2011}.
Although finite systems can only show precursors of the QPT
behaviour, they are also important for the development of the
concept.

It was shown in Sec. \ref{subsec:IV.C.3}
that at the value
$\omega_L^\mathrm{sph} = (\omega_z^2 -\omega_0^2)^{1/2}$ the magnetic
field gives rise to the {\it spherical symmetry} $(\omega_z/\Omega=1)$ in
the {\it axially-symmetric} two-electron QD (with $\omega_z > \omega_0$).
A natural question arises how to detect such a
transition looking on the {\it ground state density distribution}
only. The related question is, if such a transition occurs, what
are the concomitant structural changes?

To this end we employ the entanglement measure based on the
linear entropy of reduced density matrices (cf \cite{ColemanYukalov_2000})
\begin{equation}
{\cal E} = 1 - 2\,\mathrm{Tr}[{\rho_r^{(orb)}}^2]\,
\mathrm{Tr}[{\rho_r^{(spin)}}^2],
\label{enmes}
\end{equation}
where $\rho_r^{(orb)}$ and $\rho_r^{(spin)}$ are the
single-particle reduced density matrices in the orbital and spin
spaces, respectively. This measure is quite popular for the
analysis of the entanglement of two-fermion systems, in
particular, for two electrons confined in the parabolic potential in
the absence of the magnetic field \cite{ent1,ent2,ent3,ent4,ent5}.
Notice that the measure (\ref{enmes}) vanishes when the global (pure) state
describing the two electrons can be expressed as one
single Slater determinant.

The trace $\mathrm{Tr}[{\rho_r^{(spin)}}^2]$ of the two-electron
spin states with a definite symmetry $\chi_{S,M_S}$ has two
values: (i) $1/2$ if $M_S = 0$ (anti-parallel spins of two
electrons); (ii) $1$ if $M_S = \pm1$ (parallel spins). The
condition $M_S = S$ (see Sec.\ref{subsec:IV.C.2}, Eq.(\ref{spM})) yields
$\mathrm{Tr}[{\rho_r^{(spin)}}^2] = \hbox{$\frac{1}{2}$}(1 + |M_S|)=(3-(-1)^m)/4$.

The trace of the orbital part
\begin{eqnarray}
\label{oren}
\mathrm{Tr}[{\rho_r^{(orb)}}^2] \!\!&=&\!\! \int d\mathbf{r}_1\,
d\mathbf{r}_1^{\,\,\prime}\, d\mathbf{r}_2\,
d\mathbf{r}_2^{\,\,\prime}\, \psi(\mathbf{r}_1,\mathbf{r}_2)\,
\psi^*(\mathbf{r}_1^{\,\,\prime},\mathbf{r}_2) \nonumber
\\[-.5ex]
&&\qquad\quad \psi^*(\mathbf{r}_1,\mathbf{r}_2^{\,\,\prime})\,
\psi(\mathbf{r}_1^{\,\,\prime},\mathbf{r}_2^{\,\,\prime})
\end{eqnarray}
is more involved. Indeed, in virtue of Eq.~(\ref{expansion_3d}), it
requires cumbersome calculations of eightfold sums of terms
(integrals) obtained analytically \cite{nent}. The magnetic field dependence
of the entanglement ${\cal E}$ naturally occurs via inherent
variability of the expansion coefficients.

We choose the value
$R_W = 1.5$ (see Eq.(\ref{RW}), Sec.\ref{subsec:IV.D.2})
which corresponds for GaAs QDs to the confinement
frequency $\hbar\omega_0\approx 5.627$\,meV. The linear entropy
$\cal E$ is calculated using the basis with $n_{\max} = n_z^{\max} = 4$,
which gives $\sim 3.9\times 10^5$ terms in Eq.~(\ref{oren}).

\begin{figure}
\centerline{
\epsfxsize 1.75in \epsffile{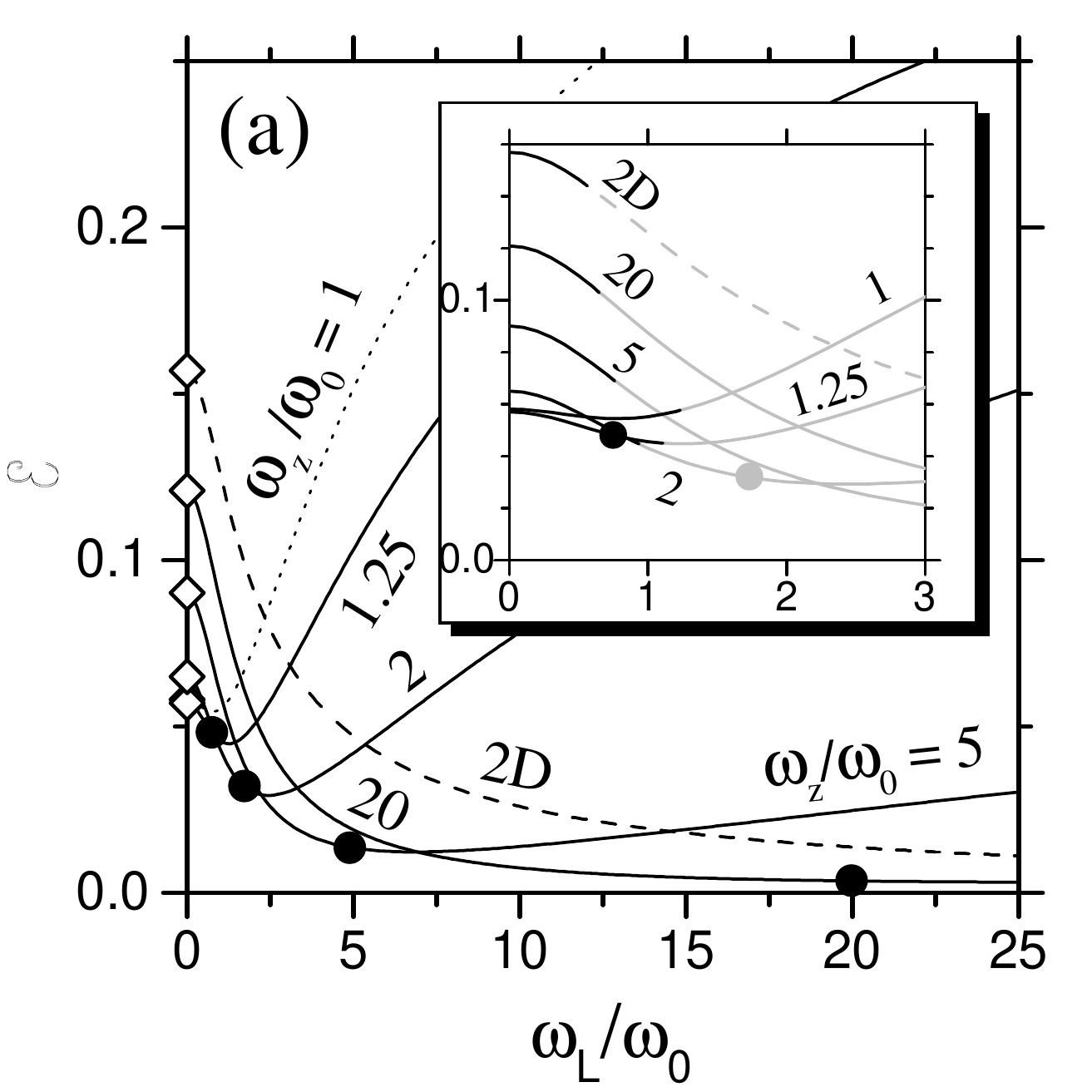}
\epsfxsize 1.75in \epsffile{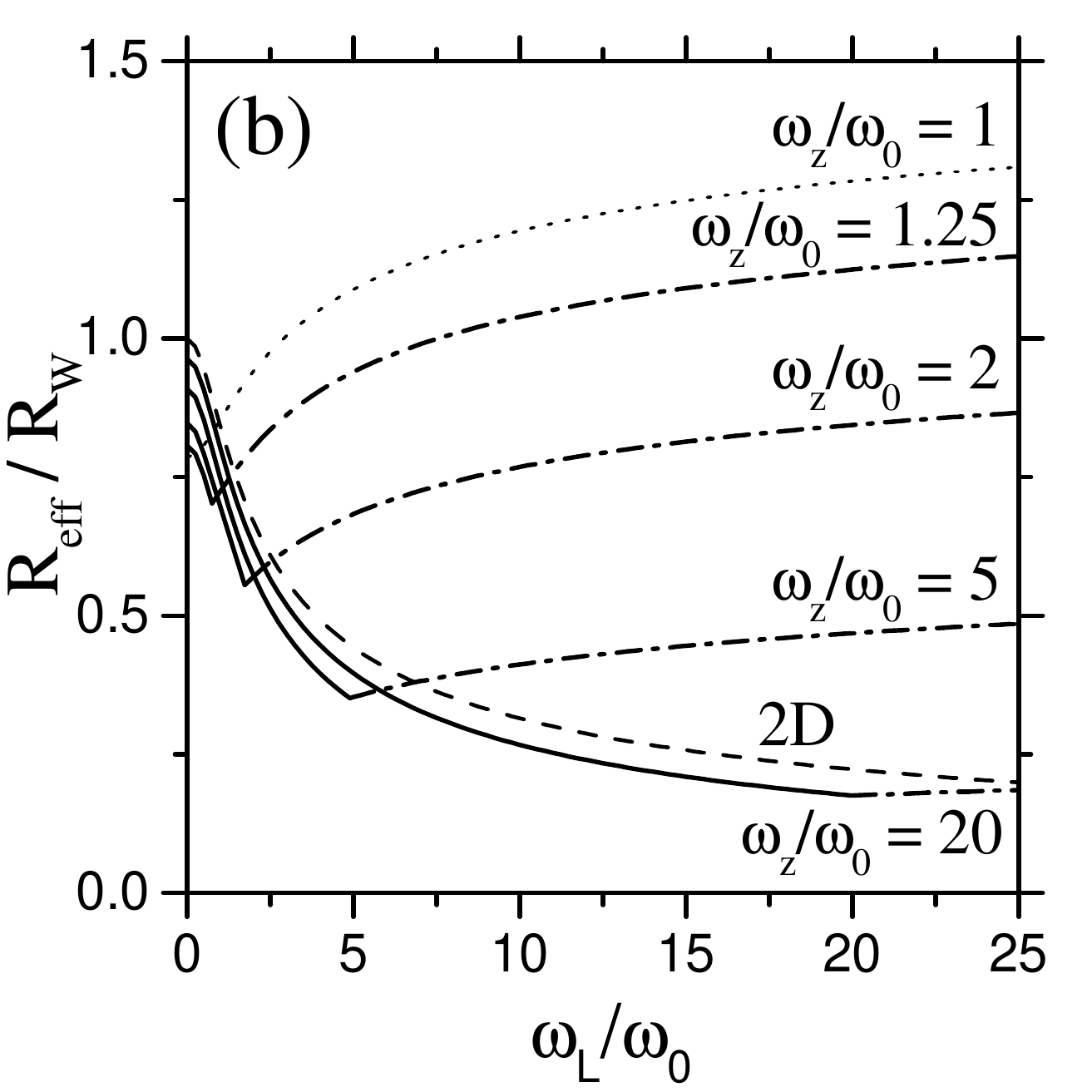}}
\caption{(a) The entanglement measure $\cal E$ of the lowest state
with $m = 0$ at $R_W = 1.5$ and various ratios $\omega_z/\omega_0$
as a function of the parameter $\omega_L/\omega_0$. The initial
(black) parts of the measure (shown in the inset) correspond to
the intervals of $\omega_L/\omega_0$, when the lowest state
$m = 0$ is the ground state. Full circles denote the values of
$\omega_L/\omega_0$, when the effective $3D$ confinements
become spherically symmetric. (b) The relative strengths of the
Coulomb interaction $R_\mathrm{eff}^\mathrm{(2D)}/R_W$ (solid line) and
$R_\mathrm{eff}^\mathrm{(1D)}/R_W$ (dash-dotted line)
for the lowest state $m = 0$ at various ratios
$\omega_z/\omega_0$ as functions of the parameter
$\omega_L/\omega_0$.}
\label{ent-oml}
\end{figure}

At zero magnetic field ($\omega_L/\omega_0 = 0$) the entanglement
of the lowest state with $m = 0$ decreases if the ratio
$\omega_z/\omega_0$ decreases from $\infty$ ($2D$ model) to $1$
(spherically symmetric $3D$ model); see open symbols (diamonds) in
Fig.~\ref{ent-oml}(a). This effect could be explained by
introducing the effective charge $k_\mathrm{eff}$, which
determines the effective electron-electron interaction
$V_C^\mathrm{eff} = k_\mathrm{eff}/\rho_{12}$ in the QD.
(see Eq.(\ref{efVC}), Sec.\ref{subsec:IV.D.3}).
In a $3D$ dot the electrons can avoid each other more efficiently than
in the $2D$ case. Consequently, the Coulomb interaction has a smaller
effect, when $\omega_0 \approx \omega_z$ (the ratio
$k_\mathrm{eff}/k \approx 0.5$), than in the anisotropic case
$\omega_0 \ll \omega_z$ ($k_\mathrm{eff}/k = 1$). Therefore, a
decreasing of the ratio $\omega_z/\omega_0$ yields the effect analogous
to the reduction of the electron-electron interaction -- a
weaker mixing of the single-particle states and, consequently,
the lowering of the entanglement.

By increasing the magnetic field from zero to
$B_\mathrm{sph}\sim \omega_L^\mathrm{sph} = \sqrt{\omega_z^2 - \omega_0^2}$
(see full circles in Fig.~\ref{ent-oml}(a)) the
entanglement decreases. Similar to the case $B=0$, one would expect
the decrease of the effective electron-electron interaction with the evolution
of the effective confinement
from the disk shape ($\Omega <\omega_z$)
to the spherical form ($\Omega = \omega_z$).
Further increase of the magnetic field yields the increase of the entanglement.
The effective confinement becomes again
anisotropic (now with $\Omega > \omega_z$).
Evidently, for $\omega_z/\omega_0 \to \infty$ (2D model) the minimum of $\cal E$
is shifted to infinity, i.e. in this case the entanglement
decreases monotonically with the increase of the field (dashed
line in Fig.~\ref{ent-oml}(a)).

The entanglement evolution can be explained by the influence of the
magnetic field on the effective strength of the electron-electron
interaction, which transforms the Wigner parameter $R_W$ to the
form $R_{mp}\equiv R_\mathrm{\Omega} = l_\Omega/a^*$.

For the quasi-$2D$ system ($\Omega \ll \omega_z$),
the influence of magnetic field on the effective strength
$R_\Omega \Rightarrow R_\mathrm{eff}^\mathrm{(2D)} =
(k_\mathrm{eff}^\mathrm{(2D)}/l_\Omega)/\hbar\Omega$ is twofold.
Here $k_\mathrm{eff}^\mathrm{(2D)}$ is determined
by Eq.(\ref{ef2d}). The magnetic field affects the
effective confinement $\hbar\Omega$ as well as the effective charge.
With the increase of the effective confinement the effective
charge $k_\mathrm{eff}^\mathrm{(2D)}/k \to 1$ and, therefore, the
effective strength decreases as $R_\mathrm{eff}^\mathrm{(2D)} \sim
1/\sqrt{\Omega}$ (see Fig.~\ref{ent-oml}(b)).

For $\Omega \gg \omega_z$ (very strong magnetic field) the
electrons are pushed laterally towards the dot's center. The
magnetic field, however, does not affect the vertical confinement.
As a consequence the electrons practically move only in the
$z$-direction, and the QD becomes a quasi-$1D$ system. In this case the
effective strength is
$R_\Omega \Rightarrow R_\mathrm{eff}^\mathrm{(1D)} =
(k_\mathrm{eff}^\mathrm{(1D)}/l_z)/\hbar\omega_z$.
Here the effective charge $k_\mathrm{eff}^\mathrm{(1D)} = \langle
|z_{12}|V_C \rangle$, and $l_z \equiv z_m=
\sqrt{\hbar/m^*\omega_z}$ is the oscillator length of
the vertical confinement. For the lowest state with $m = 0$, it can be
shown that $k_\mathrm{eff}^\mathrm{(1D)}/k = (1 +
\sqrt{\omega_z/\Omega})^{-1}$. At a very strong magnetic field the
ratio $k_\mathrm{eff}^\mathrm{(1D)}/k \to 1$, which yields the
maximal value $R_\mathrm{eff}^\mathrm{(1D)} \sim
1/\sqrt{\omega_z}$.

When $\Omega = \omega_z$ the $3D$ system is far from both
the $2D$ and the $1D$ limits. As a result,
$R_\mathrm{eff}^\mathrm{(2D)}$ and $R_\mathrm{eff}^\mathrm{(1D)}$
do not match smoothly (see Fig.~\ref{ent-oml}(b)). However, it is
clear that the effective strength reaches the minimum around this
point, i.e. when the transition from the lateral to the vertical
localization of two electrons takes place.

\begin{figure}
\centerline{
\epsfxsize 5.25in \epsffile{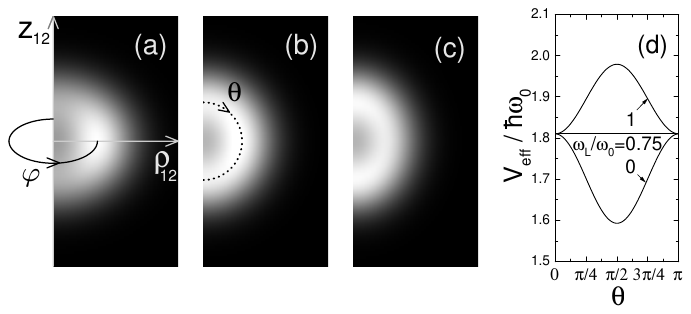}}
\caption{The ground state probability density (the relative motion
part, $m = 0$) for the near-spherical ($\omega_z/\omega_0 = 1.25$)
two-electron QD with $R_W = 1.5$ ($\hbar\omega_0 = 5.627$\,meV)
and $g^* = -0.44$ at different values of the magnetic field: (a)
$\omega_L/\omega_0 = 0$, (b) $\omega_L/\omega_0 = 0.75$, (c)
$\omega_L/\omega_0 = 1$. For the spherically symmetric case (b)
the maximum of the probability density spreads uniformly over the
spherical shell of a radius $r_0$. The panel (d) displays the
potential $V_\mathrm{eff}(\rho_{12},z_{12})$ on this sphere
($\rho_{12} = r_0\sin\theta$, $z_{12} = r_0\cos\theta$) for the
cases (a-c).} \label{fig2}
\end{figure}

In order to get a deeper insight into this transition we examine
the probability density $\vert\psi({\bf r}_{12})\vert^2$ for the
ground state, when $m = 0$. Such a ground state can be realized
for a near spherical QD with $\omega_z/\omega_0 = 1.25$.
For this ratio and $R_W=1.5$,
the first S-T transition occurs at $\omega_L/\omega_0 \approx 1.11$
(see the inset in Fig.~\ref{ent-oml}(a)). The
onset of the spherical symmetry ($\Omega = \omega_z$) takes place
at $\omega_L^{\rm sph}/\omega_0 = 0.75$, i.e., in the ground state.
For the magnetic field strengths $\omega_L < \omega_L^{\rm sph}$,
with $\Omega < \omega_z$, the density maximum forms a ring in
the $(x_{12},y_{12})$-plane (a consequence of the axial symmetry).
Fig.~\ref{fig2}(a) shows the cut of this density with the
$(\rho_{12},z_{12})$-plane at arbitrary azimuthal angle $\varphi$.
For $\omega_L = \omega_L^{\rm sph}$ the maximum of
the probability density forms a spherical shell (visible if we
rotate Fig.~\ref{fig2}(b) around $z_{12}$-axis).
For $\omega_L > \omega_L^{\rm sph}$, with $\Omega > \omega_z$,
two separate density maxima start to grow, located symmetrically
in the $z_{12}$-axis (see Fig.~\ref{fig2}(c)). In
contrast to the corresponding behaviour of the entanglement, a fuzzy
transition manifests itself in the probability density
for the chosen dot parameters. In fact, the entanglement
evolution guides us to trace a geometrical transition from the lateral
to the vertical localization of the electrons.

The probability density evolution due to the magnetic field shown in
Figs.~\ref{fig2}(a-c) can be elucidated by means of the analysis of the
effective potential
\begin{equation}
V_\mathrm{eff} = \frac{1}{2}\,\mu\,(\Omega^2
\rho_{12}^2 + \omega_z^2 z_{12}^2) + k/r_{12} +
\hbar^2m^2/(2\mu\rho_{12}^2)\,.
\end{equation}
Here, for the relative coordinate
$\mathbf{r}_{12} = \mathbf{r}_1 - \mathbf{r}_2$,
we introduce the following notations:
\begin{equation}
\rho_{12} = (x_{12}^2 +
y_{12}^2)^{1/2},\;  \varphi = \arctan(y_{12}/x_{12}),\;  r_{12} =
(\rho_{12}^2 + z_{12}^2)^{1/2}\,.
\end{equation}
The maxima of the
probability density for the ground state are directly related to
the minima of $V_\mathrm{eff}$.
For $\omega_L < \omega_L^{\rm sph}$, the potential surface has the
minimum at $\rho_{12} = \rho_0$, $z_{12} = 0$, where
$\rho_0 = (k/\mu\Omega^2)^{1/3}$, if $m = 0$. By increasing the
magnetic field to values $\omega_L > \omega_L^{\rm sph}$, this minimum
transforms to the saddle point in the $(\rho_{12},z_{12})$-plane, but
two new minima divided by this saddle (potential barrier) appear.
For $m = 0$, these minima are located at $z_{12} = \pm z_0$,
where $z_0 = (k/\mu\omega_z^2)^{1/3}$.

For weakly anisotropic (near spherical) systems it is convenient
to use the spherical coordinates $(r_{12},\theta,\varphi)$, where
the polar angle is $\theta \equiv \arctan(\rho_{12}/z_{12})$. In
these coordinates the positions of the minima are
$r_{12} = \rho_0$, $\theta = \pi/2$ for $\Omega < \omega_z$ and
$r_{12} = z_0$, $\theta = 0, \pi$ for $\Omega > \omega_z$; see the
cases $\omega_L/\omega_0 = 0$ and $1$, respectively, in
Fig.~\ref{fig2}(d). Note, that due to the axial symmetry the
azimuthal angle $\varphi$ is arbitrary. The maximum at
$\theta = \pi/2$ for $\Omega > \omega_z$ (the case $\omega_L/\omega_0 = 1$
in Fig.~\ref{fig2}(d)) corresponds to the saddle point at $z_{12} = 0$.

Applying similar to the roto-vibrational model ideas
(see Sec. \ref{subsec:IV.E.3.}), we consider small oscillations around a
minimum of the effective potential. The effective potential
can be written as the expansion up to the quadratic
terms)
\begin{equation}
V_\mathrm{eff}
\approx V_0 + \frac{1}{2}\,\omega_1^2\,q_1^2 +
\frac{1}{2}\,\omega_2^2\,q_2^2\,,
\end{equation}
where $q_1 = \Delta r$, $q_2 = r_0 \Delta\theta$ are the
normal coordinates. Here $r_0 = \rho_0$, when $\Omega < \omega_z$,
whereas $r_0 = z_0$, when $\Omega > \omega_z$. The corresponding
normal frequencies (if $m = 0$) are: (i) $\omega_1 = \sqrt{3}\,\Omega$,
$\omega_2 = (\omega_z^2 - \Omega^2)^{1/2}$ for $\Omega < \omega_z$;
and (ii) $\omega_1 = \sqrt{3}\,\omega_z$,
$\omega_2 = 2\,(\Omega^2 - \omega_z^2)^{1/2}$ for $\Omega > \omega_z$.
For the spherically symmetric case ($\Omega =
\omega_z$), one has $\omega_2 = 0$ and the minima of
$V_\mathrm{eff}$ degenerate to the sphere of radius $r_0 = \rho_0 = z_0$.
In other words, the potential $V_\mathrm{eff}$ becomes
independent on the angle $\theta$; see the case $\omega_L/\omega_0 = 0.75$
in Fig.~\ref{fig2}(d). As a consequence, the wave function
becomes spherically symmetric (Fig.~\ref{fig2}(b)).
The quantum oscillations evolve in a way similar to those
of quantum phase transitions studied for model systems \cite{Sad}.

Note that the considered geometric transformation in a two-electron quantum
dot reminds, to some extent, the orientation transition of Sec. 2.3.3.

\subsubsection{Shell structure and classical limit}
\label{subsec:IV.E.1}

Recently Maksym {\it et al.} \cite{mak2} developed a model of a vertical QD
and reproduced quite well experimental addition energies for
$N = 2,3,4$. Although the three-dimensional nature of QDs is accounted for,
the real calculations are based on the $2D$ parabolic potential.
The authors constructed the effective two-dimensional
electron-electron interaction (the screened  Coulomb interaction) which does
not depend on the magnetic field. They diagonalized the model Hamiltonian
in the many-electron basis consisting of Slater determinants.

\begin{figure}[ht]
\centerline{
\includegraphics[width=14pc]{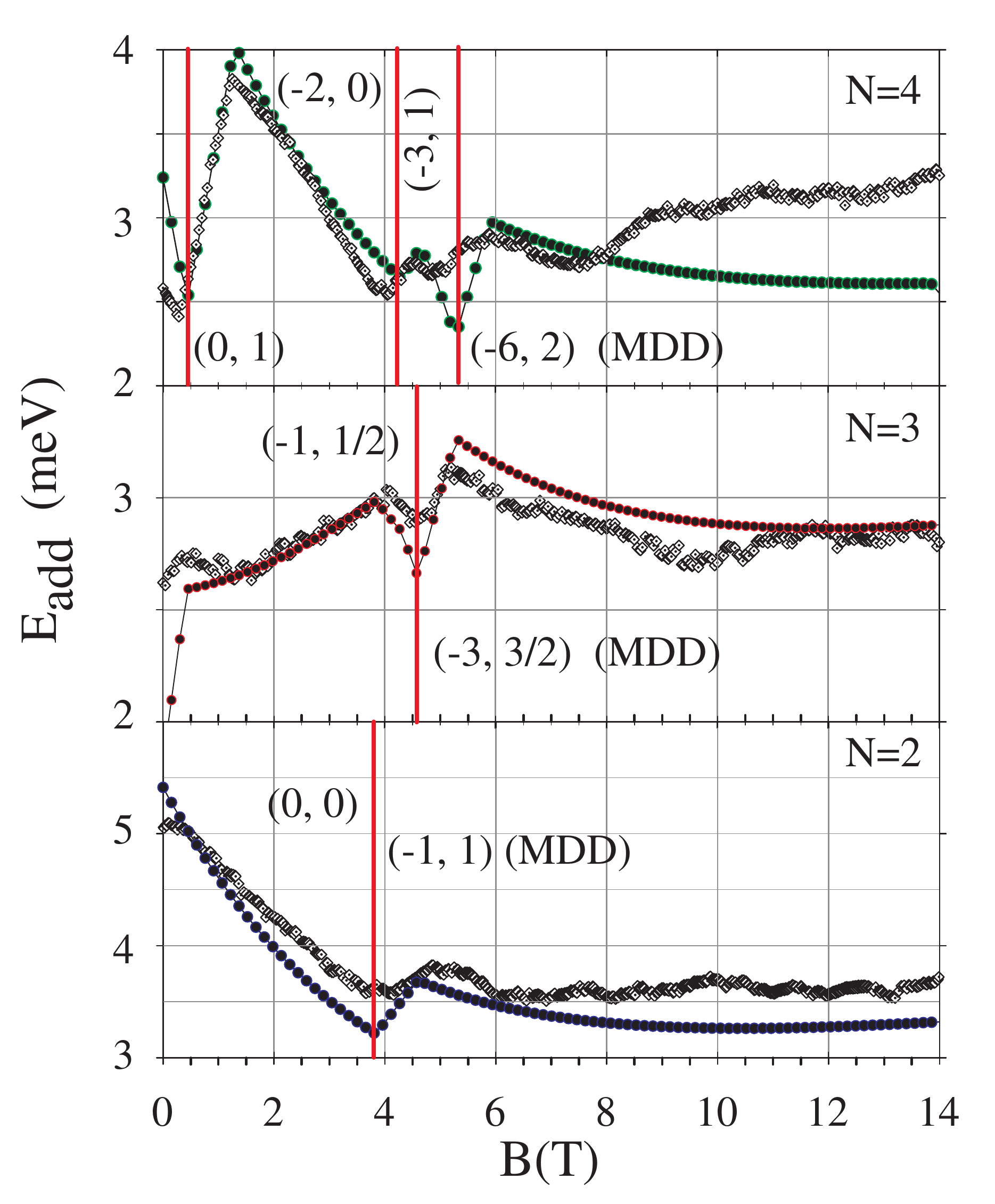} }
\caption{(Color online)
Experimental and theoretical addition energies $E_{\it add}(N)$ for $N=2,3,4$
as functions of the magnetic field $B$.
Open diamonds correspond to experimental data \cite{mak2};
full circle corresponds to the Hartree-Fock results. Ground states are labelled
by $(L_z,S_z)$. The vertical line indicates the transition between
different ground states. From \cite{ton1}.}
\label{mak09}
\end{figure}

Similar data were analyzed in a symmetry preserving Hartree-Fock
approach \cite{ton1}. The Hartree-Fock ground state is chosen as an
eigenfunction of the $z$-component
of the angular momentum $L_z$, the parity, and the spin.
To compare with the results \cite{mak2}, the following parameters for
InGaAs QDs have been used $\hbar\omega_0 = 4.5\,$meV, $m^*=0.0653$,
dielectric constant $\epsilon_r=12.7$. These values provide for the strength
of the bare Coulomb interaction the following estimate:
$R_{\it mp}(\omega_c=0)\equiv R_W \approx 1.56$.
The effective value $R_W=0.78$ is taken to reproduce the available
experimental data.

For $N=2$, it was found \cite{ton1} that the singlet $(L_z,S_z) = (0,0)$
state is replaced by fully
polarized triplet $(-1,1)$ state (MDD phase) at $B_{HF}=3.8$ T
($L_z$ and $S_z$ are the quantum numbers of the $z$-component of the
total orbital momentum and of the total spin, respectively). The experimental
transition takes place at $B_{\it exp}=4.2$ T, while the theoretical calculations
of Maksym {\it et al.} \cite{mak2} give $B_{\it theor}=3.8$ T.

For $N=3$,  the HF calculations \cite{ton1} predict that
the state $(-1,1/2)$ is the lowest state till $B_{HF}=4.72$ T.
Two electrons occupy the lowest available states with $m=0$, but with
different $S_z=\pm 1/2$ quantum numbers. The next available
state is characterized by $m=-1$,  and the largest possible
value for this configuration has $S_z=1/2$. At $B_{HF}=4.72$ T, the onset of
the MDD phase $(-3,3/2)$ occurs.

At small magnetic field $B_{HF}<0.46$ T, for $N=4$
the ground state is $(0,1)$ \cite{ton1}.
A partially polarized state is favourable due to the HF exchange
term which supports the configuration with two aligned spins.
At $B_{HF}=0.46$ T $(B_{\it exp}=0.33$ T, $B_{\it theor}=0.3$ T),
the magnetic field overcomes the exchange contribution and the HF calculations
predict $(-2,0)$ state.
This state persists till $B_{HF}<4.42$ T.
At $B_{HF}=4.42$ T $(B_{\it exp}=4.06$ T, $B_{\it theor}=4.0$ T),
the magnetic field partially polarizes the system, and, with the aid
of the exchange term, the HF calculations produce the ground state  $(-3,1)$.
At $B_{HF}\geq 5.33$ T, the onset of the MDD phase $(-6,2)$ takes place.
At $B_{HF}=14.31$ T, the MDD state is replaced by the ground state $(-10,2)$.

The HF calculations \cite{ton1} reproduced a double peak structure for $N=3$
and $N=4$ around $4-6$ T. The spike, observed for $N=3$
at $B_{HF}=3.8$ T, is due to the transition to the MDD phase in the two-electron
system. The following dip is caused by the transition from $(-1,1/2)$ to $(-3,3/2)$
for $N=3$, while the transition from $(-2,0)$ to $(-3,1)$ for $N=4$ changes only
the slope of the former transition.
The next spike, observed for $N=3$ at $B_{HF}=5.33$ T, is due to the transition
$(-3,1)\rightarrow (-6,2)$ for $N=4$. The large peak at $B_{HF}=1.3$ T for $N=4$
is due to the ground state transition $(-1,1/2)\rightarrow (-4,1/2)$
for $N=5$ (not shown). The next transition for $N=5$, that is,
$(-4,1/2)\rightarrow (-6,3/2)$ at $B_{HF}=4.80$ T, has a little effect. However,
the transition to the MDD phase $(-6,3/2)\rightarrow (-10,5/2)$ for $N=5$ is
responsible for the small spike in the addition energy of $N=4$ at $B_{HF}=5.87$ T.

Although the agreement is good in general till $B_{\it exp}\leq 6$ T, there are
some discrepancies at high magnetic field for $N=4$ case
(see Fig. \ref{mak09}). A similar problem appears in more sophisticated
calculations of \cite{mak2}.
The analysis of finite-thickness effects in two-electron QDs reveals
a magnetic-field dependence of the screening of the effective
interaction \cite{sreen}. The stronger the magnetic field, the lesser the expected screening effect in QDs.
One of the possible reasons for this discrepancy is the change of the effective
confining potential (the parabolic confinement frequency) at a high
magnetic field. It follows from the calculations \cite{ton1} that there is
a visible suppression of the electron-electron interaction in the observed
experimental data. Indeed, the experimental evidence \cite{ni2} is that the QD
has a high degree of a circular symmetry. We can conclude that, for this device,
the electron dynamics is mainly due to the confining potential.

\begin{figure}[ht]
\centerline{
\includegraphics[width=14pc]{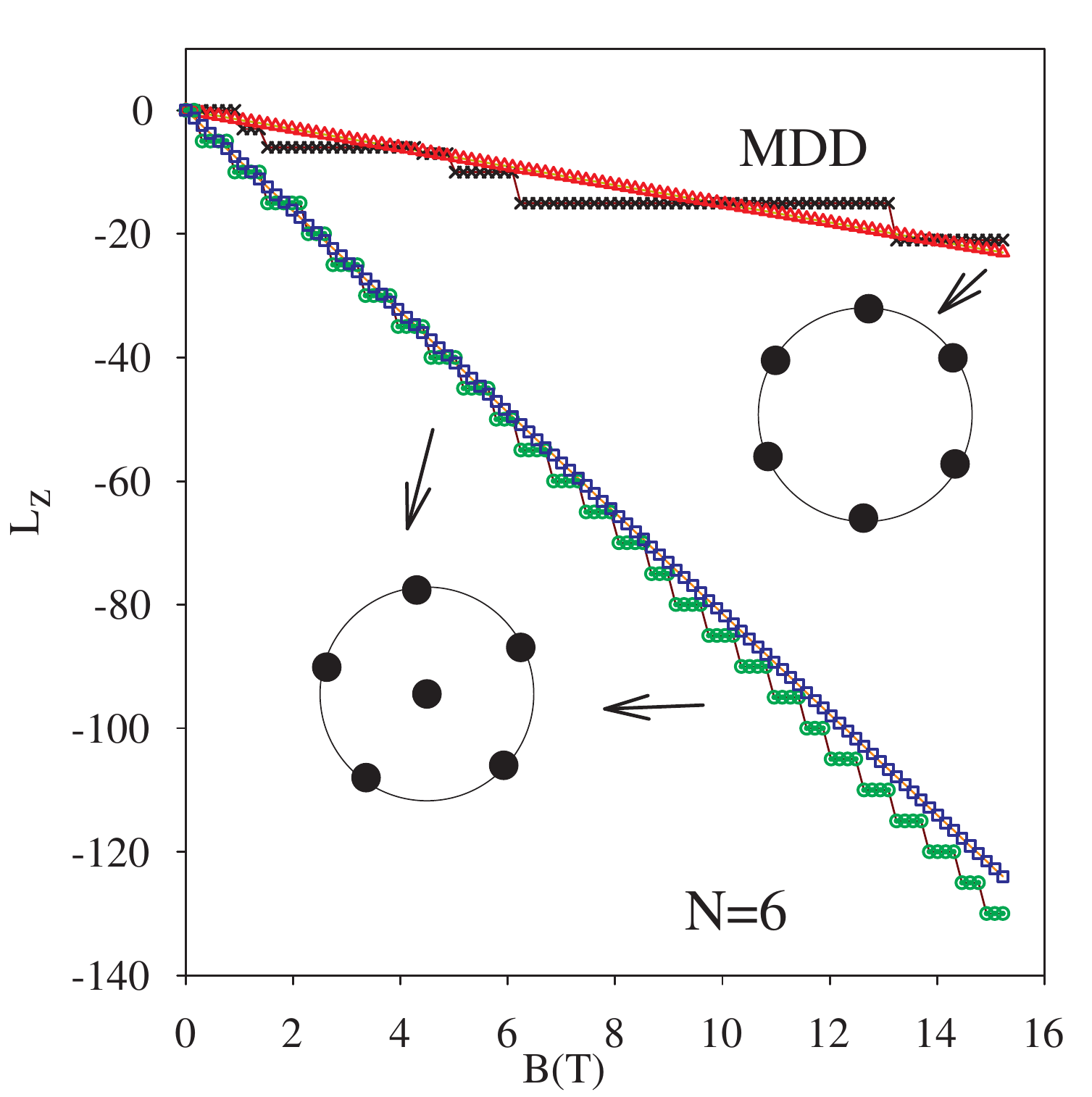} }
\caption{(Color online)
The ground state orbital momentum for six electrons as a function of magnetic
field. The symbols are: $\times$ (black) corresponds to $R_W=0.8$;
open circle (green) corresponds to $R_W=10$. In both cases the MDD phase
has $L_z=15$ (the longest plateau is for $R_W=0.8$).
For $R_W=0.8 (10)$, starting from the MDD phase,
the ground state orbital momentum follows the sequence of magic numbers
$L_z = 15, 21...$ $(L_z = 15, 20, 25,...)$. Open triangles correspond
to the values of $L_z$ from Eq. (\ref{clm}) for $R_W=0.8(10)$.
The equidistant points on the circles are associated with the equilibrium
configurations at a given $R_W$. The equilibrium
classical configurations are $N=5+1$ $(R_W=10)$. From \cite{ton1}.}
\label{angmom}
\end{figure}

The origin of the magic values for the orbital momentum, which occur after
the onset of the MDD phase
for $N=4$ $(L_z=6,10,14...)$, can be understood from the symmetry properties
of the total wave function of a few-electron QD. The classical minimum-energy
configuration of a few point charges in a parabolic potential is highly
symmetric \cite{tom}. Such configurations takes place if $N$ point charges create
equidistant nodes on the ring, with the angle $\alpha=2\pi /N$. It is
reasonable to suppose that a quantum ground state is localized around the
equilibrium classical state (see below). However, one has to take
into account that the total wave function should be antisymmetric.
Note that the rotation on the angle $\alpha$ is equivalent to a cyclic
permutation of the particle coordinates on the ring.
Since the spatial part of the total wave function $\Phi_r$ is the eigenstate
of the $L_z$ operator, one can require the fulfillment of the following symmetry
condition \cite{ruan}
\be
R_z(\alpha)\Phi_r=\exp(i\frac{2\pi}{N} L_z )\Phi_r\equiv P_{p}\Phi_r=
\varepsilon\Phi_r \; .
\ee
Here $\varepsilon=\pm$ is the parity of the permutation for the $N$ point
charges located on the ring. If the permutation is odd, then
$\exp(i\frac{2\pi}{N} L_z )\Phi_r=-\Phi_r$, and
the total orbital momentum takes the values $L_z=N(2k+1)/2$ ($k=1,2...$).
If the permutation is even, one has
$\exp(i\frac{2\pi}{N} L_z )\Phi_r=\Phi_r$.
As a result, the lowest quantum states carry the magic orbital momenta
$L_z=Nk$ ($k=1,2...$). The classical configuration ($R_W\gg 1$) for $N\ge 6$
contains a few rings \cite{loz1,loz2,bol,bed}.
In this case, the magic numbers can be determined by the number of electrons
in the external or in the internal rings starting from the orbital momentum
$L_z^{(MDD)}$ of the MDD phase. For nonzero magnetic field, this number must
obey also the condition Eq. (\ref{clm}) (see below).

The total energy for Hamiltonian (\ref{hamr}) can be
written in the following form
\bea
E_N
=\biggl \langle \sum_{i=1}^N \Bigl \lbrack \frac{p_x^2+p_y^2}{2m}+
\frac{m}{2}(\omega_0^2+\omega_L^2)r^2+\omega_L\ell_z
\Bigr \rbrack_i \biggr \rangle
+\bigl \langle V(r)+ g^* \mu_B B S_z \bigr \rangle \; ,
\eea
where $\omega_L=\omega_c/2$ and $\langle...\rangle$ means a mean field value.
Note that the potential $V(r)$ depends only on the relative distance $r$ between
particles. For a nonzero magnetic field, there is a sequence of different
energies for a fixed value of the total orbital momentum $L_z$.
Using the Hellman-Feynman theorem, we have for the lowest equilibrium state,
for a fixed value of the total orbital momentum $L_z$ at a given magnetic field,
\bea
\label{clm}
\frac{\partial E_N}{\partial \omega_L}\biggm |_{L_z=L}\!\!\!\!\!\!\!\!=\;0
\quad\Longrightarrow\quad L_z=
\Bigl \langle  \sum_{i=1}^N\ell_z\Bigr \rangle
=-m\,\omega_L\Bigl\langle
\sum_{i=1}^N r_i^2\Bigr\rangle=-\omega_L{\Im} \; .
\eea
With the moment of inertia ${\Im}\sim R_W^{2/3}$, one obtains an estimate
for the optimal sequence of values of the ground state orbital momentum $L_z$
at different values of the magnetic field $(\omega_L)$ and for different values
of the Wigner parameter $R_W$ $(\Im)$ (see Fig. \ref{angmom}).
\begin{figure}[ht]
\centerline{
\includegraphics[width=25pc]{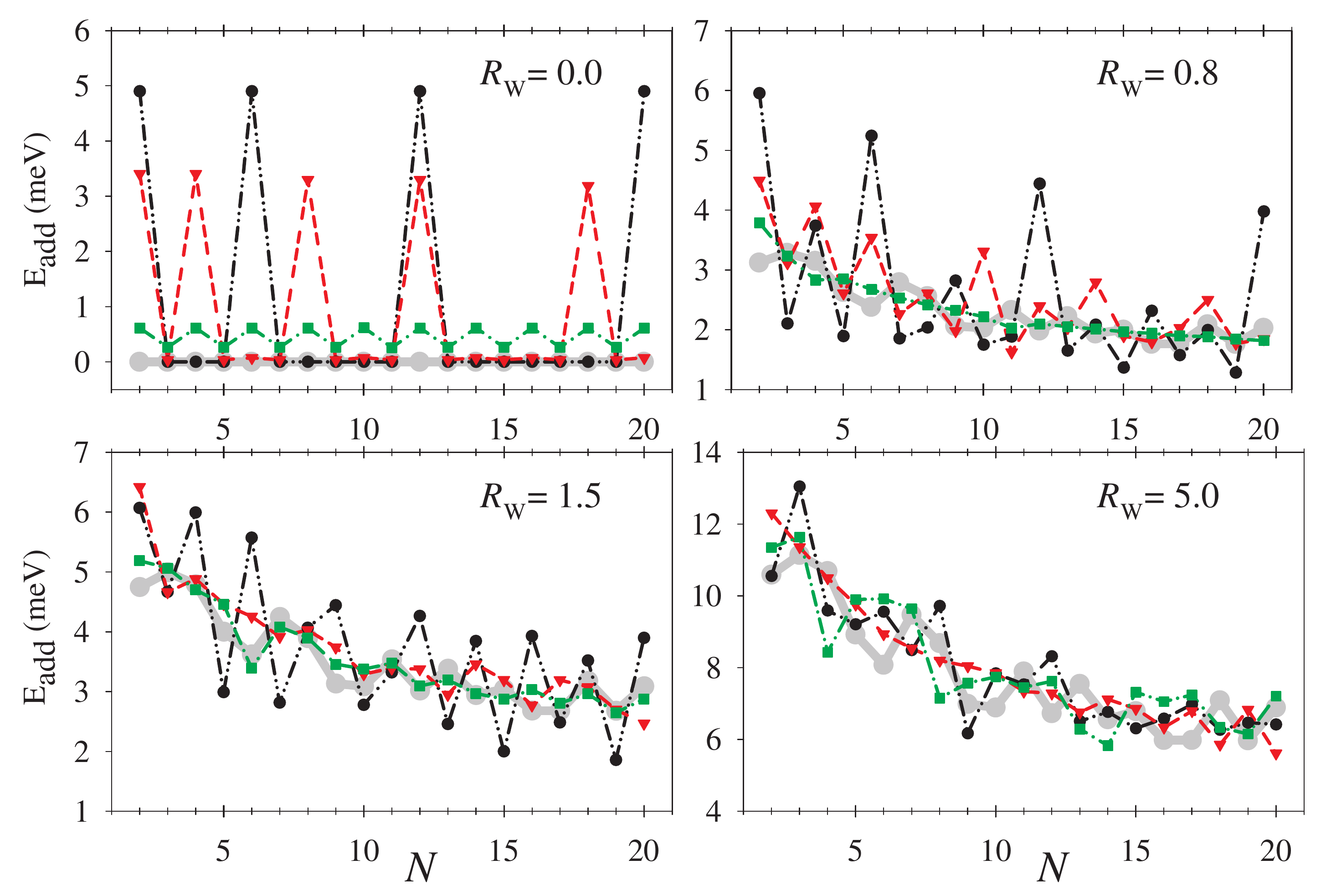} }
\caption{\label{ademig}(Color online)
Addition energy $E_{\it add}$ as a function of the magnetic field $B$ and
the strength parameter $R_W$. The values of $E_{\it add}$, at different values of
the magnetic field: dash-dotted-dash line connects the maxima of full circles
for $B=0$ T; dashed line connects the maxima of triangle-down for $B=2$ T;
dashed-dotted line connects the maxima of full squares for $B=15$ T; a thick line
displays the classical limit. From \cite{ton1}.}
\end{figure}

To trace the evolution of the shell structure,
the addition energy for the QD up to $N=20$ electrons at different values of
the strength parameter $R_W$ and the magnetic field $B$
has been calculated \cite{ton1}(see Fig.\ref{ademig}).
With the increase of the interaction strength $R_W=0.8$, the shell structure
survives only without the magnetic field. The magnetic field compresses the
energy gaps between energy levels forming the Hall liquid. At relatively
large $R_W=5$, one observes the onset of the classical limit
(Thomson model) \cite{tom,tom1}. We recall that the classical limit can also be
reached at a high magnetic field but with a relatively small $R_W$. In the
classical limit at $B=0$, the total energy (determined by Hamiltonian (\ref{rv1})
without the kinetic term) for one-ring equilibrium configuration is
\be
E_N=\frac{Na^2}{2}+\frac{R_WNS_N}{4a},\quad S_N=\sum_{j=1}^{N-1} \frac{1}{\sin{\frac{\pi}{N}j}} \; ,
\ee
where $a$ is the equilibrium radius of the electrons equispaced on the ring.
The equilibrium condition $dE_N/da=0$ enables one to define $a$ and to obtain
finally
\be
E_N^{\it cl}= (3/8)(2R_WS_N)^{2/3}N \hbar \omega_0 \; .
\ee
While the quantum
shell structure diminishes, one observes the onset of
the {\it classical shell structure} (see Fig. \ref{ademig}) due to
the optimal packing of electrons in various two-dimensional rings with
$R_W\gg 1$. In the classical limit, one has, for example, the following
sequences of rings:
$N=6\Rightarrow(1,5)$;\,$N=10\Rightarrow(2,8)$;\,$N=20\Rightarrow(1,7,12)$
\cite{tom,tom1}. A general expression for the classical energy with $N$
electrons localized on a few rings can be found in \cite{yan}.

\subsubsection{Ground state in random phase approximation}
\label{subsec:IV.E.2}

\begin{figure}[t]
\centerline{
\includegraphics[width=3.2in]{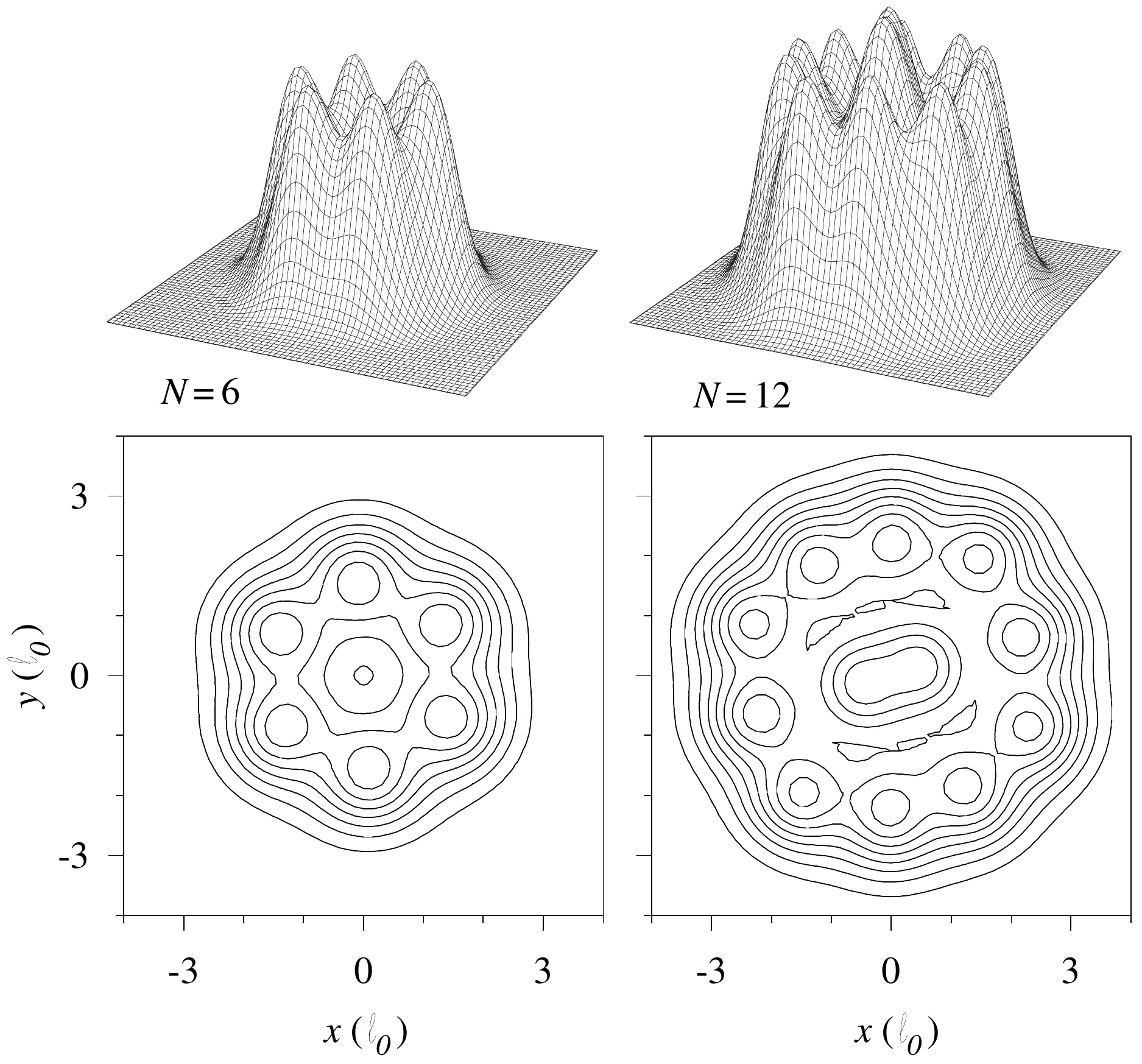} }
\caption{HF densities for the $N=6$- and 12-electron QDs at $R_W=1.89$. For
clarity, the upper plots display a 3D view of the corresponding densities.
Each line from the outermost contour line inwards corresponds, respectively,
to a density of 0.05, 0.10, 0.15 \dots,
etc in units of $\ell_0^{-2}$. From \cite{lor}.}
\label{hfQD}
\end{figure}

As discussed above, for large enough values of the ratio $R_W$ ($B=0$)
the HF field breaks the circular symmetry.
The spontaneous symmetry breaking leads to
a specific geometric distributions of electrons.
In Fig. \ref{hfQD}, the density for the $N=6$- and $12$-electron
QDs is displayed for $R_W=1.89$. Both cases show a clear
symmetry breaking; the $N=6$ having the electrons localized on a ring
while the $12$-electron dot has a central electron dimer surrounded by a
ring with 10 electrons.

The restoration of the broken symmetries
can be attained via projection techniques \cite{Ring,BR86}.
Examples demonstrating their use for the case of QDs have been
recently presented by Yannouleas and Landman \cite{yan}.
We shall focus our analysis on the RPA description
of the ground state and its connection to the excited states.

We recall that it was proved
by Thouless \cite{Th1,Th2} that, when the HF solution
corresponds to a minimum on the energy surface, the RPA equations
provide only real solutions.
To solve the RPA eigenvalue problem for QDs,
the quasi-boson approximation (QBA) developed in nuclear physics
has been used \cite{lor}.
Each particle-hole pair ($mi$) is considered as an elementary
boson, i.e., $b^\dagger_{mi}=a^\dagger_ma_i$.
A standard definition for the phonon (vibron) operator
\be
O_{\lambda}^{\dagger}=\sum_{mi}
{ \left( X_{mi}^{(\lambda)}\, a_m^{\dagger}a_i-
         Y_{mi}^{(\lambda)}\, a_i^{\dagger}a_m \right) },\quad
[O_{\lambda^\prime},O_{\lambda}^{\dagger}]=\delta_{\lambda^\prime,\lambda}
\ee
is used,
whose action on the (yet unknown) ground state $|0\rangle$
yields the excited vibrational states
$|\lambda\rangle = O_{\lambda}^{\dagger} |0\rangle \;$.

One can introduce the generalized momenta and coordinates related to the
vibron modes \cite{MW69}
\be
{\cal P}_\lambda=-i\hbar\sqrt{\frac{M_\lambda \Omega_\lambda}{2\hbar}}
(O_\lambda-O_\lambda^\dagger);\quad
{\cal X}_\lambda= \sqrt{\frac{\hbar}{2M_\lambda\Omega_\lambda}}
(O_\lambda+O_\lambda^\dagger) \; ,
\ee
which fulfill the commutation relations for the conjugate momenta and
coordinates. The number $M_\lambda$ is a mass parameter which should be
defined. The Hamiltonian, expressed in terms of the vibron operators, can
be represented through the generalized momenta and coordinates
\be
H_{RPA}=E_{HF}-\frac{1}{2}TrA+{\sum}_{\lambda}\bigg(\frac{{\cal P}_\lambda^2}{2M_\lambda}+
\frac{M_\lambda}{2}\Omega_\lambda^2{\cal X}_\lambda^2\bigg) \; ,
\label{hrpa}
\ee
where the submatrix $A$ is defined in \cite{lor}.
The operators ${\cal P}_\lambda$, ${\cal X}_\lambda$, therefore, obey the
equations of motion
\be
\label{emrpa}
[H_{RPA},{\cal P}_\lambda]=i\hbar\Omega_\lambda^2M_\lambda{\cal X}_\lambda;\quad
[H_{RPA},{\cal X}_\lambda]=-\frac{i\hbar}{M_\lambda}{\cal P}_\lambda \; .
\ee
To preserve the symmetry, broken in the mean field level and associated with a
generator (a generalized momenta ${\cal P}_0$),
one must require the fulfillment of the obvious condition:
\be
[H_{RPA},{\cal P}_0]=0 \;,
\ee
which determines a spurious (with zero energy)
RPA solution (Nambu-Goldstone mode).
The second equation (\ref{emrpa}) defines the mass parameter $M_0$.
In this case, Hamiltonian (\ref{hrpa}) transforms to the form
\be
H_{RPA}=E_{RPA}+{\sum}_{\lambda>0}\hbar\Omega_\lambda O_\lambda^\dagger O_\lambda
+\frac{{\cal P}_0^2}{2M_0} \; ,
\ee
where the third term describes {\it a collective mode} associated with
the broken symmetry mean field solution in {\it the intrinsic frame}.
The total energy in the RPA ($E_{\rm RPA}$)
can be split into the mean field contribution ($E_{\rm HF}$) and
a correction ($\Delta_{\rm RPA}$), as
\be
\label{etot}
E_{\rm RPA} = E_{\rm HF} - \Delta_{\rm RPA} \;.
\ee
The RPA correction reads \cite{Ring,BR86}
\bea
\Delta_{\rm RPA}
={\sum}_{\lambda >0} \hbar \Omega_\lambda{\sum}_{mi}|Y_{mi}^{\lambda}|^2+
\frac{\langle HF|{\cal P}_0^2|HF\rangle}{2M_0}=
\frac{1}{2}\left(\,
{\rm Tr}A-\sum_{\Omega_\lambda>0} \hbar\Omega_{\lambda} \, \right) \; .
\label{erpa}
\eea
The above Eq.\ (\ref{erpa}) includes the contribution from the vibrons
at positive frequencies $\Omega_\lambda$ and, also, from the spurious
mode. Note that Eq. \ (\ref{erpa})
is the result of a partial cancellation between two large terms. In practice,
this may cause $\Delta_{\rm RPA}$ to converge rather
slowly with a space dimension, especially for a large particle number.
For example, in heavy nuclei, the number of the
RPA eigenfrequencies is of order $10^4$ and each of them should be
determined with a high accuracy.
This problem  was reduced drastically by extending the RPA eigenvalues onto
the complex plane \cite{dan}. Using this trick, one can express the RPA
correlation $\Delta_{\rm RPA}$ as the contour integral
\be
\frac{1}{2}\left(\,\sum_{\Omega_{\lambda}>0} \hbar\Omega_{\lambda}-
 {\rm Tr}A\, \right) =
\frac{1}{4\pi i}\oint dz z\frac{\phi^\prime(z)}{\phi (z)} \; .
\label{form}
\ee
Here $\phi(z)=F(z)/F_0(z)$. The analytical function $F(z)$ of the complex
variable $z$ is defined by the general RPA eigenvalue equation
\be
F(z)=det(H-z)=0 \;,
\ee
where $H$ means the RPA matrix representation of the total Hamiltonian.
The function $F_0(z)=\lim_{V\rightarrow 0}F(z)$
is defined by poles of the RPA problem (e.g., particle-hole excitations
obtained from the HF solutions).
The integration can be done numerically.
The crucial practical advantage of the formula, Eq. (\ref{form}),
is that we are free to choose the rectangular contour sufficiently distant from
the poles such that the spectral function
\be
S(z)=\phi^\prime(z)/\phi(z)
\ee
becomes a smooth integrand. Then, the necessary grid needs not be
dense any more. In practical cases, considered
for some nuclear models \cite{dan}, the
number of integration points were reduced by a factor
$10^2$ without loss of a precision which reduces the computational
time drastically. The extension of this method for nonzero temperatures
can be found in \cite{kaneko}.

To treat the spurious mode related to rotation,
it is convenient to introduce the canonical conjugate operators
$L_z$ and
an {\em angle} operator $\Phi$; the latter being defined by the
following relations \cite{Th1,Th2,MW69}
\begin{equation}
\left[ H,{{L}_z} \right] =0,\,\,
[H,{C}]\; = \hbar {L}_z, \,\,
[{L}_z,{C}] = \hbar {\Im}_0 \; ,\;\;
\label{TV2}
\end{equation}
where $C=i{\Im}_0\Phi$ is an anti-Hermitian operator
and ${\Im}_0$ is the Thouless-Valatin moment of inertia \cite{TV62}.
The coefficients of the RPA operator
\be
L_z=\sum_{mi}{ \left( \ell_{mi}^{(z)}\, b^{\dagger}_{mi} +
         \ell_{mi}^{(z)*}\, b_{mi} \right) }
\ee
are directly given by the single-particle HF matrix elements. In contrast,
for the operator
\be
C=\sum_{mi}{ \left( c_{mi}\, b^{\dagger}_{mi} - c_{mi}^{*}\, b_{mi} \right) }\;,
\ee
one needs to solve the linear system of equations
\begin{equation}
  \label{rpasp}
  \left(
    \begin{array}{ll}
      A & B\\
      B^* & A^*
  \end{array}
  \right)
  \left(
  \begin{array}{c}
  c\\
  c^*
 \end{array}
  \right)=\hbar
  \left(
  \begin{array}{c}
  \ell^{(z)}\\
  \ell^{(z)*}
 \end{array}
  \right)\; ,
  \end{equation}
where the sub-matrix $B$ is defined in \cite{lor}.
Once these two sets of coefficients are determined,
the Thouless-Valatin moment of inertia ${\Im}_0$
may be calculated as
\be
{\Im}_0=\sum_{mi}{
\left( \ell_{mi}^{(z)*} c_{mi} + \ell_{mi}^{(z)} c_{mi}^* \right)} \; .
\ee
It was checked numerically \cite{lor} that the Thouless-Valatin moment of inertia
coincides, with a good accuracy, with the value obtained from a constrained
HF calculation for
${\cal R}={\cal H}-\lambda L_z$ as
\be
{\cal J}_0 = -\frac{d^2 \langle {\cal R}\rangle }{d\lambda^2}
= \frac{d\langle L_z\rangle}{d\lambda}
\; .
\ee
We stress that the equivalence between the two moments of inertia
can be fulfilled {\em if and only if} a self-consistent HF minimum
solution is found.

The operators ${L}_z$ and $C$, together with
the RPA phonons $O^\dagger_\lambda$, $O_\lambda$, form a complete set
for any operator linear in the bosons $b_{mi}^{\dagger}$ and $b_{mi}$.
For an arbitrary one-particle-one-hole operator the
expansion in terms of RPA excitations (for a single spurious mode)
reads as
\bea
\label{def}
{F} =
\sum_{\omega_\lambda>0}\left( [O_{\lambda},{F}]\, O_{\lambda}^{\dagger} +
                     [{F},O_{\lambda}^{\dagger}]\, O_{\lambda}\right)
+ \frac{1}{{\Im}_0}\left( [{F},{C}]\, {L}_z +
                              [{L}_z,{F}]\, {C} \right)\; .
\eea
Note that ${\Im}_0$ is nonzero only for the case in which
the symmetry is broken by the HF solution. Otherwise, the second term
in Eq. (\ref{def}) vanishes identically (see also \cite{Ring}).

In order to construct the RPA ground state for a circular-symmetric dot,
two conditions are used \cite{lor}:
\be
O_{\lambda}|0\rangle=0\;, L_z|0\rangle=0\, ,
\ee
which insure the rotational invariance of the ground state.
One looks for the solutions of the form
\be
|0\rangle=N_0\,\, e^{\cal S}\,|HF\rangle\;,\;
{\cal S}=\frac{1}{2}\sum_{mi\,nj}{Z_{mi\,nj}\, b_{mi}^{\dagger}b_{nj}^{\dagger}} \; ,
\ee
with the ${\cal S}$ operator involving the creation of two bosons \cite{Ring}.
Here $N_0$ is a normalization constant and the matrix $Z_{mi\,nj}$,
in general, is complex and symmetric in the boson indices, i.e.,
$Z_{mi\,nj} = Z_{nj\,mi}$.
Using the general identity
\be
F e^{\cal S} = e^{\cal S} \left( F + [\,F, {\cal S}\,] +
\frac{1}{2} \left[\, [\,F, {\cal S}\,], {\cal S}\, \right] +
\dots \right)
\ee
for the operators $F\equiv O_{\lambda}$ and $F\equiv L_z$, as well as
the condition of rotational invariance, and the QBA, one finds
\be
\left(\, F + [\,F, {\cal S}\,]\, \right) \, |HF\rangle = 0 .
\ee
For the case of unbroken symmetry, the above requirement leads to the
following equations for the $Z_{mi\,nj}$ coefficients:
\be
{Y_{mi}^{(\lambda)*}} =
\sum_{nj}Z_{mi\,nj}\,{X_{nj}^{(\lambda)*}}.
\ee
When the mean field breaks rotational symmetry,
we have to complement the above system of equations for phonons, which are
reduced to one equation, with the additional condition for the spurious mode
\be
\ell_{mi}^{(z)} = - \sum_{nj}Z_{mi\,nj}\,{\ell_{nj}^{(z)*}}.
\ee
Finally, with the aid of these conditions and the expansion
of the boson operators $b_{mi}^{\dagger}$ in terms of
phonons and spurious modes, Eq.\ (\ref{def}), one obtains \cite{lor}:
\begin{eqnarray}
\label{pp}
&&\langle 0| a_m^{\dagger}a_m|0\rangle =
\frac{1}{2}\sum_{i\lambda}|Y_{mi}^\lambda|^2
 + \langle 0|\Phi^2 |0\rangle \sum_{i}|\ell_{mi}^z|^2 \,,   \\
\label{hh}
&&\langle 0 | a_i^\dagger a_i |0\rangle =
1-\frac{1}{2}\sum_{m\lambda}|Y_{mi}^\lambda|^2
+ \langle 0 |\Phi^2 |0\rangle \sum_{m}|\ell_{mi}^z|^2\,\nonumber \; .
\end{eqnarray}
These equations generalize the result known in the literature \cite{Rowe}
by introducing an additional term related to the canonical variables
of the spurious mode $\{L_z, \Phi\}$.
It should be noted that the factor $1/2$ in Eqs.\ (\ref{pp})
is introduced following  \cite{14}, where the occupation
numbers were calculated using fermionic anticommutator rules, without
referring to the QBA.

The above results were used to obtain the expectation
value of any one-body operator, such as, e.g., the particle density
$$
\hat{\rho}({\bf r})=\sum_{\alpha\beta}{
\rho_{\alpha \beta}({\bf r})\,
a_{\alpha}^{\dagger}a_\beta},\quad
\rho_{\alpha \beta}({\bf r})=\varphi^*_\alpha({\bf r})\varphi_\beta({\bf r})\;,
$$
where indexes $\alpha$ and $\beta$ run over all the HF set of orbitals
(with $\varphi_{\alpha(\beta)}$ being the HF wave functions).

\begin{figure}[t]
\centerline{
\includegraphics[width=3.2in]{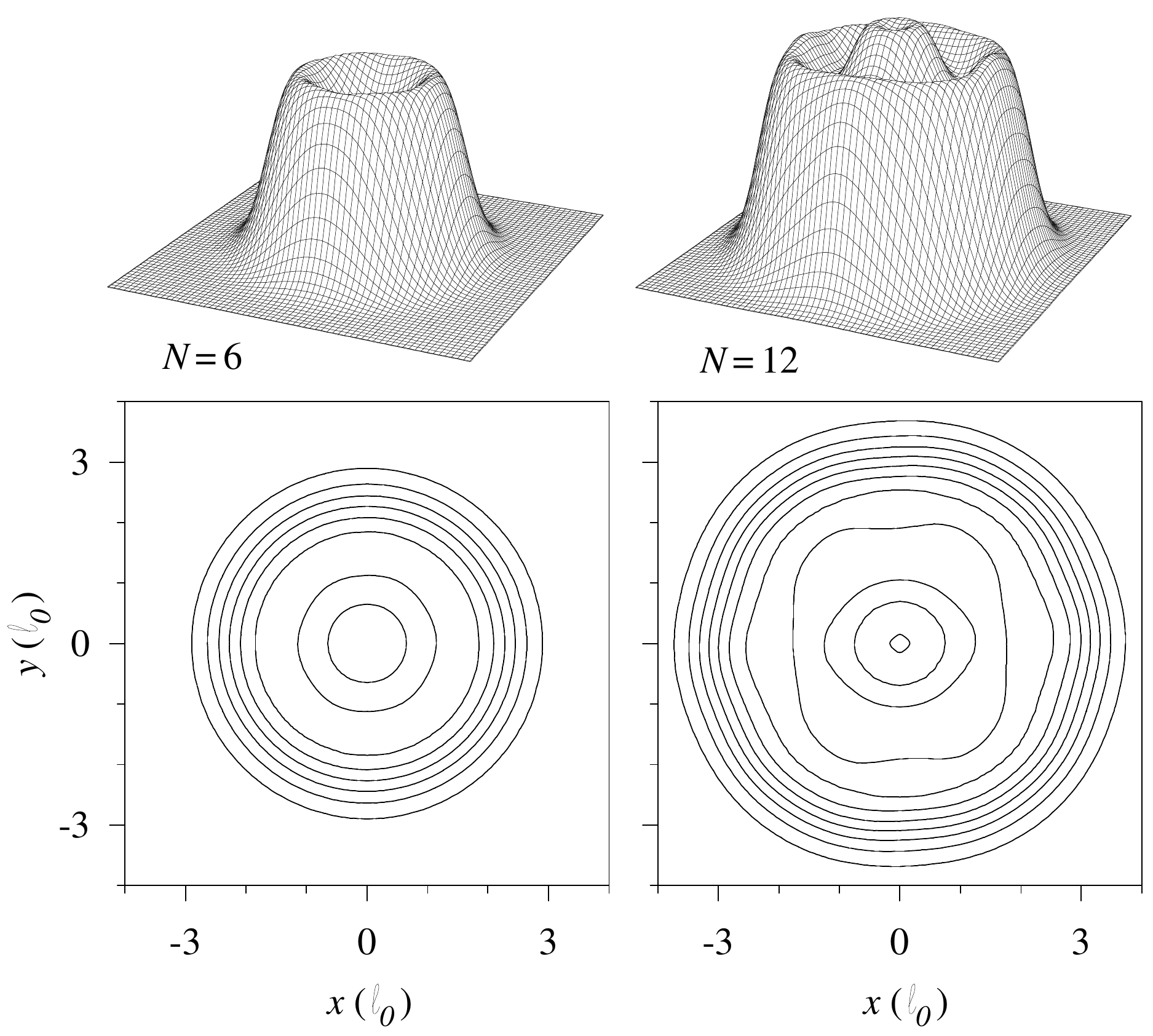} }
\caption{RPA symmetry restoration of the HF densities displayed
in Fig. \ref{hfQD}. From \cite{lor}.}
\label{rest}
\end{figure}

An excellent restoration of the circular symmetry is obtained
for the $N=6$ dot (compare Figs. \ref{hfQD},\ref{rest}). For $N=12$, the
RPA density, though more circular than the HF one, still has some residual
deformation. This can be surely attributed to incompleteness of the RPA
space considered in the numerical calculation.
We emphasize that the circular symmetry can be restored within
the RPA only if the contribution from the spurious mode is taken into
account.

The mean field  solutions, obtained within variational approaches,
are related to the classical equilibrium points
on the total energy surface of the full Hamiltonian \cite{Ring}.
Performing the RPA after the
mean field treatment (mean field+RPA approach), one takes into account
the quantized weak amplitude motion, not included
into the static mean-field solution. The quantum fluctuations
in finite systems lead not only to series of collective
excitations, corresponding to rotations and vibrations, but also give rise
to typical correlations in the ground state, which change
its properties. Hence, the calculation of
the RPA correlation energy is an important ingredient
of a consistent description of the ground state properties
of finite quantum systems within the mean field+RPA approach.
Note that the correction $\Delta_{\rm RPA}$, with a minus sign,
gives the standard correlation energy of the system. A decrease
in the ground state energy ($\Delta_{\rm RPA}>0$) implies an improvement
with respect to the mean field theory. However, since RPA
is not based on the energy minimization, it is not bound to fulfill
the variational principle and, therefore, $E_{\rm RPA}$ could be even lower
than the exact energy. In other words, the correlation energy could
be overestimated in RPA, as it has been suggested in literature
\cite{BR86}.
We conclude that self-consistent mean field
calculations, combined with the RPA analysis, could be useful to reveal
structural changes in mesoscopic systems, allowing for the detection of
quantum shape transitions. In the considered cases, the broken
mean-field solution (the Wigner molecule) can be detected through the
appearance of rotational states in a quantum spectrum of the dot.

\section{Atomic Nuclei}
\label{sec:V}

\subsection{Signatures of symmetry breaking in nuclear structure}
\label{subsec:V.A.}

The evolution of nuclear structure, with the change of proton $(Z)$ and
neutron $(N)$ numbers, is one of the fundamental issues in nuclear
physics \cite{BM,NR,heyde}. The shell structure is one of the most
important quantum features of nuclei. A remarkable stability, found in
nuclei at magic numbers 2, 8, 20, 28, 50, 82, and 126, was  a
principal step in the development of the shell model, based on the
inversion and rotational symmetries. The introduction of a
spin-orbit splitting in the nuclear shell model was crucial to
reproduce the nuclear magic numbers. The modern level of the
nuclear shell model and the results which illustrate the global
features of the approach are outlined in a review \cite{shmod05}.
Recent progress in ion beam facilities provides the technical
possibility of exploring nuclei with extreme proton-to-neutron
ratios, far from the valley of stability. Experimental results on
mass, nuclear radius, and spectroscopy, obtained in the last decade,
indicate that location and magnitude of the shell gaps in the
neighbourhood of the valley of stability may change, giving rise to
new magic numbers; see, e.g., \cite{sheros,mag1}. Evidently,
the study of the nuclear shell gaps, far from the stable isotopes, provides
a new impetus for the development of nuclear structure theory. In
particular, it requires the development of new concepts and ideas on the
properties of quasi-stationary open quantum systems. The possible
extension of the shell model for the description of weakly bound nuclei
can be found in the review \cite{mich}. The recent status of
experimental results, related to general trends in the evolution of
shell closures with the increase of $N/Z$ numbers, including nuclei
far from stability lines, is discussed in the review \cite{magnum}.
Note, that the evolution of shell gaps of neutron rich nuclei is important
for modelling the mass processing along the astrophysical rapid neutron
capture pathway.

In finite systems such as quantum dots, metallic clusters and
nuclei, the quantization of a system of fermions, moving in a common
potential, leads to a bunching of levels in the single-particle
spectrum, known as shells. When the levels of a bunch are filled, the
system is stable; an additional particle fills a level
in the next bunch at considerably higher energy and, therefore
produces a less stable system. In stable nuclei, the shell effects
are discerned by deviations of the binding energies from the smooth
variation obtained from the liquid-drop model or as oscillations in
the radial density distribution as a function of particle number and
of deformation \cite{Ring,NR}. Consequently, spherical symmetry
leads to very strong shell effects manifested in the stability of
the noble gases, nuclei and metallic clusters \cite{BM,Heer}. When
a spherical shell is only partially filled, breaking of spherical
symmetry, resulting in an energy gain, can give rise to a deformed
equilibrium shape.

Due to spin-orbit interaction between nucleons, neither spin nor
orbital angular momentum are good quantum numbers. The experimental
excitation spectra can be quite complex. However, the density of
states near the ground state is low, and one can study discrete
quantum levels by means of nuclear spectroscopy. These levels are
characterized by good quantum numbers, such as the total angular
momentum and parity $I^{\pi}$. For example, in even-even $(N,Z$
are even) nuclei the ground state has positive parity and the total
angular momentum $ I^\pi=0^+$. The study of the first low-lying
states, with positive parity and angular momentum $I^\pi=2^+$ and
$I^\pi=4^+$ with their ratio $R=E(4_1^+)/E(2_1^+)$ in various
even-even nuclei, indicates the transition from spherical to
deformed shapes, if one starts from a nucleus, that has all occupied
shells in the framework of the nuclear shell model, towards that
with a half filled next shell \cite{heyde,NR}. Indeed, a remarkably
simple phenomenology appears for these quantities (among other
empirical observables on shape-phase transitions) discussed in terms
of very simple models in a recent review \cite{pcej}.
For nuclei with either $N$ or $Z$ equal or close to the magic value,
the ratio $R\approx 2.2-2.4$, which can be interpreted as a ratio
between one and two-phonon states of a quadrupole harmonic vibrator
with a spherical ground-state. For nuclei with both $N$ and $Z$ well
away from magic numbers, $R\approx 3.33$, which corresponds to
non-spherical deformed nuclei that rotate according to the eigenvalue
expression for a quantum mechanical symmetric rotor $E\sim I(I+1)$.
Although the excitation nuclear spectra are quite complex,
experimentalists classified nuclear energy levels according to the
intensity of electromagnetic transitions of low multipolarity
$(E2,M1,E1,E3)$, which establishes a link between the states of similar
nature. In agreement with this procedure, the measured
$\gamma$-transitions, according to their multipolarity and
intensities, are arranged in rotational bands. In fact, a
transparent signature of a broken spherical symmetry in a nucleus
are the strongly enhanced electric quadrupole $(E2)$ $\gamma$-transitions
between $\Delta I=2 \hbar$ states of a rotational band. The reduced
transition probabilities, $B(E2)$, are proportional to the square of
the quadrupole moment, $Q^2$. The larger the deformation, the
stronger the $E2$ transitions between the rotational energy
levels.

Nuclear rotation has features that are rather different from the
rotation of molecules. The common practice in the analysis of the empirical
data in terms of the rotational frequency. The latter is defined by means of
the classical canonical relation
\be
\Omega=\frac{dE}{dJ}=\frac{E(I+1)-E(I-1)}{2},\quad J=I-1/2.
\ee
 In these equations, one must associate the classical angular momentum $J$
with the quantum value $I$
for a given stretched quadrupole $\gamma$-transitions $\Delta I=2$
\cite{sfrau}. Here, the term 1/2 is a quantum correction that
appears in the RPA; see discussion in \cite{ADN,mar1}. Two
characteristics of the rotational band are used in nuclear
structure, namely the kinematical ${\Im}^{(1)}$ and
dynamical ${\Im}^{(2)}$ moments of inertia:
\be
{\Im}^{(1)}=\frac{I}{\Omega},\quad {\Im}^{(2)}=\frac{dI}{d\Omega}
\approx \frac{4}{\Delta E_\gamma}.
\ee
Here, $\hbar\Omega=E_{\gamma}/2$, $E_{\gamma}$
is the $\gamma$-transition energy between two neighbouring states
that differ in two units of the angular momentum $I$, and
$\Delta E_{\gamma}$ is the difference between two consecutive
$\gamma$-transitions.

The moment of inertia ${\Im}^{(1)}$, extracted from the rotational
band at low spin region, is essentially below that of a rigid rotor, but
deviates from the value of a classical liquid drop of the same
volume. It was one of the indications of pairing correlations in
even-even nuclei noticed by Bohr {\it et al.} \cite{BMP}. The
empirically observed energy gap between the ground state and the
first single-particle excitation in even-even nuclei was the other
important fact indicating the importance of pairing correlations
in low excited states of deformed nuclei \cite{Bogolubov_1958,BMP}.
These facts led these authors to the conclusion on the similarity of the
nuclear spectrum of even-even nucleus to that of the superconducting
metal and on the existence of superfluidity in nuclei.

\begin{figure}[h]
\centerline{
\includegraphics[width=3.0in]{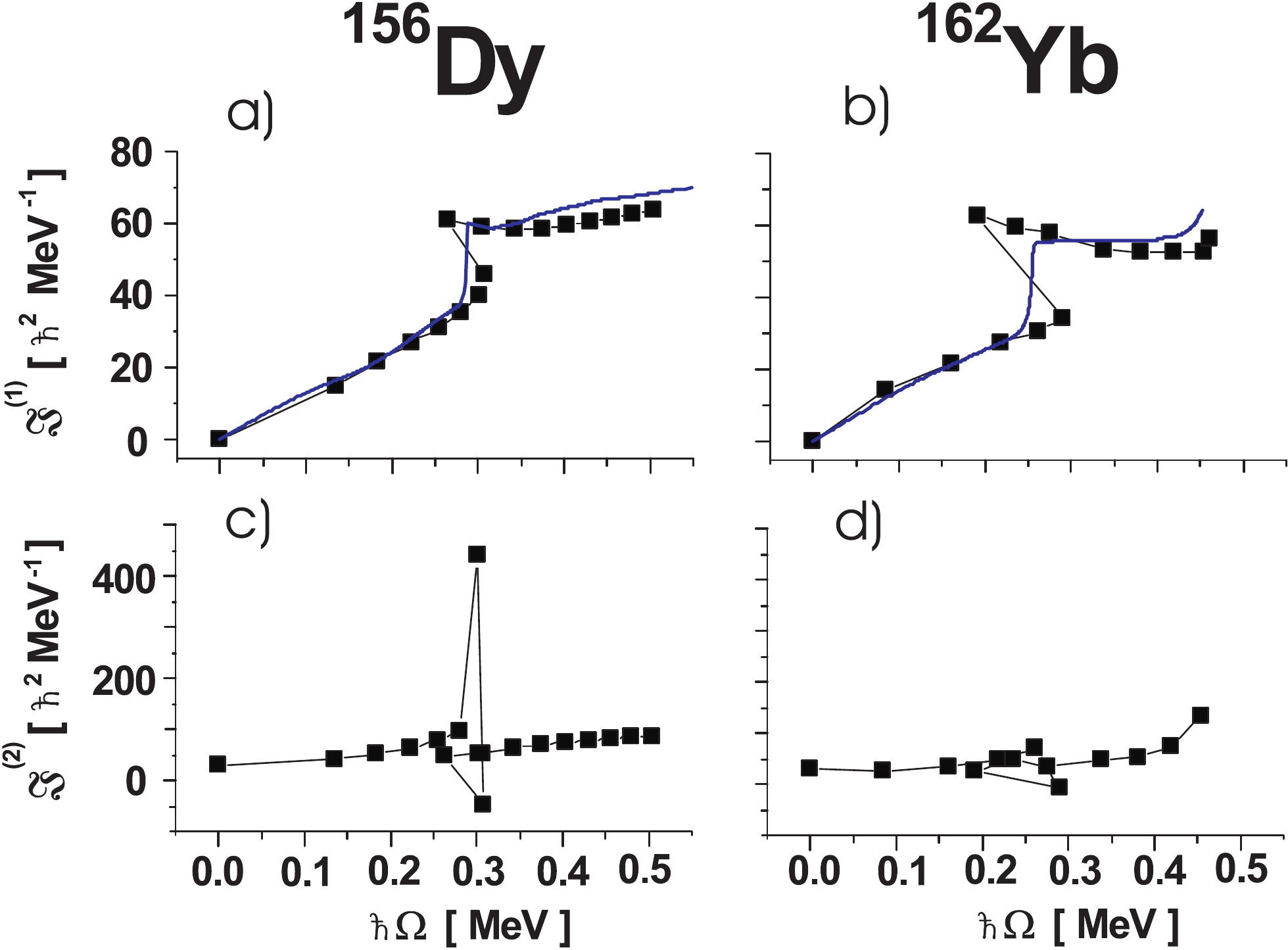} }
\caption{Rotational behaviour of the experimental, kinematical
$\Im^{(1)}$ and dynamical $\Im^{(2)}$, moments of inertia.
The experimental data denoted by black squares are taken
from \cite{brook}. The experimental
rotational frequency at the transition point is $\hbar\Omega_c
\approx 0.27, 0.32$ MeV for $^{162}$Yb and $^{156}$Dy, respectively.
The results of calculations for $\Im^{(1)}$ are connected by a solid
line. From  \cite{bb3}.}
\label{figbb}
\end{figure}

Rotational bands may terminate after a finite number of transitions
\cite{afarag}. It is accepted that the states with the lowest angular
momentum, at a given excitation energies (called {\it yrast} states),
are characterized by zero temperature. The sequence of these
states is called the {\it yrast line}. A paradigm of structural
changes in a nucleus under rotation is a {\it backbending}; the
property that is exhibited by a sudden increase of a nuclear moment
of inertia ${\Im}^{(1)}$ in the yrast rotational band at some critical
angular momentum or rotational frequency  \cite{Jonbb} (see
experimental results for
 $^{156}$Dy and $^{162}$Yb in Fig. \ref{figbb}).
This phenomenon may be explained as a result of the rotational
alignment of angular momenta of a nucleon pair occupying a high-$j$
intruder orbital near the Fermi surface along the axis of collective
rotation \cite{step}. The alignment breaks the singlet Cooper pairing
in the pair and decreases the superfluidity of a rotating nucleus.
The effect of rotation is similar to the effect of  magnetic field
on a superconductor, as was noticed long ago \cite{MV}.

While one observes a similar pattern for the backbending in the
considered nuclei (see Fig. \ref{figbb}a,b), a different response of
the nuclear field to rotation becomes more evident with the aid of the
experimental dynamical moment of inertia
$\Im^{(2)}$ as a function of the
angular frequency (see Fig. \ref{figbb} c, d). Indeed, the dynamical
moment of inertia, due to the obvious relation
\be
\Im^{(2)} = \Im^{(1)}+\Omega \frac{d\Im^{(1)}}{d\Omega}\;,
\ee
is very sensitive to
structural changes of a nuclear field. At the transition point,
$\Im^{(2)}$ wildly fluctuates with a huge amplitude in $^{156}$Dy,
whereas these fluctuation are quite mild in $^{162}$Yb. Continuing
the analogy between the backbending and the behaviour of
superconductors in a magnetic field, the dynamical moment of inertia
is similar  to a susceptibility of a sample. Its description poses a
real challenge for the microscopic approaches.

The shell structure plays a prominent role in the formation of
rotational nuclear states. It was suggested \cite{gel} that strongly
deformed nuclei could occur because of the formation of a new shell
structure. In 1986 Twin {\it et al.} \cite{twin} discovered a very
regular pattern of closely spaced $\gamma$-transitions in the
spectrum of $^{152}$Dy between the spin levels ranging from $60\hbar$
to $24\hbar$. The moment of inertia
${\Im}^{(1)}$ of the associated rotational band was close to that of
the axially symmetric rigid rotor with a $2:1$ axis ratio. This nucleus
was called superdeformed. Evidently, the high level density around the
Fermi level corresponds to less stable system. Superdeformed
rotating nuclei are among the most fascinating examples, where the
deviation from the spherical shape are a consequence of strong shell
closures giving rise to the largest level bunching, i.e., the largest
degeneracy or the lowest level density around the Fermi level. The
experimental results on superdeformed rotating nuclei are reviewed
in \cite{janss,bakt}.

The discovery of superdeformed nuclei triggered the concept of
pseudo-$SU(3)$ symmetry \cite{dra0,dra1,dra2} to explain many features
of superdeformed states described well by means of phenomenological
nuclear potentials \cite{dudek}. In the new scheme, the spin-orbit
splitting appears to be very small and the properties of the
single-particle spectrum are similar to those observed in $3D$
harmonic oscillator with rational ratios of frequencies
\cite{beng,nadob}. In this scheme, rotation stabilizes a very large
quadrupole deformation and acts like a magnetic field in QDs, giving
rise to new magic numbers (see Fig. \ref{sheff} and the following
discussion in terms of a simple shell model for QDs in
Sec. \ref{subsubsec:IV.A.1}).

Thanks to novel experimental detectors, a new frontier of
discrete-line $\gamma$-spectroscopy at very high spins has been
opened in the rare-earth nuclei; see, e.g., \cite{hspin,hspin1}
and references therein. These
nuclei can accommodate the highest values of the angular momentum,
providing one with various nuclear structure phenomena. The quest
for the manifestations of non-axial deformation is one of the driving
forces in the current high spin physics.

In recent years, rotational-like sequences of strongly enhanced $M1$
transitions between $\Delta I = 1 \hbar$ states (as opposed to the
strongly enhanced $E2$ transitions in the case, for example, of the
superdeformed band) have been observed in spherical or
near-spherical nuclei. The occurrence of regular, rotational-like
band structures may be explained by an effective interaction between
the excited high-$j$ particles and holes \cite{sfrau}. The high-$j$
particles and holes have energetically favourable alignments in the
slightly deformed potential. A consecutive alignment of these two
large vectors along the total angular momentum vector ${\bf J}$
produces rotational-like band structure. Since this process
resembles the closing of two blades of shears it was named a {\it
shears mechanism}. The coupling of the proton-particle and
neutron-hole orbitals, each with high spin $j$, results in a large
transverse component of the magnetic moment vector, $\mu_\perp$
that rotates around the total angular momentum, $I$, and creates
enhanced $M1$ transitions between the shears states. This type of
excitation has been named {\it magnetic rotation} \cite{sfrau} to
distinguish it from the well-known rotation of deformed nuclei. A
summary of the experimental evidence for the shears mechanism and of
a semi-empirical description of the observations was given in
\cite{clmac}.

One of the most important features of the nuclear many-body system is
that the nucleon density inside the nucleus is almost constant for the
most stable nuclei, and so is the binding energy per nucleon. Therefore,
it is a spatially and energetically saturated system with a
relatively sharp boundary. In virtue of a constant density with a
sharp boundary, a shape deformation can be chosen as a collective
degree of freedom. In the geometrical approach \cite{BM},  where the
concept of shape deformation is one of the basic cornerstones,
effects produced by quadrupole degrees of freedom are well
understood for various effective potentials. The need for multipole
deformations higher than the quadrupole has been recognized in
numerous calculations in order to explain experimental data \cite{sven}.
The hexadecapole deformation is essential for the understanding of
equilibrium shapes and fission process in normal and super-deformed
nuclei \cite{sven,dudek}. In some nuclei, a transition to a pear-shaped
(octupole) nuclear shape,
that breaks reflection symmetry in the intrinsic reference frame, has also
been found \cite{ahmad,Bu96}. Whether octupole degrees of freedom
are of any importance and how they are manifested,-- these questions
stimulate a noticeable fraction of experimental efforts in high spin
physics; see, e.g., \cite{garg,itemba}.

The presence of higher multipoles in the effective nuclear potential
leads, however, to a non-integrable problem from the theoretical point
of view. In fact, the
single-particle motion turns out to be chaotic. Accordingly, in the
transition from ordered to chaotic motion the quantum numbers lose
their significance, and the system behaves like a viscous fluid in a
container \cite{blocki}. Therefore, the disappearance of the shell
structure should be expected in the analogous quantum case
\cite{gutz,arv}. It was found, however, that shell structure can
appear at strong octupole deformation
\cite{arita1,arita2,H94,H95,H99}. In particular, it was shown that
for the prolate systems there is a remarkable stability against
chaos when the octupole deformation is switched on \cite{H94}. This
result is in agreement with the fact that there are more prolate
than oblate nuclei. The major conclusion of \cite{H95} is that,
albeit nonintegrable, an octupole and hexadecapole admixtures to a
quadrupole oscillator potential lead, for some values of the higher
multipole strengths, to a shell structure similar to a plain but
more deformed quadrupole potential.

A transition from quadrupole deformed shapes (regular motion) to
deformed shapes with higher multipoles (chaotic motion) can be
traced by means of tools developed in the random matrix theory. An
introduction into the basic concepts of random matrix theory and a
survey of the extant experimental information, related to the
manifestation of nuclear chaotic motion, can be found in the
recent review \cite{hans}. A deep understanding of shell structure
phenomena, in terms of classical trajectories, has been achieved by
Balian and Bloch \cite{balbl}, based on the periodic orbit theory \cite{gutz}.
According to the semiclassical theory \cite{gutz}, the
frequencies in the level density oscillations of single-particle
spectra of finite quantum systems (nuclei, quantum dots, metallic
clusters) are determined by the corresponding periods of classical
closed orbits. The short periodic orbits give the major contribution
to the gross shell structure \cite{BM,SVM,frisk,brbad}.

An important goal of nuclear structure physics is to understand
these phenomena microscopically, when one deals with nucleon degrees
of freedom and their interactions. However, the microscopic origin
of a nuclear shell model still remains a major challenge.
Collective model approaches, which express structure directly in
terms of the many-body degrees of freedom, symmetries, and their
associated quantum numbers, provide an invaluable help in the
formulation of the basic elements of the microscopic approaches to
many-body nuclear problem.
In Sec. \ref{subsec:V.B} we discuss the most popular theoretical approaches
to symmetry breaking phenomena in nuclear structure.
Sec. \ref{subsec.V.C.} reviews a few examples for the analysis of
shape phase transitions in rotating nuclei within
the cranking model+random phase approximation.

\subsection{Nuclear structure models and symmetry breaking phenomena}
\label{subsec:V.B}

Theoretical description of nuclear systems involves a number of
simplifications to tackle a problem of interpreting nuclear
structure from quark-gluon interactions. The analysis of experimental
data related to ground and excited states of different nuclei, up to the
excitation energies $\sim 20$ MeV above the yrast line with a
maximal angular momentum $I\sim 60 \hbar$, in all nuclear models,
demonstrates that treating nuclei in terms of nucleons, moving in
an effective potential, provides a reliable description. The
non-relativistic treatment of nuclei seems to be rather well
justified, if one compares the depth of the nuclear potential well
$V \le 60$ MeV with the nucleon mass $M\simeq 1000$ MeV. However,
the relativistic nature of the nucleon motion can not be excluded
from the consideration. Various versions of the relativistic models
for nuclear structure calculations have been discussed in several
reviews; see, e.g., \cite{sw1,reRMF,riRMF,meng}.
Among many applications, we can mention the
following studies related to the subject of the present review: (i) quantum
phase transitions in the ground state in the region $Z=60,62,64$,
with $N\approx 90$ \cite{qptring}, and (ii) a thorough analysis of the
time-odd mean fields (nuclear magnetism) in rotating nuclei \cite{afa1}.

\subsubsection{Geometric Collective Model}
\label{subsec:V.B.1}

The observation of rotational states in many nuclei led to models
similar to those developed in molecular physics. The theoretical
analysis of the rotational properties of many-particle systems
requires the definition of the rotating reference frame (usually
denoted as the ``body-fixed frame'' or, shortly, body frame). Note,
however, that there is no unique choice for such a system and its
definition should be based on the physical peculiarities of the
problem.

Various choices of body frames were analyzed by Eckart
\cite{eckart34:_inst_princ_ax,eckart35:_eckart_frame} within the
framework of the classical mechanics. Initially, Eckart proposed to
use the body frame defined by the principal axes of the inertia
tensor. He obtained the following expression for the rotational part
of the kinetic energy
\be
 \label{eq:1}
  K_{rot} = \frac{{\cal J}_x L_x^{2}}{2 ({\cal J}_y-{\cal J}_z)^{2}}+
\frac{{\cal J}_y L_y^{2}}{2 ({\cal J}_x-{\cal J}_z)^{2}}+
\frac{{\cal J}_z L_z^{2}}{2 ({\cal J}_x-{\cal J}_y)^{2}} \; ,
\ee
where ${\cal J}_k$ denotes the $k$-th principal moment of inertia. As is
seen, this expression diverges when some moments of inertia coincide
and it does not reproduce the kinetic energy of a rigid rotator. It
means that the Coriolis couplings in the principal axes frame are
not small for the small amplitude displacements of particles from
their equilibrium positions. Thus, the principal axes frame is not
suitable for the analysis of molecular vibrations.

Later, Eckart formulated the so-called ``Eckart condition'', which
demands that, in the body frame, the vector identity $ \sum_{i=1}^{N}
[\mathbf{r}_i \times \mathbf{r}^{(eq)}_i ] =0$ be fulfilled
\cite{eckart35:_eckart_frame}. Here, the vector
$\mathbf{r}^{(eq)}_i$ describes the equilibrium position of the
$i$-th particle. Physically, this condition means that, under small
vibrations, the components of the angular momentum in the body frame
must be proportional to the amplitude of these vibrations. Explicit
expressions for the basis vectors and the corresponding
roto-vibrational decomposition of the Hamiltonian for the Eckart
frame were recently derived in
\cite{meremianin03:_phys_rep,meremianin04:_eckart}. It was
demonstrated that, under small vibrations, the Coriolis part of the
kinetic energy, corresponding to the Eckart frame, is also small
\cite{meremianin04:_eckart}.

Eckart frame is widely used in molecular physics. Its application to
the problems of nuclear physics is limited by the observation that a
nucleus physically is more similar to a liquid drop than to a rigid
body. It is interesting that, since the quantum liquid drop, with
some equal moments of inertia, cannot rotate around the
remaining axis, the divergences of the rotational energy
(\ref{eq:1}) vanish for the liquid drop model. Therefore, the
principal axes frame can be efficiently used to describe the
rotational motion of physical systems modelled by a quantum
liquid. For example, this can be the case when trapped atoms are
considered. For QDs, in the ``Wigner crystallization'' regime,  the use
of the Eckart frame can be more adequate. Various aspects of molecular
description of QDs in the Eckart frame can be found in the
review \cite{mak}.

\begin{figure}[th]
\centerline{
\includegraphics[width=2.60in]{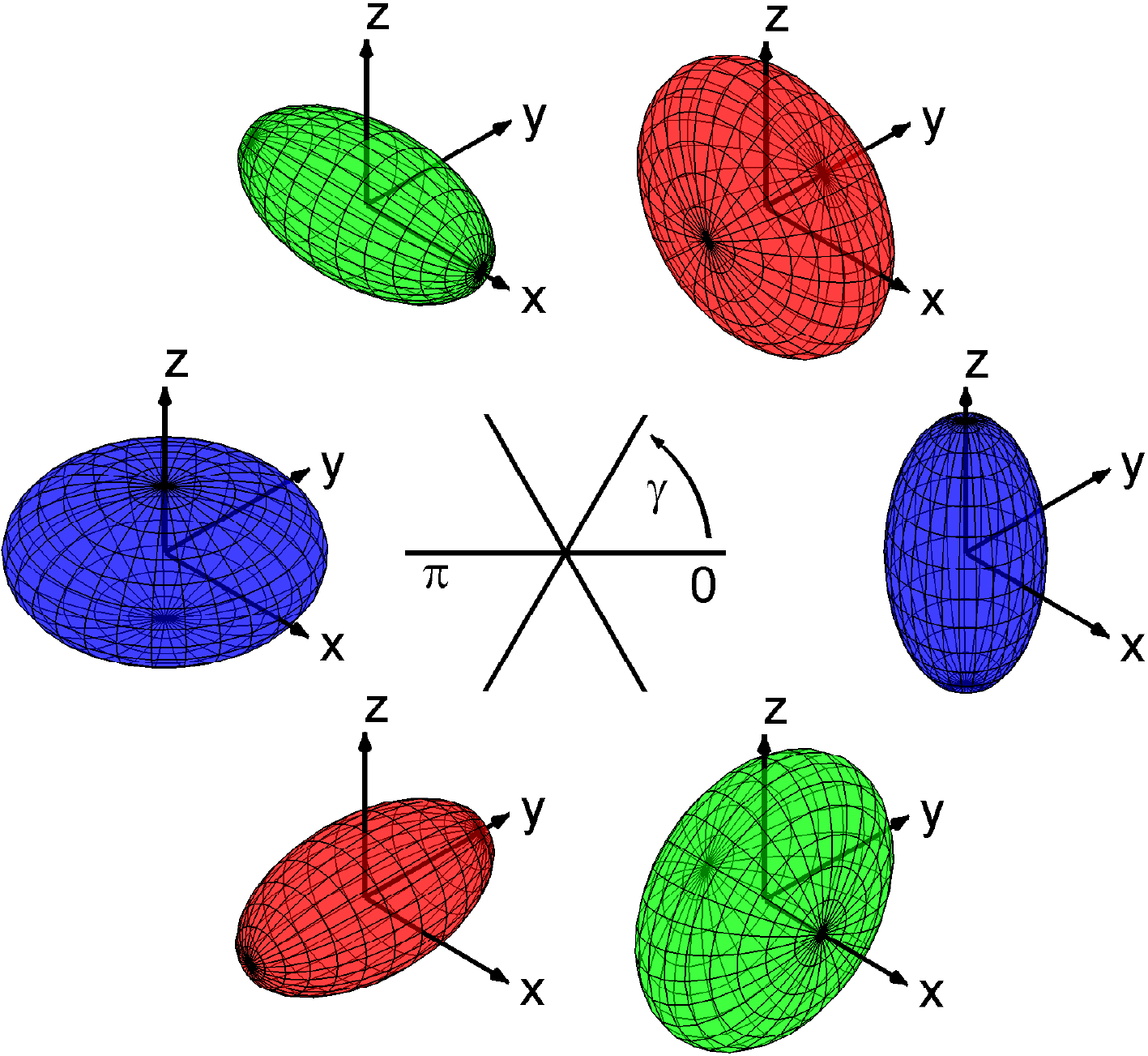} }
\caption{(Color online) Quadrupole shapes in the geometric model. The spherical
symmetry is at the center, while axially symmetric prolate and oblate
shapes are obtained along the various axes. A genuine triaxial
quadrupole deformation occurs between different axes. Adapted from
\cite{H99}.} \label{figgod}
\end{figure}

Theoretical studies of nuclear shape transitions are typically
based on the macroscopic geometric collective model introduced by Bohr
\cite{bohr}. In this model, the nucleus is considered as a droplet
and the nuclear shape is parametrized in terms of the multipole
expansion of the nuclear radius into spherical harmonics
$R=R_0(1+{\sum}_{\lambda \mu} {\alpha_{\lambda \mu}}Y_{\lambda
\mu})$. Because of the assumption that the dominant nuclear
deformation is the quadrupole one $\lambda=2$, the Bohr Hamiltonian
is parametrized in terms of two quadrupole deformation parameters
$\beta$ and $\gamma$: $\alpha_{20}=\beta \cos\gamma,\quad
\alpha_{21}=\alpha_{2-1}=0,\quad
\alpha_{22}=\alpha_{2-2}=-\beta\sin\gamma$ \cite{BM,NR}. These
parameters describe the nuclear shape in the intrinsic (body-fixed)
frame connected with the principal axes of the quadrupole tensor.
The dependence on the Euler angles $(\theta_1,\theta_2,\theta_3)$,
which describe the orientation of a nucleus in a laboratory frame, is
eliminated. Due to the symmetries of the above equations, it is
sufficient to consider the sector $\beta\ge 0,\quad 0^0\leq
\gamma\leq 60^0$ only. Prolate and oblate shapes correspond to
$\gamma=0^0$ and $60^0$, respectively. The other values of $\gamma$
describe the triaxial shapes (see Fig. \ref{figgod}).

In this model the shape phase transitions can be related to the concept
of critical-point symmetries that provide parameter-independent
predictions for excitation spectra and electromagnetic transition
rates for nuclei at the phase transition point; see details in
\cite{pcej}.

\subsubsection{Algebraic approach: Interaction Boson Model}
\label{subsec.V.B.2}

In the so-called interacting boson approximation, proposed by Arima
and Iachello \cite{ar1,ar1a,ar1b}, collective excitations of nuclei
are described in terms of bosons, which are associated with pairs of
valence fermions. To describe the quadrupole collectivity in nuclei,
one needs to consider a five-dimensional space. Although this
problem can be formulated in terms of boson variables \cite{ditmar},
it is nonlinear in terms of quadrupole bosons. By
considering the boson number as an additional degree of freedom, the
Interaction Boson Model (IBM) \cite{ibm} introduces a scalar boson as a
dynamical variable. This leads to the subsequent realization
\cite{ar2} of $s$ and $d$ bosons as pairs of nucleons coupled to the
angular momentum $\lambda=0^+$ and $\lambda=2^+$, respectively.

The finiteness of the boson number is based on the supposition
that each $d$-boson is a boson image of a quadrupole nucleon pair
occupying the levels of the valence shell in a given nucleus. Therefore,
the $d$-boson number can not exceed the half of the number of the valence
nucleons or holes. In the IBM, however, the total number of bosons
$N_b$ is taken  as a parameter. Due to this fact, the IBM can
be easily extended to thermodynamical limit $N_b \rightarrow\infty$
to study nuclear shape phase transitions at zero
temperature \cite{pcej}.
The model naturally incorporates different symmetry limits
associated with specific nuclear properties: different shapes
coincide with particular dynamical symmetries of some algebraic
structure \cite{nat}. For example, the rotational limit of this
model is associated with $SU(3)$ symmetry of axially symmetric rotor,
while the vibrational limit is associated with $U(5)$ symmetry.

The geometric model and the IBM can be related by using the
coherent state formalism \cite{gincs,diep}. The evaluation  of the
expectation value of the algebraic Hamiltonian in the coherent state
allows one to establish a correspondence between the symmetry limits of
the IBM and collective shape variables of the Bohr Hamiltonian.

The IBM does not, however, take into account the interplay between
single-particle and collective degrees of freedom in even-even
nuclei. Evidently, the shell effects are beyond the framework of
the model, though they may be crucial for the study of quantum phase
transitions in rotating nuclei, where statical and dynamical
properties are coupled. Recent review on various aspects of the IBM
and the role of dynamical symmetries, which yield benchmarks of
nuclear shapes of the ground state, can be found in \cite{pcej}.

\subsubsection{Microscopic models with effective interactions}
\label{subsec.V.B.3}

The {\it ab initio} non-relativistic models attempt to reproduce the
basic features of nuclear properties with the aid of an effective free
nucleon-nucleon interaction, which describes the nucleon-nucleon
scattering data \cite{nav}. This approach is, however, limited to
very light nuclei. Similar to trapped atoms and QDs, the nuclear many-body
problem for heavy nuclei is attacked in the mean field approach;
see, e.g., review articles \cite{good,sven,ben}.
In the nuclear shell model (SM),
one usually starts with a phenomenological single-particle model,
defines a valence space (inert core, active shells), and, with the aid
of the residual interaction, performs configuration-mixing
calculations \cite{brown}. The concept of the isospin symmetry for
proton and neutron plays an important role in the nuclear structure.
However, the first obvious difference between these particles arises
due to the electric charge of a proton. The analysis of the data for nuclei
with $N=Z$ may shed light on the evolution of nuclear structure due
to the isospin symmetry breaking under excitation. The use of
the SM to study experimental data, related to the isospin symmetry
breaking effects of different origins, in the region of nuclei between
$A\sim30$ and $A\sim60$ with increasing energy and angular momentum,
is reviewed in \cite{hs3060}. Recent development of the SM, in which
a two-body effective interaction between valence nucleons is
derived from the free nucleon-nucleon potential, is outlined in
\cite{shmod05,shmod09}. The drastic increase of the configuration
space for medium and heavy systems makes, however, the shell-model
calculations extremely difficult if not almost impossible in a
nearest future. In addition, one needs to invoke some model considerations
to interpret the SM results.

Nowadays, the concept, closely related to the density functional theory
in electronic systems, is popular in nuclear structure calculations
\cite{ben}. In this approach, one proceeds from an energy functional,
motivated from {\it ab initio} theory, but with the actual parameters
adjusted by fits to nuclear structure data. Mean-field calculations
with effective density-dependent nuclear interactions, such as Gogny
or Skyrme forces, or a relativistic mean-field approach, still do not
provide sufficiently accurate single-particle spectra for a
reliable description of experimental characteristics of low-lying
states \cite{afa}. The RPA analysis based on such mean-field
solutions is focused only on the description of various giant
resonances in non-rotating nuclei, when the accuracy of
single-particle spectra near the Fermi level is not important; see,
e.g., \cite{ben1,ben2,ves1,ves2}. Furthermore, a practical
application of the RPA for the non-separable effective forces in
rotating nuclei requires a too large configuration space and is in its
infancy. The discussion of open problems of the nuclear density functional
theory can be found, e.g., in \cite{jpg1,jpg,jpg2}. The difficulties in
using the density functional theory,
discussed in Sec. \ref{subsec:IV.C} for QDs, as well as the above
discussion, raise fundamental issues on the principles of foundation for
this approach to many-body problem of {\it finite} systems. A few
approaches to the non-relativistic density functional theory for nuclei,
based on a microscopic nuclear Hamiltonian that describes two-nucleon and
few-body scattering and bound observables, in analogy to
calculations in quantum chemistry for Coulomb systems, are reviewed
in \cite{drut}.

\subsubsection{Cranking approach}
\label{subsec.V.B.4}

To elucidate the similarities and differences between nuclei and already
discussed finite systems, trapped atoms and QDs, it is convenient to
consider a similar physical situation. Rotation, being one of the simplest
collective motions in many body systems, is a unique phenomenon which
enables one to understand the general and specific features of the
systems addressed in this review. Therefore, we will discuss, with a
few exceptions, the properties of an even-even medium and heavy nuclei
at high angular momenta, when the nucleon motion can be
substantially modified by inertial forces.  Indeed, in high spin
experiments, various symmetry breaking effects are exemplified most
clearly by rotation. On the other hand, this permits  a
semiclassical mean-field description of nuclear rotation within the
{\it cranking model} (CM) introduced by Inglis \cite{ing1,ing11}. In the
CM, for a given many-body Hamiltonian ${\cal H}=T+V$, one defines
the Routhian operator ${\cal R}$ as
\be
{\cal R} \equiv {\cal H}-{\vec \Omega} {\vec J} \label{raus} \; .
\ee
In this model, nucleons
move in an external field of single-body nature, with a constant angular
frequency $|\Omega|$. One can readily recognize that the Routhian
(\ref{raus}) is the Hamiltonian (\ref{3.72}) used for the description of
vortices created by rotation of Bose-condensed trapped atoms.

In order to justify the CM, Thouless and Valatin \cite{TV62} considered
the time-dependent Hartree-Fock solutions defined by the equation
\be
i\partial {\hat \rho}/\partial t=[{\cal R}, {\hat \rho}]
\ee
in the intrinsic frame. The moment-of-inertia tensor $\Im$ defines the
principal axes of the frame of reference:
\be
\langle {\hat J}_i\rangle=Tr {\hat J}_i{\hat \rho}={\Im}^{(1)}_i\Omega_i\quad (i=1,2,3) \; ,
\ee
where $\Im_1^{(1)},
\Im_2^{(1)}, \Im_3^{(1)}$ are the principal kinematical moments of inertia.
They have shown that the change in the average angular momentum
components in the body-fixed frame obeys the classical equation of motion
\be
d\langle {\vec J}\rangle/dt=\langle {\vec J}\times{\vec
\Omega}\rangle \; .
\ee
 For the stationary rotation
 \be
 d\langle {\vec J}\rangle/dt=0 \; ,
 \ee
one obtains the principal axis rotation.
It is usually assumed that at $\Omega=0$, the symmetry
axis is the $z$-axis. In the case of an axial symmetry (for example,
${\Im}^{(1)}_1={\Im}^{(1)}_2$), the CM  describes the uniform
rotation about an axis perpendicular to the symmetry axis. Within
the CM, one has to minimize  the energy
\be
R(\Omega)=\langle \Phi|{\cal R} |\Phi\rangle
\ee
to obtain the yrast state, where
$|\Phi\rangle$ is a Slater determinant. The angular momentum is
defined according to the semiclassical quantization
\be
J(\Omega)=\langle \Phi| {\hat J}_x |\Phi\rangle \; .
\ee
The total energy,
as a function of the angular momentum, is
\be
E(J)=R(\Omega)+\Omega J(\Omega) \; .
\ee

Additional quantum correlations can be incorporated into the CM by
means of the RPA, which describes small oscillations about the mean
field of a rotating nucleus. As is demonstrated in
Sec. \ref{subsec:IV.E.2}, the RPA, being an efficient tool to study
these quantum fluctuations (vibrational and rotational excitations),
also provides a consistent way for treating broken symmetries. Moreover,
it separates collective excitations, associated with each broken
symmetry, as the spurious RPA modes, and fixes the corresponding inertial
parameter. The alternative approach to treating broken symmetries is
based on the projection technique presented in the textbooks
\cite{Ring,BR86}. All pros and cons of this approach for the description
of high spin states are discussed in the reviews \cite{sfrau,satw}.
More technical details and recent progress
in the application of variational calculations, based on the projection
techniques, are reviewed in \cite{karl}.

During many years, various refined prescriptions have been adopted
to obtain an effective mean-field potential. It appears, however,
that it closely resembles a Woods-Saxon shape, that is, a rounded
square well that asymptotically goes to zero at a distance of a few
femtometers \cite{heyde}. This potential supplemented by a
spin-orbit interaction, nicely reproduces the famous magic numbers.
Inclusion of pairing correlations allows one to trace a transition from
the paired phase to a rigid body rotation. A harmonic oscillator
potential, supplemented by a spin-orbit interaction and $l^2$ to
mimic the surface effect of the nuclear mean-field potential (a
Nilsson potential), serves as another labor horse for the analysis of
nuclear structure effects at high spins. These potentials allow one to
construct also a self-consistent residual interaction, neglected on
the mean-field level. The RPA, with a separable multipole-multipole
interaction based on these phenomenological potentials, is an
effective tool to study low-lying collective excitations at high
spins \cite{KN,nak}. In fact, most issues addressed in this review
can be understood in the framework of the CM, using either a
three-dimensional harmonic oscillator or the Nilsson potential. In
order to keep the presentation simple and transparent, we use both these
models as an effective nuclear mean field.

Although the CM can be formulated in a self-consistent way (see,
e.g., \cite{good}), the simplest, but the most efficient version
of the CM, consists in the combination of the Strutinsky's shell
correction method \cite{strut} to account for nuclear shell structure
and a rotating liquid drop model to describe
bulk properties of nuclear matter \cite{NR}:
\be
E(\alpha,I)=E_{LDM}(\alpha,I)+\delta E_{\it shell}(\alpha,I) \; .
\ee
Here the quantum shell correction energy,
\be
\delta E_{\it shell}=\sum_{\it occ}\varepsilon _i-\tilde{E}_{\it shell} \; ,
\ee
and $\tilde{E}_{\it shell}$ is the smooth Strutinsky energy. The
single-particle energies $\varepsilon_i$ are produced by a
phenomenological nuclear potential, like Nilsson or Woods-Saxon ones,
with or without pairing interaction. $E_{LDM}$  is the deformation
energy of the rotating liquid drop with rigid-body moments of
inertia, and $\{ \alpha \}$ is a set of deformation parameters which
can be found from the minimization  of the energy functional
$E(\alpha,I)$. This approach has been, over many years, a powerful
tool for analyzing and interpreting experimental data of high spin
states. The most exciting example is the prediction of the formation
of the superdeformed high-spin states \cite{RA,neer}, which have been
discovered by Twin {\it et al.} \cite{twin}.

In this review we discuss typical, or representative, results, since
it is impossible to review all developments related to symmetry
breaking phenomena in high spin physics. To keep the discussion simple
and transparent, we will also use  the results obtained in
a fully self-consistent and analytically solvable model
\cite{TA,Ma93,NAD,wob}.

The early development of high spin nuclear physics and a mean-field
description of nuclear rotation within the {\it cranking model}, can be
found in the books \cite{Ring,szym,NR}. The experimental
data and theoretical interpretation for the properties of pear shaped
rotating nuclei are discussed in \cite{Bu96}. The
uniform rotation about an axis tilted with respect to the principal
axes of the density distribution, magnetic rotation, and band
termination is reviewed in \cite{sfrau}. The experimental
properties of the $M1$ bands, observed in near-spherical nuclei, and
the theoretical approaches, that have lead to the understanding of the
magnetic rotation, are outlined in \cite{hub}. The progress in
the development of the mean field description of high-spin states is
discussed in \cite{satw}. The basic ideas of
the cranking mean field+RPA (CRPA) approach are
outlined in \cite{KN}. The overview of pair
correlations at high spins is given in \cite{shi1}.

\subsection{Symmetry breaking in rotating nuclei}
\label{subsec.V.C.}

\subsubsection{Symmetries}
\label{subsec.V.C.1.}

For the main part of this review we assume that the Routhian
(\ref{raus}) is invariant with respect to space inversion $\hat P$.
In this case, the parity $\pi$ is a good quantum number:
\be
\hat P|\Phi\rangle=\pi|\Phi\rangle\;,
\ee
and the rotational bands are
characterized by a fixed parity $\pi$. The spin-sequences in the
rotational bands indicate broken/conserved symmetries. In
particular, the sequence of rotational levels $I^\pi$,
$(I+2)^\pi$...of a given parity is associated with a principal axis
rotation (broken rotational symmetry), but with the preserved
symmetry, called a {\it signature} $r=\pm1$, for the system with even
and $\pm i$, for the system with odd particle number \cite{BM}.
Indeed, for even-even nuclei, the cranking Hamiltonian (\ref{raus})
adheres to the $D_2$ spatial symmetry with respect to the rotation by
the angle $\pi$ around the rotational axis $x$, i.e.,
\be
[H-\Omega {\hat J}_x, {\hat R}_x]=0,\quad {\hat R}_x=\exp(-i\pi {\hat J}_x) \; .
\ee
On the other hand, the internal CM wave function is subject
to the condition
\be
{\hat R}_x|\Phi\rangle = \exp(-i\pi
J)|\Phi\rangle = \exp[-i\pi (I-1/2)]|\Phi\rangle = i \exp(-i\pi
I)|\Phi\rangle = ri|\Phi\rangle \; .
\ee
Another definition of signature $\alpha=0,1$ \cite{sfrau} is connected with
the $r$ via the relation $r=\exp (-i\pi\alpha)$. Consequently, all rotational
states can be classified by the quantum number signature $r=\exp (-i\pi\alpha)$
leading to the selection rules for the total angular momentum $I=\alpha
+ 2n$, $n=0,\pm 1, \pm 2 \ldots $ and $\alpha=0,1$. In particular,
in even-even nuclei, the yrast band, characterized by the positive
signature quantum number $r=+1 \,(\alpha = 0)$, consists of even
spins only. The members of the rotational bands, with the conserved
signature, are connected via strong, stretched $E2$
$\gamma$-transitions.

The observation of parity doublets is associated with the parity
violation in the intrinsic frame \cite{chas}. For a uniform rotation
around the principal axis $x$, the parity $\hat P$ and signature
$\hat R_x$ are broken. However, the conservation of the simplex quantum
number $\hat S_x=\hat P\hat R_x^{-1}$ \cite{svit,oct1,oct2} leads to the
alternating parity $\pi=(-1)^I$ of the bands $I^+,(I+1)^-,(I+2)^+,...$
connected by enhanced $E1$ $\gamma$-transitions.

The classification of rotational states, with respect to discrete
symmetries, such as parity, signature, and $\hat T\hat R_y(\pi)$
operations ($\hat T$ is the time reverse operation) for the
Hamiltonian $H-\Omega J_z$, is discussed thoroughly in \cite{sfrau}.
These symmetries are the special cases of the complete scheme which
classifies the mean field solutions in accordance to time-reversal
symmetry $\hat T$, spatial $\hat R_i(\pi)$, and $\hat T\hat R_i(\pi)$
$(i=x,y,z)$ symmetries \cite{dob1,dob2}.

\subsubsection{Shape transitions in rotating nuclei}
\label{sec:V.C.2.}

Nuclei are finite systems and phase transitions should be washed out
by quantum fluctuations. Nevertheless, long ago Thouless \cite{Th1}
proposed to distinguish two kinds of "phase transitions" even for
nuclei. Such phase transitions may be connected with shape
transitions, for example, from spherical to deformed or from axially
deformed to nonaxially deformed shapes. This idea was put ahead in
the analysis of shape transitions in hot rotating nuclei, based on
the cranking+Nilsson model \cite{ig2}. In this case the statistical
treatment of the finite-temperature mean-field description had
provided a justification for an application of the Landau theory for
nuclei \cite{al1}. Within this approach, simple rules for
different shape-phase transitions, induced by the variation of angular
momentum and temperature, were found \cite{al1,al2}.

With the increase of excitation energy, macroscopic effects,
i.e., the liquid drop part, dominates in the evolution process, in
particular, in the changes of the parameters of the effective mean
field. Typical excitation energies of compound states, populated in
heavy-ion reactions, are of the order of 20-30 MeV.
Thus, the shell effects cannot influence the equilibrium deformation
of such states. The $\gamma$ - decay mode of the de-excitation
process becomes prevalent when the excitation energy, measured from
the yrast line, decreases to about 7-10 MeV. According to
calculations \cite{ig2}, this region of excitations is characterized
by large fluctuations of the shape. This facilitates quadrupole
transitions from such states, that is, the de-excitation, which
carries away the angular momentum and leaves nuclei above the yrast
line. And only when the excitation energy is reduced to about 5 MeV
above the yrast line, the evolution of the nuclear shape starts to
follow the valleys in the shell component of the deformation
potential energy. At this stage, the emergence of shell
irregularities of the single-particle spectrum is essential for the
description of all physical phenomena. The effect of shell structure
could be seen quite clearly even in the region of giant resonances
$(\sim 10-20$ MeV; see for a recent review \cite{dstadta}.
The shape of the photo-absorption cross section of the giant
dipole resonance strongly depends on the type of symmetry of
the rotating effective mean field, discussed for the first time in
\cite{GR,egido}. The
experimental results and the role of shell effects for giant
dipole resonances in hot
rotating nuclei can be found, for example, in \cite{Sno,Gaar,alhex}.

Quantum phase transitions, that occur at zero temperature when
varying some non-thermal control parameters, attract a
considerable attention in recent years. The concept of a quantum phase
transition refers to a sudden change in the structure of the ground state,
as a result of varying a parameter. Since thermal fluctuations are
absent, the sole cause, responsible for the sudden changes, are
quantum fluctuations. The study of such transitions has been initiated
in condensed matter physics in order to understand the properties of
low-dimensional systems \cite{Sad} and quantum behaviour in a variety
of materials at critical points \cite{BelitzKirkVoj_2005}. In
nuclear structure, quantum phase transitions are reflected by rapid
structural changes, with varying $N$ or $Z$, or rotational frequency.

It appears that the evolution of the shell structure governs the
variation of ground-state nuclear shapes along isotopic and isotonic
chains. Shape transitions reflect the underlying modifications of
interactions between valence nucleons. Evidently, rotation acts
as an additional external field enforcing various interactions
between single-particle orbitals. It is similar to the phenomenon of
the magnetic field for QDs. The basic difference consists in the
fact that the magnetic field creates an additional confinement in
QDs, while in nuclear systems, the rotation works against the
confinement created by attractive nucleon-nucleon interactions.

The phase transition is usually detected by means of an order
parameter as a function of a control parameter. In many cases, an
order parameter is calculated with the aid of model considerations. In
particular, in rotating nuclei, one can suggest several order
parameters, like deformation parameters of a nuclear effective
potential, $\beta$ and $\gamma$, that characterize the geometrical
configuration  \cite{Ia03,Ia04}, as a function of the rotational
frequency, playing the role of a control parameter.
In order to elucidate this aspect the experimental
data in $^{156}$Dy and $^{162}$Yb (see Fig. \ref{figbb})
have been analyzed \cite{bb3,bb3b}
with the aid of the cranking Hamiltonian
\be
{\cal R} = H -\sum_{\tau =N,P} \lambda_{\tau}\hat N_{\tau}-\, \Omega \hat J_x+\,
H_{\rm int} \; .
\ee
The term  $H=H_N\,+\, H_{\rm add}$ contains the
Nilsson Hamiltonian $H_N$ and the additional term that restores the
local Galilean invariance of the Nilsson potential in the rotating
frame. The interaction includes separable monopole pairing, double
stretched quadrupole-quadrupole (QQ) and monopole-monopole terms.
The details on this model Hamiltonian can be found in \cite{bb3b}.

\begin{figure}[ht]
\centerline{
\includegraphics[width=3.0in]{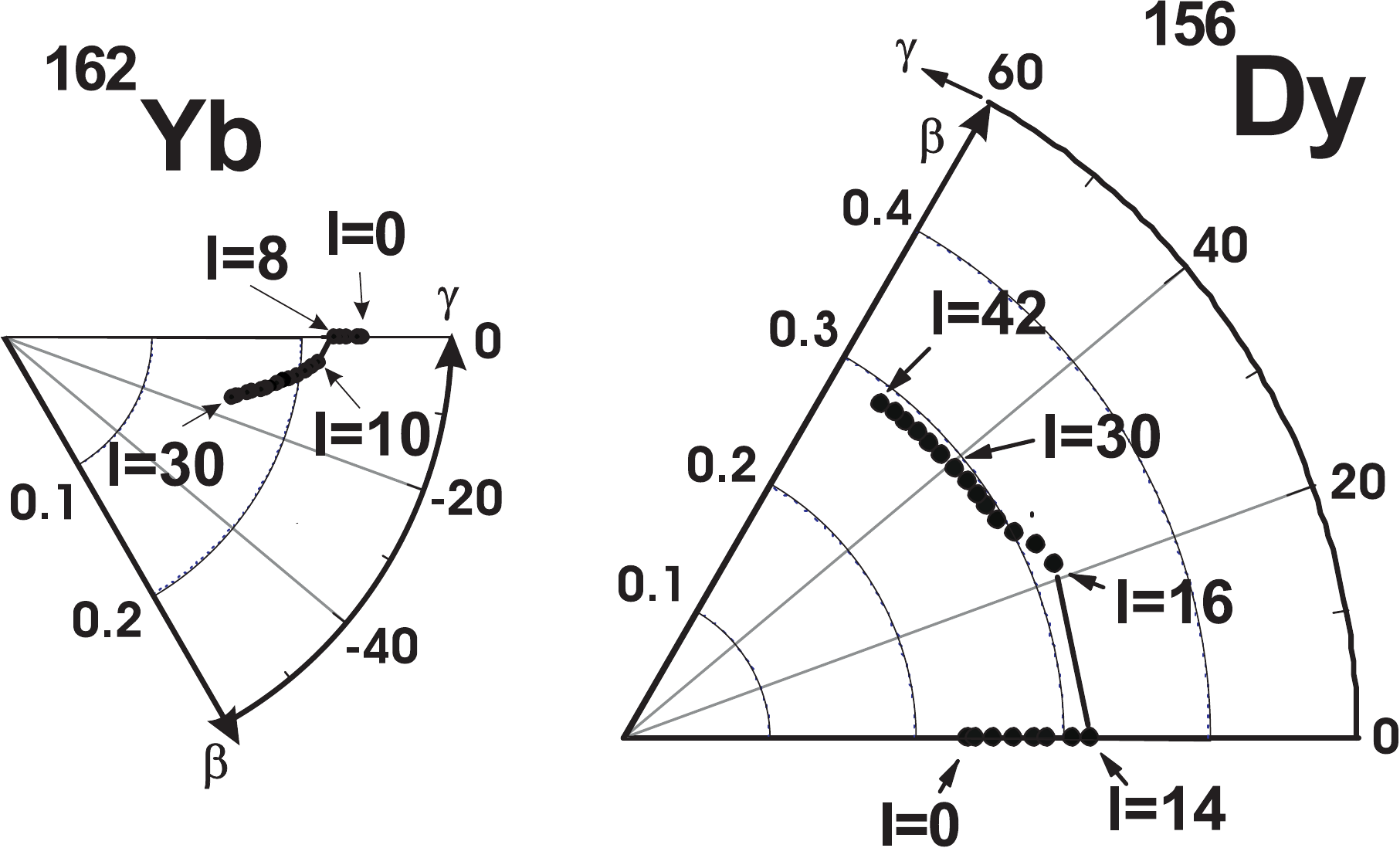} }
\caption{ Equilibrium deformations on the $\beta-\gamma$ plane, under
varying the angular momentum $I=<\,{\hat J}_x\,>-1/2$ (in units
of $\hbar$). From \cite{bb3b}.} \label{figed}
\end{figure}

At each rotational frequency, the total mean-field energy
$R=\langle \cal R \rangle$ is minimized on the mesh $\beta=0.0 - 0.6$,
$\gamma=-\pi/3 - +\pi/3$.
The results for the rotational dependence of equilibrium deformation
parameters $\beta$ and $\gamma$  (see for definitions
Sec. \ref{subsec:V.B.1}) exhibit a transition to the triaxiality
$\gamma \neq 0$ at $I=8\hbar\,\rightarrow \,I=10\hbar$ in $^{162}$Yb
and $I=14\hbar\, \rightarrow \, I=16\hbar$ in $^{156}$Dy (see
Fig. \ref{figed}). In  $^{162}$Yb, a stable, single  minimum
$E_{\Omega}(\beta,\gamma)=\langle {\cal R}\rangle$ slowly moves on
the potential energy surface $(\beta,\gamma)$ from the axially
symmetric shape to the triaxial one, with the increase of the
rotation frequency. Hereafter, $\langle ...\rangle$ means the
averaging over the mean-field states. In contrast, in $^{156}$Dy at
the vicinity of the transition point, there is a coexistence of the
axially symmetric ($\gamma=0$) and non-axial ($\gamma \neq 0$)
configurations. Slightly above the transition point, the
configuration suddenly changes from the axially symmetric into the
triaxial one. The agreement between experimental and calculated
values for the kinematical moment of inertia (Fig. \ref{figbb})
confirms the validity of the calculations.
In principle, projection methods should be used in the transition region,
since the angular momentum is not a good quantum number in the CM.
A theory of large amplitude motion would provide a superior means to
solve this problem \cite{klein,lam1}.

To elucidate the different character of the shape transition from
axially symmetric to the triaxial shape, one can
choose the deformation parameter $\gamma$ as the
order parameter that reflects the broken axial
symmetry \cite{bb3,bb3b}. Such a
choice is well justified, since the deformation parameter $\beta$
preserves its value before and after the shape transition in both
nuclei: $\beta_t \approx 0.2$ for $^{162}$Yb and $\beta_t \approx 0.31$
for $^{156}$Dy. Thus, one considers the mean-field value of the
cranking Hamiltonian,
$E_{\Omega}(\gamma;\beta_t) \equiv \langle{\cal R} \rangle$, for
different  $\Omega$ (state variable) and $\gamma$ (order parameter) at a
fixed value of $\beta_t$.

\begin{figure}[ht]
\centerline{
\includegraphics[height = 0.22\textheight]{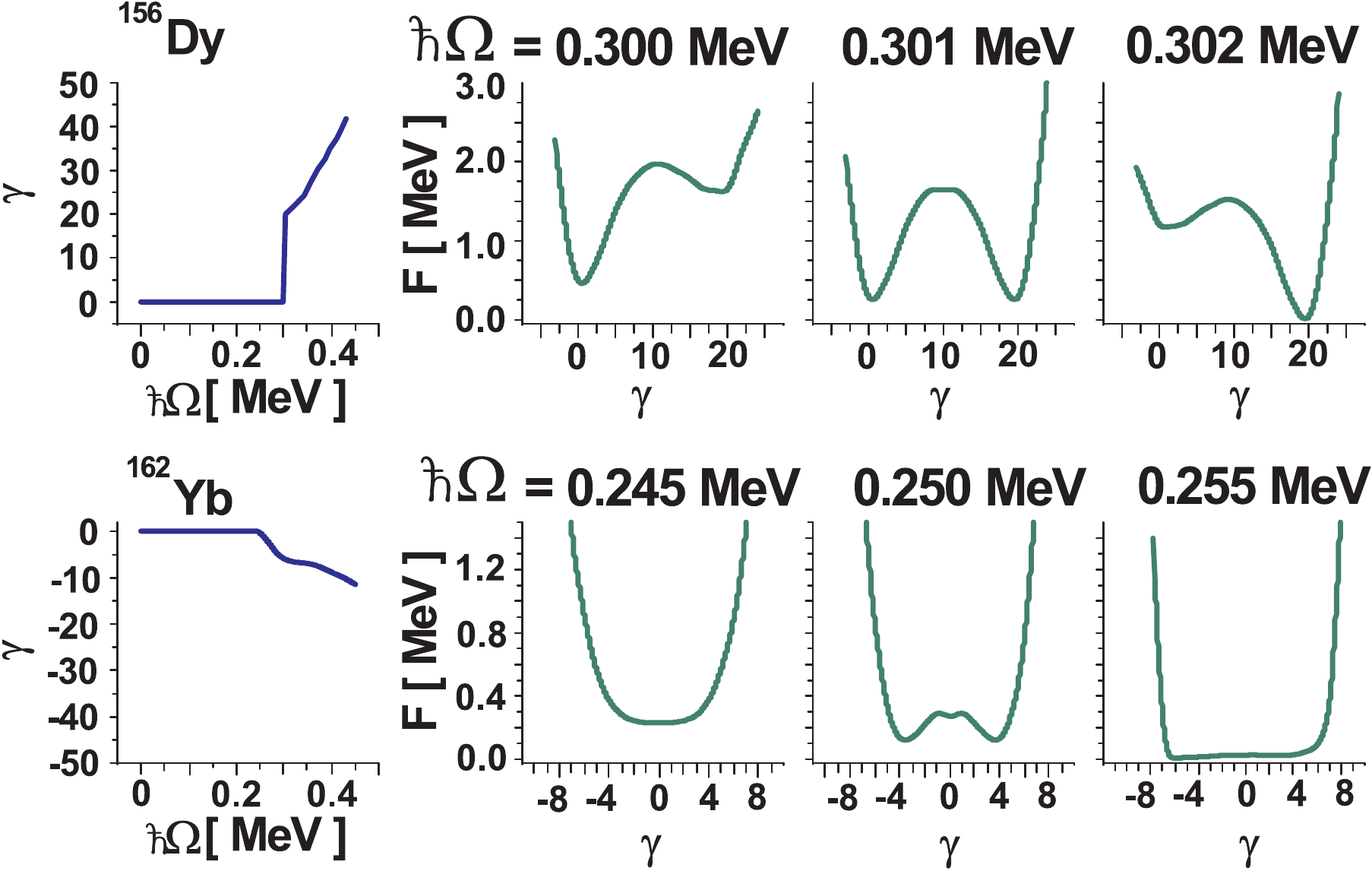} }
\caption{ (Color online) Rotational dependence of the order
parameter $\gamma$ and the energy surface sections
$F(\Omega,\gamma)=E(\gamma,\beta_t)-E_{min}$ for $^{156}$Dy (top)
and $^{162}$Yb (bottom) before and after the transition point.
The energy is counted relatively to the value
$E_{min}=E_{\Omega}(\beta_t,\gamma)$ at
$\hbar\Omega=0.255,\,\,0.302$ MeV for $^{162}$Yb and $^{156}$Dy,
respectively. From \cite{bb3b}.} \label{phatr}
\end{figure}

For $^{156}$Dy, one observes the emergence of the order parameter
$\gamma$ above the critical value $\hbar\Omega_c=0.301$ MeV of the
control parameter $\Omega$ (see a top panel in Fig. \ref{phatr}).
Below and above the transition point, there is a unique phase whose
properties are continuously connected to one of the coexisting
phases at the transition point. The order parameter changes
discontinuously as the nucleus passes through the critical point
from the axially symmetric shape to the triaxial one. The polynomial fit
of the potential landscape section at $\hbar\Omega_c=0.301$ MeV
yields the following expression
\begin{equation}
F(\Omega;\gamma) =F_0(\Omega)+ F_2(\Omega) \gamma^2 - F_3(\Omega)
\gamma^3 + F_4(\Omega) \gamma^4 \;.
\label{fitkv}
\end{equation}
One can transform this polynomial to the form
\begin{equation}
{\bar F}=\frac{F(\Omega;\gamma) -F_0(\Omega)}{\bar {F_0}} \approx
\alpha \frac{\eta^2}{2}- \frac{\eta^3}{3} +\frac{\eta^4}{4} \; ,
\label{gen}
\end{equation}
where
\begin{equation}
{\bar {F_0}}=\frac{(3F_3)^4}{(4F_4)^3},\qquad
\alpha=\frac{8F_2F_4}{9F_3^2}, \qquad \eta=\frac{4F_4}{3F_3}\gamma \; .
\end{equation}
Expression (\ref{gen}) represents the generic form of the
anharmonic model of {\it the structural first order phase
transitions} in condensed matter physics (see
Sec. \ref{subsubsec:II.A.2} and \cite{krum}). The
condition $\partial {\bar F}/\partial \eta=0$ determines the
following solutions for the order parameter $\eta$
\begin{equation}
\eta=0,\qquad \eta=\frac{1\pm\sqrt{1-4\alpha}}{2} \; .
\end{equation}
If $\alpha>1/4$, the functional ${\bar F}$ has a single minimum at
$\eta=0$. Depending on the values of $\alpha$, defined in the interval
$0<\alpha<1/4$, the functional ${\bar F}$ manifests the transition
from one stable minimum at zero order parameter via one
minimum+metastable state to the other stable minimum with the
nonzero order parameter. In particular, at the universal value of
$\alpha=2/9$ the functional ${\bar F}$ has two minimal values, with
${\bar F}=0$ at $\eta=0\rightarrow \gamma\approx0^0$ and
$\eta=2/3\rightarrow \gamma \approx27^0$ and a maximum at
$\eta=1/3\rightarrow \gamma\approx13.5^0$ (using the numerical
values for the coefficients $F_{2,3,4}$ extracted from the fit of
the potential landscape section, Eq. (\ref{fitkv})). The
correspondence between the actual value $\gamma\approx 20^0$ and the
one obtained from the generic model is quite good. Thus, the
backbending in $^{156}$Dy possesses typical features of {\it the
first order phase transition}.

In the case of $^{162}$Yb, the energy $E(\Omega;\gamma)$ and the
order parameter (Fig. \ref{phatr}) are smooth functions in the
vicinity of the transition point $\Omega_c$. This implies that the
phases with $\gamma=0$ and $\gamma\neq0$ should coincide at the
transition point. Therefore, for $\Omega$ near the
transition point $\Omega_c$, one may expand
the functional $F(\Omega,\gamma)=E_{\Omega}(\gamma,\beta_t)-E_{min}$
(see Fig. \ref{phatr}) in the form \cite{bb3b}:
\begin{equation}
F(\Omega;\gamma) = F_1(\Omega) \gamma + F_2(\Omega) \gamma^2 +
F_3(\Omega) \gamma^3 + F_4(\Omega) \gamma^4 + \ldots
\label{5}
\end{equation}
It can be shown that $F_1(\Omega)=0$ and at the transition point
$\Omega_c$ ($\gamma=0$)\;\; $F_2(\Omega_c)=F_3(\Omega_c)=0$.
Assuming that $F_3=0$ for all $\Omega$, one obtains for the order
parameter
\begin{equation}
\gamma_1=0 \,, \quad \gamma_{2,3}^2 = -
\frac{F_2(\Omega)}{2\,F_4(\Omega)} = \left\{
\begin{array}{ll}
\neq 0 & for \,\, \Omega\neq \Omega_c \\
= 0    & for \,\, \Omega = \Omega_c   \; .
\end{array}
\right. \label{par2}
\end{equation}
At the transition point $F_2(\Omega_c)=0$, it is possible to propose the
following definition of the function $F_2(\Omega)$:
\begin{equation}
F_2(\Omega) \approx \frac{dF_2(\Omega)}{d\Omega}\, \left(\Omega -
\Omega_c \right) \label{par3} \; .
\end{equation}
Thus, for the order parameter, one has
$\gamma\sim (\Omega-\Omega_c)^\nu$, with the critical exponent $\nu=1/2$,
in agreement with the classical Landau theory, where the temperature is
replaced by the rotational frequency. The observed phenomenon
resembles very much the structural phase transition discussed within
the anharmonic Landau-type model in solid state physics \cite{krum}.
One may conclude that the backbending in $^{162}$Yb is classified
as {\it the phase transition of the second order}.

Extensive experimental data have been obtained on
low-lying states of negative parity \cite{ahmad,Bu96}. For the Ra-Th
$(Z\sim 88, N\sim 134)$ and Ba-Sm $(Z\sim 56,N\sim 88)$ nuclei, low
$3^-$ states, parity doublets, and alternating parity bands with
enhanced dipole $(E1)$ $\gamma$- transitions have been found. Since
the octupole interactions are the strongest when the pairs of orbitals from
the intruder sub-shell $(l, j )$ and the normal parity sub-shell
$(l-3, j-3)$ are near the Fermi surface, the octupole correlations
are expected to become important for these nuclei \cite{BM,sven}.
The features, observed in nuclei, are similar to those familiar
from molecular physics. In molecules, a stable octupole deformation
leads to the appearance of rotational bands with alternating
parity levels connected by strong $E1$ intra-band transitions.
M\"oller and Nilsson \cite{mn} used the assumption on the symmetry
breaking of the intrinsic reflection symmetry of the single-particle
potential and demonstrated the instability with respect to the axial
octupole deformation in the Ra-Th region. The analysis was performed
within the macroscopic + microscopic approach (see Sec. \ref{subsec.V.B.4})
for the Nilsson model at zero rotational
frequency. Later, within this approach, numerous results, based on
various single-particle phenomenological potentials, have been
reported regarding the presence of a stable axial octupole
deformation in the ground state of some nuclei in the actinide
region \cite{Bu96}. The cranking macroscopic+microscopic approach
was also used (see Sec. \ref{subsec.V.B.4}) to analyze the properties of
rotating nuclei with a stable axial octupole deformation
\cite{svit,oct1,oct2}.

The potential energy surface, found in different calculations, is,
however, quite shallow. Experimental nuclear spectra show also a
pronounced parity splitting at low angular momenta. Rather dynamical
octupole effects (vibrations) are dominant over static effects
(octupole deformations) at low spins. The increase of the angular
momentum decreases the pairing correlations which reduce the octupole
interaction, since the pairing correlations couple the orbitals with
the same parity. With rotation, the density of different parity
states near the Fermi level becomes enhanced. Indeed, one
observes a smooth decrease of the parity splitting with the increase
of the rotational frequency in several nuclei. The
cranking+Hartree-Fock-Bogolubov (HFB) approach with Gogny forces
has been applied for the analysis of the data in Ba-Sm region
\cite{Ga98}. In order to compare with experimental data, the
approximate parity projection, before variation, has been developed.
Within this approach, a good agreement was obtained between the
calculated and experimental data for the energy
splitting of positive and negative parity states, as well as for the
projected $B(E1)$- transition probabilities. Note that at high spins
the differences between projected and unprojected HFB solutions
become less pronounced.

The problem of non-axial octupole deformations has attracted much
attention during recent years. The study of non-axial octupole
deformations could shed light on the tendency, for systems in the
way towards fission, to avoid superdeformation \cite{BM}. Further, a
tilted axis rotation \cite{sfrau} and non-axial octupole
deformations could be complementary parts of a signature symmetry
breaking phenomenon in rotating nuclei. Calculations, by means of the
macroscopic+microscopic method with the Woods--Saxon potential,
predict the importance of the banana-type $Y_{31}$ deformation for
highly deformed nuclei \cite{chas1}. Manifestations of pronounced
shell effects have been discovered, when non-axial octupole
deformations are added to a harmonic oscillator model \cite{H99}.
The cranking HFB calculations with Skyrme \cite{Ya01} and Gogny
\cite{Ta01} forces predict a non-axial $Y_{31}$ octupole deformation
in light nuclei at high spins. The results based on the Skyrme
interaction demonstrate the importance of a nonaxial $Y_{32}$
octupole deformation in actinide nuclei at fast rotation
\cite{Ts02}. The onset of a non-axial octupole deformation was found
in $^{162}$Yb  \cite{noct}. Notice that rotation induces the
contribution of non-axial quadrupole and octupole components of
effective nuclear potentials.

Recently it was suggested \cite{soct}
that strong octupole correlations in the mass region $A\approx 226$
might be interpreted as rotation--induced condensation of octupole
phonons having their angular momentum aligned with the rotational
axis of a triaxial nucleus. This idea was proposed for the first
time by Briancon and Mikhailov \cite{brmik}. However, another prediction
is the possible existence of tetrahedral shapes and, in particular, in the
actinide region \cite{teth1,dudtet}. Such shapes are characterized
by a triaxial octupole deformation $Y_{32}$, together with a
near-zero quadrupole deformation $Y_{20}$. This implies vanishing
in-band $E2$ $\gamma$-transitions. Evidently, the negative-parity bands,
with missing in-band $E2$ $\gamma$-transitions, are the candidates for the
rotation of a tetrahedral shape. The most favourable region for observing
tetrahedral states is in the mass region of $A\sim 160$
\cite{dudre}. Measurements of quadrupole moments of low-lying
negative-parity bands in this region have not supported the
tetrahedral hypothesis \cite{bark1}. It was found, however, that the
experimental systematics of the the dipole $(E1)$ $\gamma$-
transitions \cite{itemba} are in good agreement with the increasing triaxial
quadrupole deformation \cite{Ts02,soct}. A
lot of experimental efforts is required to understand better the
nature of octupole instability.

\subsubsection{Collective excitations as indicator of symmetry breaking}
\label{subsec:V.C.3.}

The description of rotational states is one of the oldest, yet not
fully solved, problem in nuclear structure physics. While various
microscopic models, based on the cranking approach, describe
reasonably well the kinematical moment of inertia
\be
\Im^{(1)}=-(dE/d\Omega)/\Omega
\ee
for a finite angular frequency $\Omega$, there is still a systematic
deviation \cite{amoi} of the dynamical moment of inertia
\be
\Im^{(2)}=-(d^2E/d\Omega^2) \; .
\ee
Here, the classical canonical relation $dE/d\Omega=-J$ is used (see
also Sec. \ref{subsec:V.A.}); $E$ is the total energy in the
rotating frame, defined in the CM approach as
$E=\langle {\cal R}\rangle$ or as $E=E_I-\Omega I$ from the experimental
data at high spins. Since the moments of inertia are the benchmarks for the
microscopic models of collective motion in nuclei, the understanding
of the source of the discrepancy becomes a challenge for a many-body
theory of finite Fermi systems.

The description of the moments of inertia could be improved, if quantum
oscillations around the mean solution were incorporated, as it was
suggested by Thouless and Valatin \cite{TV62}.
The effects of pairing correlations
on the moments of inertia were considered \cite{shi1}. For the
case of pairing and quadrupole vibrations, such calculations, in a
restricted configuration space (only three shells have been
included), were performed \cite{egido1}. For medium and heavy
nuclei, realistic calculations require a large configuration space
and are less reliable in the backbending region, where there are
large fluctuation of the angular momentum. If one neglects pairing
correlations (high spin limit), it is possible to calculate the
total energy, including the mean-field and the RPA correlation
energies using a self-consistent rotating harmonic oscillator
\cite{NAD}. This model provides a relatively simple, but still
realistic, frame to calculate the Thouless-Valatin moment of inertia
and the desired contributions of shape oscillations without the
usual restrictions of the configuration space.

Let us consider the mean-field part of the many-body Hamiltonian
(Routhian)  Eq. (\ref{raus}) as  ${\cal R}=\sum_i^N h_i$, where
$h=h_0+h_z$  are defined in Sec. \ref{subsubsec:IV.A.1}. The
single-particle triaxial harmonic oscillator Hamiltonian is aligned
along its principal axes (PA). For rotating nuclei we have to use
$\omega_1^2=\omega_x^2,\;\omega_2^2=\omega_y^2\;$,
$\omega_L\equiv\Omega$, and consider the rotation around $x$-axis.
In this case, all $x$-variables are to be renamed as $z$-variables.
We call this model as a PAC (a principal axis cranking). Using the
transformation (\ref{trans}), with $x$ replaced by $z$, one arrives
to the well known results \cite{val,ripka, zel}, where to simplify
notation, the $\hbar$, entering the angular momentum and the frequencies,
is suppressed:
\be
R=\omega_x\Sigma_{x}+\omega_+\Sigma_{+}+\omega_-\Sigma_{-}
\label{erot}
\ee
where the eigenmodes $\omega_{\pm}$
\bea
\label{mod11}
\omega _{\pm }^2&=&\frac{1}{2}(\omega _y^2+\omega _z^2+2\Omega^2\pm \Delta),\\
\Delta&=&[(\omega _y^2-\omega _z^2)^2+8\Omega^2(\omega _y^2+\omega _z^2)]^{1/2},\nonumber\\
\Sigma_{k}&=&\langle \Sigma_{j}^N (n_k+1/2)_j \rangle \; .
\eea
Here, $n_k=a_k^{+}a_k\; (k=x,+,-)$, where $a_k^+,\; a_k$ are the oscillator
quantum operators. The lowest levels are filled from the bottom,
which gives the ground state energy in the rotating frame. The Pauli
principle is taken into account, so that only two particles occupy a single
level. The minimization of the total energy Eq. (\ref{erot}) with
respect to all three frequencies, subject to the volume conservation
condition $\omega_x\omega_y\omega_z=\omega_0^3$, yields the
self-consistent condition Eq. (\ref{cond}) (adapted for the nuclear
rotation) at a finite rotational frequency \cite{TA}. Since all
shells are mixed, we go beyond the approximation used in
\cite{egido1} (for a cranking harmonic oscillator, see also \cite{NR}).
In this section, to distinguish the CM model
for nuclear rotation from the model of QD in the magnetic field, we
can call the condition Eq. (\ref{cond}), adapted for nuclear
rotation, as the nuclear self-consistent condition (NSCC).

It should be pointed out that, generally, the NSCC provides the
absolute minima, in comparison with the local minima obtained from
the condition of the {\it isotropic velocity distribution}
\cite{ripka,zel}
\be
\omega_x\Sigma_x=\omega_+\Sigma_+=\omega_-\Sigma_- \; ,
\ee
which describes only a rigid rotation. By means of the transformation
(\ref{trans}) (adapted for the PAC),  one can show that this
condition is equivalent to the condition
\be
\omega_x^2\langle
X^2\rangle= (\omega_y^2-\Omega^2)\langle
Y^2\rangle=(\omega_z^2-\Omega^2)\langle Z^2\rangle \; . \label{clas}
\ee
In semiclassical approximation, one can assume that
$\langle X^2\rangle=R_x^2$, $\langle Y^2\rangle=R_y^2$, and
$\langle Z^2\rangle=R_z^2$, where $R_i\; (i=x,y,z)$, are spatial extensions
of the many body system. As a result, for
$\omega_y=\omega_z=\omega_\perp$, one obtains from the condition
(\ref{clas})
\be
\frac{R_z}{R_x}=\frac{\omega_x}{\sqrt{\omega_\perp^2-\Omega^2}}
\ee
the aspect ratio, which has been used as a convenient diagnostic
tool to infer the actual angular velocity of the rotating Bose
condensates \cite{Fetter_2009}.

As it is stressed in Sec. \ref{subsec:V.A.}, because of constant
density, with a sharp boundary, in stable nuclei, the shape deformation
can be chosen to represent a collective degree of freedom. Due to short-range
attractive nuclear forces, one can conclude that the deformation of
the density distribution and that of the potential are the same
\cite{BM}. As a result, when the nuclear system undergoes collective
excitations with a change in the density distribution, it must be
accompanied by the same change in the potential
$\delta \rho=\delta V$.
Then nucleons, moving in this potential, will have to readjust
their orbits. Such readjustments result in the change in the
nucleon density distribution. Since a nucleus is a
self-sustained system, a collective motion itself is connected to the
nucleon degrees of freedom, and therefore the change in the density
thus produced must be the same density change that is caused by the
collective excitations. The self-consistency between the
particle and collective degrees of freedom is indeed the basic
ingredient responsible for inducing the effective interactions.

Based on the concept of nuclear saturation and the self-consistency
between the shape of the mean-field potential (the triaxial harmonic
oscillator) and that of a density distribution, Sakamoto and
Kishimoto \cite{sakamoto} have employed the Landau theory of Fermi liquid
\cite{LandauLif_1980} and derived the effective multipole-multipole
interaction due to shape and angular momentum oscillations; see also
\cite{sven1}. According to this result, the CRPA Hamiltonian has
the following form
\be
H_{RPA}={\cal
R}-\frac{\chi}{2}\sum_{\mu=-2}^{\mu=2}Q_\mu^+Q_\mu={\tilde H}-\Omega
L_x  \; . \label{hamRPA}
\ee
Here, the quadrupole operators
$Q_\mu=\tilde{r^2Y_{2\mu}}$ are expressed in terms of the stretched
coordinates ${\tilde q}_i=(\omega_i/\omega_0)q_i$, $(q_i=x,y,z)$.
The effective quadrupole interaction restores the rotational
invariance of the non-rotating triaxial harmonic oscillator, so that
in the frame of RPA,
\be
[{\tilde H},L_i]=0,\quad (i=x,y,z) \; .
\ee
The NSCC, Eq. (\ref{cond}), fixes the quadrupole strength
\be
\chi=(4\pi/5)m\omega_0^2 /\langle r^2\rangle,\quad
\langle r^2\rangle =\langle {\tilde x}^2 +{\tilde y}^2+ {\tilde
z}^2\rangle \; .
\ee
If the NSCC is fulfilled, in addition to the volume
conserving  constraint, the self-consistent residual interaction
does not affect the equilibrium deformation obtained from the
minimization procedure, since  $\langle\tilde Q_\mu\rangle = 0\;
(\mu=0,1,2)$.

Using the transformation (\ref{trans}) (adapted for the CHO), one
solves the RPA equations of motion
\begin{eqnarray}
[H_{RPA},{\cal X}_\lambda]=-i\omega_\lambda {\cal P}_\lambda,\quad
[H_{RPA},{\cal P}_\lambda]=i\omega_\lambda {\cal X}_\lambda,\quad
[{\cal X}_\lambda,{\cal P}_{\lambda^\prime}]=i\delta_{\lambda,\lambda^\prime}\;,
\label{RPAeq}
\end{eqnarray}
where $\omega_\lambda$ are the RPA eigenfrequencies in the rotating
frame and the associated phonon operators are
\be
O_\lambda=({\cal X}_\lambda-i{\cal P}_\lambda)/\sqrt{2} \; .
\ee
Here, ${\cal X}_\lambda=\sum_sX_s^{\lambda}{\hat f}_s$, ${\cal
P}_\lambda=i\sum_sX_s^{\lambda}{\hat g}_s$ are bilinear combinations
of the quanta $a_k^+$, $a_k$, such that
$\langle [{\hat f}_s,{\hat g}_{s^\prime}]\rangle=V_s\delta_{s,s^\prime}$,
where the quantities $V_s$ are proportional to different combinations of
$\sum_i (i=x,+,-)$. Since the mean field violates the rotational invariance,
among the RPA eigenfrequencies, there exist two spurious solutions.
One solution with zero frequency is associated with the rotation
around the $x$- axes, since
\be
[H_{RPA},L_x]=0 \; .
\ee
The other "spurious" solution at $\omega=\Omega$ corresponds to
the collective rotation \cite{marRPA}, since
\be
[H_{RPA},L_{\pm}]=[H_{RPA},L_y\pm iL_z]=\mp\Omega L_{\pm} \; .
\ee
The Hamiltonian Eq. (\ref{hamRPA}) possesses the
signature symmetry (see Sec. \ref{subsec.V.C.1.}), i.e., such that it can be
decomposed into positive and negative signature terms,
$H_{RPA}=H(+)+H(-)$, that can be separately diagonalized
\cite{marRPA,KN}.

The positive signature Hamiltonian contains the zero-frequency mode,
defined by
\be
[H(+),\phi_x]=-\frac{L_x}{\Im_{TV}},\quad
[\phi_x,L_x]=i \; ,
\ee
and allows one to determine the Thouless-Valatin
moment of inertia \cite{MW69}. Here, the angular momentum operator
$L_x=\sum_sl_s^x{\hat f}_s$ and canonically conjugated angle
$\phi_x=i\sum_s\phi_s^x{\hat g}_s$ are expressed via ${\hat f}_s$
and ${\hat g}_s$, which obey the condition
${\hat R}_x{\hat d}_s{\hat R}_x^{-1}={\hat d}_s$ $({\hat d}_s={\hat f}_s$
or ${\hat g}_s)$. The negative signature Hamiltonian contains the rotational
mode and the vibrational mode describing the wobbling motion (see below).

The Thouless-Valatin moment of inertia has been compared with the
dynamical moment of inertia $\Im^{(2)}=-d^2R/d\Omega^2$ calculated
in the mean-field approximation in the exact model \cite{NAD}.
The basic outcome is that, if the mean-field equations
are solved {\it self-consistently}, then
the mean-field dynamical moment of inertia, calculated
in the rotating frame, is {\it equivalent} to the Thouless-Valatin
moment of inertia calculated in the CRPA approach. It is the
strongest test for the validity of a microscopic model of nuclear
rotation.

Regular rotational bands, identified  in spectroscopic data, are the
most evident manifestations of an anisotropy of a
spatial nuclear density distribution. While an axial deformation of
a nuclear potential is well established, there is a long standing
debate on the existence of a triaxial deformation. A full understanding
of this degree of freedom in nuclei may give impact for other
mesoscopic systems as well. In particular, the importance of
nonaxiality is discussed recently for metallic clusters \cite{pgcl}
and atomic condensates (see \cite{bir} and references therein).

Let us consider rotation around the symmetry axis, i.e., around the $z$ axis,
similar to the rotation of the Bose-condensed trapped atoms (see
Sec. \ref{subsubsec:III.C.3}).
Without loss of generality, we consider only one type of particles
(protons or neutrons) in this model.
The eigenmodes have the
simple form (see  Appendix \ref{appa}, where
$\omega_1^2=\omega_x^2,\;\omega_2^2=\omega_y^2\;$,
$\omega_L\equiv \Omega$), and in this case the NSCC leads to the
nontrivial solution which must satisfy the equation
\be
(\omega_+\omega_--\Omega^2)=0 \; .
\ee
Setting $\omega_x=\omega_y=\omega_\perp^0$, one obtains the
bifurcation point \cite{Ma93}
\be
\Omega_{\it cr}=\omega_\perp^0/\sqrt{2} \; .
\ee
For $\Omega>\Omega_{\it cr}$, the axial symmetry is
broken and the system is driven into the domain of triaxial shape
under the PAC rotation. This situation is reminiscent of a striking
feature established experimentally in the rotating Bose condensate
\cite{mad,chev} (see also discussion in
Sec. \ref{subsubsec:III.C.1}). According to our analysis, the
dynamical instability in a nuclear system and in the rotating Bose
condensate is of similar nature, in spite of the different character
of the interaction between nucleons (attractive) and between atoms
(repulsive). The origin of this similarity is the trapping of each system
by the harmonic oscillator potential. For the nuclear system it is
the mean field and for the Bose condensate it is the effective external
magnetic field.

To understand this result, let us consider an axially deformed
system, defined by the Hamiltonian ${\tilde H}$ in the laboratory
frame that rotates about a symmetry axis $z$ with a rotational
frequency $\Omega$. The angular momentum is a good quantum number
and, consequently,
\be
[\hat{J}_z,O_K^{\dagger}]=KO_K^{\dagger } \; .
\ee
Here, the phonon $O_K^{\dagger}$ describes the vibrational state,
with $K$ being the value of the angular momentum carried by the
phonons $O_K^{\dagger}$ along the symmetry axis, that is, the $z$ axis.
Thus, one obtains
\begin{equation}
[{\cal R},O_K^{\dagger}]= [{\tilde H} -\Omega \hat{J}_z,
O_K^{\dagger }]= ({\tilde \omega}_K- K\Omega) O_K^{\dagger } \equiv
\omega_K O_K^{\dagger } \; ,
\label{man}
\end{equation}
where ${\tilde \omega}_K$ is the phonon energy of the mode $K$ in the
laboratory frame at $\Omega=0$. This equation implies that, for the
rotational frequency
\be
\Omega_{cr}={\tilde \omega}_K/K \; ,
\ee
one of the RPA frequencies $\omega_K$ vanishes in the rotating frame
\cite{Ma93,M96}. At this point of bifurcation we could expect the
symmetry breaking of the rotating mean field, due to the appearance
of the Goldstone boson related to the multipole-multipole forces
with the quantum number $K$. Indeed, solving the RPA equations
(\ref{RPAeq}) for the Hamiltonian (\ref{hamRPA}) with
$L_x\rightarrow L_z$, one obtains the RPA solution
\be
\omega_{K=2^+}=\sqrt{2}\omega_\perp^0-2\Omega
\ee
for the quadrupole phonons, with the largest projection $K=2$  \cite{wob}.
This mode is
the quadrupole excitation having two more units of angular momentum
than the vacuum state. When $\omega_{K=2^+}=0$, the transition from
non-collective rotation (around the z axis) to triaxial collective
rotation takes place, i.e., at $\Omega_{cr}=\omega_\perp^0/\sqrt{2}$, which
is just the bifurcation point of the mean field discussed above.
This bifurcation point applies for any axially symmetric system
(prolate and oblate). For the $K=1^+$ mode the RPA solution is
\be
\omega_{K=1^+}=\sqrt{\omega_\perp^2+\omega_z^2}-\Omega \; .
\ee
 The condition
$\omega_{K=1^+}=0$ yields the critical frequency at which the onset of
the nonprincipal axis (tilted) rotation should occur for the prolate
system, i.e., at
\be
\Omega_{\it cr}=\sqrt{\omega_\perp^2+\omega_z^2} \; .
\ee
However, the energy to create this mode of angular momentum 1$\hbar$
is too high, the system rather prefers the PAC rotation around the
axis perpendicular to the symmetry axis. As a consequence, this
model does not allow tilted rotations for systems which are prolate
at $\Omega=0$, i.e., when $\Sigma_1=\Sigma_2$. This frequency is
related to vibrational excitations carrying one unit of angular
momentum in an axially symmetric system, whether it is oblate or prolate.

Using the triaxial harmonic oscillator, Troudet and Arvieu
have shown that for prolate systems the PAC leads to triaxial
shapes \cite{TA}. With increasing rotational velocity, the change of the shape,
for the critical frequency $\Omega_{\it cr}^{(1)}$, leads
to an oblate shape, with the rotational axis coinciding with the
symmetry axis. For a
certain value $\Omega_{\it cr}^{(2)} >\Omega_{\it cr}^{(1)}$, the
oblate rotation may transform to a tilted rotation \cite{wob}.

For the oblate rotation around the symmetry axis $x$ one can choose
the axis $x$ as a quantization axis. In this case the projection
$\lambda$ of the angular momentum $L_x$ is a good quantum number as
\be
[\hat{J}_x,O_\lambda^{\dagger}]=\lambda O_\lambda^{\dagger } \; .
\ee
We thus obtain an equation similar to Eq. (\ref{man}) after replacing $K$
by $\lambda$. At the rotational frequency
$\Omega_{\it cr}=\tilde{\omega}/\lambda$ one of the RPA frequencies vanishes.
The bifurcation point
\bea
\Omega_{\it cr}^{(2)}=2\sqrt{\frac{\omega_x^2+\omega_ \perp^2}{3}}\cos{\frac{\psi+\pi}{3}},\\
\cos\psi=\sqrt{27}\frac{\omega_x^2\omega_\perp}{(\omega_x^2+\omega_\perp^2)^{3/2}}\frac{r-1}{r+1} \; ,
\eea
where $r=\Sigma_3/\Sigma_2$, signals on the transition from the
oblate rotation around a symmetry axis to the tilted rotation \cite{wob}.
It is worth stressing that tilted rotations do
occur for $\Omega>\Omega_{\it cr}^{(2)}$, when all three values of
$\Sigma_{i}$ differ from each other, i.e., the shape departs from
the axial symmetry. Note that there
were numerous attempts to find a stable tilted solution with the
triaxial harmonic oscillator either with a PAC or a
three-dimensional rotation. However, the solution was overlooked for
a long time; see the history in \cite{sfrau,satw}.

The analysis of specific low-lying excited states near the yrast
line could shed light on the existence of the nonaxiality. For nonaxial
shapes, one expects the appearance of low-lying vibrational states,
that may be associated with the classical wobbling motion. Such
excitations (called wobbling excitations) were suggested, first, by
Bohr and Mottelson \cite{BM} in rotating even-even nuclei, and
analyzed within simplified microscopic models
\cite{mic,D78,shi2}. According to the microscopic approach
\cite{JM79,mar2}, the wobbling excitations are the vibrational states
of the negative signature built on the positive signature yrast
(vacuum) state. Their characteristic feature are collective $E2$
transitions with $\Delta I=\pm 1 \hbar$ between these and yrast
states. First experimental evidence of such states in odd Lu nuclei
was reported only recently \cite{w1,w2,w3,w4,w5,w6}.

The CRPA provides a natural framework for the microscopic analysis
of the wobbling excitations \cite{jetp,shi3}. It can be shown that,
in the PAC regime for the triaxial nuclei, the wobbling mode is
described by the equation
\begin{equation}
\omega_{\nu=w} = \Omega \sqrt{\frac{[\Im_{x} - \Im^{\it eff}_{2}]
[\Im_{x} - \Im^{\it eff}_{3}]}{\Im^{\it eff}_{2}\Im^{\it eff}_3}} \; ,
\label{wob2}
\end{equation}
with the microscopic effective moments of inertia \cite{mar2}
\begin{equation}
\Im_{2,3}^{\it eff}=\Im_{y,z}+\Omega
S\frac{\Im_x-\Im_{y,z}-\omega_\nu ^2S/\Omega}{\Im_{z,y}+\Omega S} \; ,
\label{mmoi}
\end{equation}
depending on the RPA frequency. Here,
\be
\Im_x =
\frac{\langle \hat J_x\rangle}{\Omega}\;,\quad
S = \sum_\mu \frac{J^y_{\mu}J^z_{\mu}}{(E^{2}_{\mu}
-\omega_\nu ^2)}\;,\quad
\Im_{y,z} = \sum_\mu
\frac{E_{\mu}(J^{y,z}_\mu)^2}{(E^{2}_{\mu} - \omega_\nu ^2)}\;,
\ee
 and $E_{\mu} = \varepsilon_{i} + \varepsilon_{j}$
($E_{\bar i \bar j} = \varepsilon_{\bar i} + \varepsilon_{\bar j}$)
are two-quasiparticle energies, and $J^{y,z}_\mu$ are
two-quasiparticle matrix elements (see details in \cite{jetp}).
The first attempts to describe wobbling excitations
in the framework of the CRPA approach for odd nuclei are faced with
difficulties related to the fulfillment of the self-consistency
between the effective mean field and the RPA excitations
\cite{shi3}. The CRPA approach provides useful criteria for
identifying wobbling excitations in even-even nuclei \cite{jetp,shi2}.

\section{Metallic Grains}
\label{sec:VI}

\subsection{General properties}
\label{sec:VI.A.}

Experimental advances in sample fabrication and measurement techniques
have revealed novel aspects of superconductivity,
related to size and shape effects of finite systems.
Already fifty years ago, Anderson \cite{and1} noted that superconductivity
should  disappear in small metallic grains when the single-particle mean
level spacing $d=1/\rho$ ($\rho$ is the spectral density per spin species)
at the Fermi energy is comparable or becomes larger than the energy
gap $\tilde\Delta$ of a macroscopic superconductor.
This gap is caused by the formation of
Cooper pairs describing correlated electron pairs in time-reversed
states. Pairing correlations in superconductors
tend, therefore, to minimize the total spin of the electron system.
Ferromagnetic correlations, on the other hand, prefer to
maximize the total spin and form a macroscopic magnetic
moment. The ratio $\tilde{\Delta}/d$
provides an approximate number of electronic levels
available for the formation of Coopers pairs. Evidently, if
$\tilde{\Delta}/d\ll 1$,
then there are no active levels accessible for pair correlations and,
consequently, the ferromagnetic correlations determine the system
properties. By changing the grain size and the number of electrons one can
vary the single-particle mean level spacing in such systems and, therefore,
control the desired properties. The basic questions are: does the system
remain superconducting or transforms into normal metal under the transition
from $d < \tilde{\Delta}$ to $d\ge\tilde{\Delta}$?
What are the minimal requirements
in terms of the size and shape of a metallic grain to retain
superconducting properties? In other words, under what conditions the
gauge symmetry, related to the appearance of superconductivity, can be
considered as approximately broken? Understanding answers to these questions
is important for technological applications.

The size dependence of the critical temperature and superconducting gap
was studied for a rectangular grain \cite{parm} and for a nanoslab \cite{blthom}.
Thermodynamic properties of superconducting grains were
investigated  in \cite{scal1,scal2}. For rectangular grains, the situation
has been interpreted within the standard BCS theory of superconductivity
\cite{Bardeen_1957}. Ralph, Black, and Tinkham (RBT) \cite{ralf1,ralf2,ralf3}
have experimentally studied the excitation spectrum
of individual ultra-small metallic grains (of radii $r<5$ nm and mean level
spacings $d > 0.01$ meV), using a single-electron-tunnelling spectroscopy,
similar to the experiments with QDs (see Introduction to Sec. 4). Attaching
such a grain via oxide tunnel barriers to two leads, they constructed a
single-electron transistor having the grain as a central island, and showed
that a well-resolved, discrete excitation spectrum could indeed be extracted
from the conductance measurement. Single-electron-tunnelling spectroscopy
of ultra-small metallic grains has been used to probe superconducting pairing
correlations in Al grains \cite{ralf2,ralf3}.
The measured $d$-values (ranged from 0.02 to 0.3 meV) were much larger
than $k_BT$ for the lowest temperatures attained (around $T\simeq30$ mK),
but of the order of $\tilde\Delta$ (the bulk superconducting gap
$\tilde\Delta=0.18$ meV for Al).
The number of conduction electrons for grains of this
size was still rather large (between $10^4$ and $10^5$).
By studying the evolution of the discrete spectrum of Al grains ($r>5$ nm)
in an applied magnetic field, RBT have observed that a grain with an even
number of electrons had a distinct spectroscopic gap $(\Delta\gg d)$, but
a grain with an odd number of electrons did not \cite{ralf2,ralf3,ralf4}.
As is known from nuclear physics \cite{VG}, this is a clear evidence for
the presence of superconducting pairing correlations in these grains.

The RBT experiments initiated numerous studies of electron correlations in
metallic grains in order to understand the evolution of electron correlations
with the decrease of the sample size. A thorough overview of early experiments
is given by von Delft and Ralph \cite{dera}.
Although tunnel-spectroscopic studies of metallic grains are similar
in spirit to those of semiconductor QDs, there are a number of important
differences. They are as follows \cite{dera}:

\begin{enumerate}

\item
Since metals have much higher densities of states than semiconductors
(because the latter have smaller electron densities and effective masses),
the size of the metal samples should be much smaller
$(\leq 10$ nm) to observe discrete levels.

\item
Metallic grains have much larger charging energies than QDs. Therefore,
one can minimize the electron number fluctuations.

\item
In metallic grains, one can control the strength and type of electron
interactions in a great variety of materials through the periodic table,
including samples doped with impurities. This creates a remarkable
opportunity for studying superconductivity and itinerant ferro-magnetism.

\item
While in QDs it is possible to accomplish electrostatic control of
the tunnel barriers to the leads, for metallic grains, the tunnel
barriers are insulating oxide layers and, therefore, insensitive to the
applied voltage. This is well suited for studying nonequilibrium effects
in grains, contrary to those in QDs, where a large source-drain voltage
lowers the tunnel barrier in a poorly controlled way.

\item
Spin effects are easily probed in metallic grains by applying magnetic
field. Contrary to QDs, where the orbital effects are strongly enhanced due
to a small effective electron mass (see Secs. \ref{subsec:IV.A}
and \ref{subsubsec:IV.A.1}), spin effects
(in particular, a Zeeman splitting) are dominant in metallic grains.

\item
For the same reason, spin-orbital effects are more easily studied in
metallic grains than in QDs.
\end{enumerate}

\subsection{Pairing effects and shell structure}
\label{sec:VI.B.}

\subsubsection{Theoretical approaches}
\label{sec:VI.B.1.}

The phenomenon of superconductivity or superfluidity in nuclear and
condensed-matter systems is typically described by assuming a pairing
Hamiltonian in the BCS approximation \cite{Bardeen_1957}.
As a simplest model for a metallic grain, which
incorporates pairing interactions and a Zeeman coupling
to a magnetic field, one can adopt the following BCS Hamiltonian
\be
H_{BCS}=\sum_{j\sigma=\pm}(\varepsilon_j-\mu-\sigma h)c_{j\sigma}^{+}c_{j\sigma}-
\frac{G}{2}\sum_{ij}c_{i+}^{+}c_{i-}^{+}c_{j-}c_{j+} \;.
\label{bcs}
\ee
Here, $-\sigma h\equiv \mu_Bg\sigma B/2$ is the Zeeman energy of an electron
spin $\sigma$ in a magnetic field $B$ $(h>0)$.
To analyze experimental data for Al grains at zero magnetic field $h=0$,
von Delft {\it et al.} \cite{ralf4} proposed to consider the reduced BCS
model, assuming a completely uniform spectrum with level spacing $d$ and
$G=2\lambda d$, where $\lambda$ is regarded as a phenomenological parameter.
Braun {\it et al.} \cite{braun} studied, in the same model,
the case of nonzero magnetic field $h\neq0$; see for a review \cite{supgrain}.

One of the quantities, used for the analysis of superconducting properties
of ultra-small grains, is the condensation energy
\be
E_b^C=E_b^{GS}-\langle FS|H_{BCS}|FS\rangle \;,
\ee
where $b = 0$ for grains with even number of electrons and $b = 1$ for
those with odd number. The condensation energy is the difference between
the ground state energy of the pairing Hamiltonian (\ref{bcs}) and the energy
of the Fermi state (FS). The latter implies the Slater determinant
obtained by simply filling all levels up to the Fermi surface.
For a large number of electrons, the leading-order behaviour of $E^C$
is given by $-\tilde\Delta^2/(2d)$ \cite{duk1}, which is a standard result
for the condensation energy. Thus, $E^C$ should be independent
of the number of electrons in the system. However, a grain with the odd
number has a single electron occupying the level nearest to the Fermi energy.
The unpaired electron Pauli-blocks the scattering of other pairs into
its own singly-occupied level, i.e. it restricts the phase space available
to pair scattering and, therefore, weakens the pairing correlations.
This is the so-called "blocking effect" considered by Soloviev \cite{bef},
who discussed it extensively in the context of nuclear physics \cite{VG}.
Due to the "blocking effect", the single electron contributes through its
free energy. Furthermore, since there is one less active level at the Fermi
energy, it is harder for the pairing interaction
to overcome the gap and the total energy thus increases. This is the physical
origin of the odd-even effect in superconducting grains, similar to nuclei.
The BCS approach suggests that an abrupt crossover between the superconducting
and the normal states should occur.

As it is well known, the BCS approximation
violates particle number conservation \cite{VG,Ring,BR86}.
While this feature of the BCS approximation
has a negligible effect for macroscopic systems, it can lead to significant
errors when dealing with small or ultra-small systems. Since the fluctuations
of the particle number in BCS are of the order of $\sqrt{N}$, improvements of
the BCS theory are required for systems with $N \sim 100$ particles.
One can improve results by means of the number projection well known in
nuclear physics \cite{Ring}. Indeed, the sharp transition between the
superconducting regime and the fluctuation-dominated regime that arises
in BCS is smoothed out by the particle number projection \cite{nproj}.
Nevertheless, the number projection did not remove completely
some BCS features, in particular, for odd-electron grains.
Fernandez and Egido \cite{egidogr} proposed
a generalized BCS ansatz, which improved the BCS results and agreed with
the exact solution for the pairing Hamiltonian (\ref{bcs}).

The exact numerical solution of the pairing model for small number of
particles has been developed in nuclear physics by Kerman {\it et al.}
\cite{kerman} and Pawlikowski and Rybarska \cite{wlada}.
The basic idea of this approach
was elaborated by Richardson \cite{rich1,rich1a,rich3,rich4,rich4a,rich5,rich6}
and Richardson and Sherman \cite{rich2},
who developed in a series of papers an analytical approach to this problem.
Nowadays it is called the Richardson model. This model
was rediscovered and applied successfully to small metallic grains \cite{sdd};
see for a review \cite{dera,dukel}. The spontaneous breaking of the
$U(1)$ symmetry occurring in this model for the degenerate situation is
studied in \cite{palumbo}.

The major ingredients of this model are as follows.
As discussed above, the pairing interaction
does not affect the singly-occupied levels.
Due to  the "blocking effect", it is sufficient to diagonalize the
BCS Hamiltonian (\ref{bcs}) (in which we set $\mu=0$, $h=0$) in
the single-particle subspace $\cal U$ of empty and doubly occupied levels.
In this case, the eigenenergies are given by
\be
E_n=\sum_i^ nE_i \; .
\ee
Here $E_i$ are the parameters that characterize the eigenstate with
the number of electrons $N=2n+b$, where $b$ is the number of
singly occupied levels. Richardson's
parameters $E_i$ are found by solving the set of $n$ coupled
non-linear equations
\be
\frac{1}{G}+2\sum_{i=1,\neq m}^n\frac{1}{E_i-E_m}=
\sum_{k\in{\cal U}}\frac{1}{2\varepsilon_k-E_m},\quad m=1,...,n \; .
\ee
These equations are to be solved numerically; see details and comparison
of exact solutions with different approximations
in literature \cite{dera,dukel}.

\subsubsection{Effects of magnetic field}
\label{sec:VI.B.2.}

The superconducting gap and low-energy
excitation energies, in a rectangular grain, were computed
numerically  within the Richardson model by Gladilin {\it et al.} \cite{glad1}.
These authors have further developed an approximation
to study the magnetic response of superconductor grains
(when $N\sim 10^5$ per grain) as a function of both temperature and
magnetic field \cite{glad3}.
An approach, based on the Monte Carlo method, was suggested in \cite{belg},
which is able to simulate the reduced BCS model for any fixed number of
particles without a sign problem. Simulations can be performed at
any finite temperature and any level spacing $d$ for large
system sizes. All these calculations demonstrate that the magnetic field
attenuates the pairing correlations in a grain.
However, with increasing temperature pair correlations
may reappear for high magnetic field strengths \cite{steft}, while being
quenched at $T=0$. Nuclei, with
angular momenta along a symmetry axis,
behave in the similar way \cite{lang}. The pair correlations are
destroyed in a stepwise manner by subsequent alignment of
the angular momenta of individual nucleon orbitals along
the symmetry axis. These steps are washed out with increasing
temperature, and pair correlations appear at values of the
rotational frequency, while they are quenched at $T=0$.
A reentrance of pair correlations has been discussed, first, by
Balian, Flocard, and Veneroni \cite{BFV}, who studied the
ensembles of either even or odd numbers of particles.

\subsubsection{Shell effects}
\label{sec:VI.B.3.}

Shell effects in superconducting grains, with radial symmetry,
were studied in \cite{kres1,glad2}. Recent experiments on Al grains were
interpreted by Kresin {\it et al.} \cite{kres2} as
evidence that shell effects can drive
critical temperatures in these grains above 100 K.
Kresin and Ovchinnikov \cite{kres1} proposed that
a phase transition can occur for some metal nanoclusters
with $10^2-10^3$ valence electrons.
The effect of shell structure in larger spherical nanoparticles, together
with modifications in the effective interaction due to alteration
of the electrons wave functions, as well as a nonuniform gap
parameter, were considered in a recent study \cite{shan}.  A large and
strongly size dependent energy gap and critical temperature
were predicted for these particles.

Using the semiclassical approach \cite{brbad}, one can study
the pairing-gap fluctuations for arbitrary ballistic
potentials. Such an analysis leads to a general picture of the typical
fluctuation strength in terms of the properties of the corresponding
classical systems, namely, regular or chaotic dynamics \cite{svengr,yuzb}.
In particular, Gars\'{i}a-Gars\'{i}a {\it et al.} \cite{yuzb}
studied the dependence of low energy excitations of small superconducting
grains on their size and shape, by combining the BCS
mean field, semiclassical techniques, and leading corrections
to the mean field. They found a smooth dependence of the BCS gap function
on the excitation energy of chaotic grains. However, in the integrable
case, for certain values of the electron number $N$,
small changes in the number of electrons can substantially
modify the superconducting gap.

An increase (decrease) of the spectral density around
$E_F$ would make the pairing more (less) favourable, thereby increasing
(decreasing) the energy gap $\Delta$. As a consequence, the gap
becomes dependent on the size and the shape of the grain.
For spherical grains, one may expect that the degeneracy of
levels should decrease the pairing correlations \cite{heisel}.
Thus, for clusters, with parameters satisfying
special, but realistic, conditions, one can expect a
great strengthening of the pairing correlation and, correspondingly,
a large increase in the critical temperature \cite{kres1}.
These conditions are: the proximity of the electronic
state to a complete shell ("magic" number) and a relatively
small gap between the highest occupied shell and the
lowest free shell. Indeed, heat capacities measured for
size-selected Al clusters exhibit a jump around 200 K \cite{kres3}
which may be interpreted as a transition from a highly degenerate
electronic state near the Fermi level to a broken-symmetry superconducting
state with a high $T_c$.

\section{Summary}
\label{sec:VII}

The study of finite quantum systems is now one of the hot topics in
physics and applied sciences. This is caused by two interconnected
reasons. First of all, such systems find numerous technological
applications. This, in turn, requires to know well their properties
that are employed in all these applications. Hence, it is necessary
to have reliable theories accurately describing these systems.

In addition to the importance of technological applications, finite
quantum systems are extremely interesting by themselves. Being on
the boundary between macroscopic bulk systems and microscopic particles,
they, from one side, share the properties of large statistical systems
and, from another side, possess their own specific properties that are
absent in infinite systems. The appropriate name for such intermediate
systems is {\it mesoscopic}. And their properties is a combination of
those of large statistical systems and of small quantum objects.

The system finiteness imposes on them general features that are common
for all such systems, though they can be of rather different physical
nature. From the first glance, such physically different systems as
trapped atoms, quantum dots, atomic nuclei, and metallic grains do not
seem to have much in common. However, they do share a number of common
features, all of them being finite quantum systems.

The system finiteness immediately results in the quantization of spectra,
which, in turn, leads to the appearance of properties differing these
systems from macroscopic matter. Thus, collective excitations in finite
systems are essentially different in the short-wave and long-wave parts,
depending on the relation between the wave length and the effective system
size. Collective excitations, whose wavelength is much shorter than the
system size, behave as excitations in bulk matter, while those excitations,
whose wavelength is comparable or larger than the system size, become
discrete and essentially dependent on the system shape. These properties
are common for all finite systems, whether these are trapped atoms, quantum
dots, atomic nuclei, or metallic grains.

The system spectra are intimately connected with characteristic symmetries.
The latter changes under phase transitions. Effects of spontaneous symmetry
breaking, related to phase transitions, are different in finite systems and
in bulk matter. Strictly speaking, rigorous symmetry breaking can occur only
in infinite systems. But in mesoscopic systems, with sufficiently large
number of particles, phase transitions can be rather sharp, similar to those
in bulk matter. Therefore, it is possible to characterize such transitions
by asymptotic symmetry breaking. Respectively, one can consider such phase
transitions as Bose-Einstein condensation occurring in finite traps, or
superconducting pairing arising in trapped atomic clouds, atomic nuclei,
and metallic grains.

A specific type of transitions, typical only for finite systems, is the
geometric transition, including shape transitions and orientation transitions.
The shape transition is due to the important role of shape for the stability
of finite systems. Such transitions, happening because of instability
developing in a system, with given particle interactions, under a given shape,
are common for trapped atoms and atomic nuclei.

Rotating clouds of trapped atoms are analogous to rotating nuclei, as well
as to quantum dots with magnetic fields. Generally, the mathematical
description of different finite objects is done by so similar Hamiltonians
and methods that very often the results of a theory, developed for one
type of finite objects, can be straightforwardly extended to another type.
Since the theory of atomic nuclei is much older than the theories of trapped
atoms, quantum dots, and metallic grains, many results, obtained earlier in
nuclear theory and related to collective excitations and shape transitions,
were extended to the description of trapped atoms, quantum dots, and metallic
grains. The peculiarity of superconducting pairing, studied for atomic nuclei,
could also be investigated for trapped atoms and metallic grains.

Our aim has been to show, in the frame of one review, that finite quantum
systems of very different physical nature share a lot of common features
that invoke similar mathematical methods. We hope that comparing the content
of different chapters of the review, the reader can come to this conclusion.
Of course, since the objects, considered here, anyway, physically are quite
different, they possess a variety of particular features that could have
different interpretations and applications. It is, certainly, impossible to
discuss the whole variety of these features. We concentrated here mainly on
the common properties of finite quantum systems, related to the effects of
symmetry breaking. All additional information can be found in the cited
literature.

\section*{Acknowledgments}

The authors are grateful to late D.L. Mills for his interest and support
of our work. We appreciate many useful discussions and fruitful collaboration
with V.S. Bagnato, W.D. Heiss, N.S. Simonovi\'c, and E.P. Yukalova. We are
grateful for collaboration to D. Almehed, M. Dineykhan, F. D\"{o}nau,
M.D. Girardeau, H. Kleinert, J. Kvasil, K.P. Marzlin, E.R. Marshalek,
T. Puente, J.-M. Rost, L. Serra, and A. Tsvetkov. We also thank
K.N. Pichugin and E.N. Nazmitdinova for their help in the preparation of
the manuscript. The work of J.L.B. received some partial support from the
CUNY Faculty Research Award Program/Professional Staff Congress (CUNY FRAP/PSC).
Two of the authors (R.G.N. and V.I.Y.) acknowledge financial support from the
Russian Foundation for Basic Research.

\appendix
\section{Two-dimensional harmonic oscillator in a perpendicular
magnetic field} \label{appa}

The electronic spectrum, generated by the Hamiltonian (\ref{hamr})
without interaction, is determined by the sum $\sum_i^N h_i$ of the
single-particle harmonic oscillator Hamiltonians $h = h_0+h_z$ (see Section 2).
The properties of the Hamiltonian $h_z$ are well known
\cite{LandauLifshitz_2003}. The eigenvalue problem for the
Hamiltonian $h_0$ can be solved using the transformation
\begin{equation}
\left(\begin{array}{c}
x\\
y\\
V_x\\
V_y\\
\end{array}\right)=
\left(\begin{array}{cccc}
X_+ & X_+^{\star} & X_- & X_-^{\star} \\
Y_+ & Y_+^{\star} & Y_- & Y_-^{\star}\\
V_x^+ & {V_x^+}^{\star} & V_x^- & {V_x^-}^{\star} \\
V_y^+ & {V_y^+}^{\star} & V_y^- & {V_y^-}^{\star}\\
\end{array}\right)
\left(\begin{array}{c}
a_+ \\
a_+^{\dagger}\\
a_- \\
a_-^{\dagger}
\end{array}\right) \; .
\label{trans}
\end{equation}
Here, $a_i^+(a_i)$ is a creation(annihilation) operator of a new
mode $i=\pm$, with the following commutation relations
\begin{equation}
[a_i,a_j^+]=\delta_{i,j},\quad [a_i,a_j]=[a_i^+,a_j^+]=0 \; .
\end{equation}
One can solve the equation of motion
\begin{equation}
[a_i,h_0]= \Omega_ia_i \; \qquad (i=\pm)
\end{equation}
and express the Hamiltonian $h_0$ through the new normal modes
\begin{equation}
h_0=\hbar \Omega_+ (a_{+}^+a_{+}+1/2)+ \hbar \Omega_-
(a_{-}^+a_{-}+1/2) \; ,
\end{equation}
where
\begin{eqnarray}
\label{mod}
&&\Omega _{\pm }^2={1\over 2}(\omega _x^2+\omega _y^2+4\omega _L^2\pm \Delta) \; \\
&&\Delta=\sqrt{(\omega _x^2-\omega _y^2)^2+8\omega _L^2(\omega
_x^2+\omega _y^2)
 + 16\omega _L^4}\nonumber \; .
\end{eqnarray}
The coefficients of the transformation (\ref{trans}) can be
expressed in terms of $\omega_x,\omega_y,\omega_L$ as well:
\begin{eqnarray}
&&Y_{\pm} = iy_{\pm},\quad X_{\pm} = x_{\pm},\quad x_{\pm} =
2\omega_L\frac{\Omega_{\pm}}{\Omega_{\pm}^2-\omega_1^2+\omega_L^2}y_{\pm} \; ,\nonumber\\
&&V_x^{\pm} = -
i\omega_L\frac{\Omega_{\pm}^2+\omega_1^2-\omega_L^2}{\Omega_{\pm}^2-\omega_1^2+\omega_L^2}y_{\pm}
= -i\Omega_{\pm}\frac{\Omega_{\pm}^2-\omega_2^2-\omega_L^2}{\Omega_{\pm}^2-\omega_2^2+\omega_L^2}x_{\pm} \; ,\nonumber\\
&&V_y^{\pm} =
\Omega_{\pm}\frac{\Omega_{\pm}^2-\omega_1^2-\omega_L^2}{\Omega_{\pm}^2-\omega_1^2+\omega_L^2}y_{\pm}
= \omega_L\frac{\Omega_{\pm}^2+\omega_2^2-\omega_L^2}{\Omega_{\pm}^2-\omega_2^2+\omega_L^2}x_{\pm} \; , \nonumber\\
&&y_{\pm}^2=\pm\frac{\hbar}{2m\Omega_{\pm}}\frac{\Omega_{\pm}^2-\omega_1^2+\omega_L^2}{\Omega_{+}^2-\Omega_{-}^2} \; , \\
&&x_{\pm}^2=\pm\frac{\hbar}{2m\Omega_{\pm}}\frac{\Omega_{\pm}^2-\omega_2^2+\omega_L^2}{\Omega_{+}^2-\Omega_{-}^2} \; .
\nonumber
\end{eqnarray}
The eigenstates of the Hamiltonian $h_0$ are
\begin{equation}
|n_+n_-\rangle=\frac{1}{\sqrt{n_+!n_-!}}(a_{+}^+)^{n_+}(a_{-}^+)^{n_-}|00\rangle \; .
\label{dc}
\end{equation}
The operator $l_z$ is diagonal in this basis,
\begin{equation}
l_z=n_{-}-n_{+} \; .
\end{equation}
In the case of circular symmetry, i.e., $\omega_x=\omega_y=\omega_0$,
the eigenstate Eq. (\ref{dc}) reduces to the form of the Fock-Darwin
state (see \cite{Jac}). Quite often, it is useful to invoke the
representation of the Fock-Darwin state in cylindrical coordinates
$(\rho,\varphi)$, which has the form
\begin{equation}
\Phi_{n_rm}(\rho,\varphi)=\frac{1}{\sqrt{2\pi}}\exp^{im\varphi}R_{n_rm}(\rho) \; .
\label{fd}
\end{equation}
This state is an eigenfunction of the operator $l_z$, with an eigenvalue
$m$ and with the radius-dependent function of the form
\bea
R_{n_rm}(\rho)&=&\frac{\sqrt{2}}{l_0^B}\sqrt{\frac{n_r!}{(n_r+|m|)!}}
\biggl(\frac{\rho}{l_0^B}\biggr)^{|m|}\times\nonumber\\
&&\exp
\biggl[-\frac{\rho^2}{2{l_0^B}^2}\biggr] {\cal L}_{n_r}^{|m|}
\biggl(\frac{\rho^2}{{l_0^B}^2}\biggr) \label{fdl} \; .
\eea
Here ${l_0^B}^2=\hbar/(2m^*\Omega)$,
$\Omega=\sqrt{\omega_0^2+\omega_L^2}$, and ${\cal L}$ denotes the
Laguerre polynomials
\be
{\cal L}_{n}^{|m|}(x)={\sum}_{k=0}^n\frac{(-1)^k}{k!}
\left(\begin{array}{c}
n+|m|\\
n-k\\
\end{array}\right)x^k \; .
\ee

The pair of quantum numbers $(n,m)$ and $(n_+,n_-)$ are related as
\begin{equation}
n=n_{+}+n_{-},\quad m=n_--n_+ \; ,
\end{equation}
with $n=2n_r+|m|$. The single-particle energy in the Fock-Darwin
state is
\begin{equation}
\varepsilon(n,m)=\hbar\Omega(n+1)-
\hbar\omega_Lm=\hbar\Omega(2n_r+|m|+1)-\hbar\omega_Lm \; .
\label{fce}
\end{equation}
The quantum number $n$ is associated with the shell number $N_{\rm sh}$ for
an $N$-electron quantum dot in the $2D$ approach.

\end{document}